\documentclass{article}
\usepackage{amssymb}

\input{tcilatex}

\begin{document}

\begin{center}
\bigskip

{\LARGE QUANTUM\ ENTANGLEMENT\ FOR\bigskip }

{\LARGE SYSTEMS\ OF\ IDENTICAL\ BOSONS.\bigskip }

{\LARGE SPIN\ SQUEEZING\bigskip\ }

{\LARGE AND\ OTHER\ ENTANGLEMENT\ TESTS\ IN\bigskip\ }

{\LARGE TWO MODE SYSTEMS\bigskip\ \ }

\bigskip

B. J. Dalton$^{1,2}$, L. Heaney$^{3}$, J. Goold$^{1,4}$, B. M. Garraway$^{5}$
and Th. Busch$^{1,6}${\LARGE \bigskip }

$^{1}$\textit{Physics Department, University College Cork, Cork City, Ireland%
}

$^{2}$\textit{Centre for Atom Optics and Ultrafast Spectroscopy, Swinburne
University of Technology, Melbourne, Victoria 3122, Australia}

$^{3}$\textit{Centre for Quantum Technologies, National University of
Singapore, Singapore 117543.}

$^{4}$\textit{Clarendon Laboratory, Oxford University, Oxford OX13PU, United
Kingdom}

$^{5}$\textit{Department of Physics and Astronomy, University of Sussex,
Falmer, Brighton BN19QH, United Kingdom}

$^{6}$\textit{Quantum Systems Unit, Okinawa Institute of Science and
Technology Graduate University, Okinawa, Japan\bigskip }

\textit{\bigskip \bigskip \bigskip }
\end{center}

Email: bdalton@swin.edu.au

\pagebreak

$\mathbf{Abstract}$

Entanglement is a key quantum feature of composite systems in which the
probabilities for joint measurements on the composite sub-systems are no
longer determined from measurement probabilities on the separate
sub-systems. The topic has many aspects apart from how entanglement is
defined, including measures of how entangled a quantum state is, the linkage
between entanglement and thermodynamics as well as the technological
applications of entanglement. This review focuses on the meaning of
entanglement, the paradoxes associated with entangled states and important
measurement based tests of whether a quantum state is entangled - references
are of course given to articles on the other aspects of entanglement. In
this article entangled quantum states are specifically considered in the
context of systems of identical particles, based on the requirement that in
order to represent physical states both for the overall system and the
sub-systems which may be entangled, the density operators must satisfy the
symmetrisation principle and global and local super-selection rules that
prohibit states in which there are coherences between differing particle
numbers. These requirements and their justification are fully discussed. In
the second quantisation approach used, both the system and the sub-systems
are modes (or sets of modes) rather than particles, particles being
associated with different occupancies of the modes. The definition of
entangled states is based on first defining the non-entangled states - after
specifying which modes constitute the sub-systems. This work mainly focuses
on two mode entanglement for massive bosons. Several inequalities involving
variances and mean values of operators involving mode annihilation, creation
operators have been proposed as tests for two mode entangled states,
including the inequalities that define spin squeezing. Spin squeezing
criteria in two mode systems are examined, and spin squeezing is best
considered for principle spin operator components where the covariance
matrix is diagonal. It is shown that the presence of spin squeezing in at
least one of the spin components requires entanglement of the relevant pair
of modes. Several of the other proposed tests for entanglement, including
ones based on the sum of the variances for two spin components are
considered. All of the tests are still valid when the present concept of
entanglement based on the symmetrisation and super-selection rule criteria
is applied, but further tests have been obtained here. Sometimes the new
tests are satisfied whilst than those obtained in other papers are not.

\bigskip

\bigskip

\textbf{PACS Numbers}{\LARGE \ \ }03.65 Ud, 03.67 Bg, 03.67 Mn, 03.75 Gg

\begin{center}
\pagebreak
\end{center}

{\Large Contents\bigskip }

{\Large 1. Introduction\medskip }

{\Large 2. Entanglement - General Features}

\qquad\ \textbf{2.1 Physical States}

\textbf{\qquad 2.2 Entangled and Non-Entangled States}

\qquad \qquad \textit{2.2.1 General Considerations}

\qquad \qquad \textit{2.2.2 Local Systems and Operations}

\qquad \qquad \textit{2.2.3 Constraints on Sub-System Density Operators}

\qquad \qquad 2.2.4 \textit{Classical Entanglement}

\qquad \textbf{2.3 Separate, Joint Measurements. Reduced Density Operator}

\qquad \qquad \textit{2.3.1 Joint Measurements on Sub-Systems}

\qquad \qquad \textit{2.3.2 Single Sub-System Measurements. Reduced Density
Operator}

\qquad \qquad \textit{2.3.3 Mean Value and Variance}

\qquad \qquad \textit{2.3.4 Conditional Probabilities}

\qquad \qquad \textit{2.3.5 Conditional Mean and Variance}

\qquad \textbf{2.4 Non-Entangled States}

\qquad \qquad \textit{2.4.1 Non-Entangled States - Joint Measurements on
Sub-Systems}

\qquad \qquad \textit{2.4.2 Non-Entangled States - Single Sub-System
Measurements}

\qquad \qquad \textit{2.4.3 Non-Entangled States - Conditional Probability}

\qquad \qquad \textit{2.4.4 Non-Entangled States - Mean Values and
Correlations}

\qquad \textbf{2.5 Hidden Variable Theory}

\qquad \qquad \textit{2.5.1 HVT - Mean Values and Correlations}

\qquad \qquad \textit{2.5.2 HVT - GHZ State}

\qquad \textbf{2.6 Paradoxes}

\qquad \qquad \textit{2.6.1 EPR Paradox}

\qquad \qquad \textit{2.6.2 Schrodinger Cat Paradox}

\qquad \textbf{2.7 Bell Inequalities}

\qquad \qquad \textit{2.7.1 HVT Result}

\qquad \qquad \textit{2.7.2 Non-Entangled State Result}

\qquad \qquad \textit{2.7.3 Bell Inequality Violation and Entanglement}

\qquad \textbf{2.8 Non-Local Correlations}

\qquad \qquad \textit{2.8.1 HVT Result}

\qquad \qquad \textit{2.8.2 Non-Entangled State Result}

\qquad \qquad \textit{2.8.3 Weak Correlation Violation and Entanglement}%
{\Large \medskip }

{\Large 3. Identical Particles and Entanglement}

\qquad \textbf{3.1 Identical Particles and Entanglement}

\qquad \qquad \textit{3.1.1 Symmetrization Principle and Second Quantization}

\qquad \qquad \textit{3.1.2 Sub-Systems and Modes}

\qquad \qquad \textit{3.1.3 Multi-Mode Sub-Systems}

\qquad \textbf{3.2 Super-Selection Rules - General}

\qquad \qquad \textit{3.2.1 Global Particle Number SSR}

\qquad \qquad \textit{3.2.2 Examples of Global SSR Compliant States}

\qquad \qquad \textit{3.2.3 SSR and Conservation Laws}

\qquad \qquad \textit{3.2.4 SSR Justification - No Suitable Phase Reference}

\qquad \qquad \textit{3.2.5 SSR Justification - Physics Considerations}

\qquad \qquad \textit{3.2.6 SSR Justification - Galilean Frames ?}

\qquad \qquad \textit{3.2.7 SSR and Photons}

\qquad \textbf{3.3 Reference Frames and SSR Violation}

\qquad \qquad \textit{3.3.1 Linking SSR and Reference Frames}

\qquad \qquad \textit{3.3.2 Coherent Superposition of an Atom and a Molecule
?}

\qquad \qquad \textit{3.3.3 Detection of SSR Violating States}

\qquad \textbf{3.4 Super-Selection Rules - Separable States}

\qquad \qquad \textit{3.4.1 Local Particle Number SSR}

\qquad \qquad \textit{3.4.2 Local SSR Justification - Independent Local
Phase References}

\qquad \qquad \textit{3.4.3 Local SSR Justification -Physics Considerations}

\qquad \qquad \textit{3.4.4 Local SSR Justification - Joint Measurements}

\qquad \qquad \textit{3.4.5 States that Violate Local and Global SSR}

\qquad \qquad \textit{3.4.6 States that are Global but not Local SSR
Compliant}

\qquad \qquad \textit{3.4.7 General Form of Non-Entangled States}

\qquad \textbf{3.5 Bipartite Systems}

\qquad \qquad \textit{3.5.1 Two Single Modes}\textbf{\ - }\textit{Coherence
Terms}

\qquad \qquad \textit{3.5.2 Two Pairs of Modes - Coherence Terms\medskip }

{\Large 4. Spin Squeezing}

\qquad \textbf{3.1 Spin Operators, Bloch Vector and Covariance Matrix}

\qquad \qquad \textit{3.1.1 Spin Operators}

\qquad \qquad \textit{3.1.2 Bloch Vector and Covariance Matrix}

\qquad \textbf{3.2 New Spin Opertors and Principle Spin Fluctuations}

\qquad \textbf{3.3 Spin Squeezing - New Spin Operators}

\qquad \qquad \textit{3.3.1 Heisenberg Uncertainty Principle - Spin Squeezing%
}

\qquad \qquad \textit{3.3.2 Alternative Spin Squeezing Criteria}

\qquad \qquad \textit{3.3.3 Planar Spin Squeezing}

\qquad \textbf{3.4 Rotation Operators and New Modes}

\qquad \qquad \textit{3.4.1 Rotation Operators}

\qquad \qquad \textit{3.4.2 New Mode Operators}

\qquad \qquad \textit{3.4.3 New Modes}

\qquad \textbf{3.5 Old and New Modes - Coherence Terms}

\qquad \textbf{3.6 Quantum Correlation Functions and Spin
Measurements\medskip }

{\Large 5. Spin Squeezing Test for Entanglement}

\qquad \textbf{4.1 Spin Squeezing Requires Entanglement - New Modes}

\qquad \qquad \textit{4.1.1 Case of }$\widehat{\mathit{J}}_{x}$ \textit{and }%
$\widehat{\mathit{J}}_{y}$

\qquad \qquad \textit{4.1.2 Case of }$\widehat{\mathit{J}}_{z}$

\qquad \qquad \textit{4.1.3 No Spin Squeezing for Separable States}

\qquad \qquad \textit{4.1.4 Spin Squeezing Tests for Entanglement}

\qquad \qquad \textit{4.1.5 Inequality for }$|\left\langle \widehat{\mathit{J%
}}_{z}\right\rangle |$

\qquad \textbf{4.2 Spin Squeezing Requires Entanglement - Original Modes}

\qquad \textbf{4.3 Entangled States that are Non Spin Squeezed}

\qquad \textbf{4.4 Non-Entangled States that are Non Spin Squeezed}

\qquad \textbf{4.5 Entangled States that are Spin Squeezed}

\qquad \qquad \textit{4.5.1 Relative Phase Eigenstate}

\qquad \qquad \textit{4.5.2 New Spin Operators}

\qquad \qquad \textit{4.5.3 Bloch Vector and Covariance Matrix}

\qquad \qquad \textit{4.5.4 New Mode Operators}

\qquad \textbf{4.6 Bloch Vector Entanglement Test\medskip }

{\Large 6. Other Proposed Tests for Entanglement}

\qquad \textbf{5.1 Hillary et al 2006}

\qquad \qquad \textit{5.1.1 Hillery Spin Variance Entanglement Test}

\qquad \qquad \textit{5.1.2 Validity of Hillery Test for Local SSR
Compatible States}

\qquad \qquad \textit{5.1.3 Non-Applicable Entanglement Test Involving }$%
|\left\langle \widehat{\mathit{S}}_{z}\right\rangle |$

\qquad \textbf{5.2 Hillary et al 2009}

\qquad \qquad \textit{5.2.1 Hillery Strong Correlation Entanglement Test}

\qquad \qquad \textit{5.2.2 Correlation Test for Local SSR Compatible States}

\qquad \qquad \textit{5.2.3 Applications of Correlation Tests for
Entanglement}

\qquad \textbf{5.3 Sorensen et al 2001}

\qquad \qquad \textit{5.3.1 Sorensen Spin Squeezing Entanglement Test}

\qquad \qquad \textit{5.3.2 Revising Sorensen Spin Squeezing Test -
Localised Modes}

\qquad \qquad \textit{5.3.3 Revising Sorensen Spin Squeezing Test -
Separable State of Single Modes}

\qquad \qquad \textit{5.3.4 Revising Sorensen Spin Squeezing Test -
Separable State of Pairs of Modes with Single Boson Occupancy}

\qquad \textbf{5.4 Sorensen and Molmer 2001}

\qquad \textbf{5.5 Duan et al 2000}

\qquad \textbf{5.6 He et al 2012}

\qquad \qquad \textit{5.6.1 Correlation Tests for Entanglement}

\qquad \qquad \textit{5.6.2 Spin Squeezing Tests for Entanglement\medskip }

{\Large 7. Experiments on Spin Squeezing\medskip }

{\Large 8. Discussion and Summary of Key Results\bigskip }

\textbf{References}

\textbf{Acknowledgements\bigskip }

{\Large Appendices\bigskip }

{\Large A. Projective Measurements and Conditional Probabilities}

\qquad \textbf{A.1 Projective Measurements}

\qquad \textbf{A.2 Conditional Probabilities}

\qquad \textbf{A.3 Conditional Mean and Variance\medskip }

{\Large B. Inequalities}

\qquad \textbf{B.1 Integral Inequality}

\qquad \textbf{B.2 Sum Inequality\medskip }

{\Large C. Particle and Mode Entanglement\medskip }

{\Large D. Reference Frames and Super-Selection Rules}

\qquad \textbf{D.1 Two Observers with Different Reference Frames}

\qquad \textbf{D.2 Symmetry Groups}

\qquad \textbf{D.3 Relationships - Situation A}

\qquad \textbf{D.4 Relationships - Situation B}

\qquad \textbf{D.5 Dynamical and Measurement Considerations}

\qquad \textbf{D.6 Nature of Reference Frames}

\qquad \textbf{D.7 Relational Description of Phase References}

\qquad \textbf{D.8 Irreducible Matrix Representations and SSR}

\qquad \textbf{D.9 Non-Entangled States\medskip }

{\Large E. Super-Selection Rule Violations ?}

\qquad \textbf{E.1 Preparation of a Coherent Superposition of an Atom and a
Molecule ?}

\qquad \qquad \textit{E.1.1 Hamiltonian}

\qquad \qquad \textit{E.1.2 Initial State}

\qquad \qquad \textit{E.1.3 Implicated Reference Frame}

\qquad \qquad \textit{E.1.4 Process - Alice and Charlie Descriptions}

\qquad \qquad \textit{E.1.5 Interference Effects Without SSR Violation}

\qquad \qquad \textit{E.1.6 Conclusion}

\qquad \textbf{E.2 Detection of Coherent Superposition of a Vacuum and
One-Boson State ?\medskip }

{\Large F. Non-Physical Two Mode States\medskip }

{\Large G. Classical Entanglement}

\textbf{\qquad \qquad G.1 Classical Ket Vector Formalism and Entangled States%
}

\textbf{\qquad \qquad G.2 Circular Polarization States}

\textbf{\qquad \qquad G.3 Quantum Treatment}

\textbf{\qquad \qquad G.4 Several Light Beams}

\textbf{\qquad \qquad G.5 Classically Entangled States - Joint Measurements
on Sub-Systems}

\textbf{\qquad \qquad G.6 Classical Entanglement and Bell Inequality}{\Large %
\medskip }

{\Large H. Derivation of Sorensen et al Results\medskip }

{\Large I. Revised Sorensen et al Tests}

\qquad \textbf{I.1 Variance }$\left\langle \Delta \widehat{S}%
_{x}^{2}\right\rangle $\textbf{\ for Single Mode Sub-Systems}

\qquad \textbf{I.2 Variance }$\left\langle \Delta \widehat{S}%
_{x}^{2}\right\rangle $\textbf{\ for Two Mode Sub-Systems\medskip }

{\Large J. Heisenberg Uncertainty Principle Results\bigskip }

\textbf{Figure Captions}

\textbf{Figures}

\pagebreak

\section{Introduction}

\label{Section- - Intrduction}

Since the EPR parodox of Einstein et al \cite{Einstein35a} on the conflict
between quantum theory and local realism, the famous cat paradox of
Schrodinger \cite{Schrodinger35a} in which the cat could be thought of as
being simultaneously dead and alive, and the derivation by Bell et al \cite%
{Bell65a}, \cite{Clauser69a} of inequalities based on treating measured
quantities via a classical hidden variable theory which certain quantum
quantum systems violated, \emph{entanglement} has been recognised as being
one of the key features that distinguishes quantum physics from classical
physics. It is a feature that arises in the context of \emph{composite}
quantum systems composed of distinct components or \emph{sub-systems} and is
distinct from other features of quantum physics such as \emph{quantization}
for measurements of physical quantities, \emph{probabilistic} \emph{outcomes}
for such measurements, \emph{uncertainty principles} involving pairs of
physical quantities and so on. Such sub-systems are usually associated with
sub-sets of the physical quantities applying to the overall system, and in
general more than one choice of sub-systems can be made. The formalism of
quantum theory treats \emph{pure states} for systems made up of two or more
distinct sub-systems via tensor products of sub-system states, and since
these product states exist in a Hilbert space, it follows that linear
combinations of such products could also represent possible pure quantum
states for the system. Such \emph{quantum superpositions} which cannot be
expressed as a \emph{single} product of sub-system states are known as \emph{%
entangled} (or \emph{non-separable}) states. The concept of entanglement can
also be extended to \emph{mixed states}, where quantum states for the system
and the sub-systems are specified by density operators. The detailed
definition of entangled states is set out in Section \ref{Section -
Entanglement}. This definition is based on first carefully defining the 
\emph{non-entangled} (or \emph{separable}) states such that \emph{all}
non-entangled states must be possible physical states, and in addition these
states must be \emph{preparable} via processes involving \emph{separate
operations} on each sub-system after which correlated sub-system quantum
states are combined in accordance with \emph{classical probabilities}. Thus,
although the sub-system states retain their quantum natures the combination
resulting in the overall system state is formed classically rather than
quantum mechanically. This overall process then involves \emph{local
operations} on the distinct sub-systems and \emph{classical communication} (%
\emph{LOCC})\ to prepare a general non-entangled state. The entangled states
are then just the physical states which are not non-entangled states. The
general idea that in all composite systems the non-entangled states all
involve LOCC preparation processes was first suggested by Werner \cite%
{Werner89a}. The notion of \emph{physical states}, the nature of the systems
and \emph{sub-systems} involved and the specific features required in the
definition of non-entangled states when \emph{identical particles} are
involved is discussed in detail in Section \ref{Section - Identical
Particles and Entanglement}. Entangled states underlie a number of effects
that cannot be interpreted in terms of classical physics, including \emph{%
spin squeezing}, non-local measurement correlations - such as for the
Einstein-Podolski-Rosen \emph{(EPR) paradox} and violations of \emph{Bell
Inequalities}. More recently, entangled states have been recognised as a
resource that can be used in various \emph{quantum technologies} for
applications such as teleportation, quantum cryptography, quantum computing
and so on. Recent expositions on the effects of entanglement and its role in
quantum information science include \cite{Peres93a}, \cite{Nielsen00a}, \cite%
{Vedral07a}, \cite{Barnett09a}, \cite{Reid09a}, \cite{Reid12a}.

It would be pointless to characterise states as entangled unless such states
have some important properties. The key requirement is that entangled states
exhibit a novel \emph{quantum feature} that is only found in \emph{composite}
systems. As will be seen in SubSection\ref{SubSection - Joint Measurements},
separable states are such that the \emph{joint probability} for measurements
of all physical quantities associated with the sub-systems can be found from
separate measurement probabilities obtained from the sub-system density
operators and the overall classical probability for creating particular
products of sub-system states. Entangled states do not exhibit this feature
of separable probabilities, and it is this key \emph{non-separability feature%
} that led Schrodinger to call these states "entangled". Hidden variable
theories (HVT) (see SubSection\ref{SubSection - Hidden Variable Theory})
applied to quantum systems - which are essentially classical in nature -
also have the same separability feature for joint probabilities as quantum
separable states, though of course the basic concepts are different. The
fact that only entangled states do not exhibit the feature of separable
probabilities shown in classical HVT highlights entanglement being a \emph{%
non-classical feature} found \emph{only} in composite systems.

It is now generally recognised that entanglement is a \emph{relative}
concept (\cite{Simon02a}, \cite{Hines03a}, \cite{TerraCunha07a}), \cite%
{Vedral07a}, \cite{Horodecki09a}, \cite{Guhne09a} and not only depends on
the quantum state under discussion but also on which \emph{sub-systems} are
being considered as entangled (or non-entangled). A quantum state may be
entangled for one choice of the sub-systems but may be non-entangled if
another choice of sub-systems is made, an example being for the hydrogen
atom \cite{TerraCunha07a} where energy eigenstates are non-entangled if the
sub-systems refer to the centre of mass and the relative position of the
electron from the proton, but which would be entangled if the sub-systems
were the positions of the electron and proton. An example involving two
different choices of single particle states in a two mode Bose condensate is
given in Section \ref{Section - Relationship Spin Squeezing & Entanglement}.

For a general quantum state various \emph{measures} of entanglement have
been defined - see \cite{Vedral07a}, \cite{Barnett09a}, \cite{Horodecki09a}, 
\cite{Guhne09a}, \cite{Tichy11a}, \cite{Modi12a}, \cite{Amico08a}, for
details of these, and are aimed at quantifying entanglement to determine
which states are more entangled than others. This is important since
entanglement is considered as a resource needed in various quantum
technologies. Calculations based on such measures of entanglement confirm
that for some choices of sub-systems the quantum state is entangled, for
others it is non-entangled. For two mode pure states the \emph{entanglement
entropy} - being the difference between the entropy for the pure state
(zero) and that associated with the reduced density operator for either of
the two sub-systems - is a useful entanglement measure. As entropy and
information changes are directly linked \cite{Vedral07a}, \cite{Barnett09a},
this measure is of importance to \emph{quantum information science}. Another
entanglement measure is \emph{particle entanglement entropy}, defined by
Wiseman et al \cite{Wiseman03a}, \cite{Dowling06b}, \cite{Tichy11a} for
identical particle systems and based on projecting the quantum state onto
states with definite particle numbers.

Although not directly relatable to the various quantitative measures of
entanglement, the results for certain measurements can play the role of
being \emph{signatures} or \emph{witnesses} or \emph{tests} of entanglement 
\cite{Horodecki09a}, \cite{Guhne09a}, \cite{Tichy11a}. These are in the form
of \emph{inequalities} for \emph{variances} and \emph{mean values} for
certain physical quantities, which are consequent on the inequalities that
would apply for non-entangled quantum states. If such inequalities are \emph{%
violated} then it can be concluded that the state is \emph{entangled} for
the relevant sub-systems. It cannot be emphasised enough that these tests
provide \emph{sufficiency conditions} for establishing that a state is
entangled. The failure of a test does \emph{not} guarantee non-entanglement
- sufficiency does not imply \emph{necessity}. The violation of a \emph{Bell
inequality} is an example of such a signature of entanglement, and the
demonstration of \emph{spin squeezing} is regarded as another. However, the
absence of spin squeezing (for example) does not guarantee non-entanglement,
as the case of the \emph{NOON} state in SubSection \ref{SubSection - Ent
State Non Squeezed} shows. A significant number of such inequalities have
now been proposed and such signatures of entanglement are the primary focus
of the present paper, which is aimed at identifying which of these
inequalities really do identify entangled states, especially in the context
of \emph{two mode} systems of \emph{identical bosons}.

At present there is\emph{\ no clear linkage} between quantitative measures
of entanglement (such as entanglement entropy) and the quantities used in
conjunction with the various entanglement tests (such as the relative spin
fluctuation in spin squeezing experiments). Results for experiments
demonstrating such non-classical effects cannot yet be used to say much more
than the state \emph{is} entangled, whereas ideally these experiments should
determine \emph{how} entangled the state is. Again we emphasise that neither
the entanglement tests nor the entanglement measures are being used to \emph{%
define} entanglement. Entanglement is defined first as being the quantum
states that are non-separable, the tests for and measures of entanglement
are \emph{consequential} on this definition.

This paper deals with identical particles - bosons or fermions. In the \emph{%
second quantisation} approach used here the system is regarded as a set of 
\emph{quantum fields}, each of which may be considered as a collection of
single particle states or \emph{modes}. Hence both the system and
sub-systems will be specified via the modes that are involved, so here the
sub-systems in terms of which non-entangled (and hence entangled) states are
defined are \emph{modes} or \emph{sets} of modes, not particles \cite%
{Simon02a}, \cite{Hines03a}, \cite{TerraCunha07a}), \cite{Vedral07a}, \cite%
{Benatti10a}, \cite{Benatti11a}. In this approach, \emph{particles} will be
described via the \emph{occupancies} of the various modes, so that
situations with differing numbers of particles will be treated as differing
quantum \emph{states} of the same system, not as different systems - as in
the \emph{first quantisation} approach. Note that the choice of modes is 
\emph{not unique} - original sets of orthogonal one particle states (modes)
may be replaced by other orthogonal sets. An example is given in Section \ref%
{Section - Spin Squeezing}. Modes can often be categorised as \emph{localised%
} modes, where the corresponding single particle wavefunction is confined to
a restricted spatial region, or may be categorised as \emph{delocalised}
modes, where the opposite applies. Single particle harmonic oscillator
states are an example of localised modes, momentum states are an example of
delocalised modes. This distinction is significant when phenomena such as
EPR violations and teleportation are considered.

Although \emph{multi-mode} systems are also considered, in this paper we
mainly focus on \emph{two mode} systems of identical \emph{bosonic} atoms,
where the atoms at most occupy only two single particle states or modes. For
bosonic atoms this situation applies in two mode interferometry, where if a
single hyperfine component is involved the modes concerned may be two
distinct spatial modes, such as in a double well magnetic or optical trap,
or if two hyperfine components are involved in a single well trap each
component has its own spatial mode. Large numbers of bosons may be involved
since there is no restriction on the number of bosons that can occupy a
bosonic mode. For fermionic atoms each hyperfine component again has its own
spatial mode. However, if large numbers of \emph{fermionic} atoms are
involved then as the Pauli exclusion principle only allows each mode to
accommodate one fermion, it follows that a large number of modes must
considered and two mode systems would be restricted to at most two fermions.
Consideration of multi-mode entanglement for large numbers of fermions is 
\emph{outside} the scope of the present paper (see \cite{Lunkes05a} for a
treatment of this), and unless otherwise indicated the focus will be on 
\emph{bosonic modes}. The paper focuses on identical bosonic \emph{atoms} -
whether the paper also applies to \emph{photons} is less clear and will be
discussed below. .

The work presented here begins with the \emph{fundamental issue} of how an
entangled state should be \emph{defined} in the context of systems involving 
\emph{identical particles}. To reiterate - in the commonly used \emph{%
mathematical approach} for defining entangled states, this requires \emph{%
first} defining a general \emph{non-entangled} state, all \emph{other}
states therefore being entangled. We adhere to the original definition of
Werner \cite{Werner89a} in which the separable states are those that can be
prepared by LOCC. This approach is adopted by other authors, see for example 
\cite{Bartlett06b}, \cite{Jones07a}, \cite{Masanes08a}. However, in other
papers - see for example \cite{Verstraete03a}, \cite{Schuch04a} so-called 
\emph{separable non-local} states are introduced in which LOCC is \emph{not}
required (see SubSection.\ref{SubSection - Two Mode Coherent State Mixture}
for an example). It is contended here that the \emph{density operators} both
for the overall system states and for the sub-system states of non-entangled
states must represent \emph{physical states} and in some other work
(discussed below) this is not the case. A key feature required of all
physical states for systems involving \emph{identical particles}, entangled
or not is that they satisfy the \emph{symmetrization principle}. This places
restrictions both on the form of the overall density operator and also on
what can be validly considered to be a sub-system. In particular this rules
out \emph{individual} identical particles being treated as sub-systems, as
is done in some papers (see below). In addition, \emph{super-selection rules 
}(SSR)\emph{\ }\cite{Wick52a} only allow density operators which have \emph{%
zero coherences} between states with \emph{differing total numbers} of
particles to represent valid \emph{physical states}, and this will be taken
into account for \emph{all} physical states of the overall system, entangled
or not. This is referred to as the \emph{global particle number
super-selection rule} \ In \emph{non-entangled} or \emph{separable} states
the density operator is a sum over products of sub-system density operators,
each product being weighted by its probability of ocurring (see below for
details). For the non-entangled or separable states, a so-called\emph{\
local particle number} \emph{super-selection rule} will \emph{also} be
applied to the density operators describing each of the \emph{sub-systems}.
These sub-system density operators must then have have zero coherences
between states with differing numbers of \emph{sub-system particles}. This
additional restriction excludes density operators as defining non-entangled
states when the sub-system density operators do not conform to the local
particle number super-selection rule. Consequently, density operators where
the local particle number SSR does not apply would be regarded as entangled
states. This viewpoint is discussed in papers by Bartlett et al \cite%
{Bartlett06b}, \cite{Bartlett07a} as one of several approaches for defining
entangled states. However, other authors such as \cite{Verstraete03a}, \cite%
{Schuch04a} state on the contrary that states when the sub-system density
operators do \emph{not} conform to the local particle number super-selection
rule \emph{are} still separable, others such as \cite{Hillery06a}, \cite%
{Hillery09a} do so by implication. So in this paper we are advocating a 
\emph{different definition }to some \emph{other} \emph{definitions} of
entanglement in identical particle systems,.the consequence being that the
set of entangled states is now much \emph{larger}. This is a \emph{key idea}
in this paper - not only should super-selection rules on particle numbers be
applied to the the \emph{overall} physical state, entangled or not, but it 
\emph{also} should be applied to the density operators that describe states
of the modal \emph{sub-systems} involved in the general definition of \emph{%
non-entangled} states.

The detailed reasons for adopting this viewpoint are set out below. As will
be seen, the local particle number super-selection rule restriction \emph{%
firstly} depends on the \emph{fundamental requirement} that for \emph{all}
composite systems - whether identical particles are involved or not -
non-entangled states are only those that can be prepared via processes that
involve only \emph{local operations} and \emph{classical communication} (%
\emph{LOCC}). The requirement that the sub-system density operators in
identical particle cases satisfy the local particle number SSR is \emph{%
consequential} on the sub-system states being possible physical sub-system
states. As mentioned before, the general definition of non-entangled states
based on LOCC preparation processes was first suggested by Werner \cite%
{Werner89a}. Apart from the papers by Bartlett et al \cite{Bartlett06b}, 
\cite{Bartlett07a} we are not aware that this LOCC/SSR based criteria for
non-entangled states has been invoked previously for identical particle
systems, indeed the opposite approach has been proposed \cite{Verstraete03a}%
, \cite{Schuch04a}. However, the idea of considering whether sub-system
states should satisfy the local particle number SSR has been presented in
several papers -\emph{\ }\cite{Verstraete03a}, \cite{Schuch04a}, \cite%
{Bartlett06b}, \cite{Bartlett07a}, \cite{Vaccaro08a}, \cite{White09a}, \cite%
{Paterek11a}, mainly in the context of pure states for bosonic systems,
though in these papers the focus is on issues other than the definition of
entanglement - such as quantum communication protocols \cite{Verstraete03a},
multicopy distillation \cite{Bartlett06b}, mechanical work and accessible
entanglement \cite{Vaccaro08a}, \cite{White09a} and Bell inequality
violation \cite{Paterek11a}. The consequences for entanglement of applying
this super-selection rule requirement to the sub-system density operators
are quite \emph{significant}, and in the present paper important \emph{new} 
\emph{entanglement tests} are determined. Not only can it immediately be
established that spin squeezing \emph{requires} entangled states, but though
several of the other inequalities (see below) that have been used as
signatures of entanglement are still valid, \emph{additional tests} can be
obtained which only apply to entangled states that are defined to conform to
the symmetrisation principle and the super-selection rules.

It is worth emphasising that requiring the sub-system density operators
satisfy the local particle number SSR means that there are less states than
otherwise would be the case which are classed as non-entangled, and \emph{%
more states} will be regarded as \emph{entangled}. It is therefore not
surprising that additional tests for entanglement will result. If \emph{%
further restrictions} are placed on the sub-system density operator - such
as requiring them to correspond to a fixed number of bosons again there will
be more states regarded as entangled, and even more entanglement tests will
apply. A particular example is given in SubSection \ref{SubSection -
Sorensen 2001}, where the sub-systems are restricted to one boson states.

The \emph{symmetrisation} requirement for systems involving identical
particles is well established since the work of Dirac. There are two types
of justification for applying the \emph{super-selection rules} for systems
of identical particles. The first approach is based on simple considerations
and will be outlined here. The second approach is more sophisticated and
involves linking the absence or presence of SSR to whether or not there is a
suitable \emph{reference frame} in terms of which the quantum state is
described \cite{Aharonov67a}, \cite{Bartlett03a}, \cite{Sanders03a}, \cite%
{Verstraete03a}, \cite{Schuch04a}, \cite{Kitaev04a}, \cite{van Enk05a}, \cite%
{Bartlett06a}, \cite{Bartlett07a}, \cite{Vaccaro08a}, \cite{White09a}, \cite%
{Tichy11a}. This approach will be described in SubSection \ref{SubSection -
Super-Selection Rule} and Appendix \ref{Appendix - Reference Frames and SSR}%
, the key idea being that SSR are a consequence of considering the
description of a quantum state by an external observer (Charlie) whose phase
reference frame has an \emph{unknown phase difference} from that of an
observer ((Alice) more closely linked to the system being studied. Thus,
whilst Alice's description of the quantum state may violate the SSR, the
description of the \emph{same} quantum state by Charlie will not. In the
main part of this paper the density operator $\widehat{\rho }$ used to
describe the various quantum states will be that of the external observer
(Charlie). Note that if the relationship between the phase references is 
\emph{known}, then the SSR can be challenged (see SubSection \ref{SubSection
- Challenges to SSR} and Appendix \ref{Appendix - Reference Frames and SSR}%
). Returning to the more simple reasons referred to for invoking the
superselection rule to exclude quantum superposition states with differing
numbers of identical particles (both massive and otherwise), these may be
summarised as:

1. No way is known for creating such states.

2. No way is known for measuring all the properties of such states, even if
they existed.

3. Coherence and interference effects can be understood without invoking the
existence of such states.

4. The stability of such states against decoherence processes may not be
great, so even if they could be created, they could rapidly change to other
states. However, decoherence time scales that are not too short would be
acceptable, so this last reason is of lesser importance.

Invoking the physical existence of states that as far as we know cannot be
made or measured, and for which there are no known physical effects that
require their presence seems a rather unnecessary feature to add to the
non-relativistic quantum physics of many body systems or to quantum optics,
and considerations based on the general principle of simplicity (Occam's
razor) would suggest not doing so until a clear physical justification for
including them is found. Furthermore, experiments can be carried out on each
of the mode sub-systems considered as a \emph{separate} system, and
essentially the \emph{same reasons} that justify applying the
super-selection rule to the overall system also apply to the separate mode
sub-systems in the context of defining \emph{non-entangled states}. Hence,
unless it can be justified to ignore the super-selection rule for the
overall system it would be \emph{inconsistent} not to apply it to the
sub-system as well. The \emph{onus} is on those who wish to ignore the
super-selection rule for the separate modes to justify why it is being
applied to the overall system. In addition, \emph{joint} \emph{measurements}
on \emph{all} the sub-systems can be carried out, and the interpretation of
the measurement probabilities requires the density operators for the
sub-system \ states to be physically based. The general application of
super-selection rules has however been challenged (see SubSection \ref%
{SubSection - Super-Selection Rule}) on the basis that super-selection rules
are not a fundamental requirement of quantum theory, but are restrictions
that could be lifted if there is a suitable system that acts as a \emph{%
reference} for the coherences involved. In Section \ref{Section - Identical
Particles and Entanglement} and related Appendix \ref{Appendix -
Super-Selection Rule Violations ?} an analysis of these objections to the
super-selection rule is presented, and in Appendix \ref{Appendix - Reference
Frames and SSR} we see that the approach based on phase reference frames
does indeed justify the application of the SSR both to the general quantum
states for multi-mode systems of identical particles and to the sub-system
states for non-entangled states of these systems.

The other focus of this paper is on \emph{spin squeezing}. Heisenberg
Uncertainy Principle inequalities involving spin operators \cite%
{Wodkiewicz85a} and the consequent property of spin squeezing have been
well-known in quantum optics for many years. The importance of spin
squeezing in quantum metrology is discussed in the paper by Kitagawa et al 
\cite{Kitagawa93a} for general spin systems. It was suggested in this paper
that correlations between the individual spins was needed to produce spin
squeezing, though no quantitative proof was presented and the more precise
concept of entanglement was not mentioned. For the case of two mode systems
the earliest paper linking spin squeezing to entanglement is that of
Sorensen et al \cite{Sorensen01a}, which considers a system of identical
bosonic atoms, each of which can occupy one of two internal states. This
paper states that spin squeezing requires the quantum state to be entangled,
with a proof given in the Appendix. A consideration of how such spin
squeezing may be generated via collisional interactions is also presented.
The paper by Sorensen et al is often referred to as establishing the link
between spin squeezing and entanglement - see for example Micheli et al \cite%
{Micheli03a}, Toth et al \cite{Toth07a}, Hyllus et al \cite{Hyllus12a}.
However, the paper by Sorensen et al \cite{Sorensen01a} is based on a
definition of non-entangled states in which the sub-systems are the
identical particles, and this is inconsistent with the symmetrization
principle. The present paper establishes the link between spin squeezing and
entanglement based on a definition of entanglement consistent with the
system and sub-system density operators representing physical states.

It is also important to consider which \emph{components} of the spin
operator vector are squeezed, and this issue is also considered in the
present paper. In the context of the present second quantisation approach to
identical particle systems the three spin operator components for two mode
systems are expressed in terms of the annihilation, creation operators for
the two chosen modes. Spin squeezing can be defined (see Section \ref%
{Section - Spin Squeezing}) in terms of the variances of these spin
operators, however the \emph{covariance matrix} for the three spin operators
will in general have off-diagonal elements, and spin squeezing is better
defined in terms of rotated spin operators referred to as \emph{principal
spin operators} for which the covariance matrix is \emph{diagonal}. The
principal spin operators are related to new mode annihilation, creation
operators in the same form as for the original spin operators, where the 
\emph{new modes} are two orthogonal linear combinations of the originally
chosen modes. In discussing the relationship between spin squeezing and
entanglement, the modes which may be entangled are generally those
associated with the definition of the spin operators.

The plan of this paper is as follows. In Section \ref{Section - Entanglement}
the key definitions of entangled states are covered, and a detailed
discussion on why the symmetrisation principle and the super-selection rule
is invoked in discussed in Section \ref{Section - Identical Particles and
Entanglement}. Challenges to the necessity of the super-selection rule are
outlined, with arguements against such challenges dealt with in Appendices %
\ref{Appendix - Reference Frames and SSR}, \ref{Appendix - Super-Selection
Rule Violations ?}. and \ref{Appendix - Non-Physical Two Mode States}.
Classical entanglement is discussed in Appendix \ref{Appendix - Classical
Entanglement}. The next Section \ref{Section - Spin Squeezing} \ sets out
the definitions of spin squeezing and in the following Section \ref{Section
- Relationship Spin Squeezing & Entanglement} it is shown that spin
squeezing is a signature of entanglement, both for the principle spin
operators with entanglement of the two new modes and for the original spin
operators with entanglement of the original modes. A number of other tests
for entanglement proposed by other authors are considered in Section \ref%
{Section - Criteria for Spin Squeezing Based on Non-Physical States}, with
details of these treatments set out in Appendices \ref{Appendix - Sorensen
Results}, \ref{Appendix - Revised Sorensen}, \ref{Appendix - Heisenberg
Uncertainty Principle Results}. Two key mathematical inequalities are
derived in Appendix \ref{Appendix - Inequalities}. The final Section \ref%
{Section - Discussion & Summary of Key Results} summarises and discusses the
key results.

\pagebreak

\section{Entanglement - General Features}

\label{Section - Entanglement}

\subsection{Physical States}

The standard quantum theory notions of \emph{physical} \emph{systems} that
can exist in various \emph{states} and have associated \emph{quantitie}s on
which \emph{measurements} can be made are presumed in this paper. The
measuring system made be also treated via quantum theory, but there is
always some component that behaves \emph{classically,} so that quantum
fluctuations in the quantity recorded by the \emph{observer} are small. The
term \emph{physical state} refers to a state that can either be prepared via
a process consistent with the laws of quantum physics and on which
measurements can be then performed and the probabilistic results predicted
from this state (\emph{prediction}), or a state whose existence can be
inferred from later quantum measurements (\emph{retrodiction}). In quantum
theory, physical states are \emph{represented} by \emph{density operators}
for mixed states or \emph{state vectors} for pure states, which must satisfy
symmetrisation and other basis requirements in accordance with the laws of
quantum theory. The quantum state, the system it is associated with and the
quantities that can be measured are viewed here as entities that are viewed
as being \emph{both} ontological and epistimological. The observer is
important, but there is actually something out there to be studied. In
addition to those associated with physical states, other density operators
and state vectors may be introduced for \emph{mathematical} convenience. For
physical states, the density operator is determined from either the
preparation process or inferred from the measurement process, and in general
it is a statistical mixture of density operators for possible preparation
processes. Measurement itself constitutes a possible preparation process.
Following preparation, further experimental processes may change the
physical state and dynamical equations give the time evolution of the
density operator between preparation and measurement, the simplest situation
being where measurement takes place immediately after preparation. A full
discussion of the predictive and retrodictive aspects of the density
operator is given in papers by Pegg et al \cite{Pegg02a}, \cite{Pegg05a}.
Whilst there are often different mathematical forms for the density operator
that lead to the same predictive results for subsequent measurements,
applying the results of the measurements to retrodictively determines the 
\emph{preferred form} of the density operator that is consistent with the 
\emph{available} preparation and measurement operators. An example is given
in \cite{Pegg05a}.

\subsection{Entangled and Non-Entangled States}

\label{SubSection - Entangled and Non-Entangled States}

\subsubsection{General Considerations}

Here the commonly applied \emph{physically-based approach} to mathematically
defining entangled states will be described \cite{Barnett09a}. The
definition involves vectors and density operators that represent states than
can be prepared in real experiments, so the mathematical approach is to be 
\emph{physically based}. The concept of quantum entanglement involves \emph{%
composite systems} made up of component \emph{sub-systems} each of which are 
\emph{distinguishable} from the other sub-systems, and where each could
constitute a stand-alone quantum system. This means the each sub-system will
have its own set of physically realisable quantum states - mixed or pure -
which could be prepared independently of the quantum states of the other
sub-systems. As will be seen, the requirement that sub-systems be
distinguishable and their states be physically preparable will have
important consequences, especially in the context of identical particle
systems.\textbf{\ }The formal definition of what is meant by an entangled
state starts with the pure states, described via a vector in a Hilbert
space. The formalism of quantum theory allows for \emph{pure states} for
composite systems made up of two or more distinct sub-systems via tensor
products of sub-system states 
\begin{equation}
\left\vert \Phi \right\rangle =\left\vert \Phi _{A}\right\rangle \otimes
\left\vert \Phi _{B}\right\rangle \otimes \left\vert \Phi _{C}\right\rangle
...  \label{Eq.NonEntangledPureState}
\end{equation}%
Such products are called \emph{non-entangled} or \emph{separable} states.
However, since these product states exist in a Hilbert space, it follows
that linear combinations of such products of the form%
\begin{equation}
\left\vert \Phi \right\rangle =\tsum\limits_{\alpha \beta \gamma
..}C_{\alpha \beta \gamma ..}\left\vert \Phi _{A}^{\alpha }\right\rangle
\otimes \left\vert \Phi _{B}^{\beta }\right\rangle \otimes \left\vert \Phi
_{C}^{\gamma }\right\rangle .  \label{Eq.EntangledPureState}
\end{equation}%
could also represent possible pure quantum states for the system. Such \emph{%
quantum superpositions} which cannot be expressed as a \emph{single} product
of sub-system states are known as \emph{entangled} (or \emph{non-separable})
states.

The concept of entanglement can be extended to \emph{mixed states}, which
are described via density operators in the Hilbert space. If $A$, $B$,
...are the sub-systems with $\widehat{\rho }_{R}^{A}$, $\widehat{\rho }%
_{R}^{B}$, being density operators the sub-systems $A$, $B$, .then a \emph{%
general non-entangled }or \emph{separable} state is one where the overall
density operator $\widehat{\rho }$ can be written as the weighted sum of
tensor products of these sub-system density operators in the form \cite%
{Werner89a} 
\begin{equation}
\widehat{\rho }=\sum_{R}P_{R}\,\widehat{\rho }_{R}^{A}\otimes \widehat{\rho }%
_{R}^{B}\otimes \widehat{\rho }_{R}^{C}\otimes ...
\label{Eq.NonEntangledState}
\end{equation}%
with $\sum_{R}P_{R}=1$ and $P_{R}\geq 0$ giving the probability that the
specific product state $\widehat{\rho }_{R}=\widehat{\rho }_{R}^{A}\otimes 
\widehat{\rho }_{R}^{B}\otimes \widehat{\rho }_{R}^{C}\otimes ..$occurs. 
\emph{Entangled }states (or \emph{non-separable} states) are those that
cannot be written in this form, so in this approach knowing what the term
entangled state refers to is based on \emph{first} knowing what the general
form is for a non-entangled state. The density operator $\widehat{\rho }%
=\left\vert \Phi \right\rangle \left\langle \Phi \right\vert $ for the pure
state in (\ref{Eq.EntangledPureState}) is not of the form (\ref%
{Eq.NonEntangledState}), as there are cross terms of the form $C_{\alpha
\beta \gamma ..}C_{\theta \lambda \eta ..}^{\ast }(\left\vert \Phi
_{A}^{\alpha }\right\rangle \left\langle \Phi _{A}^{\theta }\right\vert
)\otimes (\left\vert \Phi _{B}^{\beta }\right\rangle \left\langle \Phi
_{B}^{\lambda }\right\vert )\otimes ..$involved.

The concepts of separability and entanglement based on the Eqs. (\ref%
{Eq.NonEntangledPureState}) and (\ref{Eq.NonEntangledState}) for
non-entangled states do not however just rest on the mathematical forms
alone. Implicitly there is the \emph{assumption} that separable quantum
states described by the two expressions can actually be created in \emph{%
physical processes}. The sub-systems involved must therefore be \emph{%
distinguishable quantum systems} in their own right, and the sub-system
states $\left\vert \Phi _{A}\right\rangle ,\left\vert \Phi _{B}\right\rangle
,..$or $\widehat{\rho }_{R}^{A},\widehat{\rho }_{R}^{B},..$ must also be 
\emph{possible physical states} for the sub-systems. We will return to these
requirements later. The issue of the physical preparation of non-entangled
(separable) states starting from some uncorrelated fiducial state for the
separate sub-systems was introduced by Werner \cite{Werner89a}, and
discussed further by Bartlett et al (see \cite{Bartlett06b}, Section IIB).
This involves the ideas of \emph{local operations} and \emph{classical
communication (LOCC) }dealt with in the next SubSection.

The key requirement is that entangled states exhibit a novel \emph{quantum
feature} that is only found in \emph{composite} systems. Separable states
are such that the \emph{joint probability} for measurements of all physical
quantities associated with the sub-systems can be found from separate
measurement probabilities obtained from the sub-system density operators $%
\widehat{\rho }_{R}^{A}$, $\widehat{\rho }_{R}^{B}$, etc and the overall
classical probability $P_{R}$ (see SubSection\ref{SubSection - Joint
Measurements}). This feature of separable probabilities is absent in
entangled states, and because of this key \emph{non-separability feature}
Schrodinger called these states "entangled". The separability feature for
the joint probabilities is essentially a classical feature and applies in
hidden variable theories (HVT) (see SubSection\ref{SubSection - Hidden
Variable Theory}) applied to quantum systems - as well as to quantum
separable states. The fact that only entangled states do not exhibit the
feature of separable probabilities shown in classical HVT highlights
entanglement being a \emph{non-classical feature} found only in composite
systems.

An alternative \emph{operational approach} to defining entangled states
focuses on whether or not they exhibit certain non-classical features such
as Bell Inequality violation or whether they satisfy certain mathematical
tests such as having a non-negative partial transpose\cite{Peres96a} \cite%
{Horodecki09a}, and a \emph{utilititarian approach} focuses or whether
entangled states have technological applications such as in various quantum
information protocols. As will be seen in SubSection \ref{SubSection - SSR
Separate Modes}, the particular definition of entangled states based on
their non-creatability.via LOCC essentially coincides with the approach used
in the present paper. Wiseman et al \cite{Wiseman07a}, \cite{Jones07a} and
Reid et al \cite{Reid12a} discuss the concept of a \emph{heirarchy} of \emph{%
entangled states}, with states exhibiting \emph{Bell nonlocality} being a
subset of states for which there is \emph{EPR steering}, which in turn is a
subset of the \emph{entangled states}, the latter being defined as states
whose density operators cannot be written as in Eq. (\ref%
{Eq.NonEntangledState}) though without further consideration if additional
properties are required for the sub-system density operators. The
operational approach could lead into a quagmire of differing interpretations
of entanglement dependng on which non-classical feature is highlighted, and
the utilitarian approach implies that all entangled states have a
technological use, which is by no means the case. For these reasons, the
present mathematical approach based on the quantities involved representing
physical sub-system states is generally favoured \cite{Barnett09a}. It is
also compatible with later classifying entangled states in a heirarchy.

\subsubsection{Local Systems and Operations}

As pointed out by Vedral \cite{Vedral07a}, one reason for calling states
such as in Eqs.(\ref{Eq.NonEntangledPureState}) and (\ref%
{Eq.NonEntangledState}) separable is associated with the idea of performing
operations on the separate sub-systems that do not affect the other
sub-systems. Such operations on such \emph{local systems} are referred to as 
\emph{local operations} and include unitary operations $\widehat{U}_{A}$, $%
\widehat{U}_{B}$, that change the states via $\widehat{\rho }%
_{R}^{A}\rightarrow \widehat{U}_{A}\widehat{\rho }_{R}^{A}\widehat{U}%
_{A}^{-1}$, $\widehat{\rho }_{R}^{B}\rightarrow \widehat{U}_{B}\widehat{\rho 
}_{R}^{B}\widehat{U}_{B}^{-1}$, etc as in a time evolution, and could
include processes by which the states $\widehat{\rho }_{R}^{A}$, $\widehat{%
\rho }_{R}^{B}$, are separately prepared from suitable initial states.

We note that performing local operations on a separable state only produces
another separable state, not an entangled state. Such local operations are
obviously faciltated in experiments if the sub-systems are essentially \emph{%
non-interacting} - such as when they are spatially \emph{well-separated},
though this does not have to be the case. The local systems and operations
could involve sub-systems whose quantum states and operators are just in
different parts of Hilbert space, such as for cold atoms in different
hyperfine states even when located in the same spatial region. Note the
distinction between \emph{local} and \emph{localised}. As described by
Werner \cite{Werner89a}, if one observer (Alice) is associated with
preparing separate sub-system $A$ in a physical state $\widehat{\rho }%
_{R}^{A}$ via local operations with a probability $P_{R}$, a second observer
(Bob) could be then advised via a \emph{classical communication} channel to
prepare sub-system $B$ in state $\widehat{\rho }_{R}^{B}$ via local
operations. After repeating this process for different choices $R$ of the
correlated pairs of sub-system states, the overall quantum state prepared by
both observers via this local operation and classical communication protocol 
\emph{(}$\emph{LOCC)}$ would then be the bipartite non-entangled state $%
\widehat{\rho }=\sum_{R}P_{R}\,\widehat{\rho }_{R}^{A}\otimes \widehat{\rho }%
_{R}^{B}$. Multipartite non-entangled states of the form (\ref%
{Eq.NonEntangledState}) can also be prepared via LOCC protocols involving
further observers. As will be seen, the separable or non-entangled states
are just those that can be prepared by LOCC protocols.

\subsubsection{Constraints on Sub-System Density Operators}

A key issue however is whether density operators $\widehat{\rho }$ and $%
\widehat{\rho }_{R}^{A}$, $\widehat{\rho }_{R}^{B}$, in Eq. (\ref%
{Eq.NonEntangledState}) always represent possible \emph{physical states},
even if the operators $\widehat{\rho }$ and $\widehat{\rho }_{R}^{A}$, $%
\widehat{\rho }_{R}^{B}$, etc satisfy all the standard mathematical
requirements for density operators - Hermitiancy, positiveness, trace equal
to unity, trace of density operator squared being not greater than unity. In
this paper it will be argued that there are further requirements not only on
the overall density operator, but also on those for the individual
sub-systems that are imposed by \emph{symmetrisation} and \emph{%
super-selection} rules.

\subsubsection{Classical Entanglement}

\label{SubSubSection - Classical Entanglement}

In addition to quantum entanglement there is a body of work (see \cite%
{Gisin91a}, \cite{Spreeuw98a}, \cite{Borges10a})\ dealing with so-called%
\textbf{\ }\emph{classical entanglement}\textbf{. }Here the\textbf{\ }\emph{%
states}\textbf{\ }of\textbf{\ }\emph{classical}\textbf{\ }systems - such as
a classical EM field - are represented via a formalism involving\textbf{\ }%
\emph{linear vector spaces}\textbf{\ }and classical entanglement is defined
mathematically. \ A discussion of classical entanglement is given for
completeness in Appendix \ref{Appendix - Classical Entanglement}. Although
there are some formal similarities with quantum entanglement - and even Bell
type inequalites which can be violated, there are key features that is not
analogous to that for composite quantum systems. In the end, classical and
quantum entanglement are fundamentally different when the physics of the two
different types of system - one classical and deterministic, the other
quantum and probabilistic are taken into account rather than just focusing
on similarities in the mathematical formalisms. In particular, the key
feature of quantum entanglement relating to joint measurement probailities
is quite different to the corresponding one for classical entanglement.

\subsection{Separate and Joint Measurements, Reduced Density Operator}

\label{SubSection - Joint Measurements}

In this SubSection we consider separate and joint measurements on systems
involving several sub-systems and introduce results for probabilities, mean
values for measurements on one of the sub-systems which are conditional on
the results for measurements on another of the sub-systems. This will
require consideration of quantum theoretical \emph{conditional probabilities}%
. The measurements involved will be assumed for simplicity to be von Neumann 
\emph{projective measurements} for physical quantities represented by
Hermitian operators $\widehat{\Omega }$, which project the quantum state
into subspaces for the eigenvalue $\lambda _{i}$ that is measured, the
subspaces being associated with Hermitian, idempotent \emph{projectors} $%
\widehat{\Pi }_{i}$ whose sum over all eigenvalues is unity. These concepts
are treated in several quantum theory textbooks, for example \cite{Peres93a}%
, \cite{Isham95a}. For completeness, an account setting out the key results
is presented in Appendix \ref{Appendix - Projective Measurements}.

\subsubsection{Joint Measurements on Sub-Systems}

For situations involving distinct sub-systems measurements can be carried
out on all the sub-systems and the results expressed in terms of the \emph{%
joint probability} for various outcomes. If $\widehat{\Omega }_{A}$ is a
physical quantity associated with sub-system $A$, with eigenvalues $\lambda
_{i}^{A}$ and with $\widehat{\Pi }_{i}^{A}$ the projector onto the subspace
with eigenvalue $\lambda _{i}^{A}$, $\widehat{\Omega }_{B}$ is a physical
quantity associated with sub-system $B$, with eigenvalues $\lambda _{j}^{B}$
and with $\widehat{\Pi }_{j}^{B}$ the projector onto the subspace with
eigenvalue $\lambda _{j}^{B}$ etc., then the \emph{joint probability} $%
P_{AB..}(i,j,..)$ that measurement of $\widehat{\Omega }_{A}$ leads to
result $\lambda _{i}^{A}$ , measurement of $\widehat{\Omega }_{B}$ leads to
result $\lambda _{j}^{B}$ ,etc is given by%
\begin{equation}
P_{AB..}(i,j,..)=Tr(\widehat{\Pi }_{i}^{A}\,\widehat{\Pi }_{j}^{B}\,...%
\widehat{\rho })  \label{Eq.JointProb}
\end{equation}%
This joint probability depends on the full density operator $\widehat{\rho }$
representing the physical state as well as on the quantities being measured.
Here the projectors (strictly $\widehat{\Pi }_{i}^{A}\otimes \widehat{1}%
^{B}\otimes ..$, $\widehat{1}^{A}\otimes \widehat{\Pi }_{j}^{B}\otimes ..$,
etc) commute, so the order of measurements is immaterial. An alternative
notation in which the physical quantities are also specified is $P_{AB..}(%
\widehat{\Omega }_{A},i;\widehat{\Omega }_{B},j;..)$.

\subsubsection{Single Measurements on Sub-Systems and Reduced Density
Operator}

The \emph{reduced density operator} $\widehat{\rho }_{A}$ for sub-system $A$
given by 
\begin{equation}
\widehat{\rho }_{A}=Tr_{B,C,...}(\widehat{\rho })
\label{Eq.ReducedDensityOpr}
\end{equation}%
and enables the results for measurements on sub-system $A$ to be determined
for the situation where the results for all joint measurements involving the
other sub-systems are \emph{discarded}. The probability $P_{A}(i)$ that
measurement of $\widehat{\Omega }_{A}$ leads to result $\lambda _{i}^{A}$
irrespective of the results for meaurements on the other sub-systems is
given by%
\begin{eqnarray}
P_{A}(i) &=&\tsum\limits_{j,k,..}P_{AB..}(i,j,..)  \nonumber \\
&=&Tr(\widehat{\Pi }_{i}^{A}.\widehat{\rho })  \label{Eq.SeparateProbModeA0}
\\
&=&Tr_{A}(\widehat{\Pi }_{i}^{A}\,\widehat{\rho }_{A})
\label{Eq.SeparateProbModeA}
\end{eqnarray}%
using $\tsum\limits_{j}\widehat{\Pi }_{j}^{B}=\widehat{1}$, etc. Hence the
reduced density operator $\widehat{\rho }_{A}$ plays the role of specifying
the physical state for mode $A$ considered as a separate sub-system, even if
the original state $\widehat{\rho }$ is entangled. An alternative notation
in which the physical quantity is also specified is $P_{A}(\widehat{\Omega }%
_{A},i)$.

\subsubsection{Mean Value and Variance}

The \emph{mean value} for measuring a physical quantity $\widehat{\Omega }%
_{A}$ will be given by 
\begin{eqnarray}
\left\langle \widehat{\Omega }_{A}\right\rangle &=&\sum_{\lambda
_{i}^{A}}\lambda _{i}^{A}P_{A}(i)  \nonumber \\
&=&Tr_{A}(\widehat{\Omega }^{A}\,\widehat{\rho }_{A})  \label{Eq.Mean}
\end{eqnarray}%
where we have used $\widehat{\Omega }^{A}=\sum_{\lambda _{i}^{A}}\lambda
_{i}^{A}\widehat{\Pi }_{i}^{A}$.

The \emph{variance} of measurements of the physical quantity $\widehat{%
\Omega }_{A}$ will be given by 
\begin{eqnarray}
\left\langle (\Delta \widehat{\Omega }^{A})^{2}\right\rangle
&=&\sum_{\lambda _{i}^{A}}(\lambda _{i}^{A}-\left\langle \widehat{\Omega }%
_{A}\right\rangle )^{2}P_{A}(i)  \nonumber \\
&=&Tr_{A}(\left( \widehat{\Omega }^{A}-\left\langle \widehat{\Omega }%
_{A}\right\rangle \right) ^{2}\,\widehat{\rho }_{A})  \label{Eq.Variance}
\end{eqnarray}%
so both the mean and variance only depend on the reduced density operator $%
\widehat{\rho }_{A}$.

On the other hand the \emph{mean value} of a \emph{product} of sub-system
operators $\widehat{\Omega }_{A}\otimes \widehat{\Omega }_{B}\otimes 
\widehat{\Omega }_{C}\otimes ...$, where $\widehat{\Omega }_{A}$, $\widehat{%
\Omega }_{B}$, $\widehat{\Omega }_{C}$, .. are Hermitian operators
representing physical quantities for the separate sub-systems, is given by%
\begin{eqnarray}
\left\langle \widehat{\Omega }_{A}\otimes \widehat{\Omega }_{B}\otimes 
\widehat{\Omega }_{C}\otimes .\right\rangle &=&\sum_{\lambda
_{i}^{A}}\sum_{\lambda _{j}^{B}}...\lambda _{i}^{A}\lambda
_{j}^{B}...P_{AB..}(i,j,..)  \nonumber \\
&=&Tr\left( \widehat{\Omega }_{A}\otimes \widehat{\Omega }_{B}\otimes 
\widehat{\Omega }_{C}\otimes .\right) \widehat{\rho }
\label{Eq.MeanProductGeneral}
\end{eqnarray}%
which involves the overall system density operator, as expected.

\subsubsection{Conditional Probabilities}

Treating the case of two sub-systems for simplicity we can use Bayes theorem
(see Appendix \ref{Appendix - Projective Measurements}, Eq.(\ref{Eq.BayesThm}%
)) to obtain expressions for \emph{conditional probabilities} \cite%
{Barnett09a}. The conditional probability that if measurement of $\widehat{%
\Omega }_{B}$ associated with sub-system $B$ leads to eigenvalue $\lambda
_{j}^{B}$ then measurement of $\widehat{\Omega }_{A}$ associated with
sub-system $A$ leads to eigenvalue $\lambda _{i}^{A}$ is given by 
\begin{equation}
P_{AB}(i|j)=Tr(\widehat{\Pi }_{i}^{A}\,\widehat{\Pi }_{j}^{B}\,\widehat{\rho 
})/Tr(\widehat{\Pi }_{j}^{B}\widehat{\rho })  \label{Eq.CondProbGeneralCase}
\end{equation}%
In general, the overall density operator is required to determine the
conditional probability. An alternative notation in which the physical
quantities are also specified is $P_{AB}(\widehat{\Omega }_{A},i|\widehat{%
\Omega }_{B},j)$.

As shown in Appendix \ref{Appendix - Projective Measurements} the
conditional probability is given by 
\begin{equation}
P_{AB}(i|j)=Tr(\widehat{\Pi }_{i}^{A}\widehat{\rho }_{cond}(\widehat{\Omega }%
_{B},\lambda _{j}^{B}))  \label{Eq.ConProbGC2}
\end{equation}%
where 
\begin{equation}
\widehat{\rho }_{cond}(\widehat{\Omega }_{B},\lambda _{j}^{B})=\widehat{\Pi }%
_{j}^{B}\,\widehat{\rho }\,\widehat{\Pi }_{j}^{B}/Tr(\widehat{\Pi }_{j}^{B}%
\widehat{\rho })  \label{Eq.CondDensityOprMeastB}
\end{equation}%
is the so-called \emph{conditioned density operator}, corresponding the
quantum state produced following the measurement of $\widehat{\Omega }_{B}$
that obtained the result $\lambda _{j}^{B}$. The conditional probability
result is the same as%
\begin{equation}
P_{AB}(i|j)=Tr(\widehat{\Pi }_{i}^{A}\widehat{\rho }_{cond}(\widehat{\Omega }%
_{B},\lambda _{j}^{B}))  \label{Eq.CondProbGeneralCase2}
\end{equation}%
which is the same as the expression (\ref{Eq.SeparateProbModeA0}) with $%
\widehat{\rho }$ replaced by $\widehat{\rho }_{cond}(\widehat{\Omega }%
_{B},\lambda _{j}^{B})$. This is what would be expected for a conditioned
measurement probability.

Also, if the measurement results for $\widehat{\Omega }_{B}$ are not
recorded the conditioned density operator now becomes 
\begin{eqnarray}
\widehat{\rho }_{cond}(\widehat{\Omega }_{B}) &=&\sum_{\lambda
_{j}^{B}}P_{B}(j)\widehat{\rho }_{cond}(\widehat{\Omega }_{B},\lambda
_{j}^{B})  \nonumber \\
&=&\sum_{\lambda _{j}^{B}}\widehat{\Pi }_{j}^{B}\,\widehat{\rho }\,\widehat{%
\Pi }_{j}^{B}  \label{Eq.UnrecordedCondDensOpr}
\end{eqnarray}%
This is still different to the original density operator $\widehat{\rho }$
because a measurement of $\widehat{\Omega }_{B}$ has occured, even if we
dont know the outcome. However, the measurement probability for $\widehat{%
\Omega }_{A}$ is now%
\begin{eqnarray}
P_{AB}(i|Any\,j) &=&Tr(\widehat{\Pi }_{i}^{A}\widehat{\rho }_{cond}(\widehat{%
\Omega }_{B}))  \nonumber \\
&=&Tr(\widehat{\Pi }_{i}^{A}\widehat{\rho })  \label{Eq.NoSignalling0} \\
&=&P_{A}(i)  \label{Eq.NoSignalling}
\end{eqnarray}%
where we have used the cyclic properties of the trace, $\left( \widehat{\Pi }%
_{j}^{B}\right) ^{2}=\widehat{\Pi }_{j}^{B}$ and $\sum_{\lambda _{j}^{B}}%
\widehat{\Pi }_{j}^{B}=\widehat{1}$. The results in Eqs. (\ref%
{Eq.NoSignalling0}) and (\ref{Eq.NoSignalling}) are the same as the
measurement probability for $\widehat{\Omega }_{A}$ if no measurement for $%
\widehat{\Omega }_{B}$ had taken place at all. This is perhaps not
surprising, since the record of the latter measurements was discarded.
Another way of showing this result is that Bayes Theorem tells us that $%
\sum_{j}P_{AB}(i|j)P_{B}(j)=\sum_{j}P_{AB}(i,j)=P_{A}(i)$, since $%
\sum_{j}P_{AB}(i,j)$ is the probability that measurement of $\widehat{\Omega 
}_{A}$ will lead to $\lambda _{i}^{A}$ and measurement of $\widehat{\Omega }%
_{B}$ will lead to any of the $\lambda _{j}^{B}$. This result is called the
no-signalling theorem \cite{Barnett09a}.

\subsubsection{Conditional Mean and Variance}

As explained in Appendix \ref{Appendix - Projective Measurements}, to
determine the \emph{conditioned mean value} of $\widehat{\Lambda }$ after
measurement of $\widehat{\Omega }$ has led to the eigenvalue $\lambda _{i}$
we use $\widehat{\rho }_{cond}(\widehat{\Omega },i)$ rather than $\,\widehat{%
\rho }$ in the mean formula $\left\langle \widehat{\Lambda }\right\rangle
=Tr(\widehat{\Lambda }\widehat{\rho })$ and the result is given in terms of
the conditional probability $P(\widehat{\Lambda }j|\widehat{\Omega }i)$.
Here we refer to two commuting observables and include the operators in the
notation to avoid any misinterpretation. Hence%
\begin{eqnarray}
\left\langle \widehat{\Lambda }\right\rangle _{\widehat{\Omega },i} &=&Tr(%
\widehat{\Lambda }\widehat{\rho }_{cond}(\widehat{\Omega },i))  \nonumber \\
&=&\dsum\limits_{j}\mu _{j}\,P(\widehat{\Lambda },j|\widehat{\Omega },i)
\label{Eq.CondMean0}
\end{eqnarray}

For the \emph{conditioned variance} of $\widehat{\Lambda }$ after
measurement of $\widehat{\Omega }$ has led to the eigenvalue $\lambda _{i}$
we use $\widehat{\rho }_{cond}(\widehat{\Omega },i)$ rather than $\,\widehat{%
\rho }$ and the conditioned mean $\left\langle \widehat{\Lambda }%
\right\rangle _{\widehat{\Omega },i}$rather than $\left\langle \widehat{%
\Lambda }\right\rangle $ in the variance formula $\left\langle \Delta 
\widehat{\Lambda }^{2}\right\rangle =Tr((\widehat{\Lambda }-\left\langle 
\widehat{\Lambda }\right\rangle )^{2}\widehat{\rho })$. Hence%
\begin{eqnarray}
\left\langle \Delta \widehat{\Lambda }^{2}\right\rangle _{\widehat{\Omega }%
,i} &=&Tr((\widehat{\Lambda }-\left\langle \widehat{\Lambda }\right\rangle _{%
\widehat{\Omega },i})^{2}\widehat{\rho }_{cond}(\widehat{\Omega },i)) 
\nonumber \\
&=&\dsum\limits_{j}(\mu _{j}-\left\langle \widehat{\Lambda }\right\rangle _{%
\widehat{\Omega },i})^{2}\,P(\widehat{\Lambda },j|\widehat{\Omega },i)
\label{Eq.CondVariance0}
\end{eqnarray}

If we weighted the conditioned mean by the probability $P(\widehat{\Omega }%
,i)$ that measuring $\widehat{\Omega }$ has led to the eigenvalue $\lambda
_{i}$ and summed over the possible outcomes $\lambda _{i}$ for the $\widehat{%
\Omega }$ measurement, then we obtain the mean for measurements of $\widehat{%
\Lambda }$ after un-recorded measurements of $\widehat{\Omega }$ have
occured. From Bayes theorem $\tsum\limits_{i}P(\widehat{\Lambda },j|\widehat{%
\Omega },i)P(\widehat{\Omega },i)=P(\widehat{\Lambda },j)$ so this gives the 
\emph{unrecorded mean} $\left\langle \widehat{\Lambda }\right\rangle _{%
\widehat{\Omega }}$ as 
\begin{eqnarray}
\left\langle \widehat{\Lambda }\right\rangle _{\widehat{\Omega }}
&=&\tsum\limits_{i}\left\langle \widehat{\Lambda }\right\rangle _{\widehat{%
\Omega },i}P(\widehat{\Omega },i)  \nonumber \\
&=&\dsum\limits_{j}\mu _{j}\,P(\widehat{\Lambda },j)  \nonumber \\
&=&\left\langle \widehat{\Lambda }\right\rangle  \label{Eq.UnrecordedMean}
\end{eqnarray}%
which is the usual mean value for measurements of $\widehat{\Lambda }$ when
no measurements of $\widehat{\Omega }$ have occured. Note that no such
similar result occurs for the \emph{unrecorded} \emph{variance }$%
\left\langle \Delta \widehat{\Lambda }^{2}\right\rangle _{\widehat{\Omega }}$%
\begin{eqnarray}
\left\langle \Delta \widehat{\Lambda }^{2}\right\rangle _{\widehat{\Omega }}
&=&\tsum\limits_{i}\left\langle \Delta \widehat{\Lambda }^{2}\right\rangle _{%
\widehat{\Omega },i}P(\widehat{\Omega },i)  \nonumber \\
&\neq &\left\langle \Delta \widehat{\Lambda }^{2}\right\rangle
\label{Eq.UnrecordedVariance}
\end{eqnarray}

\pagebreak

\subsection{Non-Entangled States}

\label{SubSection - Non-Entangled States}

In this SubSection we will set out the key results for measurements on
non-entangled states.

\subsubsection{Non-Entangled States - Joint Measurements on Sub-Systems}

In the case of the general \emph{non-entangled state} we find \ that the
joint probability is 
\begin{equation}
P_{AB..}(i,j,..)=\sum_{R}P_{R}\,P_{A}^{R}(i)P_{B}^{R}(j)..
\label{Eq.JointProbNonEntState}
\end{equation}%
where%
\begin{equation}
P_{A}^{R}(i)=Tr(\widehat{\Pi }_{i}^{A}\,\widehat{\rho }_{R}^{A})\qquad
P_{B}^{R}(j)=Tr(\widehat{\Pi }_{j}^{B}\,\widehat{\rho }_{R}^{B})\qquad ..
\label{Eq.SepProb}
\end{equation}%
are the probabilities for measurement results for $\widehat{\Omega }_{A}$, $%
\widehat{\Omega }_{B}$, ..on the separate sub-systems with density operators 
$\,\widehat{\rho }_{R}^{A}$, $\widehat{\rho }_{R}^{B}$, etc and the overall
joint probability is given by the products of the probalbilities $%
P_{A}^{R}(i)$, $P_{B}^{R}(j)$, ..for the measurement results $\lambda
_{i}^{A}$, $\lambda _{j}^{B}$, ..for physical quantities $\widehat{\Omega }%
_{A}$, $\widehat{\Omega }_{B}$, ..if the sub-systems are in the states $%
\widehat{\rho }_{R}^{A}$, $\widehat{\rho }_{R}^{B}$, etc. These products are
then weighted by the probability $P_{R}\,$\ that the system is prepared in
the particular product state $\widehat{\rho }_{R}^{A}\otimes \widehat{\rho }%
_{R}^{B}\otimes \widehat{\rho }_{R}^{C}\otimes ..$to determine the overall
joint probability $P_{AB..}(i,j,..)$. The overall probability is of a \emph{%
classical} form. Obviously this joint probability depends on the sub-system
density operators $\widehat{\rho }_{R}^{A}$, $\widehat{\rho }_{R}^{B}$, etc.

In the simple non-entangled\textbf{\ }\emph{pure state}\textbf{\ }situation
in Eq.(\ref{Eq.NonEntangledPureState}) the joint probabilty only involves a
single product of sub-system probabilities\textbf{\ }%
\begin{equation}
P_{AB..}(i,j,..)=P_{A}(i)P_{B}(j)..  \label{Eq.JointProbPureNonEntState}
\end{equation}%
where\textbf{\ }%
\begin{equation}
P_{A}(i)=\left\langle \Phi _{A}\right\vert \widehat{\Pi }_{i}^{A}\,\left%
\vert \Phi _{A}\right\rangle \qquad P_{B}(j)=\left\langle \Phi
_{B}\right\vert \widehat{\Pi }_{j}^{B}\,\left\vert \Phi _{B}\right\rangle
\qquad ..  \label{Eq.SubSystProb}
\end{equation}%
just give the probabilities for measurements in the separate sub-systems.

This \emph{key result} (\ref{Eq.JointProbNonEntState}) showing that the
joint measurement probability for a separable state only depends on \emph{%
separate measurement probabilities} for the sub-systems, together with the
classical probability for preparing correlated product states of the
sub-systems, does \emph{not} apply for entangled states. Hence the \emph{key
quantum feature} for \emph{composite systems} of non-separability for joint
measurement probabilites applies only to entangled states. It is the \emph{%
absence} of the feature that joint probabilities depend on sub-system
measurement probabilities that led Schrodinger to call non-separable states
entangled. This strange quantum feature of entangled states has been
regarded as particularly unusual when the sub-systems are\textbf{\ }\emph{%
spatially well-separated} (or non-local) and is linked to quantum paradoxes
such as EPR.\textbf{\ }Measurements on sub-system $A$ of physical quantity $%
\widehat{\Omega }_{A}$ affect the results of measurements of $\widehat{%
\Omega }_{B}$ at the same time on a distant sub-system $B$, even if the
choice of measured quantity $\widehat{\Omega }_{B}$ is unkown to the
experimenter measuring $\widehat{\Omega }_{A}$. As will be shown below, a
similar result to (\ref{Eq.JointProbNonEntState}) also occurs in \emph{%
hidden variable theory} - a classical theory - so non-separability for joint
measurements resulting from entanglement is a truly \emph{non-classical
feature} of composite systems.

\subsubsection{Non-Entangled States - Single Sub-System Measurements}

For the general non-entangled state, the reduced density operator for
sub-system $A$ is given by 
\begin{equation}
\widehat{\rho }_{A}=\sum_{R}P_{R}\,\widehat{\rho }_{R}^{A}
\label{Eq.ReducedDensityOprNonEntState}
\end{equation}%
A key feature of a non-entangled state is that the results of a measurement
on any \emph{one} of the sub-systems is \emph{independent} of the states for
the \emph{other} subsystems. From Eqs.(\ref{Eq.SeparateProbModeA}) and (\ref%
{Eq.ReducedDensityOprNonEntState}) the probability $P_{A}(i)$ that
measurement of $\widehat{\Omega }_{A}$ leads to result $\lambda _{i}^{A}$ is
given by 
\begin{equation}
P_{A}(i)=\sum_{R}P_{R}\,P_{A}^{R}(i)  \label{Eq.MeasProbNonEntState}
\end{equation}%
where the reduced density operator is given by Eq. (\ref%
{Eq.ReducedDensityOprNonEntState}) for the non-entangled state in Eq. (\ref%
{Eq.NonEntangledState}). This result only depends on the reduced density
operator $\widehat{\rho }_{A}$, which represents a state for sub-system $A$
and which is a statistical mixture of the sub-system states $\,\widehat{\rho 
}_{R}^{A}$, with a probability $P_{R}$ that is the \emph{same} for all
sub-systems. The result for the measurement probability $P_{A}(i)$ is just
the statistical average of the results that would apply if sub-system $A$
were in possible states $\,\widehat{\rho }_{R}^{A}$. For all quantum states
the final expression for the measurement probability $P_{A}(i)$ only
involves a trace of quantities $\widehat{\Pi }_{i}^{A}$, $\widehat{\rho }%
_{A} $ that apply to sub-system $A$, but for a non-entangled state the
reduced density operator $\widehat{\rho }_{A}$ is given by an expression (%
\ref{Eq.ReducedDensityOprNonEntState}) that does \emph{not} involve density
operators for the other sub-systems. Thus for a non-entangled state, the
probability $P_{A}(i)$ is \emph{independent} of the states $\widehat{\rho }%
_{R}^{B}$, $\widehat{\rho }_{R}^{C}$, associated with the other sub-systems.
Analogous results apply for measurements on the other sub-systems.

\subsubsection{Non-Entangled States - Conditional Probability}

For a general non-entangled bipartite mixed state the conditional
probability is given by%
\begin{equation}
P_{AB}(i|j)=\sum_{R}P_{R}\,P_{A}^{R}(i)P_{B}^{R}(j)/\sum_{R}P_{R}%
\,P_{B}^{R}(j)  \label{Eq.CondProbNonEntangledState}
\end{equation}%
which in general depends on $\widehat{\Omega }_{B}$ associated with
sub-system $B$ and the eigenvalue $\lambda _{j}^{B}$. This may seem
surprising for the case where $A$ and $B$ are localised sub-systems which
are well separated. It implies that even for separable states a measurement
result for sub-system $B$ will affect the result for a totally unrelated
measurement on sub-system $A$ which is a long distance away. This is an
example of "spooky action at a distance". However, it should be remembered
that the general separable state is still a correlated state, so each
sub-system density operator $\widehat{\rho }_{R}^{B}$ for sub-system $B$ is
matched with a corresponding density operator $\widehat{\rho }_{R}^{A}$ for
sub-system $A$. It is therefore not surprising that the measurement results
for $A$ are not independent of those for $B$.

However, for a non-entangled pure state where $\widehat{\rho }=\widehat{\rho 
}^{A}\otimes \widehat{\rho }^{B}$ we find that 
\begin{equation}
P_{AB}(i|j)=P_{A}(i)  \label{Eq.CondProbNonEntPureState}
\end{equation}%
where $P_{A}(i)=Tr(\widehat{\Pi }_{i}^{A}\widehat{\rho }^{A})$. For
separable pure states the conditional probability is independent of $%
\widehat{\Omega }_{B}$ associated with sub-system $B$ and the eigenvalue $%
\lambda _{j}^{B}$.

Also of course $\sum_{j}P_{AB}(i|j)P_{B}(j)=P_{A}(i)$ is true for separable
states since it applies to general bipartite states. Hence if the
measurement results for $\widehat{\Omega }_{B}$ are discarded then the
probability distribution for measurements on $\widehat{\Omega }_{A}$ will be
determined from the conditioned density operator $\widehat{\rho }_{cond}(%
\widehat{\Omega }_{B})$ and just result in $P_{A}(i)$ - as in shown in Eq.(%
\ref{Eq.NoSignalling}) for any quantum state.

\subsubsection{Non-Entangled States - Mean Values and Correlations}

For non-entangled states as in Eq. (\ref{Eq.NonEntangledState}) the mean
value for measuring a physical quantity $\widehat{\Omega }_{A}\otimes 
\widehat{\Omega }_{B}\otimes \widehat{\Omega }_{C}\otimes ...$, where $%
\widehat{\Omega }_{A}$, $\widehat{\Omega }_{B}$, $\widehat{\Omega }_{C}$, ..
are Hermitian operators representing physical quantities for the separate
sub-systems can be obtained from Eqs.(\ref{Eq.NonEntangledState}) and (\ref%
{Eq.MeanProductGeneral}) and is given by 
\begin{equation}
\left\langle \widehat{\Omega }_{A}\otimes \widehat{\Omega }_{B}\otimes 
\widehat{\Omega }_{C}\otimes .\right\rangle =\sum_{R}P_{R}\,\left\langle 
\widehat{\Omega }_{A}\right\rangle _{R}^{A}\,\left\langle \widehat{\Omega }%
_{B}\right\rangle _{R}^{B}\,\left\langle \widehat{\Omega }_{C}\right\rangle
_{R}^{C}...  \label{Eq.MeanValueNonEntState}
\end{equation}%
where 
\begin{equation}
\left\langle \widehat{\Omega }_{K}\right\rangle _{R}^{K}=Tr(\widehat{\Omega }%
_{K}\,\widehat{\rho }_{R}^{K}),\qquad (K=A,B,..)
\label{Eq.SubSysMeanValueSepState}
\end{equation}%
is the mean value for measuring $\widehat{\Omega }_{K}$ in the $K$
sub-system when its density operator is $\widehat{\rho }_{R}^{K}$. Since the
overall mean value is not equal to the product of the separate mean values,
the measurements on the sub-systems are said to be \emph{correlated}.
However, for the general non-entangled state as the mean value is just the
products of mean values weighted by the probability of preparing the
particular product state - which involves a LOCC\ protocal, as we have seen
- the correlation is \emph{classical} rather than \emph{quantum} \cite%
{Barnett09a}. In the case of a single product state where $\widehat{\rho }=%
\widehat{\rho }^{A}\otimes \widehat{\rho }^{B}\otimes \widehat{\rho }%
^{C}\otimes ..$we have $\left\langle \widehat{\Omega }_{A}\otimes \widehat{%
\Omega }_{B}\otimes \widehat{\Omega }_{C}\otimes .\right\rangle
=\left\langle \widehat{\Omega }_{A}\right\rangle ^{A}\,\left\langle \widehat{%
\Omega }_{B}\right\rangle ^{B}\,\left\langle \widehat{\Omega }%
_{C}\right\rangle ^{C}...$which is just the product of mean values for the
separate sub-systems, and in this case the measurements on the sub-systems
are said to be \emph{uncorrelated}. For entangled states however the last
result for $\left\langle \widehat{\Omega }_{A}\otimes \widehat{\Omega }%
_{B}\otimes \widehat{\Omega }_{C}\otimes .\right\rangle $ does not apply,
and the correlation is strictly quantum.

\subsection{Hidden Variable Theory}

\label{SubSection - Hidden Variable Theory}

In a general \emph{local hidden variable theory} physical quantities
associated with the sub-systems are denoted $\Omega _{A}$, $\Omega _{B}$
etc, which are real numbers not operators. Their values are assumed to be $%
\lambda _{i}^{A}$, $\lambda _{j}^{B}$ etc - the same as in quantum theory,
since HVT does not challenge the quantization feature. In the \emph{realist}
viewpoint of HVT all the physical quantities have definite values at any
time, these values being determined from a set of hidden variables $\xi $.
Measurement is \emph{not} required for the values for physical quantities to
be created, as in quantum theory. However, in a so-called "fuzzy" hidden
variable theory \cite{Reid03a} (see also Section 7.1 of \cite{Vedral07a})
the values for $\Omega _{A}$, $\Omega _{B}$ etc are determined
probabilisticly from the hidden variables. For particular hidden variables $%
\xi $ the probability that $\Omega _{A}$ has value $\lambda _{i}^{A}$ will
be given by $P_{A}(i,\xi )$, for particular hidden variables $\xi $ the
probability that $\Omega _{B}$ has value $\lambda _{j}^{B}$ will be given by 
$P_{B}(j,\xi )$, etc and the \emph{HVT joint probability} will be given by 
\begin{equation}
P_{AB..}(i,j,..)=\tint d\xi \,P(\xi )\,P_{A}(i,\xi )P_{B}(j,\xi )...
\label{Eq.JointProbLHVT}
\end{equation}%
Here $P(\xi )d\xi $ is the probability that the hidden variables are in the
range $d\xi $ around $\xi $, the HV being assumed continuous - which is not
a requirement. The probabilities satisfy the usual sum rules for all
outcomes giving one, thus $\sum_{i}P_{A}(i,\xi )=1$, etc., $\int d\xi
\,P(\xi )=1$.

The \emph{formal similarity} of the HVT expression for the joint probability
and that for the case of quantum separable states given in Eq.(\ref%
{Eq.JointProbNonEntState}) is noticable. Although the conceptual basis of
the various factors is quite different, it is always possible to \emph{%
describe} any quantum separable state via a HVT. The different $R$ for the
separable state can be regarded as equivalent to hidden variables $\xi $,
with $P_{R}\Rightarrow P(\xi )$ and $\tsum\limits_{R}$ $\Rightarrow \tint
d\xi \,$. The HVT classical probabilities $P_{A}(i,\xi )$, $P_{B}(j,\xi ).$%
would be given by the quantum probabilities $P_{A}^{R}(i)=Tr(\widehat{\Pi }%
_{i}^{A}\,\widehat{\rho }_{R}^{A}),.P_{B}^{R}(j)=Tr(\widehat{\Pi }_{j}^{B}\,%
\widehat{\rho }_{R}^{B}),.$respectively. There is of course no independent
fully developed classical HVT that can predict the $P_{A}(i,\xi )$, $%
P_{B}(j,\xi ).$etc. However, as we will see both the HVT and the quantum
separable state predictions are consistent with Bell Inequalities, and it
requires a quantum entangled state to demonstrate violations. Naturally it
follows that quantum entangled states cannot be described via a HVT.

\subsubsection{HVT- Mean Values and Correlation}

The actual values that would be assigned to the physical quantities $\Omega
_{A}$, $\Omega _{B}$ etc will depend on the hidden variables but can be
taken as the mean values of the possible values $\lambda _{i}^{A}$,$\lambda
_{i}^{A}$ etc. We denote these mean values as $\left\langle \Omega _{A}(\xi
_{A})\right\rangle $, $\left\langle \Omega _{B}(\xi _{B})\right\rangle $ etc
where%
\begin{equation}
\left\langle \Omega _{K}(\xi _{K})\right\rangle =\tsum\limits_{\lambda
_{k}^{K}}\lambda _{k}^{K}\,P_{K}(k,\xi _{K})\qquad (K=A,B,..)
\label{Eq.SubSysMeanValueHVT}
\end{equation}%
These expressions my be compared to Eq.(\ref{Eq.SubSysMeanValueSepState})
for the mean values of physical quantities $\widehat{\Omega }_{A}$, $%
\widehat{\Omega }_{B}$ etc in quantum separable states.

We can then obtain an expression for the mean value in HVT of the physical
quantity $\Omega _{A}\times \Omega _{B}\times \Omega _{C}\times ...$, where $%
\Omega _{A}$, $\Omega _{B}$, etc. are physical quantities for the separate
sub-systems. This is obtained from Eqs.(\ref{Eq.JointProbLHVT}) and (\ref%
{Eq.SubSysMeanValueHVT}) and is given by 
\begin{equation}
\left\langle \Omega _{A}\times \Omega _{B}\times \Omega _{C}\times
.\right\rangle _{HVT}=\tint d\xi \,P(\xi )\,\left\langle \Omega _{A}(\xi
_{A})\right\rangle \,\left\langle \Omega _{B}(\xi _{B})\right\rangle
\,\left\langle \Omega _{C}(\xi _{C})\right\rangle ..  \label{Eq.MeanValueHVT}
\end{equation}%
This may be compared to Eq.(\ref{Eq.MeanValueNonEntState}) for the mean
value of the physical quantity $\widehat{\Omega }_{A}\otimes \widehat{\Omega 
}_{B}\otimes \widehat{\Omega }_{C}\otimes $ in quantum separable states.

\subsubsection{HVT- GHZ State}

The GHZ state \cite{Greenberger03a} is an entangled state of three
sub-systems $A$, $B$ and $C$, each of which is associated with two quantum
states $\left\vert +1\right\rangle $ and $\left\vert -1\right\rangle $. Each
sub-system has three physical quantities, which are Pauli spin operators $%
\widehat{\sigma }_{x}$, $\widehat{\sigma }_{y}$ and $\widehat{\sigma }_{z}$.
The quantum states $\left\vert +1\right\rangle $ and $\left\vert
-1\right\rangle $ are eigenstates of $\widehat{\sigma }_{z}$ with
eigenvalues $+1$ and $-1$ respectively. Note that the eigenvalues of the
other two Pauli spin operators are also $+1$ and $-1$. The GHZ\ state is
defined by%
\begin{equation}
\left\vert \Psi \right\rangle _{GHZ}=(\left\vert +1\right\rangle
_{A}\left\vert +1\right\rangle _{B}\left\vert +1\right\rangle
_{C}+\left\vert -1\right\rangle _{A}\left\vert -1\right\rangle
_{B}\left\vert -1\right\rangle _{C})/\sqrt{2}  \label{Eq.GHZState}
\end{equation}

The GHZ state provides a clear example of an entangled quantum state which
cannot be described via hidden variable theory. In a non-fuzzy version of
HVT each of the nine physical quantities $\sigma _{x}^{A}$, $\sigma _{y}^{A}$%
, $\sigma _{z}^{A}$, $\sigma _{x}^{B}$, $\sigma _{y}^{B}$, $\sigma _{z}^{B}$%
, $\sigma _{x}^{C}$, $\sigma _{y}^{C}$, $\sigma _{z}^{C}$ will be associated
with hidden variables that directly specify the values $+1$ and $-1$ that
each one of these physical quantities may have. We denote these hidden
variables as $M_{\alpha }^{K}$, where $K=A,B,C$ and $\alpha =x,y,z$ and we
have $M_{\alpha }^{K}=+1$ or $-1$. With this direct specification of the
physical values Eq.(\ref{Eq.SubSysMeanValueHVT}) just becomes $\left\langle
\sigma _{\alpha }^{K}(M^{K})\right\rangle =M_{\alpha }^{K}$ and Eq.(\ref%
{Eq.MeanValueHVT}) becomes $\left\langle \sigma _{\alpha }^{A}\times \sigma
_{\beta }^{B}\times \sigma _{\gamma }^{C}.\right\rangle _{HVT}=M_{\alpha
}^{A}$ $M_{\beta }^{B}$ $M_{\gamma }^{C}$ $.$We can then derive a
contradiction with quantum theory regarding the HVT description of the GHZ
state.

Firstly, using the Pauli spin matrices for the $\left\vert +1\right\rangle $
and $\left\vert -1\right\rangle $ basis states 
\begin{equation}
\left[ \widehat{\sigma }_{x}\right] =\left[ 
\begin{tabular}{ll}
$0$ & $1$ \\ 
$1$ & $0$%
\end{tabular}%
\right] \quad \left[ \widehat{\sigma }_{y}\right] =\left[ 
\begin{tabular}{ll}
$0$ & $-i$ \\ 
$i$ & $0$%
\end{tabular}%
\right] \quad \left[ \widehat{\sigma }_{x}\right] =\left[ 
\begin{tabular}{ll}
$1$ & $0$ \\ 
$0$ & $-1$%
\end{tabular}%
\right]  \label{Eq.PauliMatrices}
\end{equation}%
it is straightforward to show that the GHZ state satisfies three eigenvalue
equations 
\begin{eqnarray}
\widehat{\sigma }_{x}^{A}\widehat{\sigma }_{y}^{B}\widehat{\sigma }%
_{y}^{C}\,\left\vert \Psi \right\rangle _{GHZ} &=&(-1)\left\vert \Psi
\right\rangle _{GHZ}  \nonumber \\
\widehat{\sigma }_{y}^{A}\widehat{\sigma }_{x}^{B}\widehat{\sigma }%
_{y}^{C}\,\left\vert \Psi \right\rangle _{GHZ} &=&(-1)\left\vert \Psi
\right\rangle _{GHZ}  \nonumber \\
\widehat{\sigma }_{y}^{A}\widehat{\sigma }_{y}^{B}\widehat{\sigma }%
_{x}^{C}\,\left\vert \Psi \right\rangle _{GHZ} &=&(-1)\left\vert \Psi
\right\rangle _{GHZ}  \label{Eq.GHZEigenV}
\end{eqnarray}%
Hence in HVT the three quantities $\sigma _{x}^{A}\sigma _{y}^{B}\sigma
_{y}^{C}$, $\sigma _{y}^{A}\sigma _{x}^{B}\sigma _{y}^{C}$ and $\sigma
_{y}^{A}\sigma _{y}^{B}\sigma _{x}^{C}$ must all have value $-1$ in the GHZ
state, so that as the values for these quantities are just the products of
the values for each of the factors we get three equations 
\begin{equation}
M_{x}^{A}\,M_{y}^{B}M_{y}^{C}=-1\quad M_{y}^{A}\,M_{x}^{B}M_{y}^{C}=-1\quad
M_{y}^{A}\,M_{y}^{B}M_{x}^{C}=-1  \label{Eq.HVTResultsGHZ}
\end{equation}%
Secondly, if we apply all three operators $\widehat{\sigma }_{x}^{A}\widehat{%
\sigma }_{x}^{B}\widehat{\sigma }_{x}^{C}$to the GHZ state we find another
eigenvalue equation 
\begin{equation}
\widehat{\sigma }_{x}^{A}\widehat{\sigma }_{x}^{B}\widehat{\sigma }%
_{x}^{C}\,\left\vert \Psi \right\rangle _{GHZ}=(+1)\left\vert \Psi
\right\rangle _{GHZ}  \label{Eq.GHZEigenV2}
\end{equation}%
which leads to 
\begin{equation}
M_{x}^{A}\,M_{x}^{B}M_{x}^{C}=+1  \label{Eq.HVTResultsGHZ2}
\end{equation}%
However, if we multiply the three equations in Eq.(\ref{Eq.HVTResultsGHZ})
together and use $(M_{y}^{K})^{2}=+1$ we find that $M_{x}^{A}%
\,M_{x}^{B}M_{x}^{C}=-1$, in direct contradiction to the last equation. Thus
the assignment of hidden variables for all the physical quantities $\sigma
_{\alpha }^{K}$ fails to describe the GHZ state. As we will see in the next
SubSection, there are tests involving the violation of Bell Inequalities
that are satisfied by some entangled states which demonstrate the failure of
more general local HVT to describe such states.

\subsection{Paradoxes}

\label{SubSection - Paradoxes}

The EPR and Schrodinger Cat paradoxes figured prominently in early
discussions about entanglement. Both paradoxes involve \emph{composite
systems }and the consideration of quantum states which are entangled \ Both
these paradoxes reflect the conflict between \emph{quantum theory}, in which
the values for physical quantities only take on definite values when
measurement occurs and \emph{classical theory}, in which the values for
physical quantities always exist even when measurement is not involved. The
latter viewpoint is referred to as \emph{realism}. Quantum theory is also
probabalistic so although the possible outcomes for measuring a physical
quantity can be determined prior to measurement, the actual outcome in a
given quantum state for the measured outcome is only known as a \emph{%
probability}. However, from the realist viewpoint, quantum theory is \emph{%
incomplete} and a \emph{future theory} based around \emph{hidden variables}
would determine the actual values of the physical quantities, as well as the
quantum probabilities that particular values will be found via measurement.

Whilst the EPR and Schrodinger Cat paradoxes are of historical interest and
have provoked much debate, it was the formulation of the \emph{Bell
inequalities} (which are described in the next SubSection \ref{SubSection -
Bell Inequalities}) and the conditions under which they could be violated
that provided the first clear case of where the predictions of quantum
theory could differ from those of hidden variable theories. It then became
possible to carry out actual experiments to distinguish these two
fundamentally different theories. The actual experimental evidence is
consistent with quantum theory and rules out hidden variable theories.

\subsubsection{EPR Paradox}

In the original version of the EPR paradox Einstein et al \cite{Einstein35a}
considered a two-particle system $A$, $B$ in which the particles were
associated with \emph{positions} $\widehat{x}_{A}$, $\widehat{x}_{B}$ and 
\emph{momenta} $\widehat{p}_{A}$, $\widehat{p}_{B}$. They envisaged a
quantum state in which the pairs of physical quantities $\widehat{x}_{A}$, $%
\widehat{x}_{B}$ or $\widehat{p}_{A}$, $\widehat{p}_{B}$ had highly
correlated values - measured or otherwise. To be specific, one may consider
a simultaneous eigenstate of the two commuting operators $\widehat{x}_{A}-%
\widehat{x}_{B}$ and $\widehat{p}_{A}+\widehat{p}_{B}$, where $(\widehat{x}%
_{A}-\widehat{x}_{B})\left\vert \Phi \right\rangle =2x\left\vert \Phi
\right\rangle $ and $(\widehat{p}_{A}+\widehat{p}_{B})\left\vert \Phi
\right\rangle =0\left\vert \Phi \right\rangle $. Thus if $A$ had a mean
momentum $p$ then $B$ would have a mean momentum $-p$. Alternatively, if $A$
had a mean position $x$ then $B$ would have a mean position $-x$. Then if
the eigenvalue $2x$ is very large the two particles will be well-separated
(in quantum theory their spatial wave functions would be localised in
separate spatial regions) so that if the position of $B$ was measured then
the position of $A$ would be immediately known, even if the particles were
light years apart. If the momentum of $B$ was measured instead then the
momentum of $A$ would immediately be known. From the realist point of view
both $A$ and $B$ always have definite positions and momenta, even if these
are not known. It would seem then that measurements of position and momentum
on particle $B$ would lead to a knowledge of the position and momentum at a
far distant particle $A$, perhaps with an accuracy that would violate the
Heisenberg Uncertainty Principle (HUP). Alternatively, the position of $B$
might be measured along with the momentum of $A$, these being commuting
observables. But as we have seen, knowing the position of $B$ leads to a
knowledge of the position at far-distant particle $A$, so the outcome is
that both the position and momentum of $A$ are known, again with the
possibility of violaing the HUP. Thus a somewhat paradoxical situation would
seem to arise. Einstein stated that this did not demonstrate that quantum
theory was wrong, only that it was incomplete.

Discussions of the \emph{EPR paradox} \cite{Einstein35a} in terms of hidden
variable theories has been given by numerous authors (see \cite{Barnett09a}, 
\cite{Vedral07a}, \cite{Reid03a}, \cite{Reid09a} for example). The recent
papers and reviews by Reid et al give a full account taking into
consideration the "fuzzy" version of HVT and determining the predictions for
the conditional variances for $x_{A}$ and $p_{A}$ based both on separable
quantum states and states described via HVT. This treatment successfully
quantifies the somewhat qualitative considerations described in the previous
paragraph. If the position for particle $B$ is measured and the result is $x$%
, then the original density operator $\widehat{\rho }$ for the two particle
system is changed into the conditional density operator $\widehat{\rho }%
_{cond}(\widehat{x}_{B},x)=\widehat{\Pi }_{x}^{B}\,\widehat{\rho }\,\widehat{%
\Pi }_{x}^{B}/Tr(\widehat{\Pi }_{x}^{B}\widehat{\rho })$, where $\widehat{%
\Pi }_{x}^{B}=(\left\vert x\right\rangle \left\langle x\right\vert )_{B}$ is
the projector onto the eigenvector $\left\vert x\right\rangle _{B}$ (the
eigenvalues $x$ are assumed for simplicity to form a quasi-continuum).
Similarly, if the momentum for particle $B$ is measured and the result is $p$%
, then the original density operator $\widehat{\rho }$ for the two particle
system is changed into the conditional density operator $\widehat{\rho }%
_{cond}(\widehat{p}_{B},p)=\widehat{\Pi }_{p}^{B}\,\widehat{\rho }\,\widehat{%
\Pi }_{p}^{B}/Tr(\widehat{\Pi }_{p}^{B}\widehat{\rho })$, where $\widehat{%
\Pi }_{p}^{B}=(\left\vert p\right\rangle \left\langle p\right\vert )_{B}$ is
the projector onto the eigenvector $\left\vert p\right\rangle _{B}$ (the
eigenvalues $p$ are assumed for simplicity to form a quasi-continuum). Here
we outline the discussion based on quantum separable states.

For\emph{\ separable states} the \emph{conditional probability} that
measurement of $\widehat{x}_{A}$ on sub-system $A$ leads to eigenvalue $%
x_{A} $ given that measurement of $\widehat{x}_{B}$ on sub-system $B$ leads
to eigenvalue $x_{B}$ is obtained from Eq.(\ref{Eq.CondProbNonEntangledState}%
) as 
\begin{equation}
P(\widehat{x}_{A},x_{A}|\widehat{x}_{B},x_{B})=\sum_{R}P_{R}\,P_{A}^{R}(%
\widehat{x}_{A},x_{A})P_{B}^{R}(\widehat{x}_{B},x_{B})/\sum_{R}P_{R}%
\,P_{B}^{R}(\widehat{x}_{B},x_{B})
\end{equation}%
where 
\begin{equation}
P_{A}^{R}(\widehat{x}_{A},x_{A})=Tr_{A}(\widehat{\Pi }_{x_{A}}^{A}\widehat{%
\rho }_{R}^{A})\qquad P_{B}^{R}(\widehat{x}_{B},x_{B})=Tr_{B}(\widehat{\Pi }%
_{x_{B}}^{B}\widehat{\rho }_{R}^{B})
\end{equation}%
are the probabilities for position measurements in the separate sub-systems.
The probability that measurement of $\widehat{x}_{B}$ on sub-system $B$
leads to eigenvalue $x_{B}$ is 
\begin{equation}
P(\widehat{x}_{B},x_{B})=\sum_{R}P_{R}\,P_{B}^{R}(\widehat{x}_{B},x_{B})
\end{equation}

The \emph{mean} result for measurement of $\widehat{x}_{A}$ for this \emph{%
conditional} measurement is from Eq.(\ref{Eq.CondMean0}) 
\begin{eqnarray}
\left\langle \widehat{x}_{A}\right\rangle _{\widehat{x}_{B},x_{B}}
&=&\dsum\limits_{x_{A}}x_{A}\,P(\widehat{x}_{A},x_{A}|\widehat{x}_{B},x_{B})
\nonumber \\
&=&\sum_{R}P_{R}\,\left\langle \widehat{x}_{A}\right\rangle _{R}P_{B}^{R}(%
\widehat{x}_{B},x_{B})/P(\widehat{x}_{B},x_{B})
\end{eqnarray}%
where 
\begin{equation}
\left\langle \widehat{x}_{A}\right\rangle
_{R}=\dsum\limits_{x_{A}}x_{A}P_{A}^{R}(\widehat{x}_{A},x_{A})
\end{equation}%
is the \emph{mean} result for measurement of $\widehat{x}_{A}$ when the
sub-system is in state $\widehat{\rho }_{R}^{A}$.

The \emph{conditional variance} for measurement of $\widehat{x}_{A}$ for the
conditional measurement of $\widehat{x}_{B}$ on sub-system $B$ which led to
eigenvalue $x_{B}$ is from Eq.(\ref{Eq.CondVariance0}) 
\begin{eqnarray}
\left\langle \Delta \widehat{x}_{A}^{2}\right\rangle _{\widehat{x}%
_{B},x_{B}} &=&\dsum\limits_{x_{A}}(x_{A}-\left\langle \widehat{x}%
_{A}\right\rangle _{\widehat{x}_{B},x_{B}})^{2}\,P(\widehat{x}_{A},x_{A}|%
\widehat{x}_{B},x_{B})  \nonumber \\
&=&\sum_{R}P_{R}\,\left\langle \Delta \widehat{x}_{A}^{2}\right\rangle _{%
\widehat{x}_{B},x_{B}}^{R}P_{B}^{R}(\widehat{x}_{B},x_{B})/P(\widehat{x}%
_{B},x_{B})
\end{eqnarray}%
where 
\[
\left\langle \Delta \widehat{x}_{A}^{2}\right\rangle _{\widehat{x}%
_{B},x_{B}}^{R}=\dsum\limits_{x_{A}}(x_{A}-\left\langle \widehat{x}%
_{A}\right\rangle _{\widehat{x}_{B},x_{B}})^{2}\,P_{A}^{R}(\widehat{x}%
_{A},x_{A}) 
\]%
is a variance for measurement of $\widehat{x}_{A}$ for when the sub-system
is in state $\widehat{\rho }_{R}^{A}$ but now with the fluctuation about the
mean $\left\langle \widehat{x}_{A}\right\rangle _{\widehat{x}_{B},x_{B}}$
for measurements conditional on measuring $\widehat{x}_{B}$.

However, for each sub-system state $R$ the quantity $\left\langle \Delta 
\widehat{x}_{A}^{2}\right\rangle _{\widehat{x}_{B},x_{B}}^{R}$ is \emph{%
minimised} if $\left\langle \widehat{x}_{A}\right\rangle _{\widehat{x}%
_{B},x_{B}}$ is replaced by the unconditioned mean $\left\langle \widehat{x}%
_{A}\right\rangle _{R}$ just determined from $\widehat{\rho }_{R}^{A}$. Thus
we have an inequality%
\begin{equation}
\left\langle \Delta \widehat{x}_{A}^{2}\right\rangle _{\widehat{x}%
_{B},x_{B}}^{R}\geq \left\langle \Delta \widehat{x}_{A}^{2}\right\rangle ^{R}
\label{Eq.VarianceIneqEPR}
\end{equation}%
where 
\begin{equation}
\left\langle \Delta \widehat{x}_{A}^{2}\right\rangle
^{R}=\dsum\limits_{x_{A}}(x_{A}-\left\langle \widehat{x}_{A}\right\rangle
)^{2}\,P_{A}^{R}(\widehat{x}_{A},x_{A})
\end{equation}%
is the \emph{normal variance} for measurement of $\widehat{x}_{A}$ for when
the sub-system is in state $\widehat{\rho }_{R}^{A}$.

Now if the measurements of $\widehat{x}_{B}$ are \emph{unrecorded} then the 
\emph{conditioned variance} is 
\begin{eqnarray}
\left\langle \Delta \widehat{x}_{A}^{2}\right\rangle _{\widehat{x}_{B}}
&=&\tsum\limits_{x_{B}}\left\langle \Delta \widehat{x}_{A}^{2}\right\rangle
_{\widehat{x}_{B},x_{B}}P(\widehat{x}_{B},x_{B})  \nonumber \\
&=&\tsum\limits_{x_{B}}\sum_{R}P_{R}\,\left\langle \Delta \widehat{x}%
_{A}^{2}\right\rangle _{\widehat{x}_{B},x_{B}}^{R}P_{B}^{R}(\widehat{x}%
_{B},x_{B})
\end{eqnarray}%
which in view of inequality (\ref{Eq.VarianceIneqEPR}) satisfies 
\begin{eqnarray}
\left\langle \Delta \widehat{x}_{A}^{2}\right\rangle _{\widehat{x}_{B}}
&\geq &\tsum\limits_{x_{B}}\sum_{R}P_{R}\,\left\langle \Delta \widehat{x}%
_{A}^{2}\right\rangle ^{R}\,P_{B}^{R}(\widehat{x}_{B},x_{B})  \nonumber \\
&=&\sum_{R}P_{R}\,\left\langle \Delta \widehat{x}_{A}^{2}\right\rangle ^{R}
\end{eqnarray}%
using $\tsum\limits_{x_{B}}P_{B}^{R}(\widehat{x}_{B},x_{B})=1$. Thus the
variance for measurement of position $\widehat{x}_{A}$ conditioned on
unrecorded measurements for position $\widehat{x}_{B}$ satisfies an
inequality that only depends on the variances for measurements of $\widehat{x%
}_{A}$ in the possible sub-system $A$ states $\widehat{\rho }_{R}^{A}$.

Now exactly the same treatment can be carried out for the variance of \emph{%
momentum} $\widehat{p}_{A}$ conditioned on unrecorded measurements of
measurements for momentum $\widehat{p}_{B}$. We have with 
\begin{eqnarray*}
\left\langle \Delta \widehat{p}_{A}^{2}\right\rangle _{\widehat{p}_{B}}
&=&\tsum\limits_{p_{B}}\left\langle \Delta \widehat{p}_{A}^{2}\right\rangle
_{\widehat{p}_{B},p_{B}}P(\widehat{p}_{B},p_{B}) \\
\left\langle \Delta \widehat{p}_{A}^{2}\right\rangle _{\widehat{p}%
_{B},p_{B}} &=&\dsum\limits_{p_{A}}(p_{A}-\left\langle \widehat{p}%
_{A}\right\rangle _{\widehat{p}_{B},p_{B}})^{2}\,P(\widehat{p}_{A},p_{A}|%
\widehat{p}_{B},p_{B}) \\
\left\langle \widehat{p}_{A}\right\rangle _{\widehat{p}_{B},p_{B}}
&=&\dsum\limits_{p_{A}}p_{A}\,P(\widehat{p}_{A},p_{A}|\widehat{p}_{B},p_{B})
\end{eqnarray*}%
the inequality 
\begin{equation}
\left\langle \Delta \widehat{p}_{A}^{2}\right\rangle _{\widehat{p}_{B}}\geq
\sum_{R}P_{R}\,\left\langle \Delta \widehat{p}_{A}^{2}\right\rangle ^{R}
\label{Eq.InequaltiyEPR2}
\end{equation}%
with 
\begin{equation}
\left\langle \Delta \widehat{p}_{A}^{2}\right\rangle
^{R}=\dsum\limits_{p_{A}}(p_{A}-\left\langle \widehat{p}_{A}\right\rangle
)^{2}\,P_{A}^{R}(\widehat{p}_{A},p_{A})
\end{equation}%
is the normal variance for measurement of $\widehat{p}_{A}$ for when the
sub-system is in state $\widehat{\rho }_{R}^{A}$.

We now multiply the two conditional variances, which it is important to note
were associated with two \emph{different conditioned states} based on two
different measurements - position and momentum - carried out on sub-system $%
B $. 
\begin{equation}
\left\langle \Delta \widehat{x}_{A}^{2}\right\rangle _{\widehat{x}%
_{B}}\left\langle \Delta \widehat{p}_{A}^{2}\right\rangle _{\widehat{p}%
_{B}}\geq \sum_{R}P_{R}\,\left\langle \Delta \widehat{x}_{A}^{2}\right%
\rangle ^{R}\sum_{S}P_{S}\,\left\langle \Delta \widehat{p}%
_{A}^{2}\right\rangle ^{S}
\end{equation}%
However, from the general inequality in Eq.(\ref{Eq.SumInequality0})%
\begin{equation}
\tsum\limits_{R}\,P_{R}\,C_{R}\,\tsum\limits_{R}\,P_{R}\,D_{R}\geq \left(
\tsum\limits_{R}\,P_{R}\,\sqrt{C_{R}D_{R}}\right) ^{2}
\end{equation}%
we then have%
\begin{eqnarray}
\left\langle \Delta \widehat{x}_{A}^{2}\right\rangle _{\widehat{x}%
_{B}}\left\langle \Delta \widehat{p}_{A}^{2}\right\rangle _{\widehat{p}_{B}}
&\geq &\left( \tsum\limits_{R}\,P_{R}\,\sqrt{\left\langle \Delta \widehat{x}%
_{A}^{2}\right\rangle ^{R}\left\langle \Delta \widehat{p}_{A}^{2}\right%
\rangle ^{R}}\right) ^{2}  \nonumber \\
&=&\left( \tsum\limits_{R}\,P_{R}\,\sqrt{\left\langle \Delta \widehat{x}%
_{A}^{2}\right\rangle ^{R}}\times \sqrt{\left\langle \Delta \widehat{p}%
_{A}^{2}\right\rangle ^{R}}\right) ^{2}
\end{eqnarray}%
But we know from the HUP that for \emph{any} given state $\widehat{\rho }%
_{R}^{A}$ that $\left\langle \Delta \widehat{x}_{A}^{2}\right\rangle
^{R}\left\langle \Delta \widehat{p}_{A}^{2}\right\rangle ^{R}\geq \frac{1}{4}%
\hbar ^{2}$, so for the conditioned variances associated with a separable
state 
\begin{equation}
\left\langle \Delta \widehat{x}_{A}^{2}\right\rangle _{\widehat{x}%
_{B}}\left\langle \Delta \widehat{p}_{A}^{2}\right\rangle _{\widehat{p}%
_{B}}\geq \frac{1}{4}\hbar ^{2}  \label{Eq.HUPResultSepStates}
\end{equation}%
showing that for a separable state the conditioned variances involving
position and momentum measurements on sub-system $B$ still satisfy the HUP.
Thus if the EPR violations are to occur then the state must be \emph{%
entangled}.

In \cite{Reid03a} an analogous treatment based on \emph{hidden variable
theory} also shows that the HUP is satisfied for the conditioned variances.
The details of this treatment will not be given here, but the formal
similarity of expressions for conditional probabilities in HVT and for
separable states indicates the steps involved.

An effect related to the EPR paradox is \emph{EPR Steering}. As we have
seen, the measurement of the position for particle $B$ changes the density
operator and consequently the probability distributions for measurements on
particle $A$ will now be determined from the conditional probabilities, such
as $P_{AB}(\widehat{x}_{A},x_{A}|\widehat{x}_{B},x_{B})$ or $P_{AB}(\widehat{%
p}_{A},p_{A}|\widehat{x}_{B},x_{B}).$Thus measurements on $B$ are said to
steer the results for measurements on $A$. Steering will of course only
apply if the measurement results for $\widehat{x}_{B}$ are recorded, and not
discarded. A discussion of EPR Steering (see \cite{Reid09a}) is beyond the
scope of this article.

\subsubsection{Schrodinger Cat Paradox}

The Schrodinger Cat Paradox \cite{Schrodinger35a}, \cite{Rinner08a} relates
to composite systems where one sub-system (the cat) is macroscopic and the
other sub-system is microscopic (the radioactive atom). Schrodinger
envisaged a state in which an alive cat and an undecayed atom existed at an
initial time, and because the decayed atom would be associated with a dead
cat, the system after a time corresponding to the half-life for radioactive
decay would be described in quantum theory via the entangled state 
\begin{equation}
\left\vert \Psi \right\rangle =\frac{1}{\sqrt{2}}(\left\vert e\right\rangle
_{Atom}\left\vert Alive\right\rangle _{Cat}+\left\vert g\right\rangle
_{Atom}\left\vert Dead\right\rangle _{Cat})  \label{Eq.SchrodingerCat}
\end{equation}%
in an obvious notation. The combined system is in an enclosed box, and
opening the box constitutes a measurement on the system. According to
quantum theory if the box was opened at this time there would be a
probability of $1/2$ of finding the atom undecayed and the cat alive, with
the same probability for finding a decayed atom and a dead cat. From the
realist viewpoint the cat should be \emph{either} dead \emph{or} it should
be alive irrespective of whether the box is opened or not, and it is a
paradox that in the quantum theory description of the state prior to
measurement the cat is in some sense \emph{both} dead \emph{and} alive. This
paradox is made worse because the cat is a macroscopic system - how could a
cat be either dead or alive at the same time, it must be one or the other?
From the quantum point of view in which the actual values of physical
quantities only appear when measurement occurs, the Schrodinger cat presents
no paradox. The two possible values signifying the health of the cat are
"alive" and "dead", and these values are found with a probability of $1/2$
when measurement takes place on opening the box, and this would entirely
explain the results if such an experiment were to be performed.

In recent times, experiments based on a \emph{Rydberg atom} in a \emph{%
microwave cavity} \cite{Haroche} involving states such as (\ref%
{Eq.SchrodingerCat}) have been performed showing that entanglement can occur
between macroscopic and microscopic systems, and it is even possible to
prepare states analogous to $\frac{1}{\sqrt{2}}(\left\vert
Alive\right\rangle _{Cat}+\left\vert Dead\right\rangle _{Cat})$ in the
macroscopic system itself. In such experiments the different macroscopic
states are large amplitude coherent states of the cavity mode. Coherent
states are possible for \emph{microwave photons} as they are created from 
\emph{classical currents} wiith \emph{well-defined phases}. A coherent
superposition of an alive and dead cat within the cat sub-system itself can
be created by measurement. The entangled state in (\ref{Eq.SchrodingerCat})
can also be written as%
\begin{eqnarray}
\left\vert \Psi \right\rangle &=&\frac{1}{\sqrt{2}}\{\frac{1}{\sqrt{2}}%
\left( \left\vert e\right\rangle _{Atom}+\left\vert g\right\rangle
_{Atom}\right) \frac{1}{\sqrt{2}}(\left\vert Alive\right\rangle
_{Cat}+\left\vert Dead\right\rangle _{Cat})  \nonumber \\
&&+\frac{1}{\sqrt{2}}\left( \left\vert e\right\rangle _{Atom}-\left\vert
g\right\rangle _{Atom}\right) \frac{1}{\sqrt{2}}(\left\vert
Alive\right\rangle _{Cat}-\left\vert Dead\right\rangle _{Cat})\}
\label{Eq.CatNewBasis}
\end{eqnarray}%
so that measurement on the atom for an observable in which the superposition
states $\frac{1}{\sqrt{2}}\left( \left\vert e\right\rangle _{Atom}\pm
\left\vert g\right\rangle _{Atom}\right) $ are the eigenstates would result
in the cat being in the corresponding.superposition states $\frac{1}{\sqrt{2}%
}(\left\vert Alive\right\rangle _{Cat}\pm \left\vert Dead\right\rangle
_{Cat})$ of an alive and dead cat.

\subsection{Bell Inequalities}

\label{SubSection - Bell Inequalities}

Violations of Bell's Inequalities represent situations where neither hidden
variable theory nor quantum theory based on separable states can account for
the result, and therefore provide a clear case where an entangled quantum
state is involved.

\subsubsection{Hidden Variable Theory Result}

A key feature of entangled states is that they are associated with \emph{%
violations of Bell inequalities} \cite{Bell65a} and hence can exhibit this
particular \emph{non-classical} feature. The Bell inequalities arise in
attempts to restore a \emph{classical} interpretation of quantum thory via
hidden variable treatments, where actual values are assigned to all
measureable quantities - including those which in quantum theory are
associated with non-commuting Hermitian operators. In this case we consider
two different physical quantities $\Omega _{A}$ for sub-system $A$, which
are listed $A_{1}$, $A_{2}$, etc , and two $\Omega _{B}$ for sub-system $B$,
which are listed $B_{1}$, $B_{2}$, etc$.$ The corresponding quantum
Hermitian operators $\widehat{\Omega }_{A}$, $\widehat{\Omega }_{B}$, etc
are $\widehat{A}_{1}$, $\widehat{A}_{2}$ and$,\widehat{B}_{1}$, $\widehat{B}%
_{2}$. The Bell inequalities involve the \emph{mean value} $\left\langle
A_{i}\times B_{j}\right\rangle _{HVT}$ of the product of observables $A_{i}$
and $B_{j}$ for subsystems $A$, $B$ respectively, for which there are two
possible measured values, $+1$ and $-1$. In hidden variable theory.the mean
values $\left\langle A_{i}\times B_{j}\right\rangle _{HVT}$ are given by 
\begin{equation}
\left\langle A_{i}\times B_{j}\right\rangle _{HVT}=\dint d\xi \,P(\xi
)\,\left\langle A_{i}(\xi )\right\rangle \,\left\langle B_{j}(\xi
)\right\rangle  \label{Eq.HiddenVarMean}
\end{equation}%
where $\left\langle A_{i}(\xi )\right\rangle $ and $\left\langle B_{j}(\xi
)\right\rangle $ are the values are assigned to $A_{i}$ and $B_{j}$ when the
hidden variables are $\xi $, and $P(\xi )$ is the hidden variable
probability distribution function. If the corresponding quantum Hermitian
operators are such that their eigenvalues are $+1$ and $-1$ - as in the case
of Pauli spin operators - then the only possible values for $\left\langle
A_{i}(\xi )\right\rangle $ and $\left\langle B_{j}(\xi )\right\rangle $ are $%
+1$ and $-1$, since HVT does not conflict with quantum theory regarding
allowed values for physical quantities. However, hidden variable theory
predicts certain inequalities for the mean values of products of physical
quantities for the two sub-systems.

The form given by Clauser et al \cite{Clauser69a} for \emph{Bell's inequality%
} is 
\begin{equation}
|S|\leq 2  \label{Eq.BellInequality}
\end{equation}
where 
\begin{equation}
S=\left\langle A_{1}\times B_{1}\right\rangle _{HVT}+\left\langle
A_{1}\times B_{2}\right\rangle _{HVT}+\left\langle A_{2}\times
B_{1}\right\rangle _{HVT}-\left\langle A_{2}\times B_{2}\right\rangle _{HVT}
\label{Eq.ClassicalBellQuantity}
\end{equation}%
The minus sign can actually be attached to any one of the four terms.

Following the proof of the Bell inequalities in \cite{Barnett09a} we have%
\begin{eqnarray}
\left\langle A_{2}\times B_{1}\right\rangle _{HVT}-\left\langle A_{2}\times
B_{2}\right\rangle _{HVT} &=&\dint d\xi \,P(\xi )\,(\left\langle A_{2}(\xi
)\right\rangle \,\left\langle B_{1}(\xi )\right\rangle -\left\langle
A_{2}(\xi )\right\rangle \,\left\langle B_{2}(\xi )\right\rangle )  \nonumber
\\
&=&\dint d\xi \,P(\xi )\,(\left\langle A_{2}(\xi )\right\rangle
\,\left\langle B_{1}(\xi )\right\rangle (1\pm \left\langle A_{1}(\xi
)\right\rangle \,\left\langle B_{2}(\xi )\right\rangle )  \nonumber \\
&&-\dint d\xi \,P(\xi )\,(\left\langle A_{2}(\xi )\right\rangle
\,\left\langle B_{2}(\xi )\right\rangle (1\pm \left\langle A_{1}(\xi
)\right\rangle \,\left\langle B_{1}(\xi )\right\rangle )  \nonumber \\
&&
\end{eqnarray}%
Now all the quantities $\left\langle A_{i}(\xi )\right\rangle $, $%
\left\langle B_{j}(\xi )\right\rangle $ are either $+1$ or $-1$, so the
expressions $(1\pm \left\langle A_{1}(\xi )\right\rangle \,\left\langle
B_{2}(\xi )\right\rangle )$ and $(1\pm \left\langle A_{1}(\xi )\right\rangle
\,\left\langle B_{1}(\xi )\right\rangle )$ are never negative. Taking the
modulus of the left side leads to an equality%
\begin{eqnarray}
&&\left\vert \left\langle A_{2}\times B_{1}\right\rangle _{HVT}-\left\langle
A_{2}\times B_{2}\right\rangle _{HVT}\right\vert  \nonumber \\
&\leq &\dint d\xi \,P(\xi )\,(\left\vert \left\langle A_{2}(\xi
)\right\rangle \right\vert \,\left\vert \left\langle B_{1}(\xi
)\right\rangle \right\vert (1\pm \left\langle A_{1}(\xi )\right\rangle
\,\left\langle B_{2}(\xi )\right\rangle )  \nonumber \\
&&+\dint d\xi \,P(\xi )\,(\left\vert \left\langle A_{2}(\xi )\right\rangle
\right\vert \,\left\vert \left\langle B_{2}(\xi )\right\rangle \right\vert
(1\pm \left\langle A_{1}(\xi )\right\rangle \,\left\langle B_{1}(\xi
)\right\rangle )  \nonumber \\
&=&\dint d\xi \,P(\xi )\,(1\pm \left\langle A_{1}(\xi )\right\rangle
\,\left\langle B_{2}(\xi )\right\rangle )+\dint d\xi \,P(\xi )\,(1\pm
\left\langle A_{1}(\xi )\right\rangle \,\left\langle B_{1}(\xi
)\right\rangle )  \nonumber \\
&=&2\pm (\dint d\xi \,P(\xi )\,\left\langle A_{1}(\xi )\right\rangle
\,\left\langle B_{2}(\xi )\right\rangle +\dint d\xi \,P(\xi )\,\left\langle
A_{1}(\xi )\right\rangle \,\left\langle B_{1}(\xi )\right\rangle )  \nonumber
\\
&=&2\pm (\left\langle A_{1}\times B_{2}\right\rangle _{HVT}+\left\langle
A_{1}\times B_{1}\right\rangle _{HVT})
\end{eqnarray}%
Hence since $\left\vert \left\langle A_{1}\times B_{2}\right\rangle
_{HVT}+\left\langle A_{1}\times B_{1}\right\rangle _{HVT}\right\vert
=+(\left\langle A_{1}\times B_{2}\right\rangle _{HVT}+\left\langle
A_{1}\times B_{1}\right\rangle _{HVT})$ or $-(\left\langle A_{1}\times
B_{2}\right\rangle _{HVT}+\left\langle A_{1}\times B_{1}\right\rangle
_{HVT}) $ we have%
\begin{equation}
\left\vert \left\langle A_{2}\times B_{1}\right\rangle _{HVT}-\left\langle
A_{2}\times B_{2}\right\rangle _{HVT}\right\vert \pm \left\vert \left\langle
A_{1}\times B_{2}\right\rangle _{HVT}+\left\langle A_{1}\times
B_{1}\right\rangle _{HVT}\right\vert \leq 2  \label{Eq.BellIneq0}
\end{equation}%
But since $\left\vert X-Y\right\vert \leq \left\vert X\right\vert
+\left\vert Y\right\vert $ we see that from the $+$ version of the last
inequality that 
\begin{equation}
\left\vert \left\langle A_{2}\times B_{1}\right\rangle _{HVT}-\left\langle
A_{2}\times B_{2}\right\rangle _{HVT}+\left\langle A_{1}\times
B_{2}\right\rangle _{HVT}+\left\langle A_{1}\times B_{1}\right\rangle
_{HVT}\right\vert \leq 2  \label{Eq.BellInequality1}
\end{equation}%
This is a Bell inequality. Interchanging $A_{2}\leftrightarrow A_{1}$ and
repeating the derivation gives $\left\vert \left\langle A_{1}\times
B_{1}\right\rangle _{HVT}-\left\langle A_{1}\times B_{2}\right\rangle
_{HVT}+\left\langle A_{2}\times B_{2}\right\rangle _{HVT}+\left\langle
A_{2}\times B_{1}\right\rangle _{HVT}\right\vert \leq 2$, which is another
Bell inequality. Interchanging $B_{1}\leftrightarrow B_{2}$ and repeating
the derivation gives $\left\vert \left\langle A_{2}\times B_{2}\right\rangle
_{HVT}-\left\langle A_{2}\times B_{1}\right\rangle _{HVT}+\left\langle
A_{1}\times B_{1}\right\rangle _{HVT}+\left\langle A_{1}\times
B_{2}\right\rangle _{HVT}\right\vert \leq 2$, and interchanging $%
A_{2}\leftrightarrow A_{1}$ and $B_{1}\leftrightarrow B_{2}$ and repeating
the derivation gives $\left\vert \left\langle A_{1}\times B_{2}\right\rangle
_{HVT}-\left\langle A_{1}\times B_{1}\right\rangle _{HVT}+\left\langle
A_{2}\times B_{1}\right\rangle _{HVT}+\left\langle A_{2}\times
B_{2}\right\rangle _{HVT}\right\vert \leq 2$. Thus the minus sign can be
attached to any one of the four terms.

\subsubsection{Non-Entangled State Result}

It can be shown that the Bell inequalities \emph{also} \emph{always} occur
for \emph{non-entangled} states (see Section 7.3 of the book by Vedral \cite%
{Vedral07a}). For Bell's inequalities we consider Hermitian operators $%
\widehat{A}_{i}$ and $\widehat{B}_{j}$ for subsystems $A$, $B$ respectively,
for which there are two eigenvalues $+1$ and $-1$, where examples of the
operators are given by the components $\widehat{A}_{i}=a_{i}\cdot \widehat{%
\sigma }_{A}$ and $\widehat{B}_{j}=b_{j}\cdot \widehat{\sigma }_{B}$ of
Pauli spin operators $\widehat{\sigma }_{A}$ and $\widehat{\sigma }_{B}$
along directions with unit vectors $a_{i}$ and $b_{j}$. The corresponding
quantum theory quantity for the Bell inequality is%
\begin{equation}
S=E(\widehat{A}_{1}\otimes \widehat{B}_{1})+E(\widehat{A}_{1}\otimes 
\widehat{B}_{2})+E(\widehat{A}_{2}\otimes \widehat{B}_{1})-E(\widehat{A}%
_{2}\otimes \widehat{B}_{2})  \label{Eq.BellQuantity}
\end{equation}%
where in quantum theory the mean value is given by $E(\widehat{A}_{i}\otimes 
\widehat{B}_{j})=\left\langle \widehat{A}_{i}\otimes \widehat{B}%
_{j}\right\rangle =Tr(\widehat{\rho }\,\widehat{A}_{i}\otimes \widehat{B}%
_{j})$. For the general bipartite non-entangled state given by \ref%
{Eq.NonEntangledState} it is easy to show that%
\begin{equation}
S=\sum_{R}P_{R}\,\left( \left\langle \widehat{A}_{1}\right\rangle
_{R}^{A}\left\langle \widehat{B}_{1}+\widehat{B}_{2}\right\rangle
_{R}^{B}+\left\langle \widehat{A}_{2}\right\rangle _{R}^{A}\left\langle 
\widehat{B}_{1}-\widehat{B}_{2}\right\rangle _{R}^{B}\right)
\label{Eq.BellQNonEntState}
\end{equation}%
where $\left\langle \widehat{A}_{i}\right\rangle _{R}^{A}=Tr(\widehat{A}%
_{i}\,\widehat{\rho }_{R}^{A})$ and $\left\langle \widehat{B}%
_{j}\right\rangle _{R}^{B}=Tr(\widehat{B}_{j}\,\widehat{\rho }_{R}^{B})$ are
the expectation values of $\widehat{A}_{i}$ and $\widehat{B}_{j}$ for the
sub-systems $A$, $B$ in states $\widehat{\rho }_{R}^{A}$ and $\widehat{\rho }%
_{R}^{B}$ respectively. Now $\left\langle \widehat{A}_{i}\right\rangle
_{R}^{A}$ and $\left\langle \widehat{B}_{j}\right\rangle _{R}^{B}$ must lie
in the range $-1$ to $+1$, so that $\left\langle \widehat{B}_{1}\pm \widehat{%
B}_{2}\right\rangle _{R}^{B}$ must each lie in the range $-2$ to $+2$. Hence%
\begin{eqnarray}
|S| &\leq &\sum_{R}P_{R}\,\left( |\left\langle \widehat{A}_{1}\right\rangle
_{R}^{A}|\,|\left\langle \widehat{B}_{1}+\widehat{B}_{2}\right\rangle
_{R}^{B}|+|\left\langle \widehat{A}_{2}\right\rangle
_{R}^{A}|\,|\left\langle \widehat{B}_{1}-\widehat{B}_{2}\right\rangle
_{R}^{B}|\right)  \nonumber \\
&\leq &\sum_{R}P_{R}\,\left( |\left\langle \widehat{B}_{1}+\widehat{B}%
_{2}\right\rangle _{R}^{B}|\,+\,|\left\langle \widehat{B}_{1}-\widehat{B}%
_{2}\right\rangle _{R}^{B}|\right)  \nonumber \\
&\leq &2  \label{Eq.BellInequalityNonEntState}
\end{eqnarray}%
since to obtain $|\left\langle \widehat{B}_{1}+\widehat{B}_{2}\right\rangle
_{R}^{B}|=2$ requires $\left\langle \widehat{B}_{1}\right\rangle
_{R}^{B}=\left\langle \widehat{B}_{2}\right\rangle _{R}^{B}=\pm 1$ and then $%
|\left\langle \widehat{B}_{1}-\widehat{B}_{2}\right\rangle
_{R}^{B}|=|\left\langle \widehat{B}_{1}\right\rangle _{R}^{B}-\left\langle 
\widehat{B}_{2}\right\rangle _{R}^{B}|=0$, or to obtain $|\left\langle 
\widehat{B}_{1}-\widehat{B}_{2}\right\rangle _{R}^{B}|=2$ requires $%
\left\langle \widehat{B}_{1}\right\rangle _{R}^{B}=-\left\langle \widehat{B}%
_{2}\right\rangle _{R}^{B}=\pm 1$ and then $|\left\langle \widehat{B}_{1}+%
\widehat{B}_{2}\right\rangle _{R}^{B}|=|\left\langle \widehat{B}%
_{1}\right\rangle _{R}^{B}+\left\langle \widehat{B}_{2}\right\rangle
_{R}^{B}|=0$.

\subsubsection{Bell Inequality Violation and Entanglement}

It follows that for a general two mode non-entangled state $|S|$ cannot
violate the Bell inequality limit of $2$. Thus, the violation of Bell
inequalities proves that the quantum state must be entangled for the
sub-systems involved, so Bell inequality violations are a test of
entanglement. For entangled states such as the one boson Bell state $%
\left\vert \Psi _{-}\right\rangle $ (see \cite{Barnett09a}, Section 2.5) 
\begin{equation}
\left\vert \Psi _{-}\right\rangle =\frac{1}{\sqrt{2}}(\left\vert
0\right\rangle _{A}\otimes \left\vert 1\right\rangle _{B}-\left\vert
1\right\rangle _{A}\otimes \left\vert 0\right\rangle _{B})
\label{Eq.BellState}
\end{equation}%
the Bell inequality can be violated for the choice where $a_{1}$, $a_{2}$
and $b_{1}$, $b_{2}$ are orthogonal and $a_{1}$, $a_{2}$ are parallel to $%
b_{1}-b_{2}$, $b_{1}+b_{2}$ respectively (see \cite{Barnett09a}, Section
5.1). Furthermore, such a quantum state cannot be described via a hidden
variable theory, since Bell inequalities are always satisfied using a hidden
variable theory. Experiments have been carried out in optical systems
providing strong evidence for the existence of quantum states that violate
Bell inequalities (see \cite{Horodecki09a} for references to experiments).
Such violation of Bell inequalities is clearly a \emph{non-classical}
feature, since the experiments rule out hidden variable theory. As Bell
inequalities do not occur for separable states, the experimental observation
of a Bell inequality indicates the presence of an entangled state.

\subsection{Non-local Correlations}

\label{SubSection - Non-local Correlations}

Another feature of entangled states is that they are associated with \emph{%
strong correlations} for \emph{observables} associated with \emph{localised
sub-systems} that are \emph{well-separated}, a particular example being 
\emph{EPR} \emph{correlations} between non-commuting observables. Entangled
states can exhibit this particular \emph{non-classical} feature, which again
cannot be accounted for via a hidden variable theory.

\subsubsection{Hidden Variable Theory}

Consider two operators $\widehat{\Omega }_{A}$ and $\widehat{\Omega }_{B}$
associated with sub-systems $A$ and $B$. These would be Hermitian if
observables are involved, but for generality this is not required. In a
hidden variable theory these would be associated with functions $\Omega
_{C}(\xi )$ $(C=A,B)$ of the hidden variables $\xi $, with the Hermitean
adjoints $\widehat{\Omega }_{C}^{\dag }$ being associated with the complex
conjugates $\Omega _{C}^{\ast }(\xi )$. In hidden variable theory \emph{%
correlation functions} are given by the following mean values 
\begin{eqnarray}
\left\langle \Omega _{A}^{\ast }\times \Omega _{B}\right\rangle _{HVT}
&=&\dint d\xi \,P(\xi )\,\Omega _{A}^{\ast }(\xi )\Omega _{B}(\xi ) 
\nonumber \\
\left\langle \Omega _{A}^{\ast }\Omega _{A}\times \Omega _{B}^{\ast }\Omega
_{B}\right\rangle _{HVT} &=&\dint d\xi \,P(\xi )\,\Omega _{A}^{\ast }(\xi
)\Omega _{A}(\xi )\,\,\Omega _{B}^{\ast }(\xi )\Omega _{B}(\xi )
\label{Eq.CorrelnFunctions}
\end{eqnarray}%
satisfy the following \emph{correlation inequality}%
\begin{equation}
|\,\left\langle \Omega _{A}^{\ast }\times \Omega _{B}\right\rangle
_{HVT}\,|^{2}\leq \left\langle \Omega _{A}^{\ast }\Omega _{A}\times \Omega
_{B}^{\ast }\Omega _{B}\right\rangle _{HVT}  \label{Eq.CorrelnInequal}
\end{equation}

This result is based on the inequality%
\begin{equation}
\tint d\xi \,P(\xi )C(\xi )\geq \left( \tint d\xi \,P(\xi )\sqrt{C(\xi )}%
\right) ^{2}  \label{Eq.IntegralInequalityB}
\end{equation}%
for real, positive functions $C(\xi ),P(\xi )$ and where $\dint d\xi \,P(\xi
)=1$, and which is proved in Appendix \ref{Appendix - Inequalities}. In the
present case we have $C(\xi )=\Omega _{A}^{\ast }(\xi )\Omega _{A}(\xi
)\,\,\Omega _{B}^{\ast }(\xi )\Omega _{B}(\xi )$, which is real, positive. A
violation of the inequality in Eq. (\ref{Eq.CorrelnInequal}) is an
indication of strong correlation between sub-systems $A$ and $B$.

\subsubsection{Non-Entangled State Result}

It can be shown that the correlation inequalities are \emph{always}
satisfied for non-entangled states. In quantum theory the correlation
functions are given by $\left\langle \widehat{\Omega }_{A}^{\dag }\otimes 
\widehat{\Omega }_{B}\right\rangle =Tr(\widehat{\rho }\,\widehat{\Omega }%
_{A}^{\dag }\otimes \widehat{\Omega }_{B})$ and $\left\langle \widehat{%
\Omega }_{A}^{\dag }\widehat{\Omega }_{A}\otimes \widehat{\Omega }_{B}^{\dag
}\widehat{\Omega }_{B}\right\rangle =Tr(\widehat{\rho }\,\widehat{\Omega }%
_{A}^{\dag }\widehat{\Omega }_{B}\otimes \widehat{\Omega }_{B}^{\dag }%
\widehat{\Omega }_{B})$. For a non-entangled state of sub-systems $A$ and $B$
we have%
\begin{eqnarray}
\left\langle \widehat{\Omega }_{A}^{\dag }\otimes \widehat{\Omega }%
_{B}\right\rangle &=&\sum_{R}P_{R}\,\left\langle \widehat{\Omega }_{A}^{\dag
}\right\rangle _{R}^{A}\left\langle \widehat{\Omega }_{B}\right\rangle
_{R}^{B}  \nonumber \\
\left\langle \widehat{\Omega }_{A}^{\dag }\widehat{\Omega }_{A}\otimes 
\widehat{\Omega }_{B}^{\dag }\widehat{\Omega }_{B}\right\rangle
&=&\sum_{R}P_{R}\,\left\langle \widehat{\Omega }_{A}^{\dag }\widehat{\Omega }%
_{A}\right\rangle _{R}^{A}\left\langle \widehat{\Omega }_{B}^{\dag }\widehat{%
\Omega }_{B}\right\rangle _{R}^{B}  \label{Eq.QuantumCorrelFns}
\end{eqnarray}

Now 
\begin{equation}
|\left\langle \widehat{\Omega }_{A}^{\dag }\otimes \widehat{\Omega }%
_{B}\right\rangle |\leq \sum_{R}P_{R}\,|\left\langle \widehat{\Omega }%
_{A}^{\dag }\right\rangle _{R}^{A}|\;|\left\langle \widehat{\Omega }%
_{B}\right\rangle _{R}^{B}|  \label{Eq.ResultE1}
\end{equation}%
since the modulus of a sum is always less than the sum of the moduli. Using $%
\left\langle \left( \widehat{\Omega }_{C}^{\dag }-\left\langle \widehat{%
\Omega }_{C}^{\dag }\right\rangle \right) \left( \widehat{\Omega }%
_{C}-\left\langle \widehat{\Omega }_{C}\right\rangle \right) \right\rangle
\geq 0$ with $(C=A,B)$, we obtain the Schwarz inequality - which is true for
all states - $\left\langle \widehat{\Omega }_{C}^{\dag }\widehat{\Omega }%
_{C}\right\rangle \geq \left\langle \widehat{\Omega }_{C}^{\dag
}\right\rangle \left\langle \widehat{\Omega }_{C}\right\rangle
=|\left\langle \widehat{\Omega }_{C}\right\rangle |^{2}=|\left\langle 
\widehat{\Omega }_{C}^{\dag }\right\rangle |^{2}$, and hence 
\begin{equation}
|\left\langle \widehat{\Omega }_{A}^{\dag }\otimes \widehat{\Omega }%
_{B}\right\rangle |\leq \sum_{R}P_{R}\,\sqrt{\left\langle \widehat{\Omega }%
_{A}^{\dag }\widehat{\Omega }_{A}\right\rangle _{R}^{A}}\;\sqrt{\left\langle 
\widehat{\Omega }_{B}^{\dag }\widehat{\Omega }_{B}\right\rangle _{R}^{B}}
\label{Eq.ResultE2}
\end{equation}

Next we use the inequality 
\begin{equation}
\tsum\limits_{R}\,P_{R}\,C_{R}\geq \left( \tsum\limits_{R}\,P_{R}\,\sqrt{%
C_{R}}\right) ^{2}  \label{Eq.SumInequalityB}
\end{equation}%
for real, positive functions $C_{R},P_{R}$ and where $\sum_{R}P_{R}=1$. This
inequality, which was used in the paper by Hillery et al \cite{Hillery06a},
is proved in Appendix \ref{Appendix - Inequalities}. In the present case we
have $C_{R}=\left\langle \widehat{\Omega }_{A}^{\dag }\widehat{\Omega }%
_{A}\right\rangle _{R}^{A}\left\langle \widehat{\Omega }_{B}^{\dag }\widehat{%
\Omega }_{B}\right\rangle _{R}^{B}$ so that 
\begin{equation}
|\left\langle \widehat{\Omega }_{A}^{\dag }\otimes \widehat{\Omega }%
_{B}\right\rangle |^{2}\leq \sum_{R}P_{R}\,\left\langle \widehat{\Omega }%
_{A}^{\dag }\widehat{\Omega }_{A}\right\rangle _{R}^{A}\left\langle \widehat{%
\Omega }_{B}^{\dag }\widehat{\Omega }_{B}\right\rangle _{R}^{B}=\left\langle 
\widehat{\Omega }_{A}^{\dag }\widehat{\Omega }_{A}\otimes \widehat{\Omega }%
_{B}^{\dag }\widehat{\Omega }_{B}\right\rangle  \label{Eq.ResultE3}
\end{equation}

Thus for a \emph{non-entangled state} we obtain the \emph{correlation
inequality}%
\begin{equation}
|\left\langle \widehat{\Omega }_{A}^{\dag }\otimes \widehat{\Omega }%
_{B}\right\rangle |^{2}=|\left\langle \widehat{\Omega }_{A}\otimes \widehat{%
\Omega }_{B}^{\dag }\right\rangle |^{2}\leq \left\langle \widehat{\Omega }%
_{A}^{\dag }\widehat{\Omega }_{A}\otimes \widehat{\Omega }_{B}^{\dag }%
\widehat{\Omega }_{B}\right\rangle  \label{Eq.QuantumCorrenInequality}
\end{equation}%
where the general result $\left\langle \widehat{\Omega }_{A}^{\dag }\otimes 
\widehat{\Omega }_{B}\right\rangle =\left\langle \widehat{\Omega }%
_{A}\otimes \widehat{\Omega }_{B}^{\dag }\right\rangle ^{\ast }$ has been
used. Thus non-entangled states have correlation functions that are
consistent with hidden variable theory.

\subsubsection{Weak Correlation Violation and Entanglement}

Hence if it is found that the correlation inequality is violated $%
|\left\langle \widehat{\Omega }_{A}^{\dag }\otimes \widehat{\Omega }%
_{B}\right\rangle |^{2}=|\left\langle \widehat{\Omega }_{A}\otimes \widehat{%
\Omega }_{B}^{\dag }\right\rangle |^{2}>\left\langle \widehat{\Omega }%
_{A}^{\dag }\widehat{\Omega }_{A}\otimes \widehat{\Omega }_{B}^{\dag }%
\widehat{\Omega }_{B}\right\rangle $ then the state must be entangled, so
the correlation inequality violation is also a test for entanglement. Again
entangled states have features that cannot be explained via hidden variable
theory, so entangled states are clearly non-classical. ..

\section{Identical Particles and Entanglement}

\label{Section - Identical Particles and Entanglement}

We now take into account the situation where systems of \emph{identical
particles} are involved. This requires us to give special consideration to
the requirement that physical states in such cases must conform to the \emph{%
symmetrisation principle} and that the nature of the possible sub-systems
must take into account that entanglement requires the specification of \emph{%
sub-system}s that are \emph{distinguishable} from each other.

\label{SubSection - Symmetrization Principle}

\subsubsection{Symmetrisation Principle and Second Quantization}

Whether \emph{entangled} or \emph{not} the physical states for systems of 
\emph{identical particles} must conform to the \emph{symmetrisation principle%
}, whereby the overall density operator has to be invariant under \emph{%
permutation operators}. Problems arise regarding how to define non-entangled
states for systems of identical particles. The basic issue is how first to
distinguish what are meaningful \emph{sub-systems} for identical particle
systems. Some authors (\cite{Sorensen01a}, \cite{Pezze09a}.. ) consider
states of the form 
\begin{equation}
\widehat{\rho }=\sum_{R}P_{R}\,\widehat{\rho }_{R}^{1}\otimes \widehat{\rho }%
_{R}^{2}\otimes \widehat{\rho }_{R}^{3}\otimes ...
\label{Eq.NonEntStateIdenticalAtoms}
\end{equation}%
to be non-entangled states, where $\widehat{\rho }_{R}^{i}$ is a density
operator for particle $i$. However such a state would not in general be
physical, since the symmetrisation principle would be violated unless the $%
\widehat{\rho }_{R}^{i}$ were related. For example, consider the state for
two identical bosonic atoms given by 
\begin{equation}
\widehat{\rho }=P_{\sigma \xi }\,\widehat{\sigma }^{1}\otimes \widehat{\xi }%
^{2}+P_{\theta \eta }\,\widehat{\theta }^{1}\otimes \widehat{\eta }^{2}
\label{Eq.NonEntangStateTwoBosons}
\end{equation}%
and apply the permutation $\widehat{P}=\widehat{P}(1\leftrightarrow 2)$. The
invariance of $\widehat{\rho }$ in general requires $\,\widehat{\sigma }=%
\widehat{\xi }$ and $\widehat{\theta }=\widehat{\eta }$, giving $\widehat{%
\rho }=P_{\sigma }\,\widehat{\sigma }^{1}\otimes \widehat{\sigma }%
^{2}+P_{\theta }\,\widehat{\theta }^{1}\otimes \widehat{\theta }^{2}$. This
is a statistical mixture of two states, one with both atoms in state $%
\widehat{\sigma }$, the other with atoms in state $\widehat{\theta }$. Of
course if the atoms were all different (atom $1$ a Rb$^{87}$ atom, atom $2$
a Na$^{23\text{ }}$atom, ..) then the expression (\ref%
{Eq.NonEntangStateTwoBosons}) would be a valid non-entangled state, but
there the atomic sub-systems are distinguishable and symmetrisation is not
required. What is distinguishable for systems of identical bosons is not the
individual particles themselves - which do not carry labels, boson $1$,
boson $2$, etc. - but the \emph{single particle states }or \emph{modes} that
the bosons may occupy. For bosonic atoms with several hyperfine components,
each component will have its own set of modes. The same would apply to
fermionic atoms. For photons the modes may be specified via wave vectors and
polarisations. Although the quantum pure states can be specified via
symmetrized products of single particle states occupied by specific
particles using a \emph{first quantization} approach, it is more convenient
to use \emph{second quantization}. Here, a basis set for the quantum states
of such sub-systems are the \emph{Fock states} $\left\vert
n_{a}\right\rangle $ ($n_{a}=0,1,2,..$) etc, which specify the number of
identical particles occupying the mode $A$, etc., so in this approach the
mode is the sub-system and the Fock states give different physical states
for this sub-system. Symmetrization is built into the definition of the Fock
states. If the atoms were fermions rather than bosons the Pauli exclusion
principle would of course restrict $n_{a}=0,1$ only. Thus in this second
quantization approach situations with differing numbers of identical
particles are different \emph{states}, not different \emph{systems}. The
overall system will be associated with physical states with density
operators and state vectors in Fock space, which includes states with total
numbers of identical particles ranging from zero in the vacuum state right
up to infinity.

\subsubsection{Sub-Systems and Modes}

The point of view in which the possible \emph{sub-systems} \emph{\ }$A$, $B$%
, etc are \emph{modes} (or \emph{sets} of modes) rather than \emph{particles}
has been adopted by several authors (\cite{Simon02a}, \cite{Hines03a}, \cite%
{TerraCunha07a}), \cite{Vedral07a}, \cite{Benatti10a}, \cite{Benatti11a} and
will be the approach used here. What are or are not entangled are \emph{modes%
} not \emph{particles}. Overall, the system is a collection of modes, not
particles. Particles are associated with mode occupancies, and therefore
related to specifying the quantum states of the system, rather than the
system itself. Note that in this approach states where there is only a \emph{%
single atom} may still be entangled states - for example with two spatial
modes $A,B$ the states which are a quantum superposition of the atom in each
of these modes, such as the Bell state $(\left\vert 1_{a}\right\rangle
\left\vert 0_{b}\right\rangle +\left\vert 0_{a}\right\rangle \left\vert
1_{b}\right\rangle )/\sqrt{2}$ are entangled states. For entangled states
associated with the EPR paradox or for quantum teleportation, the mode
functions may be \emph{localised} in well-separated spatial regions - spooky
action at a distance - but spatially overlapping mode functions apply in
other situations. Furthermore, as well as being distinguishable the modes
can act as \emph{separate systems}, with other modes being ignored. For
interacting bosonic atoms this is much harder to accomplish experimentally
than for the case of photons, where the relatively slow processes in which
photons are destroyed in one EM field mode and created in another may
require the presence of atoms as intermediaries. Two bosonic atoms in one
mode may collide rapidly disappear into other modes. However, atomic boson
interactions can be made very small via \emph{Feshbach resonance} methods.
Near absolute zero the basic physics of a BEC in a single trap potential is
describable via a \emph{one mode theory}. Hence with $A$, $B$, .. signifying
distinct modes, the general non-entangled state is given in Eq. (\ref%
{Eq.NonEntangledState}) though the present paper mainly involves only two
modes.

It is useful to clarify the meaning of entanglement used in the present
paper via a simple example. Consider a situation in which there are two
distinct single particle states (modes) designated as $\left\vert
u\right\rangle $ and $\left\vert v\right\rangle $. These states are chosen
to be orthogonal. We consider a system with $N=2$ particles, which may be 
\emph{identical} and are labeled $1$ and $2$, or they may be \emph{%
distinguishable} and labeled $\alpha $ and $\beta $.

For the case of the \emph{identical} particles we consider \emph{pure states}
for two \emph{bosons} and for two \emph{fermions}, which are written in
terms of \emph{first quantization} as 
\begin{eqnarray}
\left\vert \Psi \right\rangle _{boson} &=&\frac{1}{\sqrt{2}}(\left\vert
u(1)\right\rangle \otimes \left\vert v(2)\right\rangle +\left\vert
v(1)\right\rangle \otimes \left\vert u(2)\right\rangle )
\label{Eq.TwoBosonsTwoModes} \\
\left\vert \Psi \right\rangle _{fermion} &=&\frac{1}{\sqrt{2}}(\left\vert
u(1)\right\rangle \otimes \left\vert v(2)\right\rangle -\left\vert
v(1)\right\rangle \otimes \left\vert u(2)\right\rangle )
\label{Eq.TwoFermionsTwoModes}
\end{eqnarray}%
and clearly satisfy the symmetrization principle. The question is: are these
entangled states? They certainly look entangled because both are sums of the
products of two state vectors. In the textbook by Peres (\cite{Peres93a},
see pp126-128) it is stated that "two particles of the same type are always
entangled". Peres obviously considers such entanglement is a result of \emph{%
symmetrization}. However, noting that there is only one typre of particle
involved and there are two modes that can be occupied, in \emph{second
quantization} the state in both the fermion and boson cases is 
\begin{equation}
\left\vert \Psi \right\rangle =\left\vert 1\right\rangle _{u}\otimes
\left\vert 1\right\rangle _{v}  \label{Eq.SecondQnTwoIdentTwoModes}
\end{equation}%
which is a separable state for modes $u,v$, and not a (mode) entangled
state. On the other hand, the boson state $(\left\vert u(1)\right\rangle
\otimes \left\vert u(2)\right\rangle +\left\vert v(1)\right\rangle \otimes
\left\vert v(2)\right\rangle )/\sqrt{2}$ is an entangled state, written in
second quantization as $(\left\vert 2\right\rangle _{u}\otimes \left\vert
0\right\rangle _{v}+\left\vert 0\right\rangle _{u}\otimes \left\vert
2\right\rangle _{v})/\sqrt{2}$. There is no analogous state for fermions due
to the Pauli principle.

Now consider the case where the particles are \emph{distinguishable} \ Pure
states analogous to the previous ones are given in first quantization as 
\begin{equation}
\left\vert \Psi \right\rangle =\frac{1}{\sqrt{2}}(\left\vert u(\alpha
)\right\rangle \otimes \left\vert v(\beta )\right\rangle \pm \left\vert
v(\alpha )\right\rangle \otimes \left\vert u(\beta )\right\rangle )
\label{Eq.TwoDistParticlesTwoModes}
\end{equation}%
which are not required to satisfy the symmetrization principle since the
particles are not identical. Each may be either a boson or a fermion. The
question is: are these entangled states? Noting that for each particle there
are two modes that could be occupied, in second quantization the state would
be written as \ 
\begin{equation}
\left\vert \Psi \right\rangle =\frac{1}{\sqrt{2}}((\left\vert 1\right\rangle
_{u}\otimes \left\vert 0\right\rangle _{v})_{\alpha }\otimes (\left\vert
0\right\rangle _{u}\otimes \left\vert 1\right\rangle _{v})_{\beta }\pm
(\left\vert 0\right\rangle _{u}\otimes \left\vert 1\right\rangle
_{v})_{\alpha }\otimes (\left\vert 1\right\rangle _{u}\otimes \left\vert
0\right\rangle _{v})_{\beta })  \label{Eq.SecondQnTwoDistTwoModes}
\end{equation}%
These are both entangled states.

It is worth noting that these examples illustrate the general point that
just the \emph{mathematical form} of the state vector or the density
operator alone is not enough to determine whether a separable or an
entangled state is involved. The meaning of the factors involved also has to
be taken into account. Failure to realise this may lead to states being
regarded as separable which should not be (see SubSection \ref{SubSection -
SSR Separate Modes} for further examples). In the case just presented
involving a system with two particles, the quantities $\left\vert
u(1)\right\rangle $ or $\left\vert v(2)\right\rangle $ do not specify valid
sub-system states when the particles are identical, so the forms given by (%
\ref{Eq.TwoBosonsTwoModes}) and (\ref{Eq.TwoFermionsTwoModes}) do not
represent entangled states. On the other hand, the quantities $\left\vert
u(\alpha )\right\rangle $ or $\left\vert v(\beta )\right\rangle $ do specify
valid sub-system states when the particles are distinguishable, so the forms
given by (\ref{Eq.TwoDistParticlesTwoModes}) do represent entangled states.

The approach of Wiseman et al \cite{Wiseman03a} to defining an \emph{%
entanglement measure} in the case of identical particle systems seems to be
completely compatible with the \emph{entanglement definition} used in the
present paper. The identical particles are divided between two observers $A$
and $B$, and a general $N$ particle normalised pure state $\left\vert \Psi
_{AB}\right\rangle $ for the overall system is considered. A so-called \emph{%
entropy of particle entanglement} is defined via the expression $%
E_{P}(\left\vert \Psi _{AB}\right\rangle
)=\tsum\limits_{n}P_{n}E_{M}(\left\vert \Psi _{AB}^{(n)}\right\rangle )$,
where $\left\vert \Psi _{AB}^{(n)}\right\rangle =\widehat{\Pi }%
_{n}\left\vert \Psi _{AB}\right\rangle $ (un-normalised) is the state $%
\left\vert \Psi _{AB}\right\rangle $ projected onto the sub-space where
there are $n$ identical particles associated with $A$ and $N-n$ with $B$. $%
P_{n}$ is the probability that there will be $n$ particles associated with $%
A $, given by $P_{n}=Tr(\widehat{\Pi }_{n}\left\vert \Psi _{AB}\right\rangle
\left\langle \Psi _{AB}\right\vert )$. The (mode) entropy for the state $%
\left\vert \Psi _{AB}^{(n)}\right\rangle $ is $E_{M}(\left\vert \Psi
_{AB}^{(n)}\right\rangle )=S(\widehat{\rho }_{A}^{n})$ where $S(\widehat{%
\rho })=-Tr(\widehat{\rho }\,\log _{2}\widehat{\rho })$ is the von Neumann
entropy and $\widehat{\rho }_{A}^{n}=Tr_{B}(\left\vert \Psi
_{AB}^{(n)}\right\rangle \left\langle \Psi _{AB}^{(n)}\right\vert
)/\left\langle \Psi _{AB}^{(n)}\,|\,\Psi _{AB}^{(n)}\right\rangle $ is the
density operator for $A$ when the system is in state $\left\vert \Psi
_{AB}^{(n)}\right\rangle $. In the case considered by Wiseman et al, $A$ and 
$B$ really refer to two sub-systems (Alice and Bob's collections of qubits)
each of which consists of a number of modes. Entanglement therefore deals
with two sub-systems that are themselves collections of modes. If in the
present approach we restrict ourselves to pure $N$ particle states, then the
possible separable states consistent with our local particle number SSR
requirement are just product states of the form $\left\vert \Phi
_{A}^{(n)}\right\rangle \otimes \left\vert \Phi _{B}^{(N-n)}\right\rangle $.
where $\left\vert \Phi _{A}^{(n)}\right\rangle $ is a normalised $n$
particle state for sub-system $A$ and $\left\vert \Phi
_{B}^{(N-n)}\right\rangle $ is a normalised $N-n$ particle state for
sub-system $B$. For such a separable state of sub-systems $A$ and $B$ $%
\widehat{\Pi }_{n}(\left\vert \Phi _{A}^{(n)}\right\rangle \otimes
\left\vert \Phi _{B}^{(N-n)}\right\rangle )=\left\vert \Phi
_{A}^{(n)}\right\rangle \otimes \left\vert \Phi _{B}^{(N-n)}\right\rangle $, 
$P_{n}=1$, $E_{M}(\left\vert \Psi _{AB}^{(n)}\right\rangle )=0$ so the
entropy of particle entanglement is zero, as would be expected. The most
general pure state could always be written in terms of orthonormal states as 
$\tsum\limits_{n}\tsum\limits_{i,j}C_{ij}^{n}\left\vert \Phi
_{Ai}^{(n)}\right\rangle \otimes \left\vert \Phi _{Bj}^{(N-n)}\right\rangle $%
, and in our terms this is an entangled state of sub-systems $A$ and $B$ if
more than one $C_{ij}^{n}$ is non-zero. We see that $\left\vert \Psi
_{AB}^{(n)}\right\rangle =\tsum\limits_{i,j}C_{ij}^{n}\left\vert \Phi
_{Ai}^{(n)}\right\rangle \otimes \left\vert \Phi _{Bj}^{(N-n)}\right\rangle $%
, $P_{n}=\tsum\limits_{i,j}\left\vert C_{ij}^{n}\right\vert ^{2}\leq 1$, $%
\widehat{\rho }_{A}^{n}=(\tsum\limits_{i,k}\tsum
C_{ij}^{n}\,(C_{kj}^{n})^{\ast }\left\vert \Phi _{Ai}^{(n)}\right\rangle
\left\langle \Phi _{Ak}^{(n)}\right\vert )/P_{n}$. If more than one $%
C_{ij}^{n}$ is non-zero the entropy of particle entanglement is non-zero, so
Wiseman et al would regard this general state as entangled, just as we do.
In this paper no quantitative measure of entanglement has been specifically
proposed, so the entropy of particle entanglement proposed by Wiseman et al 
\cite{Wiseman03a} is consistent with our work.

Note however that a different concept of entanglement - \emph{particle
entanglement} - has also been applied to identical particle systems \cite%
{Hyllus12a}. This is not the same as mode entanglement so tests and measures
for particle entanglement will differ from those for mode entanglement. For
completeness a brief description highlighting the difference between mode
and particle entanglement is presented in Appendix \ref{Appendix - Particle
and Mode Entanglement}. A further discussion about the distinction is given
in \cite{Amico08a}.

\subsubsection{Multi-Mode Sub-Systems}

As well as the simple case where the sub-systems are all \emph{individual}
modes, the concept of entanglement may be \emph{extended} to situations
where the sub-systems are \emph{sets of modes}, rather than individual
modes, In this case entanglement or non-entanglement will be of these
distinct sets of modes. Such a case in considered in Subsection \ref%
{SubSection - Sorensen 2001}, where \emph{pairs} of modes associated with
distinct lattice sites are considered as the sub-systems. Another example is
treated in He et al \cite{He12a}, which involves a double well potential
with each well associated with two bosonic modes, these pairs of modes being
the two sub-systems. Entanglement criteria for the mode pairs based on local
spin operators associated with each potential well are considered (sse
SubSection \ref{SubSection - He 2012}). A further example is treated by
Heaney et al \cite{Heaney10a}, again involving \ four modes associated with
a double well potential. As in the previous example, each mode pair is
associated with the same well in the potential, but here a Bell entanglement
test was obtained for pairs of modes in the different wells. The concept of 
\emph{entanglement of sets of modes} is a straightforward extension of the
basic concept of entanglement of individual modes.

\subsection{Super-Selection Rule}

\label{SubSection - Super-Selection Rule}

As well as the symmetrisation principle there is a further requirement that
systems of identical particles must satisfy known as \emph{super-selection
rules}. These rules restrict the physical states of such systems to those in
which the \emph{coherences} between states with differing numbers of
particles are zero. This applies at the global level for the overall
physical state, but also - as will be discussed in a later sub-section - to
the sub-system states involved in the definition of separable or
non-entangled states. The justfication of the SSR at both the global and
local level will be considered both in terms of simple physics arguements
and in terms of reference frames. Examples of SSR and non-SSR compliant
states will be given, both for overall states and for separable states. The
validity of the SSR for the case of massive bosons or fermions is generally
accepted, but in the case of photons there is doubt regarding their
applcability -as will be discussed below.

\subsubsection{Global Particle Number SSR}

The question of what physical states - entangled or not - are possible in
the \emph{non-relativistic quantum physics} of a system of identical \emph{%
bosonic} particles - such as bosonic \emph{atoms} or \emph{photons} - has
been the subject of much discussion. Whether \emph{entangled} or \emph{not}
it is generally accepted that there is a \emph{super-selection rule} that
prohibits \emph{quantum superposition states} of the form 
\begin{equation}
\left\vert \Phi \right\rangle =\tsum\limits_{N=0}^{\infty }C_{N}\,\left\vert
N\right\rangle \qquad \widehat{\rho }=\tsum\limits_{N=0}^{\infty }\left\vert
C_{N}\right\vert ^{2}\,\left\vert N\right\rangle \left\langle N\right\vert
+\tsum\limits_{N=0}^{\infty }\tsum\limits_{M=0}^{\infty }(1-\delta
_{N,M})C_{N}\,C_{M}^{\ast }\,\left\vert N\right\rangle \left\langle
M\right\vert  \label{Eq.ForbiddenStates}
\end{equation}%
being \emph{physical} states when they involve Fock states $\left\vert
N\right\rangle $ with differing total numbers $N$ of particles. The density
operator for such a state would involve \emph{coherences} between states
with differing $N$. Although such superpositions - such as the \emph{Glauber
coherent state }$\left\vert \alpha \right\rangle $, where $C_{N}=\exp
(-|\alpha |^{2}/2)\,\alpha ^{N}/\sqrt{N!}$ - do have a useful \emph{%
mathematical} role, they do \emph{not} represent actual physical states
according to the super-selection rule. The papers by Sanders et al \cite%
{Sanders03a} and Cable et al \cite{Cable05a} \ are examples of applying the
SSR for optical fields, but also using the mathematical features of coherent
states to treat phenomena such as interference between independent lasers.
The super-selection rule indicates that the most \emph{general physical state%
} for a system of identical bosonic particles can only be of the form%
\begin{eqnarray}
\widehat{\rho } &=&\tsum\limits_{N=0}^{\infty }\tsum\limits_{\Phi }P_{\Phi
N}\,\left( \left\vert \Phi _{N}\right\rangle \left\langle \Phi
_{N}\right\vert \right)  \nonumber \\
\left\vert \Phi _{N}\right\rangle &=&\tsum\limits_{i}C_{i}^{N}\,\left\vert
N\,i\right\rangle  \label{Eq.PhysicalState}
\end{eqnarray}%
where $\left\vert \Phi _{N}\right\rangle $ is a quantum superposition of
states $\left\vert N\,i\right\rangle $ each of which involves exactly $N$
particles, and where different states with the same $N$ are designated as $%
\left\vert N\,i\right\rangle $. This state $\widehat{\rho }$ is a
statistical mixture of states, each of which contains a specific number of
particles. Such a SSR is referred to as a \emph{global} SSR, as it applies
to the system as a whole. Mathematically, the global particle number SSR can
be expressed as 
\begin{equation}
\lbrack \widehat{N},\widehat{\rho }]=0  \label{Eq.GloabalSSR}
\end{equation}%
where $\widehat{N}$ is the \emph{total number} operator.

\subsubsection{Examples of Global Particle Number SSR Compliant States}

Examples of a state vector $\left\vert \Phi _{N}\right\rangle $ for an
entangled pure state \cite{Hines03a} and a density operator $\widehat{\rho }$
for a non-entangled mixed \cite{Goold09a} state for a \emph{two mode bosonic}
system, both of which are possible physical states are 
\begin{eqnarray}
\left\vert \Phi _{N}\right\rangle
&=&\tsum\limits_{k=0}^{N}C(N,k)\,\left\vert k\right\rangle _{A}\otimes
\left\vert N-k\right\rangle _{B}  \label{Eq.EntangledTwoModePureState} \\
\widehat{\rho } &=&\tsum\limits_{k=0}^{N}P(k)\,\left\vert k\right\rangle
_{A}\left\langle k\right\vert _{A}\,\otimes \left\vert N-k\right\rangle
_{B}\left\langle N-k\right\vert _{B}
\label{Eq.NonEntangledTwoModeMixedState}
\end{eqnarray}%
The entangled pure state is a superposition of product states with $k$
bosons in mode $A$ and the remaining $N-k$ bosons in mode $B$. Every term in
the superposition is associated with the same total boson number $N$. The
non-entangled mixed state is a statistical mixture of product states also
with $k$ bosons in mode $A$ and the remaining $N-k$ bosons in mode $B$.
Every term in the statistical mixture is associated with the same total
boson number $N$. For the case of a \emph{two mode fermionic} system the
Pauli exclusion principle restricts the number of possible fermions to two,
with at most one fermion in each mode. Expressions for a state with exactly $%
N=2$ fermions are%
\begin{eqnarray}
\left\vert \Phi _{2}\right\rangle &=&\left\vert 1\right\rangle _{A}\otimes
\left\vert 1\right\rangle _{B}  \label{Eq.TwoModeFermionStateVector} \\
\widehat{\rho } &=&\left\vert 1\right\rangle _{A}\left\langle 1\right\vert
_{A}\,\otimes \left\vert 1\right\rangle _{B}\left\langle 1\right\vert _{B}
\label{Eq.TwoModeFermionDensOpr}
\end{eqnarray}%
Neither state is entangled and both are the same pure state since $\widehat{%
\rho }=\left\vert \Phi _{2}\right\rangle \left\langle \Phi _{2}\right\vert $%
. Although the super-selection rules and symmetrisation principle also
applies to fermions, as indicated in the Introduction this paper is focused
on bosonic systems, and it will be assumed that the modes are bosonic unless
indicated otherwise.

The \emph{Bell states} for $N=2$ bosons provide important examples of four
mode pure quantum states that are compliant with the global particle number
SSR. The modes are designated $A+,A-,B+,B-$ and the Fock states are in
general $\left\vert n_{A+},n_{A-},n_{B+},n_{B-},\right\rangle $. The Bell
states may be written 
\begin{eqnarray}
\left\vert \Psi _{\sin glet}\right\rangle &=&\frac{1}{\sqrt{2}}(\left\vert
1,0,0,1\right\rangle -\left\vert 0,1,1,0\right\rangle )\equiv \frac{1}{\sqrt{%
2}}(\left\vert A+\right\rangle \otimes \left\vert B-\right\rangle
-\left\vert A-\right\rangle \otimes \left\vert B\right\rangle )  \nonumber \\
\left\vert \Psi _{triplet,+1}\right\rangle &=&\left\vert
1,0,1,0\right\rangle \equiv \left\vert A+\right\rangle \otimes \left\vert
B+\right\rangle  \nonumber \\
\left\vert \Psi _{triplet,0}\right\rangle &=&\frac{1}{\sqrt{2}}(\left\vert
1,0,0,1\right\rangle +\left\vert 0,1,1,0\right\rangle )\equiv \frac{1}{\sqrt{%
2}}(\left\vert A+\right\rangle \otimes \left\vert B-\right\rangle
+\left\vert A-\right\rangle \otimes \left\vert B\right\rangle )  \nonumber \\
\left\vert \Psi _{triplet,-1}\right\rangle &=&\left\vert
0,1,0,1\right\rangle \equiv \left\vert A-\right\rangle \otimes \left\vert
B-\right\rangle  \label{Eq.BellStates}
\end{eqnarray}%
where the second forms may be more familiar. Of these states $\left\vert
\Psi _{\sin glet}\right\rangle $ and $\left\vert \Psi
_{triplet,0}\right\rangle $ are entangled, whilst $\left\vert \Psi
_{triplet,+1}\right\rangle $ and $\left\vert \Psi _{triplet,-1}\right\rangle 
$ are separable.

\subsubsection{Super-Selection Rules and Conservation Laws}

It is important to realise that such super-selection rules are \emph{%
additional constraints} to those imposed by \emph{conservation laws}. For
example, the conservation law on total particle number \emph{only} leads to
the requirement on the superposition state $\left\vert \Phi \right\rangle $
that the $|C_{N}|^{2}$ are time independent, it does \emph{not} require only
one $C_{N}$ being non-zero. Super-selection rules are broad in their scope,
forbidding quantum superpositions of states of systems with differing
charge, differing baryon number and differing statistics. Thus a combined
system of a hydrogen atom and a helium ion does not exist in quantum states
that are linear combinations of hydrogen atom states and helium ion states -
the super-selection rules on both charge and baryon number preclude such
states. The basis physical states for such a combined system would involve
symmetrised tensor products of hydrogen atom and helium ion states, not
linear combinations - symmetrisation being required because the system
contains two identical electrons. On the other hand, super-selection rules
do not prohibit quantum superpositions of states of systems with differing
energy, angular or linear momenta - other physical quantities that may also
be conserved. Thus in a hydrogen atom quantum superpositions of states with
differing energy and angular momentum quantum numbers are allowed physical
states.

\subsubsection{SSR Justification and No Suitable Phase Reference}

There are two types of justification for applying the super-selection rules
for systems of identical particles. The first approach is based on simple
considerations and will be outlined below in this subsection. The second
approach \cite{Aharonov67a}, \cite{Bartlett03a}, \cite{Sanders03a}, \cite%
{Kitaev04a}, \cite{van Enk05a}, \cite{Bartlett06a}, \cite{Bartlett07a}, \cite%
{Vaccaro08a}, \cite{White09a}, \cite{Tichy11a} is more sophisticated and
involves linking the absence or presence of SSR to whether or not there is a
suitable \emph{reference frame} in terms of which the quantum state is
described, and is outlined in the next subsection and Appendix \ref{Appendix
- Reference Frames and SSR}. The key idea is that SSR are a consequence of
considering the description of a quantum state by an external observer
(Charlie) whose phase reference frame has an unknown phase difference from
that of an observer ((Alice) more closely linked to the system being
studied. Thus, whilst Alice's description of the quantum state may violate
the SSR, the description of the \emph{same} quantum state by Charlie will
not. In the main part of this paper the density operator $\widehat{\rho }$
used to describe the various quantum states will be that of the external
observer (Charlie).

\subsubsection{SSR Justication and Physics Considerations}

A number of \emph{straightforward reasons} have been given in the
Introduction for why it is appropriate to apply the superselection rule to
exclude quantum superposition states of the form (\ref{Eq.ForbiddenStates})
as physical states for systems of identical particles, and these will now be
considered in more detail.

Firstly, no way is known for creating such states. The Hamiltonian for such
a system commutes with the total boson number operator, resulting in the $%
\left\vert C_{N}\right\vert ^{2}$ remaining constant, so the quantum
superposition state would need to have existed initially. In the simplest
case of non-interacting bosonic atoms, the Fock states are also energy
eigenstates, such Fock states involve total energies that differ by energies
of order the rest mass energy $mc^{2}$, so a coherent superposition of
states with such widely differing energies would at least seem unlikely in a 
\emph{non-relativistic theory, }though for massless photons this would not
be an issue as the energy differences are of order the photon energy $\hbar
\omega $. The more important question is: Is there a non-relativistic
quantum process could lead to the creation of such a state? Processes such
as the dissociation of $M$ diatomic molecules into up to $2M$ bosonic atoms
under Hamiltonian evolution involve entangled atom-molecule states of the
form 
\begin{equation}
\left\vert \Phi \right\rangle =\tsum\limits_{m=0}^{M}C_{m}\,\left\vert
M-m\right\rangle _{mol}\otimes \left\vert 2m\right\rangle _{atom}
\label{Eq.MolecDissn}
\end{equation}%
but the reduced density operator for the bosonic atoms is 
\begin{equation}
\widehat{\rho }_{atoms}=\tsum\limits_{m=0}^{M}\left\vert C_{m}\right\vert
^{2}\,\left( \left\vert 2m\right\rangle \left\langle 2m\right\vert \right)
_{atom}  \label{Eq.RDOAtoms}
\end{equation}%
which is a statistical mixture of states with differing atom numbers with no
coherence terms between such states. Such statistical mixtures are valid
physical states, corresponding to a lack of a priori knowledge of how many
atoms have been produced. To obtain a quantum superposition state for the
atoms \emph{alone}, the atom-molecule state vector would need to evolve at
some time into the form 
\begin{equation}
\left\vert \Phi \right\rangle =\tsum\limits_{m=0}^{M}B_{m}\,\left\vert
M-m\right\rangle _{mol}\otimes \tsum\limits_{n=0}^{M}A_{2n}\left\vert
2n\right\rangle _{atom}  \label{Eq.ProposedEvolvedState}
\end{equation}%
where the separate atomic system is in the required quantum superposition
state. However if such a state existed there would be terms with at least
one non-zero coefficient $B_{m}A_{2n}$ involving product states $\left\vert
M-m\right\rangle _{mol}\otimes \left\vert 2n\right\rangle _{atom}$ with $%
n\neq m$ if the state $\left\vert \Phi \right\rangle $ is not just in the
entangled form (\ref{Eq.MolecDissn}). However, the presence of such a term
would mean that the conservation law involving the number of molecules plus
two times the number of atoms was violated. This is impossible, so such an
evolution is not allowed.

Secondly, no way is known for measuring all the properties of such states,
even if they existed. If a state such as (\ref{Eq.ForbiddenStates}) did
exist then the amplitudes $C_{N}$ would oscillate with frequencies that
differ by relativistic frequencies of order $mc^{2}/\hbar $, even if
boson-boson interactions were included \ To distinguish the phases of the $%
C_{N}\,$\ in order to verify the existence of the state, \emph{measurement
operators} would need to include terms that also oscillate at relativistic
frequencies, and no such measurement operators are known.

Thirdly, there is no need to invoke the existence of such states in order to
understand coherence and interference effects..It is sometimes thought that
states involving quantum superpositions of number states are needed for
discussing \emph{coherence} and \emph{interference properties} of BECs, and
some papers describe the state via the Glauber coherent states. However, as
Leggett \cite{Leggett01a} has pointed out (see also Bach et al \cite{Bach04a}%
, Dalton and Ghanbari \cite{Dalton12a}), a highly occupied number state for
a single mode with $N$ bosons has coherence properties of high order $n$, as
long as $n\ll N$. The introduction of a Glauber coherent state is \emph{not}
required to account for coherence effects. Even the well-known presence of
spatial interference patterns produced when two independent BECs are
overlapped can be accounted for via treating the BECs as Fock states. The
interference pattern is built up as a result of successive boson position
measurements \cite{Javainainen96a}, \cite{Sanders03a}, \cite{Cable05a}.

Fourthly, the stability of such states against decoherence processes may not
be great, so even if they could be created, they could rapidly change to
other states. However, decoherence time scales that are not too short would
be acceptable. Although BECs are created in high vacuum experiments and are
well isolated from the external environment in magnetic or dipole traps,
they are not entirely free from decoherence effects because the bosons do
interact with each other. Even in a single mode case boson-boson collisions
can cause dephasing effects. These could be shown via the decay of the
coherence $\left\langle \widehat{a}\right\rangle $. However, it may turn out
that the lifetime of a coherent state in a single mode BEC is quite long -
in the case of photons the lifetime could be as long as the inverse
Townes-Schawlow line width, perhaps of order $10^{3}$s (see below). If a
coherent superposition state could be created with a non-zero coherence,
this may last long enough to carry out further experiments, so this fourth
reason for discarding coherent superposition states is relatively
unimportant though further studies of their lifetimes would be of some
theoretical interest. .

\subsubsection{SSRJustification and Galilean Frames ?}

Finally, in addition to the previous reasons there is an arguement based on
the requirement that the dynamical equations for such non-relativistic
quantum systems should be invariant under a \emph{Galilean transformation}
which has been proposed \cite{Stenholm02a} as a proof of the super-selection
rule for atom number. This approach is linked to the reference frame based
justification of SSR (see Appendix \ref{Appendix - Reference Frames and SSR}%
). However, whilst the paper shows that under a Galilean transformation -
corresponding to describing the system from the point of view of an observer
moving with a constant velocity $\mathbf{v}$ with respect to the original
observer, and where the two observers have identical clocks - the terms in a
superposition state with different numbers $N$ of massive bosons would
oscillate like $\exp i\left( \frac{1}{2}Nm\mathbf{v}^{2}t\right) /\hbar $,
and may be expected if the \emph{same} quantum state is described by a
moving observer. This feature alone does not seem to require the
super-selection rule, since here the moving observer's reference frame has a
well-defined velocity with respect to that attached to the system. However,
the moving observer's reference frame may actually have an unknown relative
velocity, in which case a twirling operation resulting in the elimination of
number state coherences could be involved (see Appendix \ref{Appendix -
Reference Frames and SSR}). This will be not be considered further at this
stage.

On the other hand, an approach of this kind involving \emph{rotation symmetry%
} would seem to rule out such states as quantum superpositions of a boson
(spin $0$) and a fermion.(spin $1/2$). Let such a state be prepared in the
form $(\left\vert F\right\rangle +\left\vert B\right\rangle )/\sqrt{2}.$%
Consider an observer whose cartesian reference frame is $X,Y,Z$. This is a
classical system that can be rotated in space. If the observer rotates with
his frame through $2\pi $ about any axis they are then back in the same
position, but the observer now sees the state as $(-\left\vert
F\right\rangle +\left\vert B\right\rangle )/\sqrt{2}.$ This state is
apparently orthogonal to the one observed before the rotation, and this is
paradoxical since the observer would be in the same position. Thus there is
a super-selection rule excluding states such as $(\left\vert F\right\rangle
+\left\vert B\right\rangle )/\sqrt{2}.$ A similar argument based on the 
\emph{time reversal} anti-unitary operator was given by Wick et al \cite%
{Wick52a}.

\subsubsection{SSR and Photons}

Though this paper is focused on massive bosonic atoms the question is
whether similar considerations also apply to the optical quantum EM field,
which involve \emph{massless} bosons - \emph{photons}. Here the situation is
not so clear.

In the case of photons, Molmer \cite{Molmer97a} has argued that the physical
state for a single mode optical laser field operating well above threshold
is not a Glauber coherent state, and the density operator would be a
statistical mixture of the form (\ref{Eq.PhysicalState}), with $\left\vert
\Phi _{N}\right\rangle =\left\vert N\right\rangle $ and $P_{\Phi N}=\exp (-%
\overline{N})\,\overline{N}^{N}/N!$. Here the density operator is a
statistical mixture of photon number states with Poisson distribution, or
equivalently a statistical mixture of coherent states $\left\vert \alpha
\right\rangle $ with $\alpha =\sqrt{\overline{N}}\exp (i\phi )$ and all
phases $\phi $ having equal probability. Some of the same general reasons
for applying the super-selection rule to systems of identical massive bosons
also apply here, though the details differ. For the free quantum EM\ field
there is a conservation law for the photon number in each mode, so in this
case again $|C_{N}|^{2}$ would be time independent. However, for photons the 
$C_{N}$ would oscillate with frequencies that only differ by
non-relativistic frequencies of order $\hbar \omega $, so the arguement
against coherent states based on this feature do not apply. In terms of
preparing states, in the case of the single mode optical laser the field is
generated via interactions with incoherently pumped atoms, there is no well
defined optical phase that can be imposed on the process, and the quantum
theory for such laser processes predicts a quantum state that is a
statistical mixture of photon number states. In the case of the optical
laser field coherent states are not physical unless there are optical
reference fields with a well-defined phase that could be used to determine
the phases associated with the expansion coefficients. This may now becoming
possible with the development of atomic clocks based on optical atomic
transitions that may supercede atomic clocks based on atomic transitions at
microwave frequencies. Optical interference and coherence effects can also
be explained without invoking Glauber coherent states, as \cite{Molmer97a}
and others such as \cite{Sanders03a} have shown. However, if coherent states
could be created they might be relatively stable. In the optical laser field
case, phase loss via diffusion is related to the laser linewidth, and this
can be reduced to the Townes-Schawlow limit that varies inversely as the
mean photon number - which is large. The Townes-Schawlow linewidth can be as
small as $10^{-3}$ hz, corresponding to a phase diffusion time of $10^{3}$
s. An alternative approach is presented by Wiseman et al \cite{Wiseman02a}, 
\cite{Wiseman02b}, in which the optical laser is treated via a master
equation, but where monitoring of the laser environment (difficult!) is
required to determine whether certain pure state ensembles - such as those
involving coherent states - are physically realisable. The conclusion
reached is that for finite self energy the coherent state ensemble is not
physically realisable, the closest ensemble being that involving squeezed
states, though for zero self energy coherent state ensembles are obtained.

Another approach to the question (see next sub-section and SubSection \ref%
{AppendixSubSection - Situation B} in Appendix \ref{Appendix - Reference
Frames and SSR}) involves the consideration of phase reference frames. The
quantum state of a single mode laser may be described as a Glauber coherent
state by an observer (Alice) with one reference frame, but would be
described as a statistical mixture of photon number states by another
observer (Charlie) with a different reference frame whose phase reference is
completely unrelated to the previous one. However, this arguement against
the presence of coherent state in Charlie's viewpoint would be overcome if
phase references at optical frequencies are developed.

\subsection{Reference Frames and Violations of Superselection Rules}

\label{SubSection - Challenges to SSR}

Challenges to the requirement for physical states to be consistent with
super-selection rules have occured since the 1960's when Aharonov and
Susskind \cite{Aharonov67a} suggested that coherent superpositions of
different charge eigenstates could be created. It is argued that
super-selection rules are not a fundamental requirement of quantum theory,
but the restrictions involved could be lifted if there is a suitable system
that acts as a \emph{reference} for the coherences involved - \cite%
{Aharonov67a}, \cite{Bartlett03a}, \cite{Sanders03a}, \cite{Kitaev04a}, \cite%
{van Enk05a}, \cite{Bartlett06a}, \cite{Bartlett07a}, \cite{Vaccaro08a}, 
\cite{White09a}, \cite{Tichy11a} provide discussions regarding reference
systems and SSR.

\subsubsection{Linking SSR and Reference Frames}

The discussion of the super-selection rule issue in terms of reference
systems is quite complex and too lengthy to be covered in the body of this
paper. However, in view of the wide use of the reference frame approach a
full outline is presented in Appendix \ref{Appendix - Reference Frames and
SSR}. The key idea is that there are two observers - Alice and Charlie - who
are describing the same quantum state in terms of their own reference
systems. The reference systems are \emph{macroscopic systems} in states
where the behaviour is essentially \emph{classical}, such as large magnets
that can be used to define \emph{cartesian axes} or BEC in Glauber coherent
states that are introduced to define a \emph{phase reference}. The
relationship between the two reference systems is represented by a \emph{%
group} of \emph{unitary transformation operators} listed as $\widehat{T}(g)$%
, where the particular transformation (translation or rotation of cartesian
axes, phase change of phase references, ..) that changes Alice's reference
system into Charlie's is denoted by $g$. Alice is the \emph{internal}
observer, closely linked to the system under study and describes the quantum
state via her density operator, whereas Charlie is the \emph{external}
observer whose specification of the \emph{same} quantum state via his
density operator is of most interest. There are two cases of importance, 
\emph{Situation A} - where the relationship between Alice's and Charlie's
reference frame is is \emph{known} and specified by a \emph{single}
parameter $g$, and \emph{Situation B} - where on the other hand the
relationship between frames is completely \emph{unknown}, all possible
transformations $g$ must be given equal weight. Situation A is not
associated with SSR, whereas Situation B leads to SSR. The relationship
between Alice's and Charlie's density operators is given in terms of the
transformation operators (see Eq. (\ref{Eq.AliceCharlieStatesSitnA}) for
Situation A and Eq. (\ref{Eq.AliceCharlieStatesSitnB}) for Situation B). In
Situation B there is often a qualitative change between Alice's and
Charlie's description of the same quantum state, with pure states as
described by Alice becoming mixed states when described by Charlie. It is
Situation B with the \emph{U(1)} transformation group - for which \emph{%
number operators} are the \emph{generators} - that is of interest for the 
\emph{single} or \emph{multi-mode} systems involving \emph{identical bosons}
on which the present paper focuses. An example of the qualitative change of
behaviour for the single mode case is that \emph{if} it is \emph{assumed}
that Alice could prepare the system in a Glauber coherent pure state - which
involves SSR breaking coherences between differing number states - then
Charlie would describe the same state as a Poisson statistical mixture of
number states - which is consistent with the operation of the SSR. Thus the
SSR applies in terms of external observer Charlie's description of the
state. This is how the dispute on whether the state for single mode laser is
a coherent state or a statistical mixture is resolved - the two descriptions
apply to different observers - Alice and Charlie. On the other hand there
are quantum states such as Fock states and Bell states which are described
the same way by both Alice and Charlie, even in Situation B. The general
justification of the SSR for Charlie's density operator description of the
quantum state in Situation B is derived in terms of the \emph{irreducible
representations} of the transformation group, there being no coherences
between states associated with differing irreducible representations (see
Eq. (\ref{Eq.CharlieDensOprSSRForm})). For the particular case of the \emph{%
U(1)} transformation group the irreducible representations are associated
with the total \emph{boson number} for the system or sub-system, hence the
SSR that prohibits coherences between states where this number differs.
Finally, it is seen that if Alice describes a general non-entangled state of
sub-systems - which being separable have their own reference frames - then
Charlie will also describe the state as a non-entangled state and with the
same probability for each product state (see Eqs. (\ref%
{Eq.AliceDensOprNonEntState}) and (\ref{Eq.CharlieDensOprNonEntState})). For
systems involving \emph{identical bosons} Charlie's description of the
sub-system density operators will only involve density operators that
conform to the SSR. This is in accord with the key idea of the present paper.

\subsubsection{Coherent Superposition of Atom and Molecule ?}

Based around the reference frame approach Dowling et al \cite{Dowling06a}
and Terra Cunha et al \cite{TerraCunha07a} propose processes using a BEC as
a reference system that would create a coherent superposition of an atom and
a molecule, or a boson and a fermion \cite{Dowling06a}. Dunningham et al 
\cite{Dunningham11a} consider a scheme for observing a superposition of a
one boson state and the vacuum state. Obviously if super-selection rules can
be overcome in these instances, it might be possible to \emph{produce}
coherent superpositions of Fock states with differing particle numbers such
as Glauber coherent states, though states with $\overline{N}$ $\symbol{126}$ 
$10^{8}$ would presumably be difficult to produce. However, detailed
considerations of such papers indicate that the states actually produced in
terms of Charlie's description are statistical mixtures consistent with the
super-selection rules rather than coherent superpositions, which are only
present in Alice's description of the state. Also, although coherence and
interference effects are demonstrated, these can also be accounted for
without invoking the presence of coherent superpositions that violate the
super-selection rule. As the paper by Dowling et al \cite{Dowling06a}
entitled "Observing a coherent superposition of an atom and a molecule." is
a good example of where the super-selection rules are challenged, the key
points are described in Appendix \ref{Appendix - Super-Selection Rule
Violations ?}. Essentially the process involves one atom $A$ interacting
with a BEC of different atoms $B$ leading to the creation of one molecule $%
AB $, with the BEC being depleted by one $B$ atom. There are three stages in
the process, the first being with the interaction that turns separate atoms $%
A$ and $B$ into the molecule $AB$ turned on at Feshbach resonance for a time 
$t$ related to the interaction strength and the mean number of bosons in the
BEC reference system, the second being free evolution at large Feshbach
detuning $\Delta $ for a time $\tau $ leading to a phase factor $\phi
=\Delta \,\tau $, the third being again with the interaction turned on at
Feshbach resonance for a further time $t$. However, it is pointed out in
Appendix \ref{Appendix - Super-Selection Rule Violations ?} that Charlie's
description of the state produced for the atom plus molecule system is
merely a statistical mixture of a state with one atom and no molecules and a
state with no atom and one molecule, the mixture coefficients depending on
the phase $\phi $ imparted during the process. However a coherent
superposition is seen in Alice's description of the final state, though this
is not surprising since a SSR violating initial state was assumed. The
feature that in Charlie's description of the final state no coherent
superposition of an atom and a molecule is produced in the process is not
really surprising, because of the averaging over phase differences in going
from Alice's reference frame to Charlie's. It is the dependence on the phase 
$\phi $ imparted during the process that demonstates coherence (Ramsey
interferometry) effects, but it is shown in Appendix \ref{Appendix -
Super-Selection Rule Violations ?} that exactly the same results can be
obtained via a treatment in which states which are coherent superpositions
of an atom and a molecules are never present, the initial BEC state being
chosen as a Fock state. In terms of the description by an external observer
(Charlie) the claim of violating the super-selection rule has not been
demonstrated via this particular process.

\subsubsection{Detection of SSR Violating States}

Whether such super-selection rule violating states can be \emph{detected}
has also not been justified. For example, consider the state given by a
superposition of a one boson state and the vacuum state (as discussed in 
\cite{Dunningham11a}). We consider an interferometric process in which one
mode $A$ for a two mode BEC interferometer is initially in the state $\alpha
\left\vert 0\right\rangle +\beta \left\vert 1\right\rangle $, and the other
mode $B$ is initially in the state $\left\vert 0\right\rangle $ - thus $%
\left\vert \Psi (i)\right\rangle =(\alpha \left\vert 0\right\rangle +\beta
\left\vert 1\right\rangle )_{A}\otimes \left\vert 0\right\rangle _{B}$ in
the usual occupancy number notation, where $|\alpha |^{2}+|\beta |^{2}=1$.
The modes are first coupled by a beam splitter, then a free evolution stage
occurs for time $\tau $ associated with a phase difference $\phi =\Delta
\tau $ (where $\Delta =\omega _{B}-\omega _{A}$ is the mode frquency
difference), the modes are then coupled again by the beam splitter and the
probability of an atom being found in modes $A$, $B$ finally being measured.
The probabilities of finding one atom in modes $A$, $B$ respectively are
found to only depend on $|\beta |^{2}$ and $\phi $. Details are given in
Appendix \ref{Appendix - Super-Selection Rule Violations ?}. There is no
dependence on the relative phase between $\alpha $ and $\beta $, as would be
required if the superposition state $\alpha \left\vert 0\right\rangle +\beta
\left\vert 1\right\rangle $ is to be specified. Exactly the same detection
probabilities are obtained if the initial state is the mixed state $\widehat{%
\rho }(i)=|\alpha |^{2}(\left\vert 0\right\rangle _{A}\left\langle
0\right\vert _{A}\otimes \left\vert 0\right\rangle _{B}\left\langle
0\right\vert _{B})+|\beta |^{2}(\left\vert 1\right\rangle _{A}\left\langle
1\right\vert _{A}\otimes \left\vert 0\right\rangle _{B}\left\langle
0\right\vert _{B})$, in which the vacuum state for mode $A$ occurs with a
probability $|\alpha |^{2}$ and the one boson state for mode $A$ occurs with
a probability $|\beta |^{2}$. In this example the proposed coherent
superposition associated with the super-selection rule violating state would
not be detected in this interferometric process, nor in the more elaborate
scheme discussed in \cite{Dunningham11a}.

\subsection{Super-Selection Rule - Separate Sub-Systems}

\label{SubSection - SSR Separate Modes}

In this sub-section the important case of SSR in separable states will be
dealt with, since this is key to understanding what entangled states are
allowed in systems involving identical particles.

\subsubsection{Local Particle Number SSR}

We now consider the role of the super-selection rule for the case of \emph{%
non-entangled} states. The global super-selection rule on \emph{total
particle number} has restricted the physical quantum state for a system of
identical bosons to be of the form (\ref{Eq.PhysicalState}). Such states may
or may not be entangled states of the modes involved. The question is - do
similar restrictions involving the \emph{sub-system particle number} apply
to the modes, considered as \emph{separate} sub-systems in the definition of
non-entangled states ? The viewpoint in this paper is that this is so. Note
that applying the SSR on the separate sub-system density operators $\widehat{%
\rho }_{R}^{A}$, $\widehat{\rho }_{R}^{B}$, .. is \emph{only} in the context
of non-entangled states. Such a SSR is referred to as a \emph{local }SSR, as
it applies to each of the separate sub-systems. Mathematically, the local
particle number SSR can be expressed as 
\begin{equation}
\lbrack \widehat{N}_{X},\widehat{\rho }_{R}^{X}]=0  \label{Eq.LocalSSR}
\end{equation}%
where $\widehat{N}_{X}$ is the \emph{number} operator for sub-system $%
X=A,B,..$.The SSR restriction is based on the proposition that the density
operators $\widehat{\rho }_{R}^{A}$, $\widehat{\rho }_{R}^{B}$, .. for the
separate sub-systems $A$, $B$, ..should themselves represent possible \emph{%
physical states} for each of the sub-systems, considered as a \emph{separate
system} and thus be required to satisfy the super-selection rule that
forbids quantum superpositions of Fock states with differing boson numbers.
It is contended that expressions for the non-entangled quantum state $%
\widehat{\rho }$ in which $\widehat{\rho }_{R}^{A}$, $\widehat{\rho }%
_{R}^{B} $, $\widehat{\rho }_{R}^{C}$.. were \emph{not} physical states for
the sub-systems would only be of mathematical interest.

Applying the local particle number SSR to the sub-system density operators
for non-entangled states is discussed in papers by Bartlett et al \cite%
{Bartlett06b}, \cite{Bartlett07a} as one of several \emph{operational
approaches} for defining entangled states. However, other authors such as 
\cite{Verstraete03a}, \cite{Schuch04a} state on the contrary that states
when the sub-system density operators do \emph{not} conform to the local
particle number super-selection rule \emph{are} still separable, others such
as \cite{Hillery06a}, \cite{Hillery09a} do so by implication, so in this
paper we are advocating a \emph{revision} to the \emph{widely held} \emph{%
notion} of entanglement in identical particle systems,.the consequence being
that the set of entangled states is now much \emph{larger}. This is a \emph{%
key idea} in this paper - not only should super-selection rules on particle
numbers be applied to the the \emph{overall} physical state, entangled or
not, but it \emph{also} should be applied to the density operators that
describe states of the modal \emph{sub-systems} involved in the general
definition of \emph{non-entangled} states. The reasons for adopting this
viewpoint are set out below. Apart from the papers by Bartlett et al \cite%
{Bartlett06b}, \cite{Bartlett07a} we are not aware that this definition of
non-entangled states has been invoked previously, indeed the opposite
approach has been proposed \cite{Verstraete03a}, \cite{Schuch04a}. However,
the idea of considering whether sub-system states should satisfy the local
particle number SSR has been presented in several papers -\emph{\ }\cite%
{Verstraete03a}, \cite{Schuch04a}, \cite{Bartlett06b}, \cite{Bartlett07a}, 
\cite{Vaccaro08a}, \cite{White09a}, \cite{Paterek11a}, mainly in the context
of pure states for bosonic systems, though in these papers the focus is on
issues other than the definition of entanglement - such as quantum
communication protocols \cite{Verstraete03a}, multicopy distillation \cite%
{Bartlett06b}, mechanical work and accessible entanglement \cite{Vaccaro08a}%
, \cite{White09a} and Bell inequality violation \cite{Paterek11a}. However,
there are a number of papers that do not apply the SSR to the sub-system
density operators, and those that do have not studied the consequences for
various entanglement tests - as is done in the present paper.

\subsubsection{Local SSR Justification and Independent Local Phase References%
}

The more elaborate justification in terms of reference frames for this SSR
requirement on non-entangled states is presented in SubSection \ref{Appendix
SubSection - Non Ent States} of Appendix \ref{Appendix - Reference Frames
and SSR}. Essentially the idea is that in the context of separable states,
each sub-system has its own \emph{independent phase reference frames}, and
those of Charlie having an unknown phase in relation to those of Alice. This
leads to the local particle number SSR.

\subsubsection{Local SSR Justication and Physics Considerations}

The more simple reasons for this assertion are analogous to those for the
overall multi-mode system and may be summarised as: absence of both a
preparation process and a measurement process for such states, the lack of
need of such states to describe single mode interference and coherence
effects. Such superposition states may also be unstable, though again this
is not a fatal problem.

Firstly, sub-system states incompatible with the SSR cannot be prepared.
Consider for example a typical preparation process. For the situation of two
modes $A,B$ physically allowed pure states $\left\vert \Phi
_{N}\right\rangle $ could be prepared which in general are entangled states
of the form 
\begin{equation}
\left\vert \Phi _{N}\right\rangle
=\tsum\limits_{k=0}^{N}A_{k}^{N}\,\left\vert k\right\rangle _{A}\left\vert
N-k\right\rangle _{B}  \label{Eq.TwoModeEntState}
\end{equation}%
so that the general mixed physical state for the two mode system is 
\begin{equation}
\widehat{\rho }=\tsum\limits_{N=0}^{\infty }\tsum\limits_{\Phi }P_{\Phi
N}\,\tsum\limits_{k=0}^{N}\tsum\limits_{l=0}^{N}A_{k}^{N}\,(A_{l}^{N})^{\ast
}\left\vert k\right\rangle _{A}\left\langle l\right\vert _{A}\otimes
\left\vert N-k\right\rangle _{B}\left\langle N-l\right\vert _{B}
\label{Eq.DensityOprTwoModeEntState}
\end{equation}%
Hence the reduced density operator - which specifies the state for mode $A$
if measurements on this mode were carried out and measurements on other
modes discarded - will be given by 
\begin{equation}
\widehat{\rho }_{A}=\tsum\limits_{N=0}^{\infty }\tsum\limits_{\Phi }P_{\Phi
N}\,\tsum\limits_{k=0}^{N}A_{k}^{N}\,(A_{k}^{N})^{\ast }\left\vert
k\right\rangle _{A}\left\langle k\right\vert _{A}
\label{Eq.ReducedDensityOprTwoModeES}
\end{equation}%
which is a statistical mixture of Fock states $\left\vert k\right\rangle
_{A} $. Thus the quantum state for mode $A$ considered separately contains
no superposition of states $\left\vert k\right\rangle _{A}$ with differing
numbers of bosons occupying mode $A$. As in the example considered in the
previous section, the evolution of $\left\vert \Phi _{N}\right\rangle $ into
a tensor product of superposition states for modes $A$ and $B$ of the form 
\begin{equation}
\left\vert \Phi _{N}\right\rangle
=\tsum\limits_{k=0}^{N}C_{k}^{N}\,\left\vert k\right\rangle _{A}\otimes
\tsum\limits_{k=0}^{N}D_{k}^{N}\left\vert N-k\right\rangle _{B}
\label{Eq.ImpossibleEvolvedState}
\end{equation}%
is not possible. The preparation of the state for mode $A$ must have
involved first preparing a physical state for the full multi-mode system -
for which the two mode state in Eq. (\ref{Eq.DensityOprTwoModeEntState}) is
a specific example - from which the state associated with a particular mode
is then determined as given by the reduced density operator. As illustrated
by the example just given, the super-selection rule on the total number of
identical bosons for the overall system produces a reduced density operator
for the sub-system in which the super-selection rule for boson number also
applies - that is the state for the sub-system does not involve quantum
superpositions of mode Fock states with differing boson numbers, it only can
involve statistical mixtures of such states.

Secondly, measurement processes may be applied to each separate mode and
again the lack of measurement systems with well defined relativistic phases
would preclude measurements that determine the rapidly varying phase
differences between the expansion coefficients.in single mode state vectors
of the form $\left\vert \Phi _{A}\right\rangle =\tsum\limits_{n=0}^{\infty
}C_{n}\,\left\vert n\right\rangle _{A}$. Invoking the existence of states
whose key properties cannot be measured is somewhat dubious.

Thirdly, experimental setups involving single mode BECs and optical systems
can be created and yet there is no need to invoke coherent superpositions of
number states to explain coherence and interferometric effects. Thus
essentially the \emph{same reasons} that justify applying the
super-selection rule to the overall many boson system also apply to the
separate mode sub-systems.

\subsubsection{Local SSR Justification and Joint Measurements}

A consideration of \emph{joint measurements} on all the sub-systems leads to
other fundamental reasons why the individual density operators $\widehat{%
\rho }_{R}^{A}$, $\widehat{\rho }_{R}^{B}$, .in the specific situation of
the general mixed non-entangled state given in Eq. (\ref%
{Eq.NonEntangledState}) must represent physical states for the sub-systems.
This state is a statistical mixture of product states $\widehat{\rho }%
_{R}^{A}\otimes \widehat{\rho }_{R}^{B}\otimes \widehat{\rho }%
_{R}^{C}\otimes ...$- each product state being an overall state of the
system that \emph{could} have been prepared. If sub-sysystem $A$ is prepared
by one experimenter in state $\widehat{\rho }_{R}^{A}$ with probability $%
P_{R}$, classical communications to other local experimenters to prepare the
other sub-systems in states $\widehat{\rho }_{R}^{B}$, $\widehat{\rho }%
_{R}^{C}$, etc with the same probability will result in the preparation of
the overall mixed state. If such an overall product state is a physical
state, then so must be the states of the uncorrelated sub-systems involved.
Furthermore, measurements on \emph{all} the sub-systems can be carried out,
not just those on one particular sub-system $A$ - where the results for the
sub-system probabilities $P_{A}(i)$ are determined from the reduced density
operator $\widehat{\rho }_{A}$ - see Eq. (\ref{Eq.MeasProbNonEntState}). We
have seen in Eq (\ref{Eq.JointProbNonEntState}) that the joint probability $%
P_{AB..}(i,j,..)$ for measurements on all the sub-systems is determined from
the product of the individual sub-system probabilities $P_{A}^{R}(i)$, $%
P_{B}^{R}(j)$, ..associated with sub-system density operators $\widehat{\rho 
}_{R}^{A}$, $\widehat{\rho }_{R}^{B}$, ..., the overall product being
weighted by the probability $P_{R}$ that a particular product state is
prepared. The reduced density operators for all the sub-systems do \emph{not}
determine this \emph{joint probability} - what is required are the \emph{%
full set} of sub-system density operators $\widehat{\rho }_{R}^{A}$, $%
\widehat{\rho }_{R}^{B}$, ...along with the overall probability $P_{R}$ that
a particular product state is prepared. As these individual sub-system
probabilities $P_{A}^{R}(i)$, $P_{B}^{R}(j)$, ..must determine actual
possible measurements then the density operators $\widehat{\rho }_{R}^{A}$, $%
\widehat{\rho }_{R}^{B}$, ..must correspond to possible physical states for
the sub-systems, the sub-systems being modes or single particle states in
the present case. But as we have seen, the possible physical states that can
be prepared for these sub-systems are those as in Eq. (\ref%
{Eq.ReducedDensityOprTwoModeES}) which are a statistical mixture of number
states with no coherences between Fock states with differing boson numbers,
so the $\widehat{\rho }_{R}^{A}$, $\widehat{\rho }_{R}^{B}$, themselves
satisfy the super-selection rule.

\subsubsection{State that Violates Local and Global Particle Number SSR}

Finally, an objection to applying the super-selection rule to separate modes
based on emphasising only measurements on only one mode and its the reduced
density operator may be raised, and suggest that . $\widehat{\rho }_{R}^{A}$%
, $\widehat{\rho }_{R}^{B}$ etc may be allowable provided that the overall
reduced density operators comply with the super-selection rule. However, as
will be seen this is not in general possible. As shown above, measurements
on the subsystems with measurements on the other sub-systems discarded - are
determined \emph{only} from the reduced density operators $\widehat{\rho }%
_{A}=\sum_{R}P_{R}\,\widehat{\rho }_{R}^{A}$ alone. Hence it may seem that
providing the reduced density operators represent physical states then it
does not matter if the $\widehat{\rho }_{R}^{A}$, $\widehat{\rho }_{R}^{B}$, 
$\widehat{\rho }_{R}^{C}$.. do not. Indeed, for special cases we can find
density operators $\widehat{\rho }_{R}^{A}$ that are unphysical even though
the reduced density operator $\widehat{\rho }_{A}$ is physical \ One such
example is where 
\begin{eqnarray}
\widehat{\rho }_{1}^{A} &=&\left( \frac{1}{\sqrt{2}}(\left\vert
0\right\rangle _{A}+\left\vert 1\right\rangle _{A})\right) \left( \frac{1}{%
\sqrt{2}}(\left\langle 0\right\vert _{A}+\left\langle 1\right\vert
_{A})\right) \qquad P_{1}=\frac{1}{2}  \nonumber \\
\widehat{\rho }_{2}^{A} &=&\left( \frac{1}{\sqrt{2}}(\left\vert
0\right\rangle _{A}-\left\vert 1\right\rangle _{A})\right) \left( \frac{1}{%
\sqrt{2}}(\left\langle 0\right\vert _{A}-\left\langle 1\right\vert
_{A})\right) \qquad P_{2}=\frac{1}{2}  \label{Eq.NonPhysicalStates}
\end{eqnarray}%
which yields 
\begin{equation}
\widehat{\rho }_{A}=\frac{1}{2}\left( \left\vert 0\right\rangle
_{A}\left\langle 0\right\vert _{A}\right) +\frac{1}{2}\left( \left\vert
1\right\rangle _{A}\left\langle 1\right\vert _{A}\right)  \label{Eq.RDO}
\end{equation}%
This is a valid statistical mixture of two physical states for mode $A$,
namely a state with no bosons and a state with one boson, even though the
contributions $\widehat{\rho }_{1}^{A}$ and $\widehat{\rho }_{2}^{A}$ are
non physical states consisting of pure states that are each quantum
superpositions of a zero boson state and \ a one boson state - in violation
of the super-selection rule. However even a minute change in the $P_{R}$
will lead to the reduced density operators $\widehat{\rho }_{A}$, $\widehat{%
\rho }_{B}$, .. that are non physical. In the example given, changes to $%
P_{1}=0.51$ and $P_{2}=0.49$ will lead to non physical contributions $%
\left\vert 0\right\rangle _{A}\left\langle 1\right\vert _{A}$ and $%
\left\vert 1\right\rangle _{A}\left\langle 0\right\vert _{A}$ to the reduced
density operator $\widehat{\rho }_{A}$. Also, as \emph{all} the reduced
density operators must represent physical states, then the sums in $\widehat{%
\rho }_{A}=\sum_{R}P_{R}\,\widehat{\rho }_{R}^{A}$, $\widehat{\rho }%
_{B}=\sum_{R}P_{R}\,\widehat{\rho }_{R}^{B}$, ... must \emph{all} lead to
physical states. Since the probabilities $P_{R}$ depend on the preparation
process that generates the mixed non-entangled state, and may for example
depend on external parameters such as temperature, it would be extremely
unlikely for given $\widehat{\rho }_{R}^{A}$, $\widehat{\rho }_{R}^{B}$,
..that \emph{all} such sums will lead to physical states, though for \emph{%
special choices} of the mode density operators and the $P_{R}$ this can
occur. In addition, the density operators for the other modes must be chosen
so that the overall density operator is consistent with the super-selection
rule. For example in the case where there are only two modes, the density
operators $\widehat{\rho }_{1}^{B}=\left\vert 0\right\rangle
_{B}\left\langle 0\right\vert _{B}$ and $\widehat{\rho }_{2}^{B}=\left\vert
1\right\rangle _{B}\left\langle 1\right\vert _{B}$ would lead to a
physically valid reduced density operator $\widehat{\rho }_{B}=\frac{1}{2}%
\left( \left\vert 0\right\rangle _{B}\left\langle 0\right\vert _{B}\right) +%
\frac{1}{2}\left( \left\vert 1\right\rangle _{B}\left\langle 1\right\vert
_{B}\right) $ for mode $B$, but there would be terms such as $\frac{1}{4}$ $%
\left\vert 0\right\rangle _{A}\left\langle 1\right\vert _{A}\otimes
\left\vert 0\right\rangle _{B}\left\langle 0\right\vert _{B}$ in the overall
density operator, and such a term involves a coherence between an $N=0$
state and an $N=1$ state which is disallowed. Indeed, for the $\widehat{\rho 
}_{1}^{A}$, $\widehat{\rho }_{2}^{A}$ and $P_{1}$, $P_{2}$ as in Eq. (\ref%
{Eq.NonPhysicalStates}), there may be no choice for $\widehat{\rho }_{1}^{B}$
and $\widehat{\rho }_{2}^{B}$ that gives rise to an overall physical state.
In Appendix \ref{Appendix - Non-Physical Two Mode States}\ the situation
where $\widehat{\rho }_{1}^{B}$ and $\widehat{\rho }_{2}^{B}$ are associated
with two general pure orthogonal states of the form $\alpha \left\vert
0\right\rangle _{B}+\beta \left\vert 1\right\rangle _{B}$ and $-\beta ^{\ast
}\left\vert 0\right\rangle _{B}+\alpha ^{\ast }\left\vert 1\right\rangle
_{B} $ with $(|\alpha |^{2}+|\beta |^{2})=1$, is considered, and we find
that no choice of $\alpha $ and $\beta $ leads to an overall physical state
- although again the reduced density operator $\widehat{\rho }_{B}=\frac{1}{2%
}\left( \left\vert 0\right\rangle _{B}\left\langle 0\right\vert _{B}\right) +%
\frac{1}{2}\left( \left\vert 1\right\rangle _{B}\left\langle 1\right\vert
_{B}\right) $ is physical.

\subsubsection{Global but not Local Particle Number SSR Compliant States}

\label{SubSection - Two Mode Coherent State Mixture}

However, in some cases sub-system density operators can be chosen in the
context of two mode systems which comply with the global particle number SSR
but not the local particle number SSR. Such a case involving four \emph{zero
and one boson superpositions} is presented by Verstraete et al \cite%
{Verstraete03a}, \cite{Schuch04a}. The overall density operator is a
statistical mixture 
\begin{eqnarray}
\widehat{\rho } &=&\frac{1}{4}(\left\vert \psi _{1}\right\rangle
\left\langle \psi _{1}\right\vert )_{A}\otimes \left\vert \psi
_{1}\right\rangle \left\langle \psi _{1}\right\vert )_{B}+\frac{1}{4}%
(\left\vert \psi _{i}\right\rangle \left\langle \psi _{i}\right\vert
)_{A}\otimes \left\vert \psi _{i}\right\rangle \left\langle \psi
_{i}\right\vert )_{B}  \nonumber \\
&&+\frac{1}{4}(\left\vert \psi _{-1}\right\rangle \left\langle \psi
_{-1}\right\vert )_{A}\otimes \left\vert \psi _{-1}\right\rangle
\left\langle \psi _{-1}\right\vert )_{B}+\frac{1}{4}(\left\vert \psi
_{-i}\right\rangle \left\langle \psi _{-i}\right\vert )_{A}\otimes
\left\vert \psi _{-i}\right\rangle \left\langle \psi _{-i}\right\vert )_{B} 
\nonumber \\
&&  \label{Eq.VerstraeteState}
\end{eqnarray}%
where $\left\vert \psi _{\omega }\right\rangle =(\left\vert 0\right\rangle
+\omega \left\vert 1\right\rangle )/\sqrt{2}$, with $\omega =1,i,-,-i$. The $%
\left\vert \psi _{\omega }\right\rangle $ are superpositions of zero and one
boson states and consequently the local particle number SSR is violated by
each of the sub-system density operators $\left\vert \psi _{\omega
}\right\rangle \left\langle \psi _{\omega }\right\vert )_{A}$ and $%
\left\vert \psi _{\omega }\right\rangle \left\langle \psi _{\omega
}\right\vert )_{B}$. On the other hand, the global particle number SSR is
obeyed since the density operator can also be wriiten as 
\begin{eqnarray}
\widehat{\rho } &=&\frac{1}{4}(\left\vert 0\right\rangle \left\langle
0\right\vert )_{A}\otimes \left\vert 0\right\rangle \left\langle
0\right\vert )_{B}+\frac{1}{4}(\left\vert 1\right\rangle \left\langle
1\right\vert )_{A}\otimes \left\vert 1\right\rangle \left\langle
1\right\vert )_{B}  \nonumber \\
&&+\frac{1}{2}(\left\vert \Psi _{+}\right\rangle \left\langle \Psi
_{+}\right\vert )_{AB}  \label{Eq.VerstraeteState2}
\end{eqnarray}%
where $\left\vert \Psi _{+}\right\rangle _{AB}=(\left\vert 0\right\rangle
_{A}\left\vert 1\right\rangle _{B}+\left\vert 1\right\rangle _{A}\left\vert
0\right\rangle _{B})/\sqrt{2}$. This is a statistical mixture of $N=0,1,2$
boson states. Although the expression in Eq.(\ref{Eq.VerstraeteState}) is of
the form in Eq.(\ref{Eq.NonEntangledState}), the subsystem density operators 
$\left\vert \psi _{\omega }\right\rangle \left\langle \psi _{\omega
}\right\vert )_{A}$ and $\left\vert \psi _{\omega }\right\rangle
\left\langle \psi _{\omega }\right\vert )_{B}$ do not comply with the local
particle number SSR, so this paper the state would be regarded as entangled.
However, Verstraete et al \cite{Verstraete03a}, \cite{Schuch04a} regard it
as separable. They would call it separable but nonlocal. However, Eq.(\ref%
{Eq.VerstraeteState2}) indicates that the state could be prepared as a mixed
state containing two terms that comply with the local particle number SSR in
each of the sub-systems plus a term which is an entangled state of the two
sub-systems. The presence of an entangled state in such an obvious
preparation process challenges the description of the state as being
separable.

To further illustrate some of the points made about super-selection rules -
local and global - it is useful to consider a specific case also presented
by Verstraete et al \cite{Verstraete03a}, \cite{Schuch04a}. This \emph{%
mixture} of \emph{two mode coherent states }is represented by the two mode
density operator%
\begin{eqnarray}
\widehat{\rho } &=&\tint \frac{d\theta }{2\pi }\,\left\vert \alpha ,\alpha
\right\rangle \left\langle \alpha ,\alpha \right\vert  \nonumber \\
&=&\tint \frac{d\theta }{2\pi }\,\left( \left\vert \alpha \right\rangle
\left\langle \alpha \right\vert \right) _{A}\otimes \left( \left\vert \alpha
\right\rangle \left\langle \alpha \right\vert \right) _{B}
\label{Eq.TwoModeCoherentStateMixture}
\end{eqnarray}%
where $\left\vert \alpha \right\rangle _{C}$ is a one mode coherent state
for mode $C=A,B$ with $\alpha =|\alpha |\,\exp (-i\theta )$, and modes $A,B$
are associated with bosonic annihilation operators $\widehat{a}$, $\widehat{b%
}$. The magnitude $|\alpha |$ is fixed.

This density operator \emph{appears} to be that for a non-entangled state of
modes $A,B$ in the form%
\begin{equation}
\widehat{\rho }=\tsum\limits_{R}P_{R}\,\widehat{\rho }_{R}^{A}\otimes 
\widehat{\rho }_{R}^{B}  \label{Eq.NonEntState}
\end{equation}%
with $\tsum\limits_{R}P_{R}\rightarrow \tint \frac{d\theta }{2\pi }$ and $%
\widehat{\rho }_{R}^{A}\rightarrow \left( \left\vert \alpha \right\rangle
\left\langle \alpha \right\vert \right) _{A}$ and $\widehat{\rho }%
_{R}^{B}\rightarrow \left( \left\vert \alpha \right\rangle \left\langle
\alpha \right\vert \right) _{B}$. However although this choice of $\widehat{%
\rho }_{R}^{A}$, $\widehat{\rho }_{R}^{B}$ satisfy the Hermitiancy, unit
trace, positivity features they do \emph{not} conform to the requirement of
satisfying the (\emph{local}) sub-system boson number \emph{super-selection
rule}. From Eq. (\ref{Eq.TwoModeCoherentStateMixture}) we have%
\begin{eqnarray}
\left\langle n\right\vert \left( \left\vert \alpha \right\rangle
\left\langle \alpha \right\vert \right) \left\vert m\right\rangle _{A}
&=&\exp (-|\alpha |^{2})\frac{\alpha ^{n}}{\sqrt{n!}}\frac{(\alpha )^{\ast
\,m}}{\sqrt{m!}}  \nonumber \\
\left\langle p\right\vert \left( \left\vert \alpha \right\rangle
\left\langle \alpha \right\vert \right) \left\vert q\right\rangle _{B}
&=&\exp (-|\alpha |^{2})\frac{\alpha ^{p}}{\sqrt{p!}}\frac{(\alpha )^{\ast
\,q}}{\sqrt{q!}}  \label{Eq.MatrixElements}
\end{eqnarray}%
so clearly for each of the separate modes there are \emph{coherences}
between Fock states with differing boson occupation numbers. In the approach
in the present paper the density operator in Eq. (\ref%
{Eq.TwoModeCoherentStateMixture}) does \emph{not} represent a non-entangled
state. However, in the papers of Verstraete et al \cite{Verstraete03a}, \cite%
{Schuch04a}, Hillery et al \cite{Hillery06a}, \cite{Hillery09a} and others
it would represent an allowable non-entangled (separable) state. Indeed,
Verstraete et al \cite{Verstraete03a} specifically state ".., this state is 
\emph{obviously} separable, though the states $\left\vert \alpha
\right\rangle $ are incompatible with the (local) super-selection rule.".
Verstraete et al \cite{Verstraete03a} introduce the state defined in Eq. (%
\ref{Eq.TwoModeCoherentStateMixture}) as an example of a state that is
separable (in their terms) but which cannot be prepared locally, because it
is incompatible with the local particle number super-selection rule.

The \emph{mixture} of \emph{two mode coherent states} does of course satisfy
the \emph{total} or \emph{global} boson number super-selection rule. The
matrix elements between two mode Fock states are%
\begin{eqnarray}
(\left\langle n\right\vert _{A}\otimes \left\langle p\right\vert _{B})\,%
\widehat{\rho }\,(\left\vert m\right\rangle _{A}\otimes \left\vert
q\right\rangle _{B}) &=&\exp (-2|\alpha |^{2})\frac{|\alpha |^{n+m}}{\sqrt{n!%
}\sqrt{m!}}\frac{|\alpha |^{p+q}}{\sqrt{p!}\sqrt{q!}}\tint \frac{d\theta }{%
2\pi }\,\exp (-i(n-m+p-q)\theta )  \nonumber \\
&=&\exp (-2|\alpha |^{2})\frac{|\alpha |^{n+m}}{\sqrt{n!}\sqrt{m!}}\frac{%
|\alpha |^{p+q}}{\sqrt{p!}\sqrt{q!}}\,\delta _{n+p,m+q}
\label{Eq.OverallMatrixElements}
\end{eqnarray}%
These overall matrix elements are zero unless $n+p=m+q$, showing that there
are \emph{no coherences} between two mode Fock states where the total boson
number differs. The mixture of two mode coherent states has the interesting
feature of providing an example of a two mode state which satisfies the
global but not the local super-selection rule.

The \emph{reduced density operators} for modes $A,B$ are 
\[
\widehat{\rho }_{A}=\tint \frac{d\theta }{2\pi }\,\left( \left\vert \alpha
\right\rangle \left\langle \alpha \right\vert \right) _{A}\qquad \widehat{%
\rho }_{B}=\tint \frac{d\theta }{2\pi }\,\left( \left\vert \alpha
\right\rangle \left\langle \alpha \right\vert \right) _{B} 
\]%
and a straightforward calculation gives 
\[
\widehat{\rho }_{A}=\exp (-|\alpha |^{2})\dsum\limits_{n}\frac{|\alpha |^{2n}%
}{n!}\left( \left\vert n\right\rangle \left\langle n\right\vert \right)
_{A}\qquad \widehat{\rho }_{B}=\exp (-|\alpha |^{2})\dsum\limits_{p}\frac{%
|\alpha |^{2p}}{p!}\left( \left\vert p\right\rangle \left\langle
p\right\vert \right) _{B} 
\]%
which are statistical mixtures of Fock states with the expected Poisson
distribution associated with coherent states. This shows that the reduced
density operators \emph{are} consistent with the separate mode local
super-selection rule, whereas the density operators $\widehat{\rho }%
_{R}^{A}=\left( \left\vert \alpha \right\rangle \left\langle \alpha
\right\vert \right) _{A}$ , $\widehat{\rho }_{R}^{B}$ $=\left( \left\vert
\alpha \right\rangle \left\langle \alpha \right\vert \right) _{B}$ are \emph{%
not} . Later we will revisit this example in the context of entanglement
tests.

Note that if a twirling operation (see Eq.(\ref{Eq.BECDensityOpr3})) were to
be applied to mode $A$, the result would be equivalent to applying two
independent twirling operations to each mode. In this case the density
operator for each mode is a Poisson statistical mixture of number states, so
each mode has a density operator that complies with the local particle
number SSR.

\subsubsection{General Form of Non-Entangled States}

To summarise: basically the sub-systems are \emph{single modes} that the
identical bosons can occupy, the super-selection rule for identical bosons,
massive or otherwise, prohibits states which are coherent superpositions of
states with different numbers of bosons, and the only physically allowable $%
\widehat{\rho }_{R}^{A}$, $\widehat{\rho }_{R}^{B}$, ..for the separate mode
sub-systems that are themselves compatiible with the local particle number
SSR are allowed. For single mode sub-systems these can be written as
statistical mixtures of states with definite numbers of bosons in the form 
\begin{equation}
\widehat{\rho }_{R}^{A}=\sum_{n_{A}}P_{n_{A}}^{A}\left\vert
n_{A}\right\rangle \left\langle n_{A}\right\vert \qquad \widehat{\rho }%
_{R}^{B}=\sum_{n_{B}}P_{n_{B}}^{B}\left\vert n_{B}\right\rangle \left\langle
n_{B}\right\vert \qquad ..  \label{Eq.PhysicalStatesSubSys}
\end{equation}

However, in cases where the sub-systems are \emph{pairs of modes} the
density operators $\widehat{\rho }_{R}^{A}$, $\widehat{\rho }_{R}^{B}$,
..for the separate sub-systems are still required to conform to the
symmetrisation principle and the super-selection rule. The forms for $%
\widehat{\rho }_{R}^{A}$, $\widehat{\rho }_{R}^{B}$, .. are now of course
more complex, as entanglement \emph{within} the pairs of modes $A_{1}$, $%
A_{2}$ associated with sub-system $A$, the pairs of modes $B_{1}$, $B_{2}$
associated with sub-system $B$, etc is now possible within the definition
for the general non-entangled state Eq. (\ref{Eq.NonEntangledState}) for
these \emph{pairs} of modes. Within each pair of modes $A_{1}$, $A_{2}$
statistical mixtures of states with differing total numbers $N_{A}$ bosons
in the two modes are possible and the sub-system density operators are based
on states of the form given in Eq. (\ref{Eq.EntangledTwoModePureState}). We
have%
\begin{eqnarray}
\left\vert \Phi _{N_{A}}\right\rangle _{A}
&=&\tsum\limits_{k=0}^{N_{A}}C_{A\Phi }(N_{A},k)\,\left\vert k\right\rangle
_{A_{1}}\otimes \left\vert N_{A}-k\right\rangle _{A_{2}}  \nonumber \\
\widehat{\rho }_{R}^{A} &=&\dsum\limits_{N_{A}=0}^{\infty
}\dsum\limits_{\Phi }P_{\Phi N_{A}}\left\vert \Phi _{N_{A}}\right\rangle
_{A}\left\langle \Phi _{N_{A}}\right\vert _{A}
\label{Eq.PhysStateSubSystPairA}
\end{eqnarray}%
with analogous expressions for the density operators $\widehat{\rho }%
_{R}^{B} $ etc for the other pairs of modes. Note that $\left\vert \Phi
_{N_{A}}\right\rangle _{A}$ only involves quantum superpositions of states
with the same total number of bosons $N_{A\text{. }}$The expression (\ref%
{Eq.GeneralDensityOprModePair}) in SubSection \ref{SubSection - Sorensen
2001} is of this form.

\subsection{Bipartite Systems}

We now consider the bipartite case where there are just two sub-systems
involved. The simplest case is where each sub-system involves only a single
mode, such as for two modes in a double well potential when only a single
hyperfine state is involve. Another important case is where each sub-system
contains two modes, such as in the double well case where modes with two
different hyperfine states are involved.

\subsubsection{Two Single Modes - Coherence Terms}

The general non-entangled state for modes $\widehat{a}$ and $\widehat{b}$ is
given by 
\begin{equation}
\widehat{\rho }=\sum_{R}P_{R}\,\widehat{\rho }_{R}^{A}\otimes \widehat{\rho }%
_{R}^{B}
\end{equation}%
and as a consequence of the requirement that $\widehat{\rho }_{R}^{A}$ and $%
\widehat{\rho }_{R}^{B}$ are physical states for modes $\widehat{a}$ and $%
\widehat{b}$ satisying the super-selection rule, it follows that 
\begin{eqnarray}
\left\langle (\widehat{a})^{n}\right\rangle _{a} &=&Tr(\widehat{\rho }%
_{R}^{A}(\widehat{a})^{n})=0\qquad \left\langle (\widehat{a}^{\dag
})^{n}\right\rangle _{a}=Tr(\widehat{\rho }_{R}^{A}(\widehat{a}^{\dag
})^{n})=0  \nonumber \\
\left\langle (\widehat{b})^{m}\right\rangle _{b} &=&Tr(\widehat{\rho }%
_{R}^{B}(\widehat{b})^{m})=0\qquad \left\langle (\widehat{b}^{\dag
})^{m}\right\rangle _{b}=Tr(\widehat{\rho }_{R}^{B}(\widehat{b}^{\dag
})^{m})=0  \nonumber \\
&&  \label{Eq.CondNonEntStateAB}
\end{eqnarray}%
Thus coherence terms are zero. As we will see these results will limit spin
squeezing to entangled states of modes $\widehat{a}$ and $\widehat{b}$. Note
that similar results also apply when non-entangled states for the original
modes $\widehat{c}$ and $\widehat{d}$ are considered - $\left\langle (%
\widehat{c})^{n}\right\rangle _{c}=0$, etc..

\subsubsection{Two Pairs of Modes - Coherence Terms}

\label{SubSystem - Two SubSystems of Pairs}

In this case the general non-entangled state where $A$ and $B$ are pairs of
modes - $A_{1}$, $A_{2}$ associated with sub-system $A$, and modes $B_{1}$, $%
B_{2}$ associated with sub-system $B$, the overall density operator is of
the form (\ref{Eq.NonEntStateModesCD}), with $C\rightarrow A$, $D\rightarrow
B$, whilst the sub-system density operators are of the forms given in (\ref%
{Eq.PhysStateSubSystPairA}). In this case we now have in general%
\begin{eqnarray}
\left\langle (\widehat{a}_{i})^{n}\right\rangle _{A} &=&Tr(\widehat{\rho }%
_{R}^{A}(\widehat{a}_{i})^{n})\neq 0\qquad \left\langle (\widehat{a}%
_{i}^{\dag })^{n}\right\rangle _{A}=Tr(\widehat{\rho }_{R}^{A}(\widehat{a}%
_{i}^{\dag })^{n})\neq 0  \nonumber \\
\left\langle (\widehat{b}_{j})^{m}\right\rangle _{B} &=&Tr(\widehat{\rho }%
_{R}^{B}(\widehat{b}_{j})^{m})\neq 0\qquad \left\langle (\widehat{b}%
_{j}^{\dag })^{m}\right\rangle _{B}=Tr(\widehat{\rho }_{R}^{B}(\widehat{b}%
_{j}^{\dag })^{m})\neq 0  \nonumber \\
i,j &=&1,2  \label{Eq.CondNonEntStatePairs}
\end{eqnarray}%
so unlike the case where the two sub-systems are single modes, there are
non-zero coherences when they are pairs of modes.\pagebreak

\section{Spin Squeezing}

\label{Section - Spin Squeezing}

The basic concept of spin squeezing was first introduced by Kitagawa and
Ueda \cite{Kitagawa93a} for general spin systems. These include cases based
on two mode systems, such as may occur both for optical fields and for
Bose-Einstein condensates. Though focused on systems of massive identical
bosons, the treatment in this paper also applies to photons though details
will differ.

\subsection{Spin Operators, Bloch Vector and Covariance Matrix}

\label{SubSection - Spin Operators}

\subsubsection{Spin Operators}

For two mode systems with mode annihilation operators $\widehat{a}$, $%
\widehat{b}$ associated with the two single particle states $\left\vert \phi
_{a}\right\rangle $, $\left\vert \phi _{b}\right\rangle $, and where the
non-zero bosonic commutation rules are $[\widehat{e},\widehat{e}^{\dag }]=%
\widehat{1}$ ($\widehat{e}=\widehat{a}$ or $\widehat{b}$), Schwinger \emph{%
spin angular momentum operators} $\widehat{S}_{\xi }$ ($\xi =x,y,z$) are
defined as 
\begin{equation}
\widehat{S}_{x}=(\widehat{b}^{\dag }\widehat{a}+\widehat{a}^{\dag }\widehat{b%
})/2\qquad \widehat{S}_{y}=(\widehat{b}^{\dag }\widehat{a}-\widehat{a}^{\dag
}\widehat{b})/2i\qquad \widehat{S}_{z}=(\widehat{b}^{\dag }\widehat{b}-%
\widehat{a}^{\dag }\widehat{a})/2  \label{Eq.OldSpinOprs}
\end{equation}%
and which satisfy the commutation rules $[\widehat{S}_{\xi }$ $,\widehat{S}%
_{\mu }$ $]=i\epsilon _{\xi \mu \lambda }\widehat{S}_{\lambda }$ for angular
momentum operators. For bosons the square of the angular momentum operators
is given by $\widehat{S}_{x}^{2}+\widehat{S}_{y}^{2}+\widehat{S}_{z}^{2}=(%
\widehat{N}/2)(\widehat{N}/2+1)$, where $\widehat{N}=(\widehat{b}^{\dag }%
\widehat{b}+\widehat{a}^{\dag }\widehat{a})$ is the boson total number
operator, those for the separate modes being $\widehat{n}_{e}=\widehat{e}%
^{\dag }\widehat{e}$ ($\widehat{e}=\widehat{a}$ or $\widehat{b}$). The
Schwinger spin operators are the second quantization form of symmetrized one
body operators $\widehat{S}_{x}=\sum_{i}(\left\vert \phi
_{b}(i)\right\rangle \left\langle \phi _{a}(i)\right\vert +\left\vert \phi
_{a}(i)\right\rangle \left\langle \phi _{b}(i)\right\vert )/2$ ; $\widehat{S}%
_{y}=\sum_{i}(\left\vert \phi _{b}(i)\right\rangle \left\langle \phi
_{a}(i)\right\vert -\left\vert \phi _{a}(i)\right\rangle \left\langle \phi
_{b}(i)\right\vert )/2i$ ; $\widehat{S}_{z}=\sum_{i}(\left\vert \phi
_{b}(i)\right\rangle \left\langle \phi _{b}(i)\right\vert -\left\vert \phi
_{a}(i)\right\rangle \left\langle \phi _{a}(i)\right\vert )/2$ , where the
sum $i$ is over the identical bosonic particles. In the case of the two mode
EM\ field the spin angular momentum operators are related to the Stokes
parameters.

\subsubsection{Bloch Vector and Covariance Matrix}

If the density operator for the overall system is $\widehat{\rho }$ then
expectation values of the three spin operators $\left\langle \widehat{S}%
_{\xi }\right\rangle =Tr(\widehat{\rho }\widehat{S}_{\xi })$ ($\xi =x,y,z$)
define the \emph{Bloch vector}. Spin squeezing is related to the fluctuation
operators $\Delta \widehat{S}_{\xi }=\widehat{S}_{\xi }-\left\langle 
\widehat{S}_{\xi }\right\rangle $, in terms of which a real, symmetric \emph{%
covariance matrix} $C(\widehat{S}_{\xi },\widehat{S}_{\mu })$ ($\xi ,\mu
=x,y,z$) is defined \cite{Jaaskelainen06a}, \cite{Dalton12a} via%
\begin{eqnarray}
C(\widehat{S}_{\xi },\widehat{S}_{\mu }) &=&(\left\langle \Delta \widehat{S}%
_{\xi }\,\Delta \widehat{S}_{\mu }\right\rangle +\left\langle \Delta 
\widehat{S}_{\mu }\,\Delta \widehat{S}_{\xi }\right\rangle )/2  \nonumber \\
&=&\left\langle \widehat{S}_{\xi }\,\widehat{S}_{\mu }+\widehat{S}_{\mu }\,%
\widehat{S}_{\xi }\right\rangle /2-\left\langle \widehat{S}_{\xi
}\right\rangle \left\langle \widehat{S}_{\mu }\right\rangle
\label{Eq.CovMatrix}
\end{eqnarray}%
and whose diagonal elements $C(\widehat{S}_{\xi },\widehat{S}_{\xi
})=\left\langle \Delta \widehat{S}_{\xi }{}^{2}\right\rangle $ gives the
variance for the fluctuation operators. The variances for the spin operators
satisfy the three Heisenberg uncertainty principle reations $\left\langle
\Delta \widehat{S}_{x}{}^{2}\right\rangle \left\langle \Delta \widehat{S}%
_{y}{}^{2}\right\rangle \geq \frac{1}{4}|\left\langle \widehat{S}%
_{z}\right\rangle |^{2}$; $\left\langle \Delta \widehat{S}%
_{y}{}^{2}\right\rangle \left\langle \Delta \widehat{S}_{z}{}^{2}\right%
\rangle \geq \frac{1}{4}|\left\langle \widehat{S}_{x}\right\rangle |^{2}$; $%
\left\langle \Delta \widehat{S}_{z}{}^{2}\right\rangle \left\langle \Delta 
\widehat{S}_{x}{}^{2}\right\rangle \geq \frac{1}{4}|\left\langle \widehat{S}%
_{y}\right\rangle |^{2}$, and spin squeezing is usually defined via
conditions such as $\left\langle \Delta \widehat{S}_{x}{}^{2}\right\rangle <%
\frac{1}{2}|\left\langle \widehat{S}_{z}\right\rangle |$ with $\left\langle
\Delta \widehat{S}_{y}{}^{2}\right\rangle >\frac{1}{2}|\left\langle \widehat{%
S}_{z}\right\rangle |,$ for $\widehat{S}_{x}{}$ being squeezed compared to $%
\widehat{S}_{y}$ and so on. However this definition is unsatisfactory since
it ignores the presence of the off-diagonal elements of the covariance
matrix, so a better definition is required.

\subsection{New Spin Operators and Principal Spin Fluctuations}

The covariance matrix has real, non-negative eigenvalues and can be
diagonalised via an orthogonal \emph{rotation matrix} $M(-\alpha ,-\beta
,-\gamma )$ that defines \emph{new spin angular momentum operators} $%
\widehat{J}_{\xi }$ ($\xi =x,y,z$) via 
\begin{equation}
\widehat{J}_{\xi }=\sum_{\mu }M_{\xi \mu }(-\alpha ,-\beta ,-\gamma )%
\widehat{S}_{\mu }  \label{Eq.NewSpinOprsRotnMatrix}
\end{equation}%
and where 
\begin{eqnarray}
C(\widehat{J}_{\xi },\widehat{J}_{\mu }) &=&\sum_{\lambda \theta }M_{\xi
\lambda }(-\alpha ,-\beta ,-\gamma )C(\widehat{S}_{\lambda },\widehat{S}%
_{\theta })M_{\mu \theta }(-\alpha ,-\beta ,-\gamma )  \nonumber \\
&=&\delta _{\xi \mu }\left\langle \Delta \widehat{J}_{\xi
}{}^{2}\right\rangle  \label{Eq.DiagnCovMatrix}
\end{eqnarray}%
is the covariance matrix for the new spin angular momentum operators $%
\widehat{J}_{\xi }$ ($\xi =x,y,z$), and which is \emph{diagonal} with the
diagonal elements $\left\langle \Delta \widehat{J}_{x}{}^{2}\right\rangle
,\left\langle \Delta \widehat{J}_{y}{}^{2}\right\rangle $ and $\left\langle
\Delta \widehat{J}_{z}{}^{2}\right\rangle $ giving the so-called \emph{%
principal spin fluctuations}. The matrix $M(\alpha ,\beta ,\gamma )$ is
parameterised in terms of three Euler angles $\alpha ,\beta ,\gamma $ and is
given in \cite{Rose57a} (see Eq. (4.43)).

The Bloch vector and spin fluctuations are illustrated in Figure 1. In Fig 1
the Bloch vector and spin fluctuation ellipsoid is shown in terms of the
original spin operators $\widehat{S}_{\xi }$ ($\xi =x,y,z$)

\bigskip

\begin{center}
Figure 1 near here.

\bigskip
\end{center}

\subsection{Spin Squeezing for New Spin Operators}

\label{SubSection - Spin Squeezing New Spin Oprs}

\subsubsection{Heisenberg Uncertainty Principle and Spin Squeezing}

Since the new spin operators also satisfy \emph{Heisenberg uncertainty
principle} relationships 
\begin{eqnarray}
\left\langle \Delta \widehat{J}_{x}{}^{2}\right\rangle \left\langle \Delta 
\widehat{J}_{y}{}^{2}\right\rangle &\geq &\frac{1}{4}|\left\langle \widehat{J%
}_{z}\right\rangle |^{2}  \nonumber \\
\left\langle \Delta \widehat{J}_{y}{}^{2}\right\rangle \left\langle \Delta 
\widehat{J}_{z}{}^{2}\right\rangle &\geq &\frac{1}{4}|\left\langle \widehat{J%
}_{x}\right\rangle |^{2}  \nonumber \\
\left\langle \Delta \widehat{J}_{z}{}^{2}\right\rangle \left\langle \Delta 
\widehat{J}_{x}{}^{2}\right\rangle &\geq &\frac{1}{4}|\left\langle \widehat{J%
}_{y}\right\rangle |^{2}  \label{Eq.HeisenbergUncert}
\end{eqnarray}%
\emph{spin squeezing} will now be defined via condtions such as 
\begin{eqnarray}
\left\langle \Delta \widehat{J}_{x}{}^{2}\right\rangle &<&\frac{1}{2}%
|\left\langle \widehat{J}_{z}\right\rangle |\;and\;\left\langle \Delta 
\widehat{J}_{y}{}^{2}\right\rangle >\frac{1}{2}|\left\langle \widehat{J}%
_{z}\right\rangle |  \nonumber \\
\left\langle \Delta \widehat{J}_{y}{}^{2}\right\rangle &<&\frac{1}{2}%
|\left\langle \widehat{J}_{x}\right\rangle |\;and\;\left\langle \Delta 
\widehat{J}_{z}{}^{2}\right\rangle >\frac{1}{2}|\left\langle \widehat{J}%
_{x}\right\rangle |  \nonumber \\
\left\langle \Delta \widehat{J}_{z}{}^{2}\right\rangle &<&\frac{1}{2}%
|\left\langle \widehat{J}_{y}\right\rangle |\;and\;\left\langle \Delta 
\widehat{J}_{x}{}^{2}\right\rangle >\frac{1}{2}|\left\langle \widehat{J}%
_{y}\right\rangle |  \label{Eq.SpinSqueezingJXJY}
\end{eqnarray}%
for $\widehat{J}_{x}{}$being squeezed compared to $\widehat{J}_{y}$, and so
on. By convention we may choose $\left\langle \Delta \widehat{J}%
_{x}{}^{2}\right\rangle \leq \left\langle \Delta \widehat{J}%
_{y}{}^{2}\right\rangle \leq $ $\left\langle \Delta \widehat{J}%
_{z}{}^{2}\right\rangle $, so the primary spin operator of interest will be $%
\widehat{J}_{x}$ since this has the smallest fluctuation. Note that here we
have chosen principal spin fluctuations, but of course the last Heisenberg
uncertainty relations apply for \emph{any} new choice of rotated spin
operators - as occurs in the next part of this section.

\subsubsection{Alternative Spin Squeezing Criteria}

\emph{Other criteria} for spin squeezing are also used, for example in the
article by Wineland et al \cite{Wineland94a} . To focus on spin squeezing
for $\widehat{J}_{x}{}$compared to \emph{any} orthogonal spin operators we
can combine the first and third Heisenberg uncertainty principle
relationships to give%
\begin{equation}
\left\langle \Delta \widehat{J}_{x}{}^{2}\right\rangle \left( \left\langle
\Delta \widehat{J}_{y}{}^{2}\right\rangle +\left\langle \Delta \widehat{J}%
_{z}{}^{2}\right\rangle \right) \geq \frac{1}{4}\left( |\left\langle 
\widehat{J}_{y}\right\rangle |^{2}+|\left\langle \widehat{J}%
_{z}\right\rangle |^{2}\right)  \label{Eq.AdditionalHeisebergUncertainty}
\end{equation}%
Then we may define two new spin operators via%
\begin{equation}
\widehat{J}_{\perp \,1}=\cos \theta \;\widehat{J}_{y}+\sin \theta \;\widehat{%
J}_{z}\qquad \widehat{J}_{\perp \,2}=-\sin \theta \;\widehat{J}_{y}+\cos
\theta \;\widehat{J}_{z}  \label{Eq.NewOrthogSpinOprs}
\end{equation}%
where $\theta $ corresponds to a rotation angle in the $yz$ plane, and which
satisfy the standard angular momentun commutation rules $[\widehat{J}_{\perp
\,1}$ $,\widehat{J}_{\perp \,2}]=i\widehat{J}_{x}$, $[\widehat{J}_{\perp
\,2} $ $,\widehat{J}_{x}$ $]=i\widehat{J}_{\perp \,1}$, $[\widehat{J}_{x}$ $,%
\widehat{J}_{\perp \,1}]=i\widehat{J}_{\perp \,2}$. It is straightforward to
show that $\left\langle \Delta \widehat{J}_{y}{}^{2}\right\rangle
+\left\langle \Delta \widehat{J}_{z}{}^{2}\right\rangle =\left\langle \Delta 
\widehat{J}_{\perp \,1}{}^{2}\right\rangle +\left\langle \Delta \widehat{J}%
_{\perp \,2}{}^{2}\right\rangle $ and $|\left\langle \widehat{J}_{\perp
\,1}\right\rangle |^{2}+|\left\langle \widehat{J}_{\perp \,2}\right\rangle
|^{2}=$ $|\left\langle \widehat{J}_{y}\right\rangle |^{2}+|\left\langle 
\widehat{J}_{z}\right\rangle |^{2}$ so that 
\begin{equation}
\left\langle \Delta \widehat{J}_{x}{}^{2}\right\rangle \left( \left\langle
\Delta \widehat{J}_{\perp \,1}{}^{2}\right\rangle +\left\langle \Delta 
\widehat{J}_{\perp \,2}{}^{2}\right\rangle \right) \geq \frac{1}{4}\left(
|\left\langle \widehat{J}_{\perp \,1}\right\rangle |^{2}+|\left\langle 
\widehat{J}_{\perp \,2}\right\rangle |^{2}\right)
\label{Eq.NewHeisenbergUncertaintyPpl}
\end{equation}%
so that \emph{spin squeezing} for $\widehat{J}_{x}{}$compared to \emph{any
two }orthogonal spin operators such as $\widehat{J}_{\perp \,1}$ or $%
\widehat{J}_{\perp \,2}$ would be defined as 
\begin{eqnarray}
\left\langle \Delta \widehat{J}_{x}{}^{2}\right\rangle &<&\frac{1}{2}\sqrt{%
\left( |\left\langle \widehat{J}_{\perp \,1}\right\rangle
|^{2}+|\left\langle \widehat{J}_{\perp \,2}\right\rangle |^{2}\right) } 
\nonumber \\
&&and  \nonumber \\
\left\langle \Delta \widehat{J}_{\perp \,1}{}^{2}\right\rangle +\left\langle
\Delta \widehat{J}_{\perp \,2}{}^{2}\right\rangle &>&\frac{1}{2}\sqrt{\left(
|\left\langle \widehat{J}_{\perp \,1}\right\rangle |^{2}+|\left\langle 
\widehat{J}_{\perp \,2}\right\rangle |^{2}\right) }
\label{Eq.NewCriterionSpinSqueezing}
\end{eqnarray}%
This criterion would apply however the choice of rotation matrix $M(-\alpha
,-\beta ,-\gamma )$ is made, so $\Delta \widehat{J}_{x}$ does not have to
correspond to the principal spin fluctuation with the smallest variance
though obviously such a choice is preferable over some arbitrary set of new
spin operators. For spin squeezing in $\left\langle \Delta \widehat{J}%
_{x}{}^{2}\right\rangle $ we require 
\begin{equation}
\xi ^{2}=\frac{\left\langle \Delta \widehat{J}_{x}{}^{2}\right\rangle }{%
\left( |\left\langle \widehat{J}_{\perp \,1}\right\rangle
|^{2}+|\left\langle \widehat{J}_{\perp \,2}\right\rangle |^{2}\right) }<%
\frac{1}{2\sqrt{\left( |\left\langle \widehat{J}_{\perp \,1}\right\rangle
|^{2}+|\left\langle \widehat{J}_{\perp \,2}\right\rangle |^{2}\right) }}\sim 
\frac{1}{N}  \label{Eq.SpinSqueezingMeasure}
\end{equation}%
The last step is an approximation based on the assumption that the Bloch
vector lies in the $yz$ plane and close to the Bloch sphere, this situation
being the most conducive to detecting the fluctuation $\left\langle \Delta 
\widehat{J}_{x}{}^{2}\right\rangle $. In \ this situation $\sqrt{\left(
|\left\langle \widehat{J}_{\perp \,1}\right\rangle |^{2}+|\left\langle 
\widehat{J}_{\perp \,2}\right\rangle |^{2}\right) }$ is approximately $N/2$.
The condition $\xi ^{2}<1/N$ is sometimes taken as the condition for spin
squeezing \cite{Toth09a}, but it should be noted that this is approximate
and Eq. (\ref{Eq.NewCriterionSpinSqueezing}) gives the correct expression.

\subsubsection{Planar Spin Squeezing}

A special case of recent interest is that referred to as \emph{planar
squeezing} \cite{He11b} in which the Bloch vector for a suitable choice of
spin operators lies in a \emph{plane} and along one of the \emph{axes}. If
this plane is chosen to be the $xy$ plane and the $x$ axis is chosen then $%
\left\langle \widehat{J}_{z}\right\rangle =0$ and $\left\langle \widehat{J}%
_{y}\right\rangle =0$, resulting in only one Heisenberg uncertainty
principle relationship where the right side is non-zero, namely $%
\left\langle \Delta \widehat{J}_{y}{}^{2}\right\rangle \left\langle \Delta 
\widehat{J}_{z}{}^{2}\right\rangle \geq \frac{1}{4}|\left\langle \widehat{J}%
_{x}\right\rangle |^{2}$. Combining this with $\left\langle \Delta \widehat{J%
}_{x}{}^{2}\right\rangle \left\langle \Delta \widehat{J}_{y}{}^{2}\right%
\rangle \geq 0$ gives $\left( \left\langle \Delta \widehat{J}%
_{y}{}^{2}\right\rangle +\left\langle \Delta \widehat{J}_{x}{}^{2}\right%
\rangle \right) \left\langle \Delta \widehat{J}_{z}{}^{2}\right\rangle \geq 
\frac{1}{4}|\left\langle \widehat{J}_{x}\right\rangle |^{2}$. So the total
spin fluctuation \emph{in} the $xy$ plane defined as $\left\langle \Delta 
\widehat{J}_{{\LARGE \Vert }}{}^{2}\right\rangle =\left\langle \Delta 
\widehat{J}_{y}{}^{2}\right\rangle +\left\langle \Delta \widehat{J}%
_{x}{}^{2}\right\rangle $ will be squeezed compared to the total spin
fluctuation perpendicular to the $xy$ plane given by $\left\langle \Delta 
\widehat{J}_{{\LARGE \bot }}{}^{2}\right\rangle =\left\langle \Delta 
\widehat{J}_{z}{}^{2}\right\rangle $ if 
\begin{equation}
\left\langle \Delta \widehat{J}_{{\LARGE \Vert }}{}^{2}\right\rangle <\frac{1%
}{2}|\left\langle \widehat{J}_{x}\right\rangle |\;and\;\left\langle \Delta 
\widehat{J}_{{\LARGE \bot }}{}^{2}\right\rangle >\frac{1}{2}|\left\langle 
\widehat{J}_{x}\right\rangle |  \label{Eq.PlanarSpinSqg}
\end{equation}%
By minimising $\left\langle \Delta \widehat{J}_{{\LARGE \Vert }%
}{}^{2}\right\rangle $ whilst satisfying the constraints $\left\langle 
\widehat{J}_{z}\right\rangle =\left\langle \widehat{J}_{y}\right\rangle =0$
a spin squeezed state is found that satisfies (\ref{Eq.PlanarSpinSqg}) with $%
\left\langle \Delta \widehat{J}_{{\LARGE \Vert }}{}^{2}\right\rangle \;%
\symbol{126}\;J^{2/3}$, $\left\langle \Delta \widehat{J}_{{\LARGE \bot }%
}{}^{2}\right\rangle \;\symbol{126}\;J^{4/3}$, $|\left\langle \widehat{J}%
_{x}\right\rangle |\;\symbol{126}\;J$ for large $J=N/2$ \cite{He11b}. The
Bloch vector is on the Bloch sphere and condition (\ref%
{Eq.SpinSqueezingMeasure}) is also satisfied.

\subsection{Rotation Operators and New Modes}

\label{SubSection - Rotation Operators}

\subsubsection{Rotation Operators}

The new spin operators are also related to the original spin operators via a 
\emph{unitary rotation operator }$\widehat{R}(\alpha ,\beta ,\gamma )$
parameterised in terms of Euler angles so that 
\begin{equation}
\widehat{J}_{\xi }=\widehat{R}(\alpha ,\beta ,\gamma )\,\widehat{S}_{\xi }\,%
\widehat{R}(\alpha ,\beta ,\gamma )^{-1}  \label{Eq.UnitaryTransfnSpinOprs}
\end{equation}%
where 
\begin{equation}
\widehat{R}(\alpha ,\beta ,\gamma )=\widehat{R}_{z}(\alpha )\widehat{R}%
_{y}(\beta )\widehat{R}_{z}(\gamma )  \label{Eq.UnitaryRotnOpr}
\end{equation}%
with $\widehat{R}_{\xi }(\phi )=\exp (i\phi \widehat{S}_{\xi })$ describing
a rotation about the $\xi $ axis anticlockwise through an angle $\phi $.
Details for the rotation operators and matrices are set out in \cite%
{Dalton12a}. Note that Eq. (\ref{Eq.UnitaryTransfnSpinOprs}) specifies a
rotation of the vector spin operator rather than a rotation of the axes, so $%
\widehat{J}_{\xi }$ ($\xi =x,y,z$) are the components of the rotated vector
spin operator with respect to the original axes.

\subsubsection{New Mode Operators}

We can also see that the new spin operators are related to \emph{new mode
operators} $\widehat{c}$ and $\widehat{d}$ via 
\begin{equation}
\widehat{J}_{x}=(\widehat{d}^{\dag }\widehat{c}+\widehat{c}^{\dag }\widehat{d%
})/2\qquad \widehat{J}_{y}=(\widehat{d}^{\dag }\widehat{c}-\widehat{c}^{\dag
}\widehat{d})/2i\qquad \widehat{J}_{z}=(\widehat{d}^{\dag }\widehat{d}-%
\widehat{c}^{\dag }\widehat{c})/2  \label{Eq.NewSpinOprs}
\end{equation}%
where 
\begin{equation}
\widehat{c}=\widehat{R}(\alpha ,\beta ,\gamma )\,\widehat{a}\,\widehat{R}%
(\alpha ,\beta ,\gamma )^{-1}\qquad \widehat{d}=\widehat{R}(\alpha ,\beta
,\gamma )\,\widehat{b}\,\widehat{R}(\alpha ,\beta ,\gamma )^{-1}
\label{Eq.UnitaryTransfnModeOprs}
\end{equation}

For the bosonic case a straight-forward calculation gives the new mode
operators as 
\begin{eqnarray}
\widehat{c} &=&\exp (\frac{1}{2}i\gamma )\left( \cos (\frac{\beta }{2}%
)\,\exp (\frac{1}{2}i\alpha )\,\widehat{a}+\sin (\frac{\beta }{2})\,\exp (-%
\frac{1}{2}i\alpha )\,\widehat{b}\right)  \nonumber \\
\widehat{d} &=&\exp (-\frac{1}{2}i\gamma )\left( -\sin (\frac{\beta }{2}%
)\,\exp (\frac{1}{2}i\alpha )\,\widehat{a}+\cos (\frac{\beta }{2})\,\exp (-%
\frac{1}{2}i\alpha )\,\widehat{b}\right)  \nonumber \\
&&  \label{Eq.NewModeOprs}
\end{eqnarray}%
and it is easy to then check that $\widehat{c}$ and $\widehat{d}$ satisfy
the expected non-zero bosonic commutation rules are $[\widehat{e},\widehat{e}%
^{\dag }]=\widehat{1}$ ($\widehat{e}=\widehat{c}$ or $\widehat{d}$) and that
the \emph{total boson number operator} is $\widehat{N}=(\widehat{d}^{\dag }%
\widehat{d}+\widehat{c}^{\dag }\widehat{c})$. As $\widehat{N}$ is invariant
under unitary rotation operators it follows that $\widehat{J}_{x}^{2}+%
\widehat{J}_{y}^{2}+\widehat{J}_{z}^{2}=(\widehat{N}/2)(\widehat{N}/2+1)$.

\subsubsection{New Modes}

The new mode operators correspond to \emph{new single particle states} $%
\left\vert \phi _{c}\right\rangle $, $\left\vert \phi _{d}\right\rangle $
where 
\begin{eqnarray}
\left\vert \phi _{c}\right\rangle &=&\exp (-\frac{1}{2}i\gamma )\left( \cos (%
\frac{\beta }{2})\,\exp (-\frac{1}{2}i\alpha )\,\left\vert \phi
_{a}\right\rangle +\sin (\frac{\beta }{2})\,\exp (\frac{1}{2}i\alpha
)\,\left\vert \phi _{b}\right\rangle \right)  \nonumber \\
\left\vert \phi _{d}\right\rangle &=&\exp (\frac{1}{2}i\gamma )\left( -\sin (%
\frac{\beta }{2})\,\exp (-\frac{1}{2}i\alpha )\,\left\vert \phi
_{a}\right\rangle +\cos (\frac{\beta }{2})\,\exp (\frac{1}{2}i\alpha
)\,\left\vert \phi _{b}\right\rangle \right)  \nonumber \\
&&  \label{Eq.NewModes}
\end{eqnarray}%
These are two orthonormal quantum superpositions of the original single
particle states $\left\vert \phi _{a}\right\rangle $, $\left\vert \phi
_{b}\right\rangle $, and as such represent an \emph{alternative choice} of
modes that could be realised experimentally.

Eqs. (\ref{Eq.NewModeOprs}) can be inverted to give the old mode operators
via 
\begin{eqnarray}
\widehat{a} &=&\exp (-\frac{1}{2}i\alpha )\left( \cos (\frac{\beta }{2}%
)\,\exp (-\frac{1}{2}i\gamma )\,\widehat{c}-\sin (\frac{\beta }{2})\,\exp (+%
\frac{1}{2}i\gamma )\,\widehat{d}\right)  \nonumber \\
\widehat{b} &=&\exp (+\frac{1}{2}i\alpha )\left( \sin (\frac{\beta }{2}%
)\,\exp (\frac{1}{2}i\gamma )\,\widehat{c}+\cos (\frac{\beta }{2})\,\exp (-%
\frac{1}{2}i\gamma )\,\widehat{d}\right)  \nonumber \\
&&  \label{Eq.OldModeOperators}
\end{eqnarray}

\subsection{Old and New Modes - Coherence Terms}

For our two-mode case we have also seen that the original choice of modes
with annihilation operators $\widehat{a}$ and $\widehat{b}$ may be replaced
by new modes with annihilation operators $\widehat{c}$ and $\widehat{d}$.
Since the new modes are associated with new spin operators $\widehat{J}_{\xi
}$ ($\xi =x,y,z$) for which the covariance matrix is diagonal and where the
diagonal elements give the variances that are relevant for the definition of
spin squeezing, it is therefore more relevant to consider entanglement for
he case where the sub-systems are modes $\widehat{c}$ and $\widehat{d}$,
rather than $\widehat{a}$ and $\widehat{b}$. Consequently the general
non-entangled state for modes $\widehat{c}$ and $\widehat{d}$ is given by 
\begin{equation}
\widehat{\rho }=\sum_{R}P_{R}\,\widehat{\rho }_{R}^{C}\otimes \widehat{\rho }%
_{R}^{D}  \label{Eq.NonEntStateModesCD}
\end{equation}%
and as a consequence of the requirement that $\widehat{\rho }_{R}^{C}$ and $%
\widehat{\rho }_{R}^{D}$ are physical states for modes $\widehat{c}$ and $%
\widehat{d}$ satisying the super-selection rule, it follows that 
\begin{eqnarray}
\left\langle (\widehat{c})^{n}\right\rangle _{c} &=&Tr(\widehat{\rho }%
_{R}^{C}(\widehat{c})^{n})=0\qquad \left\langle (\widehat{c}^{\dag
})^{n}\right\rangle _{c}=Tr(\widehat{\rho }_{R}^{C}(\widehat{c}^{\dag
})^{n})=0  \nonumber \\
\left\langle (\widehat{d})^{m}\right\rangle _{d} &=&Tr(\widehat{\rho }%
_{R}^{D}(\widehat{d})^{m})=0\qquad \left\langle (\widehat{d}^{\dag
})^{m}\right\rangle _{d}=Tr(\widehat{\rho }_{R}^{D}(\widehat{d}^{\dag
})^{m})=0  \nonumber \\
&&  \label{Eq.CondNonEntStateCD}
\end{eqnarray}%
Thus coherence terms are zero. As we will see these results will limit spin
squeezing to entangled states of modes $\widehat{c}$ and $\widehat{d}$.

\subsection{Quantum Correlation Functions and Spin Measurements}

Finally, we note that the principal spin fluctuations can be related to 
\emph{quantum correlation functions}. For example, it is easy to show that%
\begin{eqnarray}
\left\langle \Delta \widehat{J}_{x}{}^{2}\right\rangle &=&\frac{1}{4}\left(
\left\langle (\widehat{d}^{\dag })^{2}(\widehat{c})^{2}\right\rangle
+\left\langle (\widehat{c}^{\dag })^{2}(\widehat{d})^{2}\right\rangle
+2\left\langle \widehat{d}^{\dag }\widehat{c}^{\dag }\widehat{c}\widehat{d}%
\right\rangle +\left\langle \widehat{d}^{\dag }\widehat{d}\right\rangle
+\left\langle \widehat{c}^{\dag }\widehat{c}\right\rangle \right)  \nonumber
\\
&&-\frac{1}{4}\left( \left\langle (\widehat{d}^{\dag }\widehat{c}%
\right\rangle ^{2}+\left\langle (\widehat{c}^{\dag }\widehat{d}\right\rangle
^{2}+2\left\langle (\widehat{d}^{\dag }\widehat{c}\right\rangle \left\langle
(\widehat{c}^{\dag }\widehat{d}\right\rangle \right)  \label{Eq.QCFReln}
\end{eqnarray}%
showing that $\left\langle \Delta \widehat{J}_{x}{}^{2}\right\rangle $ is
related to various first and second order quantum correlation functions.
These can be measured experimentally and are given theoretically in terms of
phase space integrals involving distribution functions to represent the
density operator and phase space variables to represent the mode
annihilation, creation operators.\pagebreak

\section{Spin Squeezing as a Test for Entanglement}

\label{Section - Relationship Spin Squeezing & Entanglement}

With the general non-entangled state now required to be such that the
density operators for the individual sub-systems must represent physical
states and conform to the super-selection rule, the consequential link
between entanglement in two mode bosonic systems and spin squeezing can now
be established. We first consider spin squeezing for the principal spin
operators $\widehat{J}_{x}$, $\widehat{J}_{y},$ $\widehat{J}_{z}$ and
entangled states of the related new modes $\widehat{c}$, $\widehat{d}$ and
then spin squeezing for the original spin operators $\widehat{S}_{x}$, $%
\widehat{S}_{y},$ $\widehat{S}_{z}$ and entangled states of the original
modes $\widehat{a}$, $\widehat{b}$. Examples of entangled states that are
not spin squeezed and states that are not entangled nor spin squeezed for
one choice of mode sub-systems, but are entangled and spin squeezed for
another choice are then presented.

\subsection{Spin Squeezing Requires Entanglement - New Modes}

Firstly, the \emph{variance} for a Hermitian operator $\widehat{\Omega }$ in
a mixed state 
\begin{equation}
\widehat{\rho }=\sum_{R}P_{R}\,\widehat{\rho }_{R}  \label{Eq.MixedState}
\end{equation}%
is always greater than or equal to the the average of the variances for the
separate components 
\begin{equation}
\left\langle \Delta \widehat{\Omega }\,^{2}\right\rangle \geq
\sum_{R}P_{R}\,\left\langle \Delta \widehat{\Omega }{}^{2}\right\rangle _{R}
\label{Eq.VarianceResult}
\end{equation}%
where $\left\langle \Delta \widehat{\Omega }\,^{2}\right\rangle =Tr(\widehat{%
\rho }\,\Delta \widehat{\Omega }\,^{2})$ with $\Delta \widehat{\Omega }=%
\widehat{\Omega }-\left\langle \widehat{\Omega }\right\rangle $ and $%
\left\langle \Delta \widehat{\Omega }\,^{2}\right\rangle _{R}=Tr(\widehat{%
\rho }_{R}\,\Delta \widehat{\Omega }_{R}\,^{2})$ with $\Delta \widehat{%
\Omega }_{R}=\widehat{\Omega }-\left\langle \widehat{\Omega }\right\rangle
_{R}$ . The proof is straight-forward and given in Ref. \cite{Hoffmann03a}.

\subsubsection{Cases of $\widehat{J}_{x}$ and $\widehat{J}_{y}$}

Next we calculate $\left\langle \Delta \widehat{J}\,_{x}^{2}\right\rangle
_{R}$, $\left\langle \Delta \widehat{J}\,_{y}^{2}\right\rangle _{R}$ and $%
\left\langle \widehat{J}_{x}\right\rangle _{R}$, $\left\langle \widehat{J}%
_{y}\right\rangle _{R}$, $\left\langle \widehat{J}_{z}\right\rangle _{R}$
for the case where $\widehat{\rho }_{R}=$ $\widehat{\rho }_{R}^{c}\otimes 
\widehat{\rho }_{R}^{d}$. From Eqs. (\ref{Eq.NewSpinOprs}) we find that%
\begin{eqnarray}
\widehat{J}\,_{x}^{2} &=&\frac{1}{4}((\widehat{d}^{\dag })^{2}(\widehat{c}%
)^{2}+\widehat{d}^{\dag }\widehat{d}\widehat{c}\widehat{c}^{\dag }+\widehat{c%
}^{\dag }\widehat{c}\widehat{d}\widehat{d}^{\dag }+(\widehat{d})^{2}(%
\widehat{c}^{\dag })^{2})  \nonumber \\
\widehat{J}\,_{y}^{2} &=&-\frac{1}{4}((\widehat{d}^{\dag })^{2}(\widehat{c}%
)^{2}-\widehat{d}^{\dag }\widehat{d}\widehat{c}\widehat{c}^{\dag }-\widehat{c%
}^{\dag }\widehat{c}\widehat{d}\widehat{d}^{\dag }+(\widehat{d})^{2}(%
\widehat{c}^{\dag })^{2})  \label{Eq.SquareNewSpinOprs}
\end{eqnarray}%
so that on taking the trace with $\widehat{\rho }_{R}$ and using Eqs. (\ref%
{Eq.CondNonEntStateCD}) we get after applying the commutation rules $[%
\widehat{e},\widehat{e}^{\dag }]=\widehat{1}$ ($\widehat{e}=\widehat{c}$ or $%
\widehat{d}$) 
\begin{eqnarray}
\left\langle \widehat{J}\,_{x}^{2}\right\rangle _{R} &=&\frac{1}{4}%
(\left\langle \widehat{d}^{\dag }\widehat{d}\right\rangle _{R}+\left\langle 
\widehat{c}^{\dag }\widehat{c}\right\rangle _{R})+\frac{1}{2}(\left\langle 
\widehat{c}^{\dag }\widehat{c}\right\rangle _{R}\left\langle \widehat{d}%
^{\dag }\widehat{d}\right\rangle _{R})  \nonumber \\
\left\langle \widehat{J}\,_{y}^{2}\right\rangle _{R} &=&\frac{1}{4}%
(\left\langle \widehat{d}^{\dag }\widehat{d}\right\rangle _{R}+\left\langle 
\widehat{c}^{\dag }\widehat{c}\right\rangle _{R})+\frac{1}{2}(\left\langle 
\widehat{c}^{\dag }\widehat{c}\right\rangle _{R}\left\langle \widehat{d}%
^{\dag }\widehat{d}\right\rangle _{R})
\label{Eq.MeanSquareNewSpinXYProdState}
\end{eqnarray}

As we also have%
\begin{eqnarray}
\left\langle \widehat{J}\,_{x}\right\rangle _{R} &=&\frac{1}{2}(\left\langle 
\widehat{d}^{\dag }\right\rangle _{R}\left\langle \widehat{c}\right\rangle
_{R}+\left\langle \widehat{c}^{\dag }\right\rangle _{R}\left\langle \widehat{%
d}\right\rangle _{R})=0  \nonumber \\
\left\langle \widehat{J}\,_{y}\right\rangle _{R} &=&\frac{1}{2i}%
(\left\langle \widehat{d}^{\dag }\right\rangle _{R}\left\langle \widehat{c}%
\right\rangle _{R}-\left\langle \widehat{c}^{\dag }\right\rangle
_{R}\left\langle \widehat{d}\right\rangle _{R})=0
\label{Eq.MeanNewSpinXYProdState}
\end{eqnarray}%
using Eqs. (\ref{Eq.CondNonEntStateCD}) and we see finally that the
variances are%
\begin{eqnarray}
\left\langle \Delta \widehat{J}\,_{x}^{2}\right\rangle _{R} &=&\frac{1}{4}%
(\left\langle \widehat{d}^{\dag }\widehat{d}\right\rangle _{R}+\left\langle 
\widehat{c}^{\dag }\widehat{c}\right\rangle _{R})+\frac{1}{2}(\left\langle 
\widehat{c}^{\dag }\widehat{c}\right\rangle _{R}\left\langle \widehat{d}%
^{\dag }\widehat{d}\right\rangle _{R})  \nonumber \\
\left\langle \Delta \widehat{J}\,_{y}^{2}\right\rangle _{R} &=&\frac{1}{4}%
(\left\langle \widehat{d}^{\dag }\widehat{d}\right\rangle _{R}+\left\langle 
\widehat{c}^{\dag }\widehat{c}\right\rangle _{R})+\frac{1}{2}(\left\langle 
\widehat{c}^{\dag }\widehat{c}\right\rangle _{R}\left\langle \widehat{d}%
^{\dag }\widehat{d}\right\rangle _{R})
\label{Eq.VariancesNewSpinXYProdState}
\end{eqnarray}%
and therefore from Eq. (\ref{Eq.VarianceResult}) 
\begin{eqnarray}
\left\langle \Delta \widehat{J}\,_{x}^{2}\right\rangle &\geq &\sum_{R}P_{R}\,%
\frac{1}{4}(\left\langle \widehat{d}^{\dag }\widehat{d}\right\rangle
_{R}+\left\langle \widehat{c}^{\dag }\widehat{c}\right\rangle _{R})+\frac{1}{%
2}(\left\langle \widehat{c}^{\dag }\widehat{c}\right\rangle _{R}\left\langle 
\widehat{d}^{\dag }\widehat{d}\right\rangle _{R})  \nonumber \\
\left\langle \Delta \widehat{J}\,_{y}^{2}\right\rangle &\geq &\sum_{R}P_{R}\,%
\frac{1}{4}(\left\langle \widehat{d}^{\dag }\widehat{d}\right\rangle
_{R}+\left\langle \widehat{c}^{\dag }\widehat{c}\right\rangle _{R})+\frac{1}{%
2}(\left\langle \widehat{c}^{\dag }\widehat{c}\right\rangle _{R}\left\langle 
\widehat{d}^{\dag }\widehat{d}\right\rangle _{R})
\label{Eq.InequalityXYVariancesNonEntState}
\end{eqnarray}

Now 
\begin{equation}
\left\langle \widehat{J}\,_{z}\right\rangle =\sum_{R}P_{R}\,\frac{1}{2}%
(\left\langle \widehat{d}^{\dag }\widehat{d}\right\rangle _{R}-\left\langle 
\widehat{c}^{\dag }\widehat{c}\right\rangle _{R}))  \label{Eq.MeanNewSpinZ}
\end{equation}%
so that 
\begin{equation}
\frac{1}{2}|\left\langle \widehat{J}\,_{z}\right\rangle |\leq \sum_{R}P_{R}\,%
\frac{1}{4}|(\left\langle \widehat{d}^{\dag }\widehat{d}\right\rangle
_{R}-\left\langle \widehat{c}^{\dag }\widehat{c}\right\rangle _{R}))|\leq
\sum_{R}P_{R}\,\frac{1}{4}(\left\langle \widehat{d}^{\dag }\widehat{d}%
\right\rangle _{R}+\left\langle \widehat{c}^{\dag }\widehat{c}\right\rangle
_{R}))  \label{Eq.InequalityMeanNewSpinZEntState}
\end{equation}%
and thus for any non-entangled state of modes $\widehat{c}$ and $\widehat{d}$
\begin{eqnarray}
&&\left\langle \Delta \widehat{J}\,_{x}^{2}\right\rangle -\frac{1}{2}%
|\left\langle \widehat{J}\,_{z}\right\rangle |  \nonumber \\
&\geq &\sum_{R}P_{R}\,\frac{1}{4}(\left\langle \widehat{d}^{\dag }\widehat{d}%
\right\rangle _{R}+\left\langle \widehat{c}^{\dag }\widehat{c}\right\rangle
_{R})+\frac{1}{2}(\left\langle \widehat{c}^{\dag }\widehat{c}\right\rangle
_{R}\left\langle \widehat{d}^{\dag }\widehat{d}\right\rangle
_{R})-\sum_{R}P_{R}\,\frac{1}{4}(\left\langle \widehat{d}^{\dag }\widehat{d}%
\right\rangle _{R}+\left\langle \widehat{c}^{\dag }\widehat{c}\right\rangle
_{R}))  \nonumber \\
&\geq &\sum_{R}P_{R}\,\frac{1}{2}(\left\langle \widehat{c}^{\dag }\widehat{c}%
\right\rangle _{R}\left\langle \widehat{d}^{\dag }\widehat{d}\right\rangle
_{R})  \nonumber \\
&\geq &0  \label{Eq.NonSpinSqResult}
\end{eqnarray}%
Similar final steps show that $\left\langle \Delta \widehat{J}%
\,_{y}^{2}\right\rangle -\frac{1}{2}|\left\langle \widehat{J}%
\,_{z}\right\rangle |\geq 0$ for any non-entangled state of modes $\widehat{c%
}$ and $\widehat{d}$.

This shows that for the general non-entangled state with modes $\widehat{c}$
and $\widehat{d}$ as the sub-systems, the variances for two of the principal
spin fluctuations $\left\langle \Delta \widehat{J}\,_{x}^{2}\right\rangle $
and $\left\langle \Delta \widehat{J}\,_{y}^{2}\right\rangle $ are \emph{both}
greater than $\frac{1}{2}|\left\langle \widehat{J}\,_{z}\right\rangle |$,
and hence there is no spin squeezing for $\widehat{J}_{x}$ or $\widehat{J}%
_{y}$. Note that as $|\left\langle \widehat{J}\,_{y}\right\rangle |=0$, the
quantity $\sqrt{\left( |\left\langle \widehat{J}_{\perp \,1}\right\rangle
|^{2}+|\left\langle \widehat{J}_{\perp \,2}\right\rangle |^{2}\right) }$ is
the same as $|\left\langle \widehat{J}\,_{z}\right\rangle |$, so the
alternative criterion in Eq. (\ref{Eq.NewCriterionSpinSqueezing}) is the
same as that in Eq. (\ref{Eq.SpinSqueezingJXJY}) which is used here.

We can extend the above to obtain further inequalities for the non-entangled
state. Using Eq. (\ref{Eq.MeanNewSpinXYProdState})%
\begin{equation}
\left\langle \widehat{J}\,_{x}\right\rangle =\sum_{R}P_{R}\left\langle 
\widehat{J}\,_{x}\right\rangle _{R}=0\qquad \left\langle \widehat{J}%
\,_{y}\right\rangle =\sum_{R}P_{R}\left\langle \widehat{J}%
\,_{y}\right\rangle _{R}=0  \label{Eq.MeanNewSpinXY}
\end{equation}%
it is easy to see that 
\begin{equation}
\left\langle \Delta \widehat{J}\,_{x}^{2}\right\rangle -\frac{1}{2}%
|\left\langle \widehat{J}\,_{y}\right\rangle |\geq 0\qquad \left\langle
\Delta \widehat{J}\,_{y}^{2}\right\rangle -\frac{1}{2}|\left\langle \widehat{%
J}\,_{x}\right\rangle |\geq 0  \label{Eq.NonSpinSqResultB}
\end{equation}%
for any non-entangled state of modes $\widehat{c}$ and $\widehat{d}$. This
completes the set of inequalities for the variances of $\widehat{J}_{x}$ and 
$\widehat{J}_{y}$.

\subsubsection{Case of $\widehat{J}_{z}$}

For the other principal spin fluctuation we find that 
\begin{equation}
\left\langle \Delta \widehat{J}\,_{z}^{2}\right\rangle _{R}=\frac{1}{4}%
(\left\langle \left( \widehat{d}^{\dag }\widehat{d}-\left\langle \widehat{d}%
^{\dag }\widehat{d}\right\rangle _{R}\right) \left( \widehat{d}^{\dag }%
\widehat{d}-\left\langle \widehat{d}^{\dag }\widehat{d}\right\rangle
_{R}\right) \right\rangle _{R}+\left\langle \left( \widehat{c}^{\dag }%
\widehat{c}-\left\langle \widehat{c}^{\dag }\widehat{c}\right\rangle
_{R}\right) \left( \widehat{c}^{\dag }\widehat{c}-\left\langle \widehat{c}%
^{\dag }\widehat{c}\right\rangle _{R}\right) \right\rangle _{R}
\label{Eq.VariancesNewSpinZProdState}
\end{equation}%
so that using (\ref{Eq.VarianceResult}) 
\begin{equation}
\left\langle \Delta \widehat{J}\,_{z}^{2}\right\rangle \geq \sum_{R}P_{R}\,%
\frac{1}{4}(\left\langle \left( \widehat{d}^{\dag }\widehat{d}-\left\langle 
\widehat{d}^{\dag }\widehat{d}\right\rangle _{R}\right) ^{2}\right\rangle
_{R}+\left\langle \left( \widehat{c}^{\dag }\widehat{c}-\left\langle 
\widehat{c}^{\dag }\widehat{c}\right\rangle _{R}\right) ^{2}\right\rangle
_{R}  \label{Eq.VariancesNewSpinZNonEntState}
\end{equation}%
From Eq. (\ref{Eq.MeanNewSpinXY}) it follows that 
\begin{eqnarray}
&&\left\langle \Delta \widehat{J}\,_{z}^{2}\right\rangle -\frac{1}{2}%
|\left\langle \widehat{J}\,_{x}\right\rangle |\,  \nonumber \\
&\geq &\sum_{R}P_{R}\,\frac{1}{4}(\left\langle \left( \widehat{d}^{\dag }%
\widehat{d}-\left\langle \widehat{d}^{\dag }\widehat{d}\right\rangle
_{R}\right) ^{2}\right\rangle _{R}+\left\langle \left( \widehat{c}^{\dag }%
\widehat{c}-\left\langle \widehat{c}^{\dag }\widehat{c}\right\rangle
_{R}\right) ^{2}\right\rangle _{R}  \nonumber \\
&\geq &0  \label{Eq.NonSpinSqResultC}
\end{eqnarray}%
Similarly $\left\langle \Delta \widehat{J}\,_{z}^{2}\right\rangle -\frac{1}{2%
}|\left\langle \widehat{J}\,_{y}\right\rangle |\,\geq 0$.

\subsubsection{No Spin Squeezing for Separable States}

So overall, we have for the general non-entangled state of modes $\widehat{c}
$ and $\widehat{d}$%
\begin{eqnarray}
\left\langle \Delta \widehat{J}_{x}{}^{2}\right\rangle &\geq &\frac{1}{2}%
|\left\langle \widehat{J}_{z}\right\rangle |\;and\;\left\langle \Delta 
\widehat{J}_{y}{}^{2}\right\rangle \geq \frac{1}{2}|\left\langle \widehat{J}%
_{z}\right\rangle |  \nonumber \\
\left\langle \Delta \widehat{J}_{y}{}^{2}\right\rangle &\geq &\frac{1}{2}%
|\left\langle \widehat{J}_{x}\right\rangle |\;and\;\left\langle \Delta 
\widehat{J}_{z}{}^{2}\right\rangle \geq \frac{1}{2}|\left\langle \widehat{J}%
_{x}\right\rangle |  \nonumber \\
\left\langle \Delta \widehat{J}_{z}{}^{2}\right\rangle &\geq &\frac{1}{2}%
|\left\langle \widehat{J}_{y}\right\rangle |\;and\;\left\langle \Delta 
\widehat{J}_{x}{}^{2}\right\rangle \geq \frac{1}{2}|\left\langle \widehat{J}%
_{y}\right\rangle |  \label{Eq.CombinedResultNonEntState}
\end{eqnarray}%
Note that the last two pairs of inequalities are trivially true for the
general non-entangled state, since $\left\langle \widehat{J}%
_{x}\right\rangle =\left\langle \widehat{J}_{y}\right\rangle =0$. This
overall result tells us that for \emph{any} non-entangled state of modes $%
\widehat{c}$ and $\widehat{d}$ we do \emph{not} have $\widehat{J}_{x}$ being
squeezed compared to $\widehat{J}_{y}$ (or vice-versa), $\widehat{J}_{y}$
being squeezed compared to $\widehat{J}_{z}$ (or vice-versa), $\widehat{J}%
_{z}$ being squeezed compared to $\widehat{J}_{x}$ (or vice-versa). That is,
there is \emph{no} \emph{spin squeezing} for the non-entangled state!

\subsubsection{Spin Squeezing Tests for Entanglement}

The key value of these results is the \emph{spin squeezing test} for \emph{%
entanglement} - \emph{if} for a given state we find that 
\begin{equation}
If\emph{\quad }\left\langle \Delta \widehat{J}\,_{x}^{2}\right\rangle <\frac{%
1}{2}|\left\langle \widehat{J}\,_{z}\right\rangle |\qquad \emph{or\quad }%
\left\langle \Delta \widehat{J}\,_{y}^{2}\right\rangle <\frac{1}{2}%
|\left\langle \widehat{J}\,_{z}\right\rangle |
\label{Eq.SpinSqueezeEntangleTest}
\end{equation}

or 
\begin{equation}
If\emph{\quad }\left\langle \Delta \widehat{J}\,_{y}^{2}\right\rangle <\frac{%
1}{2}|\left\langle \widehat{J}\,_{x}\right\rangle |\qquad \emph{or\quad }%
\left\langle \Delta \widehat{J}\,_{z}^{2}\right\rangle <\frac{1}{2}%
|\left\langle \widehat{J}\,_{x}\right\rangle |
\label{Eq.SpinSqueezeEntangleTestB}
\end{equation}

or%
\begin{equation}
If\emph{\quad }\left\langle \Delta \widehat{J}\,_{z}^{2}\right\rangle <\frac{%
1}{2}|\left\langle \widehat{J}\,_{y}\right\rangle |\qquad \emph{or\quad }%
\left\langle \Delta \widehat{J}\,_{x}^{2}\right\rangle <\frac{1}{2}%
|\left\langle \widehat{J}\,_{y}\right\rangle |
\label{Eq.SpinSqueezeEntangleTestC}
\end{equation}%
then the state \emph{must} be entangled. Thus we only need to have spin
squeezing in \emph{any} of the $\widehat{J}_{x}$, $\widehat{J}_{y}$ or $%
\widehat{J}_{z}$ to demonstrate entanglement. No particular component need
be singled out. Note that one cannot have both $\left\langle \Delta \widehat{%
J}\,_{x}^{2}\right\rangle <\frac{1}{2}|\left\langle \widehat{J}%
\,_{z}\right\rangle |$ and$\left\langle \Delta \widehat{J}%
\,_{y}^{2}\right\rangle <\frac{1}{2}|\left\langle \widehat{J}%
\,_{z}\right\rangle |$ etc. due to the Heisenberg uncertainty principle.

It is then straightforward to show that 
\begin{equation}
If\emph{\quad }\left\langle \Delta \widehat{J}\,_{x}^{2}\right\rangle <\frac{%
1}{2}\sqrt{|\left\langle \widehat{J}\,_{\perp 1}^{x}\right\rangle
|^{2}+|\left\langle \widehat{J}\,_{\perp 2}^{x}\right\rangle |^{2}}
\label{Eq.SpinSqEntangleTestD}
\end{equation}%
or%
\begin{equation}
If\emph{\quad }\left\langle \Delta \widehat{J}\,_{y}^{2}\right\rangle <\frac{%
1}{2}\sqrt{|\left\langle \widehat{J}\,_{\perp 1}^{y}\right\rangle
|^{2}+|\left\langle \widehat{J}\,_{\perp 2}^{y}\right\rangle |^{2}}
\label{Eq.SpinSqEntangleTestE}
\end{equation}%
or%
\begin{equation}
If\emph{\quad }\left\langle \Delta \widehat{J}\,_{z}^{2}\right\rangle <\frac{%
1}{2}\sqrt{|\left\langle \widehat{J}\,_{\perp 1}^{z}\right\rangle
|^{2}+|\left\langle \widehat{J}\,_{\perp 2}^{z}\right\rangle |^{2}}
\label{Eq.SpinSqEntangleTestF}
\end{equation}%
that is, if $\widehat{J}_{x}$, $\widehat{J}_{y}$ or $\widehat{J}_{z}$ are
squeezed compared to \emph{any} of their two orthogonal spin components -
then the state must be entangled. Again we only need to have spin squeezing
in \emph{any} of the $\widehat{J}_{x}$, $\widehat{J}_{y}$ or $\widehat{J}%
_{z} $ compared to \emph{any} of their two orthogonal spin components to
demonstrate entanglement.

\subsubsection{Inequality for $|\left\langle \widehat{J}\,_{z}\right\rangle
|\,$}

Of the results for a \emph{non-entangled} physical state for modes $\widehat{%
c}$ and $\widehat{d}$ we will later find it particularly important to
consider the first of (\ref{Eq.CombinedResultNonEntState})%
\begin{equation}
\left\langle \Delta \widehat{J}\,_{x}^{2}\right\rangle \geq \frac{1}{2}%
|\left\langle \widehat{J}\,_{z}\right\rangle |\quad and\quad \left\langle
\Delta \widehat{J}\,_{y}^{2}\right\rangle \geq \frac{1}{2}|\left\langle 
\widehat{J}\,_{z}\right\rangle |  \label{Eq.NonEntStateSpinSqCondn}
\end{equation}%
This is because we can show that for any quantum state%
\begin{equation}
|\left\langle \widehat{J}\,_{z}\right\rangle |\,=|\left\langle \frac{1}{2}(%
\widehat{n}_{d}-\widehat{n}_{c})\right\rangle |\,\leq \frac{1}{2}%
(|\left\langle \widehat{n}_{d}\right\rangle |\,+|\left\langle \widehat{n}%
_{c}\right\rangle |)\,=\frac{1}{2}\left\langle \widehat{N}\right\rangle
\label{Eq.GenInequalNJZ}
\end{equation}%
there is an inequality involving $|\left\langle \widehat{J}%
\,_{z}\right\rangle |\,$\ and the mean number of bosons $\left\langle 
\widehat{N}\right\rangle $ in the two mode system. Note that there \emph{may}
be entangled states for which $\left\langle \Delta \widehat{J}%
\,_{x}^{2}\right\rangle $ and $\left\langle \Delta \widehat{J}%
\,_{y}^{2}\right\rangle $ are both greater than $\frac{1}{2}|\left\langle 
\widehat{J}\,_{z}\right\rangle |$, since all that has been proven is that
for non-entangled states we must have \emph{both} $\left\langle \Delta 
\widehat{J}\,_{x}^{2}\right\rangle \geq \frac{1}{2}|\left\langle \widehat{J}%
\,_{z}\right\rangle |$ \emph{and} $\left\langle \Delta \widehat{J}%
\,_{y}^{2}\right\rangle \geq \frac{1}{2}|\left\langle \widehat{J}%
\,_{z}\right\rangle |$.

Hence we may conclude that spin squeezing in either of the principal spin
fluctuations $\widehat{J}_{x}$ , $\widehat{J}_{y}$ or $\widehat{J}_{z}$
requires the quantum state to be entangled for the modes $\widehat{c}$ and $%
\widehat{d}$ as the sub-systems, these modes being associated with the
principal spin fluctuations via Eq. (\ref{Eq.NewSpinOprs}). Although finding
spin squeezing tells us that the state is entangled, there are however no
simple relationships between the measures of entanglement and those of spin
squeezing, so the linkage is essentially a qualitative one. For general
quantum states, measures of entanglement for the specific situation of two
sub-systems (bi-partite entanglement) are reviewed in \cite{Amico08a}. \ 

\subsection{Spin Squeezing Requires Entanglement - Original Modes}

It is also of some interest to consider spin squeezing for the original spin
operators $\widehat{S}_{x}$, $\widehat{S}_{y}$, $\widehat{S}_{z}$ with the
original modes $\widehat{a}$ and $\widehat{b}$ as the sub-systems, even
though these spin operators are in general associated with a non-diagonal
covariance matrix and the concept of spin squeezing is rather problematic in
view of principal spin fluctuations not being involved. In this case the
general non-entangled state for the \emph{original} modes is given by 
\begin{equation}
\widehat{\rho }=\sum_{R}P_{R}\,\widehat{\rho }_{R}^{A}\otimes \widehat{\rho }%
_{R}^{B}  \label{Eq.NonEntangStateModesAB}
\end{equation}%
with the $\widehat{\rho }_{R}^{A}$ and $\widehat{\rho }_{R}^{B}$
representing physical states for modes $\widehat{a}$ and $\widehat{b}$, and
where results analogous to Eqs. (\ref{Eq.CondNonEntStateCD}) apply. The same
treatment applies as for spin operators $\widehat{J}_{x}$, $\widehat{J}_{y}$%
, $\widehat{J}_{z}$ with the modes $\widehat{c}$ and $\widehat{d}$ as the
sub-systems and leads to the result for a \emph{non-entangled} state of
modes $\widehat{a}$ and $\widehat{b}$ 
\begin{equation}
\left\langle \Delta \widehat{S}\,_{x}^{2}\right\rangle \geq \frac{1}{2}%
|\left\langle \widehat{S}\,_{z}\right\rangle |\quad and\quad \left\langle
\Delta \widehat{S}\,_{y}^{2}\right\rangle \geq \frac{1}{2}|\left\langle 
\widehat{S}\,_{z}\right\rangle  \label{Eq.NonSpinSqResultSXSY}
\end{equation}%
showing that neither $\widehat{S}_{x}$ or $\widehat{S}_{y}$ is spin squeezed
for the general non-entangled state for modes $\widehat{a}$ and $\widehat{b}$
given in Eq. (\ref{Eq.NonEntangStateModesAB}). We also have 
\begin{equation}
\left\langle \widehat{S}\,_{x}\right\rangle =\sum_{R}P_{R}\left\langle 
\widehat{S}\,_{x}\right\rangle _{R}=0\qquad \left\langle \widehat{S}%
\,_{y}\right\rangle =\sum_{R}P_{R}\left\langle \widehat{S}%
\,_{y}\right\rangle _{R}=0  \label{Eq.MeanOldSpinXY}
\end{equation}%
so all the results analogous to Eqs. (\ref{Eq.CombinedResultNonEntState})
also follow. Hence we may also conclude that spin squeezing in \emph{any} of
the original spin fluctuations requires the quantum state to be entangled
when the original modes $\widehat{a}$ and $\widehat{b}$ are the sub-systems.
Thus the \emph{entanglement test} is 
\begin{equation}
If\emph{\quad }\left\langle \Delta \widehat{S}\,_{x}^{2}\right\rangle <\frac{%
1}{2}|\left\langle \widehat{S}\,_{z}\right\rangle |\qquad \emph{or\quad }%
\left\langle \Delta \widehat{S}\,_{y}^{2}\right\rangle <\frac{1}{2}%
|\left\langle \widehat{S}\,_{z}\right\rangle |
\label{Eq.SpinSqEntTestOrigModes}
\end{equation}%
or%
\begin{equation}
If\emph{\quad }\left\langle \Delta \widehat{S}\,_{y}^{2}\right\rangle <\frac{%
1}{2}|\left\langle \widehat{S}\,_{x}\right\rangle |\qquad \emph{or\quad }%
\left\langle \Delta \widehat{S}\,_{z}^{2}\right\rangle <\frac{1}{2}%
|\left\langle \widehat{S}\,_{x}\right\rangle |
\label{Eq.SpinSqEntTestOrigModesB}
\end{equation}%
or%
\begin{equation}
If\emph{\quad }\left\langle \Delta \widehat{S}\,_{z}^{2}\right\rangle <\frac{%
1}{2}|\left\langle \widehat{S}\,_{y}\right\rangle |\qquad \emph{or\quad }%
\left\langle \Delta \widehat{S}\,_{x}^{2}\right\rangle <\frac{1}{2}%
|\left\langle \widehat{S}\,_{y}\right\rangle |
\label{Eq.SpinSqEntTestOrigModesC}
\end{equation}%
then we have an entangled state for the original modes $\widehat{a}$ and $%
\widehat{b}$.

Hence we have seen that spin squeezing - either of the new or original spin
operators requires entanglement of the new or original modes - the question
then is: Does entanglement automatically lead to spin squeezing? The answer
is no, since cases where the quantum state is entangled but not spin
squeezed can be found. Thus in general, spin squeezing and entanglement are 
\emph{not equivalent}.- they do not occur \emph{together} for all states.
Spin squeezing is a \emph{sufficient} condition for entanglement, it is not
a \emph{necessary} condition.

\subsection{Entangled States that are Non Spin-Squeezed}

\label{SubSection - Ent State Non Squeezed}

One such example is the generalised $N$ boson \emph{NOON\ state} defined as%
\begin{eqnarray}
\widehat{\rho } &=&\left\vert \Phi \right\rangle \left\langle \Phi
\right\vert  \nonumber \\
\left\vert \Phi \right\rangle &=&\cos \theta \,\frac{(\widehat{c}^{\dag
})^{N}}{\sqrt{N!}}\left\vert 0\right\rangle +\sin \theta \,\frac{(\widehat{d}%
^{\dag })^{N}}{\sqrt{N!}}\left\vert 0\right\rangle  \nonumber \\
&=&\cos \theta \,\left\vert \frac{N}{2},-\frac{N}{2}\right\rangle +\sin
\theta \,\left\vert \frac{N}{2},+\frac{N}{2}\right\rangle
\label{Eq.GeneralNOONState}
\end{eqnarray}%
which is an entangled state for modes $\widehat{c}$ and $\widehat{d}$ in all
cases except where $\cos \theta \,$or $\sin \theta \,$is zero. In the last
form the state is expressed in terms of the eigenstates for $(%
\underrightarrow{\widehat{J}})^{2}\,$and $\widehat{J}\,_{z}$, as detailed in 
\cite{Dalton12a}.

A straight-forward calculation gives 
\begin{eqnarray}
\left\langle \widehat{J}\,_{x}\right\rangle &=&0\qquad \left\langle \widehat{%
J}\,_{y}\right\rangle =0\qquad \left\langle \widehat{J}\,_{z}\right\rangle =-%
\frac{N}{2}\cos 2\theta  \nonumber \\
\left\langle \Delta \widehat{J}\,_{x}^{2}\right\rangle &=&\frac{N}{4}\qquad
\left\langle \Delta \widehat{J}\,_{y}^{2}\right\rangle =\frac{N}{4}\qquad
\left\langle \Delta \widehat{J}\,_{z}^{2}\right\rangle =\frac{N^{2}}{4}%
(1-\cos ^{2}2\theta )  \label{Eq.MenVarianceNOONState}
\end{eqnarray}%
for $N>1$, so that using the criteria for spin squeezing given in Eq. (\ref%
{Eq.SpinSqueezingJXJY}) we see that as $\left\langle \Delta \widehat{J}%
\,_{x}^{2}\right\rangle -\frac{1}{2}|\left\langle \widehat{J}%
\,_{z}\right\rangle |\,\geq 0$, etc, and hence spin squeezing does not occur
for this entangled state.

\subsection{Non-Entangled States that are Non Spin Squeezed}

Of course from the previous section \emph{any} non entangled state is
definitely not spin squeezed. A specific example illustrating this is the $N$
boson binomial state given by 
\begin{eqnarray}
\widehat{\rho } &=&\left\vert \Phi \right\rangle \left\langle \Phi
\right\vert  \nonumber \\
\left\vert \Phi \right\rangle &=&\frac{(-\widehat{c}^{\dag })^{N}}{\sqrt{N!}}%
\left\vert 0\right\rangle  \label{Eq.BinomialState}
\end{eqnarray}%
where $\widehat{c}$ and $\widehat{d}$ are given by Eqs. (\ref{Eq.NewModeOprs}%
) with Euler angles $\alpha =-\pi +\chi $, $\beta =-2\theta $ and $\gamma
=-\pi $, we find that 
\begin{eqnarray}
\widehat{c} &=&-\cos \theta \,\exp (\frac{1}{2}i\chi )\,\widehat{a}-\sin
\theta \,\exp (-\frac{1}{2}i\chi )\,\widehat{b}=-\widehat{a}_{1}  \nonumber
\\
\widehat{d} &=&\sin \theta \,\exp (\frac{1}{2}i\chi )\,\widehat{a}-\cos
\theta \,\exp (-\frac{1}{2}i\chi )\,\widehat{b}=-\widehat{a}_{2}
\label{Eq.SpecialNewModeOprs}
\end{eqnarray}%
where the mode operators $\widehat{a}_{1}$ and $\widehat{a}_{2}$ are as
defined in \cite{Dalton12a} (see Eqs. (53) therein). The new spin angular
momentum operators $\widehat{J}_{\xi }$ ($\xi =x,y,z$) are the same as those
defined in \cite{Dalton12a} (see Eqs. (64) therein) and in \cite{Dalton12a}
it has been shown (see Eq. (60) therein) for the same binomial state as in (%
\ref{Eq.BinomialState}) that 
\begin{eqnarray}
\left\langle \widehat{J}\,_{x}\right\rangle &=&0\qquad \left\langle \widehat{%
J}\,_{y}\right\rangle =0\qquad \left\langle \widehat{J}\,_{z}\right\rangle =-%
\frac{N}{2}  \nonumber \\
\left\langle \Delta \widehat{J}\,_{x}^{2}\right\rangle &=&\frac{N}{4}\qquad
\left\langle \Delta \widehat{J}\,_{y}^{2}\right\rangle =\frac{N}{4}\qquad
\left\langle \Delta \widehat{J}\,_{z}^{2}\right\rangle =0
\label{Eq.MeanVarianceBinomialState}
\end{eqnarray}%
(see Eqs. (162) and (176) therein). Hence the binomial state is not spin
squeezed since $\left\langle \Delta \widehat{J}\,_{x}^{2}\right\rangle
=\left\langle \Delta \widehat{J}\,_{y}^{2}\right\rangle =\frac{1}{2}%
|\left\langle \widehat{J}\,_{z}\right\rangle |.$ It is of course a \emph{%
minimum uncertainty state} with spin fluctuations at the \emph{standard
quantum limit}. Clearly, it is a non-entangled state for modes $\widehat{c}$
and $\widehat{d}$ , being the product of a number state for mode $\widehat{c}
$ with the vacuum state for mode $\widehat{d}$.

Note that from the point of view of the original modes $\widehat{a}$ and $%
\widehat{b}$, this is an entangled state. so the question is: Is it a spin
squeezed state with respect to the original spin operators $\widehat{S}_{\xi
}$ ($\xi =x,y,z$) ? The Bloch vector and variances for this binomial state
are given in \cite{Dalton12a} (see Eq. (163) in the main paper and Eq. (410)
in the Appendix). The results include:

\begin{eqnarray}
\left\langle \widehat{S}\,_{z}\right\rangle &=&-\frac{N}{2}\cos 2\theta 
\nonumber \\
\left\langle \Delta \widehat{S}\,_{x}^{2}\right\rangle &=&\frac{N}{4}(\cos
^{2}2\theta \,\cos ^{2}\chi +\sin ^{2}\chi )\qquad \left\langle \Delta 
\widehat{S}\,_{y}^{2}\right\rangle =\frac{N}{4}(\cos ^{2}2\theta \,\sin
^{2}\chi +\cos ^{2}\chi )  \nonumber \\
&&  \label{Eq.BinomStateOrigSpinOprs}
\end{eqnarray}%
This gives $\left\langle \Delta \widehat{S}\,_{x}^{2}\right\rangle
\left\langle \Delta \widehat{S}\,_{y}^{2}\right\rangle -\frac{1}{4}%
|\left\langle \widehat{S}\,_{z}\right\rangle |^{2}=\frac{1}{16}N^{2}(\cos
^{2}2\theta -1)^{2}\cos ^{2}\chi \sin ^{2}\chi \geq 0$ as required for the
Heisenberg uncertainty principle. With $\chi =0$ we have $\left\langle
\Delta \widehat{S}\,_{x}^{2}\right\rangle =\frac{N}{4}\cos ^{2}2\theta \,$\
and $\left\langle \Delta \widehat{S}\,_{y}^{2}\right\rangle =\frac{N}{4}$,
whilst $\frac{1}{2}|\left\langle \widehat{S}\,_{z}\right\rangle |=\frac{N}{4}%
|\cos 2\theta |$. As $\left\langle \Delta \widehat{S}\,_{x}^{2}\right\rangle
<\frac{1}{2}|\left\langle \widehat{S}\,_{z}\right\rangle |$ there is spin
squeezing in $\widehat{S}\,_{x}$ for this entangled state of modes $\widehat{%
a}$ and $\widehat{b}$, though not of course for the new spin operator $%
\widehat{J}\,_{x}$ since this state is non-entangled for modes $\widehat{c}$
and $\widehat{d}$. This example illustrates the need to carefully define
spin squeezing and entanglement in terms of related sets of spin operators
and modes. The same state is entangled with respect to one choice of modes -
and spin squeezing occurs, whilst it is non-entangled with respect to
another set of modes - and no spin squeezing occurs.

To summarise - with a physically based definition of non-entangled states
for bosonic systems with two modes (related to the principal spin operators
that have a diagonal covariance matrix) being the sub-systems and with a
criterion for spin squeezing that focuses on these principal spin
fluctuations, it seen that whilst non-entangled states are never spin
squeezed and therefore although entanglement is a necessary condition for
spin squeezing, it is not a sufficient one. There are entangled states that
are not spin squeezed. Furthermore, as there is no simple quantitative links
between measures of spin squeezing and those for entanglement, the mere
presence of spin squeezing only demonstrates the qualitative result that the
quantum state is entangled. Nevertheless, for high precision measurements
based on spin operators where the primary emphasis is on how much spin
squeezing can be achieved, knowing that entangled states are needed provides
an impetus for studying such states and how they might be produced.

\subsection{Entangled States that are Spin Squeezed}

\label{SubSection - Ent States that are Spin Sq}

\subsubsection{Relative Phase Eigenstate}

As an example of an entangled state that is spin squeezed we consider the
relative phase eigenstate $\left\vert \frac{N}{2},\theta _{p}\right\rangle $
for a two mode system in which there are $N$ bosons. For modes with
annihilation operators $\widehat{a}$, $\widehat{b}$ the \emph{relative phase
eigenstate} is defined as 
\begin{equation}
\left\vert \frac{N}{2},\theta _{p}\right\rangle =\frac{1}{\sqrt{N+1}}%
\dsum\limits_{k=\,-N/2}^{N/2}\exp (ik\theta _{p})\,\frac{(\widehat{a}^{\dag
})^{N/2-k}}{\sqrt{(N/2-k)!}}\frac{(\widehat{b}^{\dag })^{N/2+k}}{\sqrt{%
(N/2+k)!}}\left\vert 0\right\rangle  \label{Eq.RelativePhaseState}
\end{equation}%
where the relative phase $\theta _{p}=p(2\pi /(N+1))$ with $%
p=-N/2,-N/2+1,...,+N/2$, is an eigenvalue of the relative phase Hermitian
operator of the type introduced by Barnett and Pegg \cite{Barnett89a} (see 
\cite{Dalton12a} and references therein). Note that the eigenvalues form a
quasi-continuum over the range $-\pi $ to $+\pi $, with a small separation
between neighboring phases of $O(1/N)$. The relative phase state is
consistent with the super-selection rule and is an entangled state for modes 
$\widehat{a}$, $\widehat{b}$. The Bloch vector for spin operators $\widehat{S%
}_{x}$, $\widehat{S}_{y}$, $\widehat{S}_{z}$ is given by (see \cite%
{Dalton12a}) 
\begin{equation}
\left\langle \widehat{S}\,_{x}\right\rangle =N\,\frac{\pi }{8}\cos \theta
_{p}\qquad \left\langle \widehat{S}\,_{y}\right\rangle =-N\,\frac{\pi }{8}%
\sin \theta _{p}\qquad \left\langle \widehat{S}\,_{z}\right\rangle =0
\end{equation}%
but the covariance matrix (see Eq. (178) in \cite{Dalton12a}) is
non-diagonal.

\subsubsection{New Spin Operators}

It is more instructive to consider spin squeezing in terms of new spin
operators $\widehat{J}_{x}$, $\widehat{J}_{y}$, $\widehat{J}_{z}$ for which
the covariance matrix is diagonal. The new spin operators are related to the
original spin operators via 
\begin{eqnarray}
\widehat{J}_{x} &=&\widehat{S}_{z}  \nonumber \\
\widehat{J}_{y} &=&\sin \theta _{p}\,\widehat{S}_{x}+\cos \theta _{p}\,%
\widehat{S}_{y}  \nonumber \\
\widehat{J}_{z} &=&-\cos \theta _{p}\,\widehat{S}_{x}+\sin \theta _{p}\,%
\widehat{S}_{y}  \label{Eq.NewSpinOprsPhaseState}
\end{eqnarray}%
corresponding to the transformation in Eq. (\ref{Eq.NewSpinOprsRotnMatrix})
with Euler angles $\alpha =-\pi +\theta _{p}$, $\beta =-\pi /2$ and $\gamma
=-\pi $.

\subsubsection{Bloch Vector and Covariance Matrix}

The Bloch vector and covariance matrix for spin operators $\widehat{J}_{x}$, 
$\widehat{J}_{y}$, $\widehat{J}_{z}$ are given by (see Eqs. (180), (181) in 
\cite{Dalton12a} - note that the $C(\widehat{J}_{y},\widehat{J}_{y})$
element is incorrect in Eq. (181)) 
\begin{equation}
\left\langle \widehat{J}\,_{x}\right\rangle =0\qquad \left\langle \widehat{J}%
\,_{y}\right\rangle =0\qquad \left\langle \widehat{J}\,_{z}\right\rangle
=-N\,\frac{\pi }{8}  \label{Eq.BlochVectorPhaseState}
\end{equation}%
and%
\begin{eqnarray}
&&\left[ 
\begin{tabular}{lll}
$C(\widehat{J}_{x},\widehat{J}_{x})$ & $C(\widehat{J}_{x},\widehat{J}_{y})$
& $C(\widehat{J}_{x},\widehat{J}_{z})$ \\ 
$C(\widehat{J}_{y},\widehat{J}_{x})$ & $C(\widehat{J}_{y},\widehat{J}_{y})$
& $C(\widehat{J}_{y},\widehat{J}_{z})$ \\ 
$C(\widehat{J}_{z},\widehat{J}_{y})$ & $C(\widehat{J}_{z},\widehat{J}_{y})$
& $C(\widehat{J}_{z},\widehat{J}_{z})$%
\end{tabular}%
\right]  \nonumber \\
&\doteqdot &\left[ 
\begin{tabular}{lll}
$\frac{{\LARGE 1}}{{\LARGE 12}}N^{2}$ & $0$ & $0$ \\ 
$0$ & $\frac{1}{4}+\frac{1}{8}\ln N$ & $0$ \\ 
$0$ & $0$ & $\left( \frac{{\LARGE 1}}{{\LARGE 6}}-\frac{{\LARGE \pi }^{2}}{%
{\LARGE 64}}\right) N^{2}$%
\end{tabular}%
\right] \qquad N\gg 1  \label{Eq.CovMatrixPhaseStateNewSpins}
\end{eqnarray}%
With \ $\left\langle \Delta \widehat{J}\,_{x}^{2}\right\rangle =\frac{%
{\LARGE 1}}{{\LARGE 12}}N^{2}$, $\left\langle \Delta \widehat{J}%
\,_{y}^{2}\right\rangle =\frac{1}{4}+\frac{1}{8}\ln N$ and $\left\langle
\Delta \widehat{J}\,_{z}^{2}\right\rangle =\left( \frac{{\LARGE 1}}{{\LARGE 6%
}}-\frac{{\LARGE \pi }^{2}}{{\LARGE 64}}\right) N^{2}$ and the only non-zero
Bloch vector component being $\left\langle \widehat{J}\,_{z}\right\rangle
=-N\,\frac{\pi }{8}$ it is easy to see that $\left\langle \Delta \widehat{J}%
\,_{x}^{2}\right\rangle \left\langle \Delta \widehat{J}\,_{y}^{2}\right%
\rangle \geq \frac{1}{4}|\left\langle \widehat{J}\,_{z}\right\rangle |^{2}$
as required by the Heisenberg Uncertainty Principle. The principal spin
fluctuations in both $\widehat{J}_{x}$ and $\widehat{J}_{z}$ are comparable
to the length of the Bloch vector and no spin squeezing occurs in either of
these components. However, spin squeezing occurs in that $\widehat{J}_{y}$
is squeezed with respect to $\widehat{J}_{x}$ - $\left\langle \Delta 
\widehat{J}\,_{y}^{2}\right\rangle $ only increases as $\frac{1}{8}\ln N$
whilst $\frac{1}{2}|\left\langle \widehat{J}\,_{z}\right\rangle |$ increases
as $\frac{\pi }{16}N\,$\ for large $N$. Hence the relative phase state
satisfies the test in Eq. (\ref{Eq.SpinSqueezeEntangleTestC}) to demonstate
entanglement for modes $\widehat{c}$, $\widehat{d}$.

\subsubsection{New Modes Operators}

To confirm that the relative phase state is in fact an entangled state for
modes $\widehat{c}$, $\widehat{d}$ as well as for the original modes $%
\widehat{a}$, $\widehat{b}$, we note that the new mode operators $\widehat{c}
$, $\widehat{d}$ are given in in Eq. (\ref{Eq.NewModeOprs}) with Euler
angles $\alpha =-\pi +\theta _{p}$, $\beta =-\pi /2$ and $\gamma =-\pi $.
The old mode operators are given in Eq. (\ref{Eq.OldModeOperators}) and with
these Euler angles we have 
\begin{eqnarray}
\widehat{a} &=&-\exp (\frac{1}{2}i\theta _{p})\frac{1}{\sqrt{2}}\left( 
\widehat{c}-\widehat{d}\right)  \nonumber \\
\widehat{b} &=&-\exp (-\frac{1}{2}i\theta _{p})\frac{1}{\sqrt{2}}\left( 
\widehat{c}+\widehat{d}\right)  \nonumber \\
&&  \label{Eq.OldModeOprsPhaseState}
\end{eqnarray}%
This enables us to write the phase state in terms of new mode operators $%
\widehat{c}$, $\widehat{d}$ as 
\begin{eqnarray}
\left\vert \frac{N}{2},\theta _{p}\right\rangle &=&\frac{1}{\sqrt{N+1}}%
\left( \frac{-1}{\sqrt{2}}\right)
^{N}\dsum\limits_{k=\,-N/2}^{N/2}\dsum\limits_{r=\,-N/4+k/2}^{\,N/4-k/2}%
\dsum\limits_{s=\,-N/4-k/2}^{N/4+k/2}  \nonumber \\
&&\times \frac{1}{\sqrt{(N/2-k)!}}\frac{1}{\sqrt{(N/2+k)!}}(-1)^{N/4-k/2+r} 
\nonumber \\
&&\times \frac{(N/2-k)!}{(N/4-k/2-r)!(N/4-k/2+r)!}\frac{(N/2+k)!}{%
(N/4+k/2-s)!(N/4+k/2+s)!}  \nonumber \\
&&\times (\widehat{c}^{\dag })^{N/2-(r+s)}\,(\widehat{d}^{\dag
})^{N/2+(r+s)}\left\vert 0\right\rangle  \nonumber \\
&&  \label{Eq.PhaseStateNewModes}
\end{eqnarray}%
We see that the expression does not depend explicitly on the relative phase $%
\theta _{p}$ when written in terms of the new mode unnormalised Fock states $%
(\widehat{c}^{\dag })^{N/2-(r+s)}\,(\widehat{d}^{\dag
})^{N/2+(r+s)}\left\vert 0\right\rangle $. This pure state is a linear
combination of product states of the form $\left\vert N/2-m\right\rangle
_{c}\otimes \left\vert N/2+m\right\rangle _{d}$ for various $m$ - each of
which is an $N$ boson state and an eigenstate for $\widehat{J}\,_{z}$ with
eigenvalue $m$, and therefore is an entangled state for modes $\widehat{c}$, 
$\widehat{d}$ which is compatible with the global super-selection rule. Note
that there cannot just be a single term $m$ involved, otherwise the variance
for $\widehat{J}\,_{z}$ would be zero instead of $\left( \frac{{\LARGE 1}}{%
{\LARGE 6}}-\frac{{\LARGE \pi }^{2}}{{\LARGE 64}}\right) N^{2}$. We will
return to the relative phase state again in SubSection \ref{SubSection -
Hillery 2006}.

\subsection{Bloch Vector Entanglement Test}

We have seen for the general non-entangled states of modes $\widehat{c}$ and 
$\widehat{d}$ or of modes \ $\widehat{a}$ and $\widehat{b}$ that 
\begin{eqnarray}
\left\langle \widehat{J}\,_{x}\right\rangle &=&0\qquad \left\langle \widehat{%
J}\,_{y}\right\rangle =0  \label{Eq.BlochVectNonEntStateCD} \\
\left\langle \widehat{S}\,_{x}\right\rangle &=&0\qquad \left\langle \widehat{%
S}\,_{y}\right\rangle =0  \label{Eq.BlochVectNonEntStateAB}
\end{eqnarray}%
From Eqs. (\ref{Eq.NewSpinOprs}) and (\ref{Eq.OldSpinOprs}) these results
are equivalent to 
\begin{eqnarray}
\left\langle \widehat{d}\,\widehat{c}^{\dag }\right\rangle &=&0\qquad
\left\langle \widehat{c}\,\widehat{d}^{\dag }\right\rangle =0
\label{Eq.CondNESCD} \\
\left\langle \widehat{b}\,\widehat{a}^{\dag }\right\rangle &=&0\qquad
\left\langle \widehat{a}\,\widehat{b}^{\dag }\right\rangle =0
\label{Eq.CondNESAB}
\end{eqnarray}

Hence we find further \emph{tests} for \emph{entangled states} of modes $%
\widehat{c}$ and $\widehat{d}$ or of modes \ $\widehat{a}$ and $\widehat{b}$%
\begin{eqnarray}
|\left\langle \widehat{d}\,\widehat{c}^{\dag }\right\rangle |^{2}\,
&>&0\qquad |\left\langle \widehat{c}\,\widehat{d}^{\dag }\right\rangle
|^{2}\,>0  \label{Eq.EntangTestModesCD} \\
|\left\langle \widehat{b}\,\widehat{a}^{\dag }\right\rangle |^{2}\,
&>&0\qquad |\left\langle \widehat{a}\,\widehat{b}^{\dag }\right\rangle
|^{2}\,>0  \label{Eq.EntTestsModesAB}
\end{eqnarray}%
As we will see in Section \ref{Section - Criteria for Spin Squeezing Based
on Non-Physical States}, these tests are particular cases with $m=n=1$ of
the simpler entanglement test in Eq. (\ref{Eq.EntangTest}) that applies for
the situation in the present paper where non-entangled states are required
to satisfy the super-selection rule. \pagebreak

\section{Other Proposed Tests for Entanglement}

\label{Section - Criteria for Spin Squeezing Based on Non-Physical States}

There are a number of inequalities involving not only the variances of the
spin operators but also other quantities, that have been derived for testing
whether a state for a system of identical bosons is entangled. These are 
\emph{not} always associated with criteria for spin squeezing. Some of these
are based on the implicit assumption that the density operators $\widehat{%
\rho }_{R}^{A}$, $\widehat{\rho }_{R}^{B}$ in the expression for a
non-entangled state are \emph{not} required to conform to the
super-selection rule that prohibits quantum superpositions of single mode
states with differing numbers of bosons. These results are based in effect
on a \emph{different criterion} as to what constitutes an \emph{entangled
state}, so of course the resulting inequalities will \emph{differ} from
those that would apply if the definition of an entangled state is based on
the considerations presented here in this paper - which emphasise the
requirement that the density operators $\widehat{\rho }_{R}^{A}$, $\widehat{%
\rho }_{R}^{B}$ should represent physical states for the separate modes and
hence satisfy the super-selection rule. Other results are based on forms of
the density operator for non-entangled states that do not satisfy the
symmetrisation principle. In this Section we examine a number of such
entanglement tests and find that some are not valid, though some may be
revised as tests for entangled states defined in accord with symmetrisation
and super-selection rules.

\subsection{Hillery et al 2006}

\label{SubSection - Hillery 2006}

\subsubsection{Hillery Spin Variance Entanglement Test}

One such entanglement test is presented in the paper by Hillery and Zubairy 
\cite{Hillery06a} entitled "Entanglement conditions for two-mode states".
The paper actually dealt with EM\ field modes and the super-selection rule
was not applied, so density operators $\widehat{\rho }_{R}^{A}$, $\widehat{%
\rho }_{R}^{B}$ for photon modes allowed for coherences between states with
differing photon numbers, and hence the conditions in Eq. (\ref%
{Eq.CondNonEntStateCD}) did not apply. However, even for the situation of
EM\ field modes where massless photons are involved, it is argued here that
the super-selection rule also should be applied. Conditions involving the
variances $\left\langle \Delta \widehat{S}\,_{x}\right\rangle ^{2}$, $%
\left\langle \Delta \widehat{S}\,_{y}\right\rangle ^{2}$ can be obtained by
applying similar arguements to those in Section \ref{Section - Criteria for
Spin Squeezing Based on Non-Physical States}. It is found that for the
original spin operators $\widehat{S}_{x}$, $\widehat{S}_{y}$, $\widehat{S}%
_{z}$ and modes $\widehat{a}$ and $\widehat{b}$ 
\begin{eqnarray}
\left\langle \widehat{S}\,_{x}^{2}\right\rangle _{R} &=&\frac{1}{4}%
(\left\langle \widehat{b}^{\dag }\widehat{b}\right\rangle _{R}+\left\langle 
\widehat{a}^{\dag }\widehat{a}\right\rangle _{R})+\frac{1}{2}(\left\langle 
\widehat{a}^{\dag }\widehat{a}\right\rangle _{R}\left\langle \widehat{b}%
^{\dag }\widehat{b}\right\rangle _{R})+\frac{1}{4}(\left\langle (\widehat{b}%
^{\dag })^{2}\right\rangle _{R}\left\langle (\widehat{a})^{2}\right\rangle
_{R}+\left\langle (\widehat{b})^{2}\right\rangle _{R}\left\langle (\widehat{a%
}^{\dag })^{2}\right\rangle )  \nonumber \\
\left\langle \widehat{S}\,_{y}^{2}\right\rangle _{R} &=&\frac{1}{4}%
(\left\langle \widehat{b}^{\dag }\widehat{b}\right\rangle _{R}+\left\langle 
\widehat{a}^{\dag }\widehat{a}\right\rangle _{R})+\frac{1}{2}(\left\langle 
\widehat{a}^{\dag }\widehat{a}\right\rangle _{R}\left\langle \widehat{b}%
^{\dag }\widehat{b}\right\rangle _{R})-\frac{1}{4}(\left\langle (\widehat{b}%
^{\dag })^{2}\right\rangle _{R}\left\langle (\widehat{a})^{2}\right\rangle
_{R}+\left\langle (\widehat{b})^{2}\right\rangle _{R}\left\langle (\widehat{a%
}^{\dag })^{2}\right\rangle )  \nonumber \\
&&
\end{eqnarray}%
where terms such as $\left\langle (\widehat{b}^{\dag })^{2}\right\rangle
_{R} $ and $\left\langle (\widehat{a})^{2}\right\rangle _{R}$ previously
shown to be zero have been retained. Note that in \cite{Hillery06a} the
operators $\widehat{S}_{x}$, $\widehat{S}_{y}$, $\widehat{S}_{z}$
constructed from the EM field mode operators as in Eq. (\ref{Eq.OldSpinOprs}%
) would be related to Stokes parameters Hence 
\begin{eqnarray}
&&\left\langle \widehat{S}\,_{x}^{2}\right\rangle _{R}+\left\langle \widehat{%
S}\,_{y}^{2}\right\rangle _{R}  \nonumber \\
&=&\frac{1}{2}(\left\langle \widehat{b}^{\dag }\widehat{b}\right\rangle
_{R}+\left\langle \widehat{a}^{\dag }\widehat{a}\right\rangle
_{R})+(\left\langle \widehat{a}^{\dag }\widehat{a}\right\rangle
_{R}\left\langle \widehat{b}^{\dag }\widehat{b}\right\rangle _{R})  \nonumber
\\
&=&\frac{1}{2}(\left\langle \widehat{b}^{\dag }\widehat{b}\right\rangle
_{R}(\left\langle \widehat{a}^{\dag }\widehat{a}\right\rangle _{R}+1)+\frac{1%
}{2}(\left\langle \widehat{a}^{\dag }\widehat{a}\right\rangle
_{R}(\left\langle \widehat{b}^{\dag }\widehat{b}\right\rangle _{R}+1))
\end{eqnarray}%
where the terms $\left\langle (\widehat{b}^{\dag })^{2}\right\rangle _{R}$,
..., $\left\langle (\widehat{a}^{\dag })^{2}\right\rangle $ cancel out. This
is the same as before.

However, 
\begin{eqnarray}
\left\langle \widehat{S}\,_{x}\right\rangle _{R} &=&\frac{1}{2}(\left\langle 
\widehat{b}^{\dag }\right\rangle _{R}\left\langle \widehat{a}\right\rangle
_{R}+\left\langle \widehat{a}^{\dag }\right\rangle _{R}\left\langle \widehat{%
b}\right\rangle _{R})  \nonumber \\
\left\langle \widehat{S}\,_{y}\right\rangle _{R} &=&\frac{1}{2i}%
(\left\langle \widehat{b}^{\dag }\right\rangle _{R}\left\langle \widehat{a}%
\right\rangle _{R}-\left\langle \widehat{a}^{\dag }\right\rangle
_{R}\left\langle \widehat{b}\right\rangle _{R})
\end{eqnarray}%
is now non-zero, since the previously zero terms have again been retained.
Hence 
\begin{equation}
\left\langle \widehat{S}\,_{x}\right\rangle _{R}^{2}+\left\langle \widehat{S}%
\,_{y}\right\rangle _{R}^{2}=\left\langle \widehat{b}^{\dag }\right\rangle
_{R}\left\langle \widehat{b}\right\rangle _{R}\left\langle \widehat{a}^{\dag
}\right\rangle _{R}\left\langle \widehat{a}\right\rangle _{R}
\end{equation}%
so that we now have%
\begin{eqnarray}
&&\left\langle \Delta \widehat{S}\,_{x}^{2}\right\rangle _{R}+\left\langle
\Delta \widehat{S}\,_{y}^{2}\right\rangle _{R}  \nonumber \\
&=&\frac{1}{2}(\left\langle \widehat{b}^{\dag }\widehat{b}\right\rangle
_{R}(\left\langle \widehat{a}^{\dag }\widehat{a}\right\rangle _{R}+1)+\frac{1%
}{2}(\left\langle \widehat{a}^{\dag }\widehat{a}\right\rangle
_{R}(\left\langle \widehat{b}^{\dag }\widehat{b}\right\rangle
_{R})+1)-\left\langle \widehat{b}^{\dag }\right\rangle _{R}\left\langle 
\widehat{b}\right\rangle _{R}\left\langle \widehat{a}\right\rangle
_{R}\left\langle \widehat{a}^{\dag }\right\rangle _{R}\left\langle \widehat{a%
}\right\rangle _{R}  \nonumber \\
&=&\frac{1}{2}\left( \left\langle \widehat{b}^{\dag }\widehat{b}%
\right\rangle _{R}+\left\langle \widehat{a}^{\dag }\widehat{a}\right\rangle
_{R}\right) +\left( \left\langle \widehat{b}^{\dag }\widehat{b}\right\rangle
_{R}(\left\langle \widehat{a}^{\dag }\widehat{a}\right\rangle
_{R}-|\left\langle \widehat{a}\right\rangle _{R}|^{2}|\left\langle \widehat{b%
}^{\dag }\right\rangle _{R}|^{2}\right)  \label{Eq.ProductStateInequality}
\end{eqnarray}%
But from the Schwarz inequality - which is based on $\left\langle (\widehat{a%
}^{\dag }-\left\langle \widehat{a}^{\dag }\right\rangle )(\widehat{a}%
-\left\langle \widehat{a}\right\rangle )\right\rangle \geq 0$ for any state 
\begin{equation}
|\left\langle \widehat{a}\right\rangle _{R}|^{2}\leq \left\langle \widehat{a}%
^{\dag }\widehat{a}\right\rangle _{R}\qquad |\left\langle \widehat{b}%
\right\rangle _{R}|^{2}\leq \left\langle \widehat{b}^{\dag }\widehat{b}%
\right\rangle _{R}  \label{Eq.Schwarz}
\end{equation}%
so that 
\begin{equation}
\left\langle \Delta \widehat{S}\,_{x}^{2}\right\rangle _{R}+\left\langle
\Delta \widehat{S}\,_{y}^{2}\right\rangle _{R}\geq \frac{1}{2}(\left\langle 
\widehat{b}^{\dag }\widehat{b}\right\rangle _{R}+\left\langle \widehat{a}%
^{\dag }\widehat{a}\right\rangle _{R})
\end{equation}%
and thus from Eq. (\ref{Eq.VarianceResult}) it follows that for a general
non entangled state 
\begin{equation}
\left\langle \Delta \widehat{S}\,_{x}^{2}\right\rangle +\left\langle \Delta 
\widehat{S}\,_{y}^{2}\right\rangle \geq \sum_{R}P_{R}\frac{1}{2}\left(
\left\langle \widehat{n}_{b}\right\rangle _{R}+\left\langle \widehat{n}%
_{a}\right\rangle _{R}\right)
\end{equation}%
However, half the expectation value of the number operator is 
\begin{equation}
\frac{1}{2}\left\langle \widehat{N}\right\rangle =\frac{1}{2}\left\langle (%
\widehat{n}_{a}+\widehat{n}_{b})\right\rangle =\sum_{R}P_{R}\frac{1}{2}%
\left( \left\langle \widehat{n}_{b}\right\rangle _{R}+\left\langle \widehat{n%
}_{a}\right\rangle _{R}\right)
\end{equation}%
so for a non-entangled state%
\begin{equation}
\left\langle \Delta \widehat{S}\,_{x}^{2}\right\rangle +\left\langle \Delta 
\widehat{S}\,_{y}^{2}\right\rangle \geq \frac{1}{2}\left\langle \widehat{N}%
\right\rangle  \label{Eq.InvalidInequalityVariance}
\end{equation}%
This inequality for non-entangled states is given in \cite{Hillery06a} (see
Eq. (3)). The above proof was based on a different definition of entangled
states to that in this paper.

\subsubsection{Validity of Hillery Test for Local SSR Compatible
Non-Entangled States}

However, it turns out that the same inequality is \emph{also} valid when the
definition of entangled states is the same as in the present paper. We would
then find that $\left\langle \widehat{S}\,_{x}\right\rangle
_{R}=\left\langle \widehat{S}\,_{y}\right\rangle _{R}=0$ and hence 
\begin{equation}
\left\langle \Delta \widehat{S}\,_{x}^{2}\right\rangle _{R}+\left\langle
\Delta \widehat{S}\,_{y}^{2}\right\rangle _{R}=\frac{1}{2}\left(
\left\langle \widehat{b}^{\dag }\widehat{b}\right\rangle _{R}+\left\langle 
\widehat{a}^{\dag }\widehat{a}\right\rangle _{R}\right) +\left( \left\langle 
\widehat{b}^{\dag }\widehat{b}\right\rangle _{R}(\left\langle \widehat{a}%
^{\dag }\widehat{a}\right\rangle _{R}\right)  \label{Eq.NewProductInequality}
\end{equation}%
instead of Eq.(\ref{Eq.ProductStateInequality}). Since the term $%
\left\langle \widehat{b}^{\dag }\widehat{b}\right\rangle _{R}(\left\langle 
\widehat{a}^{\dag }\widehat{a}\right\rangle _{R}$ is always positive we find
after applying Eq. (\ref{Eq.VarianceResult}) that 
\begin{equation}
\left\langle \Delta \widehat{S}\,_{x}^{2}\right\rangle +\left\langle \Delta 
\widehat{S}\,_{y}^{2}\right\rangle \geq \frac{1}{2}\left\langle \widehat{N}%
\right\rangle  \label{Eq.InequalityVariance2}
\end{equation}%
which is the same as in Eq. (\ref{Eq.InvalidInequalityVariance}). Hence,
finding that $\left\langle \Delta \widehat{S}\,_{x}^{2}\right\rangle
+\left\langle \Delta \widehat{S}\,_{y}^{2}\right\rangle <\frac{1}{2}%
\left\langle \widehat{N}\right\rangle $ would show that the state was
entangled, irrespective of whether or not entanglement is defined in terms
of non-physical unentangled states. The Hillery et al \cite{Hillery06a} 
\emph{entanglement test} 
\begin{equation}
\left\langle \Delta \widehat{S}\,_{x}^{2}\right\rangle +\left\langle \Delta 
\widehat{S}\,_{y}^{2}\right\rangle <\frac{1}{2}\left\langle \widehat{N}%
\right\rangle  \label{Eq.HillerySpinEntTest}
\end{equation}%
is still used in recent papers, for example \cite{He12a}, \cite{He12b} which
deal with the entanglement of sub-systems each consisting of single modes $%
\widehat{a}$, $\widehat{b}$ for a double well situation (in these papers $%
\widehat{S}\,_{x}\rightarrow \widehat{J}_{AB}^{X}$, $\widehat{S}%
\,_{y}\rightarrow -\widehat{J}_{AB}^{Y}$, $\widehat{S}\,_{z}\rightarrow -%
\widehat{J}_{AB}^{Z}$).

\subsubsection{Non-Applicable Entanglement Test Involving $|\left\langle 
\widehat{S}\,_{z}\right\rangle |$}

Previously we had found for a general non-entangled state that is based on
physically valid density operators $\widehat{\rho }_{R}^{A}$, $\widehat{\rho 
}_{R}^{B}$ 
\begin{eqnarray}
\left\langle \Delta \widehat{S}\,_{x}^{2}\right\rangle -\frac{1}{2}%
|\left\langle \widehat{S}\,_{z}\right\rangle | &\geq &0  \nonumber \\
\left\langle \Delta \widehat{S}\,_{y}^{2}\right\rangle -\frac{1}{2}%
|\left\langle \widehat{S}\,_{z}\right\rangle | &\geq &0
\end{eqnarray}%
so that the sum of the variances satisfies the inequality 
\begin{equation}
\left\langle \Delta \widehat{S}\,_{x}^{2}\right\rangle +\left\langle \Delta 
\widehat{S}\,_{y}^{2}\right\rangle \geq |\left\langle \widehat{S}%
\,_{z}\right\rangle |\,  \label{Eq.CorrectInequalityVariances}
\end{equation}%
This is another correct inequality required for a non-entangled state as
defined in the present paper. It follows that if only physical states $%
\widehat{\rho }_{R}^{A}$, $\widehat{\rho }_{R}^{B}$ are allowed, the related 
\emph{entanglement test} involving $\left\langle \Delta \widehat{S}%
\,_{x}^{2}\right\rangle +\left\langle \Delta \widehat{S}\,_{y}^{2}\right%
\rangle $ would be 
\begin{equation}
\left\langle \Delta \widehat{S}\,_{x}^{2}\right\rangle +\left\langle \Delta 
\widehat{S}\,_{y}^{2}\right\rangle <|\left\langle \widehat{S}%
\,_{z}\right\rangle |\,  \label{Eq.TrueSpinEntTest}
\end{equation}%
For \emph{any} quantum state we have 
\begin{equation}
|\left\langle \widehat{S}\,_{z}\right\rangle |\,=\frac{1}{2}|\left(
\left\langle \widehat{n}_{b}\right\rangle -\left\langle \widehat{n}%
_{a}\right\rangle \right) |\,\leq \frac{1}{2}\left( \left\langle \widehat{n}%
_{b}\right\rangle +\left\langle \widehat{n}_{a}\right\rangle \right) =\frac{1%
}{2}\left\langle \widehat{N}\right\rangle  \label{Eq.GeneralInequalSZN}
\end{equation}%
which means that it is now required that $\left\langle \Delta \widehat{S}%
\,_{x}^{2}\right\rangle +\left\langle \Delta \widehat{S}\,_{y}^{2}\right%
\rangle $ be less than a quantity that is \emph{smaller} than in the
criterion in (\ref{Eq.InvalidInequalityVariance}).

However, it should be noted that \emph{all} states, entangled or otherwise,
satisfy the inequality $\left\langle \Delta \widehat{S}\,_{x}^{2}\right%
\rangle +\left\langle \Delta \widehat{S}\,_{y}^{2}\right\rangle \geq
|\left\langle \widehat{S}\,_{z}\right\rangle |$ so the inequality in (\ref%
{Eq.CorrectInequalityVariances}) - though true, is of no use in establishing
whether a state is entangled in the terms of the meaning of entanglement in
the present paper. There are \emph{no} quantum states, entangled or
otherwise that satisfy the proposed entanglement test given in Eq. (\ref%
{Eq.TrueSpinEntTest}). This general result was stated by Hillery et al \cite%
{Hillery06a}. To show this we write the Heisenberg uncertainty principle for 
$\left\langle \Delta \widehat{S}\,_{x}^{2}\right\rangle ,\left\langle \Delta 
\widehat{S}\,_{y}^{2}\right\rangle $ as $\left\langle \Delta \widehat{S}%
\,_{x}^{2}\right\rangle \left\langle \Delta \widehat{S}\,_{y}^{2}\right%
\rangle =\xi \frac{1}{4}|\left\langle \widehat{S}\,_{z}\right\rangle |^{2}$,
where $\xi \geq 1$, then 
\begin{eqnarray}
\frac{\left\langle \Delta \widehat{S}\,_{x}^{2}\right\rangle +\left\langle
\Delta \widehat{S}\,_{y}^{2}\right\rangle }{|\left\langle \widehat{S}%
\,_{z}\right\rangle |} &=&\frac{1}{2}\left( y+\frac{\xi }{y}\right)
=F(y)\qquad where\;y=\frac{\left\langle \Delta \widehat{S}%
\,_{x}^{2}\right\rangle }{\frac{1}{2}|\left\langle \widehat{S}%
\,_{z}\right\rangle |}  \nonumber \\
&&
\end{eqnarray}%
It is straightforward to show that $F(y)\geq 1$ for all $\xi ,y$. The
minimum value is $1$, which occurs for $\xi =1$ and $y=1$. Even spin
squeezed states with $\left\langle \Delta \widehat{S}\,_{x}^{2}\right\rangle
<\frac{1}{2}|\left\langle \widehat{S}\,_{z}\right\rangle |$ still have $%
\left\langle \Delta \widehat{S}\,_{x}^{2}\right\rangle +\left\langle \Delta 
\widehat{S}\,_{y}^{2}\right\rangle \geq |\left\langle \widehat{S}%
\,_{z}\right\rangle |\,$, so it is \emph{never} found that $\left\langle
\Delta \widehat{S}\,_{x}^{2}\right\rangle +\left\langle \Delta \widehat{S}%
\,_{y}^{2}\right\rangle <|\left\langle \widehat{S}\,_{z}\right\rangle |$ and
hence this latter inequality \emph{cannot} used as a test for entanglement.

Fortunately - as we have seen, showing that spin squeezing occurs via \emph{%
either} $\left\langle \Delta \widehat{S}\,_{x}^{2}\right\rangle <\frac{1}{2}%
|\left\langle \widehat{S}\,_{z}\right\rangle |$ \emph{or} $\left\langle
\Delta \widehat{S}\,_{y}^{2}\right\rangle <\frac{1}{2}|\left\langle \widehat{%
S}\,_{z}\right\rangle |$ is sufficient to establish that the state is an
entangled state for modes $\widehat{a},\widehat{b}$, with analogous results
if principle spin operators are considered. Applying the Hillery et al
entanglement test in Eq. (\ref{Eq.HillerySpinEntTest}) involving $\frac{1}{2}%
\left\langle \widehat{N}\right\rangle $ is also a valid entanglement test,
but is usually \emph{less stringent} than the spin squeezing test involving
either $\left\langle \Delta \widehat{S}\,_{x}^{2}\right\rangle <\frac{1}{2}%
|\left\langle \widehat{S}\,_{z}\right\rangle |$ \emph{or} $\left\langle
\Delta \widehat{S}\,_{y}^{2}\right\rangle <\frac{1}{2}|\left\langle \widehat{%
S}\,_{z}\right\rangle |$. For the Hillery et al entanglement test to be
satisfied at least one of $\left\langle \Delta \widehat{S}%
\,_{x}^{2}\right\rangle $ or $\left\langle \Delta \widehat{S}%
\,_{y}^{2}\right\rangle $ is required to be less than $\frac{1}{2}%
\left\langle \widehat{N}\right\rangle $, whereas for the spin squeezing test
to apply at least one of $\left\langle \Delta \widehat{S}\,_{x}^{2}\right%
\rangle $ or $\left\langle \Delta \widehat{S}\,_{y}^{2}\right\rangle $ must
be less than $\frac{1}{2}|\left\langle \widehat{S}\,_{z}\right\rangle |$.
The quantity $\frac{1}{2}|\left\langle \widehat{S}\,_{z}\right\rangle |$ is
likely to be smaller than $\frac{1}{2}\left\langle \widehat{N}\right\rangle $
- for example the Bloch vector may lie close to the $xy$ plane, so a greater
degree of reduction in spin fluctuation is needed to satisfy the spin
squeezing test for entanglement.

However, this is not always the case as the example of the \emph{relative
phase state} discussed in SubSection \ref{SubSection - Ent States that are
Spin Sq} shows. The results in the current SubSection can easily be modified
to apply to new spin operators $\widehat{J}_{x}$, $\widehat{J}_{y}$, $%
\widehat{J}_{z}$ , with entanglement being considered for new modes $%
\widehat{c}$ and $\widehat{d}$. The Hillery et al \cite{Hillery06a}
entanglement test then becomes 
\begin{equation}
\left\langle \Delta \widehat{J}\,_{x}^{2}\right\rangle +\left\langle \Delta 
\widehat{J}\,_{y}^{2}\right\rangle <\frac{1}{2}\left\langle \widehat{N}%
\right\rangle  \label{Eq.HilleryTestB}
\end{equation}%
In the case of the relative phase eigenstate we have from Eq. (\ref%
{Eq.CovMatrixPhaseStateNewSpins}) that $\left\langle \Delta \widehat{J}%
\,_{x}^{2}\right\rangle +\left\langle \Delta \widehat{J}\,_{y}^{2}\right%
\rangle =\frac{{\LARGE 1}}{{\LARGE 12}}N^{2}+\frac{1}{4}+\frac{1}{8}\ln
N\approx \frac{{\LARGE 1}}{{\LARGE 12}}N^{2}$ for large $N$. This clearly
exceeds $\frac{1}{2}\left\langle \widehat{N}\right\rangle =\frac{{\LARGE 1}}{%
{\LARGE 2}}N$, so the Hillery et al \cite{Hillery06a} test for entanglement
fails. On the other hand, as we have seen in SubSection \ref{SubSection -
Ent States that are Spin Sq} $\left\langle \Delta \widehat{J}%
\,_{y}^{2}\right\rangle <\frac{1}{2}|\left\langle \widehat{J}%
\,_{z}\right\rangle |\,\approx \,\frac{\pi }{16}N$, so the spin squeezing
test is satisfied for this entangled state of modes $\widehat{c}$ and $%
\widehat{d}$.

\subsection{Hillery et al 2009}

\label{SubSection - Hillery 2009}

\subsubsection{Hillery Strong Correlation Entanglement Test}

In a later paper entitled "Detecting entanglement with non-Hermitian
operators" Hillery et al \cite{Hillery09a} apply other inequalities for
determining entanglement derived in the earlier paper \cite{Hillery06a} but
now also to systems of massive identical bosons, while still retaining
.density operators $\widehat{\rho }_{R}^{A}$, $\widehat{\rho }_{R}^{B}$ that
contain coherences between states with differing boson numbers. In
particular, for a non-entangled state the following family of inequalities -
originally derived in \cite{Hillery06a}, is invoked. 
\begin{equation}
|\left\langle (\widehat{a})^{m}\,(\widehat{b}^{\dag })^{n}\right\rangle
|^{2}\leq \left\langle (\widehat{a}^{\dag })^{m}(\widehat{a})^{m}\,(\widehat{%
b}^{\dag })^{n}(\widehat{b})^{n}\right\rangle
\label{Eq.GeneralHilleryNonEntState}
\end{equation}%
Thus if $|\left\langle (\widehat{a})^{m}\,(\widehat{b}^{\dag
})^{n}\right\rangle |^{2}>\left\langle (\widehat{a}^{\dag })^{m}(\widehat{a}%
)^{m}\,(\widehat{b}^{\dag })^{n}(\widehat{b})^{n}\right\rangle $ then the
state is entangled.

A particular case for $n=m=1$ is the test $|\left\langle \widehat{a}\,%
\widehat{b}^{\dag }\right\rangle |^{2}>\left\langle \widehat{n}_{a}\,%
\widehat{n}_{b}\right\rangle $ for an entangled state. To put this result in
context, for a general quantum state and any operator $\widehat{\Omega }$ we
have $\left\langle \widehat{\Omega }^{\dag }\right\rangle =\left\langle 
\widehat{\Omega }\right\rangle ^{\ast }$ and $\left\langle \left( \widehat{%
\Omega }^{\dag }-\left\langle \widehat{\Omega }^{\dag }\right\rangle \right)
\left( \widehat{\Omega }-\left\langle \widehat{\Omega }\right\rangle \right)
\right\rangle \geq 0$, hence leading to the Schwarz inequality $%
|\left\langle \widehat{\Omega }\right\rangle |^{2}=|\left\langle \widehat{%
\Omega }^{\dag }\right\rangle |^{2}\leq \left\langle \widehat{\Omega }^{\dag
}\,\widehat{\Omega }\right\rangle $. Taking $\widehat{\Omega }=\widehat{a}\,%
\widehat{b}^{\dag }$ leads to the inequality $|\left\langle \widehat{a}\,%
\widehat{b}^{\dag }\right\rangle |^{2}\leq \left\langle \widehat{n}_{a}\,(%
\widehat{n}_{b}+1)\right\rangle $, whilst choosing $\widehat{\Omega }=%
\widehat{b}\,\widehat{a}^{\dag }$ leads to the inequality $|\left\langle 
\widehat{a}\,\widehat{b}^{\dag }\right\rangle |^{2}\leq \left\langle (%
\widehat{n}_{a}+1)\,\widehat{n}_{b}\right\rangle $ for \emph{all} quantum
states. In both cases the right side of the inequality is greater than $%
\left\langle \widehat{n}_{a}\,\widehat{n}_{b}\right\rangle $, so \emph{if }%
it was found that $|\left\langle \widehat{a}\,\widehat{b}^{\dag
}\right\rangle |^{2}>\left\langle \widehat{n}_{a}\,\widehat{n}%
_{b}\right\rangle $ (though of course still $\leq \left\langle \widehat{n}%
_{a}\,(\widehat{n}_{b}+1)\right\rangle $ and $\leq \left\langle (\widehat{n}%
_{a}+1)\,\widehat{n}_{b}\right\rangle $) then it could be concluded that the
state was entangled. However,. as we will see the left side $|\left\langle 
\widehat{a}\,\widehat{b}^{\dag }\right\rangle |^{2}$ actually works out to
be zero if physical states for $\widehat{\rho }_{R}^{A}$, $\widehat{\rho }%
_{R}^{B}$ are involved in defining non-entangled states, so that for a
non-entangled state defined as in the present paper the true inequality
replacing $|\left\langle \widehat{a}\,\widehat{b}^{\dag }\right\rangle
|^{2}\leq \left\langle \widehat{n}_{a}\,\widehat{n}_{b}\right\rangle $ is
just $0\leq \left\langle \widehat{n}_{a}\,\widehat{n}_{b}\right\rangle $,
which is trivially true for any quantum state. The test for entanglement
requires modification.

The derivation of the general inequality in \cite{Hillery06a}, as in Eq. (%
\ref{Eq.GeneralHilleryNonEntState}) follows directly from the inequality in
Eq. (\ref{Eq.QuantumCorrenInequality}) obtained in SubSection \ref%
{SubSection - Non-local Correlations} for a general non-entangled state of
sub-systems $A$ and $B$. If we choose $\widehat{\Omega }_{A}=(\widehat{a}%
)^{m}$ and $\widehat{\Omega }_{B}=(\widehat{b})^{n}$ then from $%
|\left\langle \widehat{\Omega }_{A}\otimes \widehat{\Omega }_{B}^{\dag
}\right\rangle |^{2}\leq \left\langle \widehat{\Omega }_{A}^{\dag }\widehat{%
\Omega }_{A}\otimes \widehat{\Omega }_{B}^{\dag }\widehat{\Omega }%
_{B}\right\rangle $ the result of Hillery et al \cite{Hillery06a} stated in
Eq. (\ref{Eq.GeneralHilleryNonEntState}) immediately follows. The Hillery et
al \cite{Hillery06a} \emph{entanglement test} is that if 
\begin{equation}
|\left\langle (\widehat{a})^{m}\,(\widehat{b}^{\dag })^{n}\right\rangle
|^{2}>\left\langle (\widehat{a}^{\dag })^{m}(\widehat{a})^{m}\,(\widehat{b}%
^{\dag })^{n}(\widehat{b})^{n}\right\rangle  \label{Eq.HilleryEntangTest}
\end{equation}%
then it may be concluded that the state is an entangled state for
sub-systems $A$ and $B$. Note that the proof of this result did \emph{not}
depend on the sub-system density operators $\widehat{\rho }_{R}^{A},\widehat{%
\rho }_{R}^{B}$ being required to satisfy SSR.

\subsubsection{Correlation Test for Local SSR Compatible Non-Entangled States%
}

However, for a non-entangled state based on \emph{physical} $\widehat{\rho }%
_{R}^{A},\widehat{\rho }_{R}^{B}$ for modes $\widehat{a}$ and $\widehat{b}$
where the SSR is satisfied we actually have 
\begin{equation}
\left\langle (\widehat{a})^{m}\,(\widehat{b}^{\dag })^{n}\right\rangle
=\tsum\limits_{R}P_{R}\,\left\langle (\widehat{a})^{m}\,(\widehat{b}^{\dag
})^{n}\right\rangle _{R}=\tsum\limits_{R}P_{R}\,\left\langle (\widehat{a}%
)^{m}\right\rangle _{R}\left\langle (\widehat{b}^{\dag })^{n}\right\rangle
_{R}=0  \label{Eq.ResultTrueNonEntState}
\end{equation}%
since from Eqs. analogous to (\ref{Eq.CondNonEntStateCD}) $\left\langle (%
\widehat{a})^{m}\right\rangle _{R}=\left\langle (\widehat{b}^{\dag
})^{n}\right\rangle _{R}=0$. Hence for a physical non-entangled state as
defined in the present paper the inequality becomes 
\begin{equation}
0\leq \left\langle (\widehat{a}^{\dag })^{m}(\widehat{a})^{m}\,(\widehat{b}%
^{\dag })^{n}(\widehat{b})^{n}\right\rangle
\label{Eq.InequalNonEntPhysState}
\end{equation}%
which is trivially true and applies for \emph{any} state, entangled or not.

Since $\left\langle (\widehat{a})^{m}\,(\widehat{b}^{\dag
})^{n}\right\rangle $ is zero for non-entangled states it follows that it is
merely necessary to show that this quantity is non-zero to establish that
the state is entangled. Hence an \emph{entanglement test} in the case of
sub-systems consisting of single modes $\widehat{a}$ and $\widehat{b}$
becomes 
\begin{equation}
|\left\langle (\widehat{a})^{m}\,(\widehat{b}^{\dag })^{n}\right\rangle
|^{2}>0  \label{Eq.EntangTest}
\end{equation}%
for a non-entangled state based on \emph{physical} $\widehat{\rho }_{R}^{A},%
\widehat{\rho }_{R}^{B}$. This is a useful criterion for entanglement in
terms the definition of entanglement in the present paper, and is different
to that of Hillery et al \cite{Hillery06a}. The Hillery et al \cite%
{Hillery06a} entanglement test $|\left\langle (\widehat{a})^{m}\,(\widehat{b}%
^{\dag })^{n}\right\rangle |^{2}>\left\langle (\widehat{a}^{\dag })^{m}(%
\widehat{a})^{m}\,(\widehat{b}^{\dag })^{n}(\widehat{b})^{n}\right\rangle $
is \emph{also} a valid test for entanglement and is actually a \emph{more
stringent test} than merely showing that $|\left\langle (\widehat{a})^{m}\,(%
\widehat{b}^{\dag })^{n}\right\rangle |^{2}>0$, since the quantity $%
|\left\langle (\widehat{a})^{m}\,(\widehat{b}^{\dag })^{n}\right\rangle
|^{2} $ is now required to be \emph{larger}. In a paper by He et al \cite%
{He12a} (see Section IIIA) the Hillery et al \cite{Hillery06a} entanglement
test $|\left\langle (\widehat{a})^{m}\,(\widehat{b}^{\dag
})^{n}\right\rangle |^{2}>\left\langle (\widehat{a}^{\dag })^{m}(\widehat{a}%
)^{m}\,(\widehat{b}^{\dag })^{n}(\widehat{b})^{n}\right\rangle $ is applied
for the case where $A$ and $B$ \emph{each} consist of \emph{one mode}
localised in each well of a double well potential. This test whilst
applicable could be replaced by the more easily satisfied test $%
|\left\langle (\widehat{a})^{m}\,(\widehat{b}^{\dag })^{n}\right\rangle
|^{2}>0$. However, as will be seen below in SubSection \ref{SubSection - He
2012}, Hillery et al \cite{Hillery06a} entanglement criterion is needed if
the sub-systems each consist of \emph{pairs of modes}, as treated in \cite%
{He11a}, \cite{He12a}.

\subsubsection{Applications of Correlation Tests for Entanglement}

As an example of applying these tests consider the \emph{mixed two mode
coherent states} described in SubSection \ref{SubSection - Two Mode Coherent
State Mixture}, whose density operator for the two mode $\widehat{a}$, $%
\widehat{b}$ system is given in Eq. (\ref{Eq.TwoModeCoherentStateMixture}).
We can now examine the Hillery et al \cite{Hillery09a} entanglement test in
Eq.(\ref{Eq.HilleryEntangTest}) and the entanglement test in Eq.(\ref%
{Eq.EntangTest}) for the case where $m=n=1$. It is straight-forward to show
that 
\begin{eqnarray}
|\left\langle \widehat{a}\,\widehat{b}^{\dag }\right\rangle |^{2} &=&|\alpha
|^{4}  \nonumber \\
\left\langle (\widehat{a}^{\dag }\widehat{a})\,(\widehat{b}^{\dag }\widehat{b%
})\right\rangle &=&|\alpha |^{4}  \label{Eq.TestResults}
\end{eqnarray}%
so that $|\left\langle \widehat{a}\,\widehat{b}^{\dag }\right\rangle
|^{2}=\left\langle (\widehat{a}^{\dag }\widehat{a})\,(\widehat{b}^{\dag }%
\widehat{b})\right\rangle $. A non-entangled state defined in terms of the
SSR requirement for the separate modes satisfies $|\left\langle \widehat{a}\,%
\widehat{b}^{\dag }\right\rangle |^{2}=0$, whilst for a non-entangled state
in which the SSR requirement for separate modes is not specifically required
merely satisfies $|\left\langle \widehat{a}\,\widehat{b}^{\dag
}\right\rangle |^{2}\leq \left\langle (\widehat{a}^{\dag }\widehat{a})\,(%
\widehat{b}^{\dag }\widehat{b})\right\rangle $. Hence the test for
entanglement of modes $A$, $B$ in the present paper $|\left\langle \widehat{a%
}\,\widehat{b}^{\dag }\right\rangle |^{2}>0$ is satisfied, whilst the
Hillery et al \cite{Hillery09a} test $|\left\langle \widehat{a}\,\widehat{b}%
^{\dag }\right\rangle |^{2}>\left\langle (\widehat{a}^{\dag }\widehat{a})\,(%
\widehat{b}^{\dag }\widehat{b})\right\rangle $ is not.

In terms of the definition of non-entangled states in the present paper, the
mixture of two mode coherent states given in Eq.(\ref%
{Eq.TwoModeCoherentStateMixture}) is an \emph{entangled state}, not a
separable state. However, in terms of the definition of non-entangled states
in other papers such as those of Hillery et al \cite{Hillery06a}, \cite%
{Hillery09a} the mixture of two mode coherent states would be a \emph{%
non-entangled} state. It is thus a useful state for providing an example of
the different outcomes of definitions where the local SSR is applied or not.

\subsection{Sorensen et al 2001}

\label{SubSection - Sorensen 2001}

\subsubsection{Sorensen Spin Squeezing Entanglement Test}

In a paper entitled "Many-particle entanglement with Bose-Einstein
condensates" Sorensen et al \cite{Sorensen01a} consider the implications for
spin squeezing for non-entangled states of the form in Eq. (\ref%
{Eq.NonEntStateIdenticalAtoms}). As discussed previously, a density operator
of this general form is not consistent with the symmetrisation principle -
having separate density operators $\widehat{\rho }_{R}^{i}$ for specific
particles $i$ in an identical particle system (such as for a BEC) is not
compatible with the indistinguishability of such particles. It is modes that
are distinguishable, not identical particles, so the basis for applying
their results to systems of identical bosons is flawed. However, they derive
an inequality for the spin variance $\left\langle \Delta \widehat{S}%
\,_{z}^{2}\right\rangle $%
\begin{equation}
\left\langle \Delta \widehat{S}\,_{z}^{2}\right\rangle \geq \frac{1}{N}%
\left( \left\langle \widehat{S}\,_{x}\right\rangle ^{2}+\left\langle 
\widehat{S}\,_{y}\right\rangle ^{2}\right)  \label{Eq.SorensenInequality}
\end{equation}%
that applies in the case of non-entangled states. Key steps in their
derivation are stated in the Appendix to \cite{Sorensen01a}, but as the
justification of these steps is not obvious for completeness the full
derivation is given in Appendix \ref{Appendix - Sorensen Results} of the
present paper. This inequality (\ref{Eq.SorensenInequality}) establishes
that if 
\begin{equation}
\xi ^{2}=\frac{\left\langle \Delta \widehat{S}\,_{z}^{2}\right\rangle }{%
\left( \left\langle \widehat{S}\,_{x}\right\rangle ^{2}+\left\langle 
\widehat{S}\,_{y}\right\rangle ^{2}\right) }<\frac{1}{N}
\label{Eq.SpinSqueezingMeasure2}
\end{equation}%
then the state is entangled, so that if the condition for spin squeezing
analogous to that in Eq. (\ref{Eq.NewCriterionSpinSqueezing}) is satisfied,
then entanglement is required if spin squeezing for $\widehat{S}\,_{z}$ to
occur. Spin squeezing is then a test for entanglement in terms of their
definition of an entangled state. Note that the condition (\ref%
{Eq.NewCriterionSpinSqueezing}) requires the Bloch vector to be in the $xy$
plane and close to the Bloch sphere of radius $N/2$.

\subsubsection{Revising Sorensen Spin Squeezing Entanglement Test -
Localised Modes}

The work of Sorensen et al really applies only when the individual spins are
distinguishable. It is possible however to modify the work of Sorensen et al 
\cite{Sorensen01a} to apply to a system of identical bosons in accordance
with the symmetrisation and super-selection rules if the index $i$ is \emph{%
re-interpreted} as specifying diffferent modes, for example modes localised
on \emph{optical lattice} sites $i=1,2,..,N$. Details are given in Appendix %
\ref{Appendix - Revised Sorensen}. With two single particle states $a,b$
available on each site (these could be two different internal atomic states
or two distinct spatial modes localised on the site) the modes would then be
labelled $\left\vert \phi _{\alpha i}\right\rangle $ with $\alpha =a,b$. The
mode orthogonality and completeness relations would then be 
\begin{eqnarray}
\left\langle \phi _{\alpha \,i}|\phi _{\beta \,j}\right\rangle &=&\delta
_{\alpha \beta }\delta _{ij}  \nonumber \\
\tsum\limits_{\alpha i}\left\vert \phi _{\alpha \,i}\right\rangle
\left\langle \phi _{\alpha \,i}\right\vert &=&\widehat{1}
\label{Eq.ModesRevisedSorensen}
\end{eqnarray}%
With the particles now labelled $K=1,2,3,...$one can define spin operators
in first quantization via 
\begin{eqnarray}
\widehat{S}_{x} &=&\sum_{K}\sum_{i}(\left\vert \phi _{b\,i}(K)\right\rangle
\left\langle \phi _{a\,i}(K)\right\vert +\left\vert \phi
_{a\,i}(K)\right\rangle \left\langle \phi _{b\,i}(K)\right\vert )/2 
\nonumber \\
\widehat{S}_{y} &=&\sum_{K}\sum_{i}(\left\vert \phi _{b\,i}(K)\right\rangle
\left\langle \phi _{a\,i}(K)\right\vert -\left\vert \phi
_{a\,i}(K)\right\rangle \left\langle \phi _{b\,i}(K)\right\vert )/2i 
\nonumber \\
\widehat{S}_{z} &=&\sum_{K}\sum_{i}(\left\vert \phi _{b\,i}(K)\right\rangle
\left\langle \phi _{b\,i}(K)\right\vert -\left\vert \phi
_{a\,i}(K)\right\rangle \left\langle \phi _{a\,i}(K)\right\vert )/2
\label{Eq.SpinOprsRevisedSorensen}
\end{eqnarray}%
In second quantization if the annihilation, creation operators for the modes 
$\left\vert \phi _{ai}\right\rangle $ ,$\left\vert \phi _{bi}\right\rangle $
are $\widehat{a}_{i}$, $\widehat{b}_{i}$ and $\widehat{a}_{i}^{\dag }$, $%
\widehat{b}_{i}^{\dag }$ respectively, then the Schwinger spin operators are
just%
\begin{eqnarray}
\widehat{S}_{x} &=&\sum_{i}(\widehat{b}_{i}^{\dag }\widehat{a}_{i}+\widehat{a%
}_{i}^{\dag }\widehat{b}_{i})/2=\sum_{i}\widehat{S}_{x}^{i}  \nonumber \\
\widehat{S}_{y} &=&\sum_{i}(\widehat{b}_{i}^{\dag }\widehat{a}_{i}-\widehat{a%
}_{i}^{\dag }\widehat{b}_{i})/2i=\sum_{i}\widehat{S}_{y}^{i}  \nonumber \\
\widehat{S}_{z} &=&\sum_{i}(\widehat{b}_{i}^{\dag }\widehat{b}_{i}-\widehat{a%
}_{i}^{\dag }\widehat{a}_{i})/2=\sum_{i}\widehat{S}_{z}^{i}
\label{Eq.SchwingerSpinOprsRevisedSorensen}
\end{eqnarray}%
It is easy to confirm that the overall spin operators $\widehat{S}_{\alpha }$
and the spin operators $\widehat{S}_{\alpha }^{i}$ for the separate \emph{%
pairs} of \emph{modes} $\left\vert \phi _{ai}\right\rangle $ ,$\left\vert
\phi _{bi}\right\rangle $ (or $\widehat{a}_{i}$, $\widehat{b}_{i}$ for
short) satisfy the same commutation rules as Sorensen et al \cite%
{Sorensen01a} have for the overall spin operators and those for the separate 
\emph{particles}. With this modification the non-entangled state in Eq. (\ref%
{Eq.NonEntStateIdenticalAtoms}) could be interpreted as being a
non-entangled state where the subsystems are actually \emph{pairs }of\emph{\
modes} $\left\vert \phi _{ai}\right\rangle $ ,$\left\vert \phi
_{bi}\right\rangle $ and the density operators $\widehat{\rho }_{R}^{i}$
would then refer to a subsystem consisting of these pairs of modes. It is to
be noted that entanglement of \emph{pairs} of modes is different to
entanglement of \emph{all separate} modes. It is an example of a special
kind of \emph{multimode entanglement} - since the modes $\left\vert \phi
_{ai}\right\rangle $ ,$\left\vert \phi _{bi}\right\rangle $ may themselves
be entangled we may have "entanglement of entanglement". In terms of the
present paper the density operators $\widehat{\rho }_{R}^{i}$ would be
restricted by the super-selection rule to statistical mixtures of states
with specific total numbers $N_{i}$ of bosons in the pair of modes $%
\left\vert \phi _{ai}\right\rangle $ ,$\left\vert \phi _{bi}\right\rangle $.
In terms of Fock states $\left\vert n_{a\,i}\right\rangle ,\left\vert
n_{b\,i}\right\rangle $ for this pair of modes the allowed quantum states
for the sub-system will be%
\begin{equation}
\left\vert \Phi _{N_{i}}\right\rangle
=\tsum\limits_{k=0}^{N_{i}}A_{k}^{N_{i}}\,\left\vert k\right\rangle
_{a\,i}\left\vert N_{i}-k\right\rangle _{b\,i}
\label{Eq.TwoModeQuantumSuperposition}
\end{equation}%
so at this stage the general mixed physical state for the two mode system 
\emph{could} be 
\begin{equation}
\widehat{\rho }_{R}^{i}=\tsum\limits_{N_{i}=0}^{\infty }\tsum\limits_{\Phi
}P_{\Phi
N_{i}}\,\tsum\limits_{k=0}^{N_{i}}\tsum\limits_{l=0}^{N_{i}}A_{k}^{N_{i}}%
\,(A_{l}^{N})^{\ast }\left\vert k\right\rangle _{a\,i}\left\langle
l\right\vert _{a\,i}\otimes \left\vert N_{i}-k\right\rangle
_{b\,i}\left\langle N_{i}-l\right\vert _{b\,i}
\label{Eq.GeneralDensityOprModePair}
\end{equation}%
This state has no coherences between states of the two mode subsystem with
differing total boson number $N_{i}$ for the pair of modes. However this is
still an entangled states for the two modes $\left\vert \phi
_{ai}\right\rangle $ ,$\left\vert \phi _{bi}\right\rangle $, so the overall
state in Eq. (\ref{Eq.NonEntStateIdenticalAtoms}) is not a non-entangled
state if the subsystems were to consist of \emph{all} the distinct modes.

\subsubsection{Revising Sorensen Spin Squeezing Entanglement Test -
Separable State of Single Modes}

It is possible however to link spin squeezing and entanglement in the case
where the sub-systems consist of \emph{all} the distinct modes. To obtain a 
\emph{fully non-entangled state} of \emph{all} the modes $\left\vert \phi
_{ai}\right\rangle $ ,$\left\vert \phi _{bi}\right\rangle $ the density
operator $\widehat{\rho }_{R}^{i}$ must then be a product of density
operators for modes $\left\vert \phi _{ai}\right\rangle $ and $\left\vert
\phi _{bi}\right\rangle $%
\begin{equation}
\widehat{\rho }_{R}^{i}=\widehat{\rho }_{R}^{a\,i}\otimes \widehat{\rho }%
_{R}^{b\,i}  \label{Eq.DensityOprModesAiBi}
\end{equation}%
giving the full density operator as 
\begin{equation}
\widehat{\rho }=\sum_{R}P_{R}\,\left( \widehat{\rho }_{R}^{a\,1}\otimes 
\widehat{\rho }_{R}^{b\,1}\right) \otimes \left( \widehat{\rho }%
_{R}^{a\,2}\otimes \widehat{\rho }_{R}^{b\,2}\right) \otimes \left( \widehat{%
\rho }_{R}^{a\,3}\otimes \widehat{\rho }_{R}^{b\,3}\right) \otimes .
\label{Eq.RevisedSorensenDensityOprNonEnt}
\end{equation}%
as is required for a general non-entangled state all $2N$ modes.
Furthermore, as previously the density operators for the individual modes
must represent possible physical states for such modes, so the
super-selection rule for atom number will apply and we have 
\begin{eqnarray}
\left\langle (\widehat{a}_{i})^{n}\right\rangle _{a\,i} &=&Tr(\widehat{\rho }%
_{R}^{a\,i}(\widehat{a}_{i})^{n})=0\qquad \left\langle (\widehat{a}%
_{i}^{\dag })^{n}\right\rangle _{a\,i}=Tr(\widehat{\rho }_{R}^{a\,i}(%
\widehat{a}_{i}^{\dag })^{n})=0  \nonumber \\
\left\langle (\widehat{b}_{i})^{m}\right\rangle _{b\,i} &=&Tr(\widehat{\rho }%
_{R}^{b\,i}(\widehat{b}_{i})^{m})=0\qquad \left\langle (\widehat{b}%
_{i}^{\dag })^{m}\right\rangle _{b\,i}=Tr(\widehat{\rho }_{R}^{b\,i}(%
\widehat{b}_{i}^{\dag })^{m})=0  \nonumber \\
&&  \label{Eq.RevisedSorensenAverages}
\end{eqnarray}

The question is whether this reformulation will lead to a useful inequality
for the spin variances such as $\left\langle \Delta \widehat{S}%
\,_{x}^{2}\right\rangle $. This issue is dealt with in Appendix \ref%
{Appendix - Revised Sorensen} and it is found that we can indeed show for
the general \emph{fully non-entangled} state (\ref%
{Eq.RevisedSorensenDensityOprNonEnt}) that 
\begin{equation}
\left\langle \Delta \widehat{S}\,_{x}^{2}\right\rangle \geq \frac{1}{2}%
|\left\langle \widehat{S}\,_{z}\right\rangle |\quad and\quad \left\langle
\Delta \widehat{S}\,_{y}^{2}\right\rangle \geq \frac{1}{2}|\left\langle 
\widehat{S}\,_{z}\right\rangle |  \label{Eq.VarianceInequalRevisedSorensen}
\end{equation}%
This shows that if there is spin squeezing in \emph{either} $\widehat{S}%
\,_{x}$ \emph{or} $\widehat{S}\,_{y}$ then the state must be entangled. Note
that this result depends on the general non-entangled state being
non-entangled for \emph{all} modes and that the density operator for each
mode $\widehat{a}_{i}$ or $\widehat{b}_{i}$ being a physical state with no
coherences between mode Fock states with differing atom numbers. In terms of
the revised interpretation of the density operator to refer to a multi-mode
system with modes $\left\vert \phi _{ai}\right\rangle $ ,$\left\vert \phi
_{bi}\right\rangle $ the statement that spin squeezing for systems of
identical massive bosons requires all the modes to be entangled is correct.
However superposition states of the form (\ref%
{Eq.TwoModeQuantumSuperposition}) that are consistent with the
super-selection rule applying to pure states of a two mode system are
precluded, and such states ought to be allowed if entanglement of \emph{pairs%
} of modes rather than of \emph{separate} modes is to be considered.

\subsubsection{Revising Sorensen Spin Squeezing Entanglement Test -
Separable State of Pairs of Modes with One Boson Occupancy}

It is also possible however to link spin squeezing and entanglement in the
case where the subsystems consist of \emph{pairs} of modes, but only if 
\emph{further restrictions} are applied. The general \emph{non-entangled}
state of the \emph{pairs} of \emph{modes} would actually be of the form 
\begin{equation}
\widehat{\rho }=\sum_{R}P_{R}\,\widehat{\rho }_{R}^{1}\otimes \widehat{\rho }%
_{R}^{2}\otimes \widehat{\rho }_{R}^{3}\otimes ...
\label{Eq.NonEntStateModePairs}
\end{equation}%
where the $\widehat{\rho }_{R}^{i}$ are now of the form given in Eq. (\ref%
{Eq.GeneralDensityOprModePair}) and no longer are density operators for the $%
i$th identical particle. Unlike in (\ref{Eq.RevisedSorensenAverages}) we now
have expectation values $\left\langle (\widehat{a}_{i})^{n}\right\rangle
_{\,i}=Tr(\widehat{\rho }_{R}^{\,i}(\widehat{a}_{i})^{n})$ etc that are
non-zero, so considerations of the link between spin squeezing and
entanglement - now entanglement of pairs of modes, willl be different.

If the density operators $\widehat{\rho }_{R}^{i}$ associated with the \emph{%
pair} of modes $\widehat{a}_{i}$, $\widehat{b}_{i}$ are all \emph{restricted}
to be associated with \emph{one boson states} then this density operator is
of the form 
\begin{eqnarray}
\widehat{\rho }_{R}^{i} &=&\rho _{aa}^{i}(\left\vert 1\right\rangle
_{ia}\left\langle 1\right\vert _{ia}\otimes \left\vert 0\right\rangle
_{ib}\left\langle 0\right\vert _{ib})+\rho _{ab}^{i}(\left\vert
1\right\rangle _{ia}\left\langle 0\right\vert _{ia}\otimes \left\vert
0\right\rangle _{ib}\left\langle 1\right\vert _{ib})  \nonumber \\
&&+\rho _{ba}^{i}(\left\vert 0\right\rangle _{ia}\left\langle 1\right\vert
_{ia}\otimes \left\vert 1\right\rangle _{ib}\left\langle 0\right\vert
_{ib})+\rho _{bb}^{i}(\left\vert 0\right\rangle _{ia}\left\langle
0\right\vert _{ia}\otimes \left\vert 1\right\rangle _{ib}\left\langle
1\right\vert _{ib})  \nonumber \\
&&  \label{Eq.DensityOprModePair}
\end{eqnarray}%
where the $\rho _{ef}^{i}$ are density matrix elements. With this
restriction the pair of modes $\widehat{a}_{i}$, $\widehat{b}_{i}$ behave
like \emph{distinguishable} two state particles, essentially the case that
Sorensen et al \cite{Sorensen01a} implicitly considered. The expectation
values for the spin operators $\widehat{S}_{x}^{i}$, $\widehat{S}_{y}^{i}$
and $\widehat{S}_{z}^{i}$ associated with the $i$th pair of modes are then 
\begin{eqnarray}
\left\langle \widehat{S}_{x}^{i}\right\rangle _{R} &=&\frac{1}{2}\left( \rho
_{ab}^{i}+\rho _{ba}^{i}\right) \qquad \left\langle \widehat{S}%
_{y}^{i}\right\rangle _{R}=\frac{1}{2i}\left( \rho _{ab}^{i}-\rho
_{ba}^{i}\right)  \nonumber \\
\left\langle \widehat{S}_{z}^{i}\right\rangle _{R} &=&\frac{1}{2}\left( \rho
_{bb}^{i}-\rho _{aa}^{i}\right)  \label{Eq.ExpnValues}
\end{eqnarray}

If in addition Hermitiancy, positivity, unit trace $Tr(\widehat{\rho }%
_{R}^{i})=1$ and $Tr(\widehat{\rho }_{R}^{i})^{2}\leq 1$ are used (see
Appendix \ref{Appendix - Sorensen Results}) then we can show that $\rho
_{bb}^{i}$ and $\rho _{aa}^{i}$ are real and positive, $\rho _{ab}^{i}=(\rho
_{ba}^{i})^{\ast }$ and $\rho _{aa}^{i}\rho _{bb}^{i}-|\rho
_{ab}^{i}|^{2}\geq 0$. The condition $Tr(\widehat{\rho }_{R}^{i})=1$ leads
to $\rho _{aa}^{i}+\rho _{bb}^{i}=1$, from which $Tr(\widehat{\rho }%
_{R}^{i})^{2}\leq 1$ follows using the previous positivity results. These
results enable the matrix elements in (\ref{Eq.DensityOprModePair}) to be
parameterised in the form%
\begin{eqnarray}
\rho _{aa}^{i} &=&\sin ^{2}\alpha _{i}\qquad \rho _{bb}^{i}=\cos ^{2}\alpha
_{i}  \nonumber \\
\rho _{ab}^{i} &=&\sqrt{\sin ^{2}\alpha _{i}\,\cos ^{2}\alpha _{i}}\,\sin
^{2}\beta _{i}\,\exp (+i\phi _{i})\qquad \rho _{ba}^{i}=\sqrt{\sin
^{2}\alpha _{i}\,\cos ^{2}\alpha _{i}}\,\sin ^{2}\beta _{i}\,\exp (-i\phi
_{i})  \nonumber \\
&&  \label{Eq.ParamTwoModeDensityME}
\end{eqnarray}%
where $\alpha _{i}$, $\beta _{i}$ and $\phi _{i}$ are real. In terms of
these quantities we then have 
\begin{eqnarray}
\left\langle \widehat{S}_{x}^{i}\right\rangle _{R} &=&\frac{1}{2}\sin
2\alpha _{i}\,\sin ^{2}\beta _{i}\,\cos \phi _{i}\qquad \left\langle 
\widehat{S}_{y}^{i}\right\rangle _{R}=\frac{1}{2}\sin 2\alpha _{i}\,\sin
^{2}\beta _{i}\,\sin \phi _{i}  \nonumber \\
\left\langle \widehat{S}_{z}^{i}\right\rangle _{R} &=&\frac{1}{2}\cos
2\alpha _{i}\,  \label{Eq.ExpnValuesB}
\end{eqnarray}%
and then a key inequality 
\begin{equation}
\left\langle \widehat{S}_{x}^{i}\right\rangle _{R}^{2}+\left\langle \widehat{%
S}_{y}^{i}\right\rangle _{R}^{2}+\left\langle \widehat{S}_{z}^{i}\right%
\rangle _{R}^{2}=\frac{1}{4}-\frac{1}{4}\sin ^{2}2\alpha _{i}\,(1-\sin
^{4}\beta _{i}\,)\leq \frac{1}{4}  \label{Eq.SorensenInequalityB}
\end{equation}%
follows. This result depends on the density operators $\widehat{\rho }%
_{R}^{i}$ being for one boson states, as in (\ref{Eq.DensityOprModePair}).
The same steps as in Sorensen et al \cite{Sorensen01a} (see Appendix \ref%
{Appendix - Sorensen Results}) leads to the result%
\begin{equation}
\left\langle \Delta \widehat{S}\,_{z}^{2}\right\rangle \geq \frac{1}{N}%
\left( \left\langle \widehat{S}\,_{x}\right\rangle ^{2}+\left\langle 
\widehat{S}\,_{y}\right\rangle ^{2}\right)
\label{Eq.SpinSqgCondModifiedSorensen}
\end{equation}%
for non-entangled \emph{pair} of modes $\widehat{a}_{i}$, $\widehat{b}_{i}$.
Thus when the interpretation is changed so that are the separate sub-systems
are these pairs of modes \emph{and} the sub-systems are in one boson states,
it follows that spin squeezing requires entanglement of all the mode pairs.

A similar proof extending the test of Sorensen et al \cite{Sorensen01a} to
appply to systems of identical bosons is given by Hyllus et al \cite%
{Hyllus12a} based on a particle entanglement approach. In their approach
bosons in differing external modes (analogous to differing $i$ here) are
treated as distinguishable, and the symmetrization principle is ignored for
such bosons.

\subsection{Sorensen and Molmer 2001}

\label{SubSection - Sorensen and Molmer 2001}

In a paper entitled "Entanglement and Extreme Spin Squeezing" Sorensen and
Molmer \cite{Sorensen01b} first consider the limits imposed by the
Heisenberg uncertainty principle on the variance $\left\langle \Delta 
\widehat{J}\,_{x}^{2}\right\rangle $ considered as a function of $%
|\left\langle \widehat{J}\,_{z}\right\rangle |$ for states where the spin
operators are chosen such that $\left\langle \widehat{J}\,_{x}\right\rangle
=\left\langle \widehat{J}\,_{y}\right\rangle =0$. Note that such spin
operators can always be chosen so that the Bloch vector does lie along the $%
z $ axis, even if the spin operators are not principal spin operators. Their
treatment is based on combining the result from the Schwarz inequality 
\begin{equation}
\left\langle \widehat{J}\,_{x}^{2}\right\rangle +\left\langle \widehat{J}%
\,_{y}^{2}\right\rangle +\left\langle \widehat{J}\,_{z}\right\rangle
^{2}\leq J(J+1)  \label{Eq.SchwarzResult}
\end{equation}%
where $J=N/2$, and the Heisenberg uncertainty principle 
\begin{equation}
\left\langle \Delta \widehat{J}\,_{x}^{2}\right\rangle \left\langle \Delta 
\widehat{J}\,_{y}^{2}\right\rangle =\xi \frac{1}{4}|\left\langle \widehat{J}%
\,_{z}\right\rangle |^{2}  \label{Eq.HUP}
\end{equation}%
where $\xi \geq 1$. In fact two inequalities can be obtained 
\begin{eqnarray}
\left\langle \Delta \widehat{J}\,_{x}^{2}\right\rangle &\geq &\frac{1}{2}%
\left\{ \left( J(J+1)-\left\langle \widehat{J}\,_{z}\right\rangle
^{2}\right) -\sqrt{\left( J(J+1)-\left\langle \widehat{J}\,_{z}\right\rangle
^{2}\right) ^{2}-\xi \left\langle \widehat{J}\,_{z}\right\rangle ^{2}}%
\right\}  \nonumber \\
&&  \label{Eq.HUPRestriction1} \\
\left\langle \Delta \widehat{J}\,_{x}^{2}\right\rangle &\leq &\frac{1}{2}%
\left\{ \left( J(J+1)-\left\langle \widehat{J}\,_{z}\right\rangle
^{2}\right) +\sqrt{\left( J(J+1)-\left\langle \widehat{J}\,_{z}\right\rangle
^{2}\right) ^{2}-\xi \left\langle \widehat{J}\,_{z}\right\rangle ^{2}}%
\right\}  \nonumber \\
&&  \label{Eq.HUPRestriction2}
\end{eqnarray}%
which restricts the region in a $\left\langle \Delta \widehat{J}%
\,_{x}^{2}\right\rangle $ versus $|\left\langle \widehat{J}%
\,_{z}\right\rangle |$ plane that applies for states that are consistent
with the Heisenberg uncertainty principle. The first of these two
inequalities is given as Eq. (3) in \cite{Sorensen01b}. For states in which $%
\widehat{J}\,_{x}$ is squeezed relative to $\widehat{J}\,_{y}$ the points in
the$\left\langle \Delta \widehat{J}\,_{x}^{2}\right\rangle $ versus $%
|\left\langle \widehat{J}\,_{z}\right\rangle |$ plane must also satisfy 
\begin{equation}
\left\langle \Delta \widehat{J}\,_{x}^{2}\right\rangle \leq \frac{1}{2}%
|\left\langle \widehat{J}\,_{z}\right\rangle |  \label{Eq.SpinSqRestriction}
\end{equation}%
Note that as $\widehat{J}\,_{z}$ is a spin angular momentum component we
always have $|\left\langle \widehat{J}\,_{z}\right\rangle |\,\leq J$, which
places an overall restriction on $|\left\langle \widehat{J}%
\,_{z}\right\rangle |$. However, for $\xi >1$ there are values of $%
|\left\langle \widehat{J}\,_{z}\right\rangle |$.which are excluded via the
Heisenberg uncertainty principle, since the quantity $\left(
J(J+1)-\left\langle \widehat{J}\,_{z}\right\rangle ^{2}\right) ^{2}-\xi
\left\langle \widehat{J}\,_{z}\right\rangle ^{2}$ then becomes negative.
This effect is seen in Figure 4.

The question is: Is it possible to find values for $\left\langle \Delta 
\widehat{J}\,_{x}^{2}\right\rangle $ and $|\left\langle \widehat{J}%
\,_{z}\right\rangle |$ in which all three inequalities are satisfied? The
answer is yes. Results showing the regions in the $\left\langle \Delta 
\widehat{J}\,_{x}^{2}\right\rangle $ versus $|\left\langle \widehat{J}%
\,_{z}\right\rangle |$ plane corresponding to the three inequalities are
shown in Figures 2 and 3 for the cases where $J=1000$ and with $\xi =1.0$
and $\xi =10.0$ respectively. The quantities for which the regions are shown
are the scaled variance and mean $\left\langle \Delta \widehat{J}%
\,_{x}^{2}\right\rangle /J$ and $|\left\langle \widehat{J}%
\,_{z}\right\rangle |/J$, with $\left\langle \Delta \widehat{J}%
\,_{x}^{2}\right\rangle $ given as a function of $|\left\langle \widehat{J}%
\,_{z}\right\rangle |$ via (\ref{Eq.HUPRestriction1}), (\ref%
{Eq.HUPRestriction2}) and (\ref{Eq.SpinSqRestriction}). The spin squeezing
region is always consistent with the second Heisenberg inequality (\ref%
{Eq.HUPRestriction2}) and for large $J=1000$ there is a large region of
overlap with the first inequality (\ref{Eq.HUPRestriction1}). For small $J$
and large $\xi $ the region of overlap becomes much smaller, as the result
in Figure 4 for $J=1$ and with $\xi =10.0$ shows. As the derivation of the
Heisenberg principle inequalities is not obvious, this is set out in
Appendix.\ref{Appendix - Heisenberg Uncertainty Principle Results}.

Sorensen and Molmer \cite{Sorensen01b} also determine the minimum for $%
\left\langle \Delta \widehat{J}\,_{x}^{2}\right\rangle =\left\langle 
\widehat{J}\,_{x}^{2}\right\rangle $ as a function of $|\left\langle 
\widehat{J}\,_{z}\right\rangle |$ for various choices of $J$, subject to the
constraints $\left\langle \widehat{J}\,_{x}\right\rangle =\left\langle 
\widehat{J}\,_{y}\right\rangle =0$. The results show again that there is a
region in the $\left\langle \Delta \widehat{J}\,_{x}^{2}\right\rangle $
versus $|\left\langle \widehat{J}\,_{z}\right\rangle |$ plane which is
compatible with spin squeezing.

So although these considerations show that the Heisenberg uncertainty
principle does not rule out spin squeezing, nothing is determined about
whether the spin squeezed states are entangled states for modes $\widehat{c}$%
, $\widehat{d}$\textbf{\ }, where the $\widehat{J}_{\alpha }$ are given as
in Eq. (\ref{Eq.NewSpinOprs}). The discussion in \cite{Sorensen01b}
regarding entanglement is based on the physically incorrect density operator
for non-entangled states of identical particles in Eq. (\ref%
{Eq.NonEntStateIdenticalAtoms}), discussed in the previous section.

\subsection{Duan et al 2000}

\label{Duan et al 2000}

A further inequality aimed at providing a signature for entanglement is set
out in the papers by Duan et al \cite{Duan00a}, Toth et al \cite{Toth03a}.
For simplicity we only set out the case for which $a=1$ in the former paper.
This inequality involves position and momentum like Hermitian operators
defined by 
\begin{eqnarray}
\widehat{x}_{A} &=&\frac{1}{\sqrt{2}}(\widehat{a}+\widehat{a}^{\dag })\qquad 
\widehat{p}_{A}=\frac{1}{\sqrt{2}i}(\widehat{a}-\widehat{a}^{\dag }) 
\nonumber \\
\widehat{x}_{B} &=&\frac{1}{\sqrt{2}}(\widehat{b}+\widehat{b}^{\dag })\qquad 
\widehat{p}_{B}=\frac{1}{\sqrt{2}i}(\widehat{b}-\widehat{b}^{\dag })
\label{Eq.PositionMtmOprs}
\end{eqnarray}%
These are essentially \emph{quadrature operators} and satisfy commutation
rules $[\widehat{x}_{A},\widehat{p}_{A}]=[\widehat{x}_{B},\widehat{p}_{B}]=i$
similar to those for position and momentum. An inequality is obtained for a
general two mode non-entangled state involving the variances for the
commuting observables $\widehat{x}_{A}+\widehat{x}_{B}$ and $\widehat{p}_{A}-%
\widehat{p}_{B}$%
\begin{equation}
\left\langle \Delta (\widehat{x}_{A}+\widehat{x}_{B})^{2}\right\rangle
+\left\langle \Delta (\widehat{p}_{A}-\widehat{p}_{B})^{2}\right\rangle \geq
2  \label{Eq.DuanInequalityNonEntState}
\end{equation}%
which could be used to establish a \emph{quadrature variance test} for
entangled states of the mode $A$ and mode $B$ sub-systems, so that if%
\begin{equation}
\left\langle \Delta (\widehat{x}_{A}+\widehat{x}_{B})^{2}\right\rangle
+\left\langle \Delta (\widehat{p}_{A}-\widehat{p}_{B})^{2}\right\rangle <2
\label{Eq.QuadratureVarianceEntTest}
\end{equation}%
then the modes are entangled. Such states are possible - consider for
example any simultaneous eigenstate of the commuting observables $\widehat{x}%
_{A}+\widehat{x}_{B}$ and $\widehat{p}_{A}-\widehat{p}_{B}$. For such a
state $\left\langle \Delta (\widehat{x}_{A}+\widehat{x}_{B})^{2}\right%
\rangle $ and $\left\langle \Delta (\widehat{p}_{A}-\widehat{p}%
_{B})^{2}\right\rangle $ are both zero, so the simultaneous eigenstates are
entangled states of modes $A$, $B$.

To confirm whether the inequality (\ref{Eq.DuanInequalityNonEntState})
applies for non-entangled states (\ref{Eq.NonEntangStateModesAB}) in which
the sub-system states $\widehat{\rho }_{R}^{A}$, $\widehat{\rho }_{R}^{B}$
are physical, the general variance result in Eq. (\ref{Eq.VarianceResult})
plus the factorisations $\left\langle \widehat{x}_{A}\widehat{x}%
_{B}\right\rangle _{R}$ $=\left\langle \widehat{x}_{A}\right\rangle
_{R}\left\langle \widehat{x}_{B}\right\rangle _{R}$ and $\left\langle 
\widehat{p}_{A}\widehat{p}_{B}\right\rangle _{R}$ $=\left\langle \widehat{p}%
_{A}\right\rangle _{R}\left\langle \widehat{p}_{B}\right\rangle _{R}$ are
first used to show that 
\begin{eqnarray}
&&\left\langle \Delta (\widehat{x}_{A}+\widehat{x}_{B})^{2}\right\rangle
+\left\langle \Delta (\widehat{p}_{A}-\widehat{p}_{B})^{2}\right\rangle 
\nonumber \\
&\geq &\sum_{R}P_{R}\,\left( \left\langle \widehat{x}_{A}{}^{2}\right\rangle
_{R}-\left\langle \widehat{x}_{A}\right\rangle _{R}^{2}+\left\langle 
\widehat{x}_{B}{}^{2}\right\rangle _{R}-\left\langle \widehat{x}%
_{B}\right\rangle _{R}^{2}+\left\langle \widehat{p}_{A}{}^{2}\right\rangle
_{R}-\left\langle \widehat{p}_{A}\right\rangle _{R}^{2}+\left\langle 
\widehat{p}_{B}{}^{2}\right\rangle _{R}-\left\langle \widehat{p}%
_{B}\right\rangle _{R}^{2}\right)  \nonumber \\
&&  \label{Eq.InequalityDuanResult}
\end{eqnarray}%
For sub-system states $\widehat{\rho }_{R}^{A}$, $\widehat{\rho }_{R}^{B}$
that are physical we have in addition $\left\langle \widehat{x}%
_{A}\right\rangle _{R}=\left\langle \widehat{x}_{B}\right\rangle
_{R}=\left\langle \widehat{p}_{A}\right\rangle _{R}=\left\langle \widehat{p}%
_{B}\right\rangle _{R}=0$. Also using $\left\langle \widehat{a}%
^{2}\right\rangle _{R}=\left\langle (\widehat{a}^{\dag })^{2}\right\rangle
_{R}=\left\langle \widehat{b}^{2}\right\rangle _{R}=\left\langle (\widehat{b}%
^{\dag })^{2}\right\rangle _{R}=0$ for physical states we find for the
remaining terms in Eq.(\ref{Eq.InequalityDuanResult}) that 
\begin{eqnarray}
\left\langle \widehat{x}_{A}{}^{2}\right\rangle _{R} &=&\frac{1}{2}\left(
\left\langle \widehat{a}^{2}\right\rangle _{R}+\left\langle (\widehat{a}%
^{\dag })^{2}\right\rangle _{R}+1+2\left\langle \widehat{a}^{\dag }\,%
\widehat{a}\right\rangle _{R}\right) \geq \frac{1}{2}  \nonumber \\
\left\langle \widehat{x}_{B}{}^{2}\right\rangle _{R} &=&\frac{1}{2}\left(
\left\langle \widehat{b}^{2}\right\rangle _{R}+\left\langle (\widehat{b}%
^{\dag })^{2}\right\rangle _{R}+1+2\left\langle \widehat{b}^{\dag }\,%
\widehat{b}\right\rangle _{R}\right) \geq \frac{1}{2}  \nonumber \\
\left\langle \widehat{p}_{A}{}^{2}\right\rangle _{R} &=&-\frac{1}{2}\left(
\left\langle \widehat{a}^{2}\right\rangle _{R}+\left\langle (\widehat{a}%
^{\dag })^{2}\right\rangle _{R}-1-2\left\langle \widehat{a}^{\dag }\,%
\widehat{a}\right\rangle _{R}\right) \geq \frac{1}{2}  \nonumber \\
\left\langle \widehat{p}_{B}{}^{2}\right\rangle _{R} &=&-\frac{1}{2}\left(
\left\langle \widehat{b}^{2}\right\rangle _{R}+\left\langle (\widehat{b}%
^{\dag })^{2}\right\rangle _{R}-1-2\left\langle \widehat{b}^{\dag }\,%
\widehat{b}\right\rangle _{R}\right) \geq \frac{1}{2}
\end{eqnarray}%
Substituting these results into Eq.(\ref{Eq.InequalityDuanResult})
establishes the validity of (\ref{Eq.DuanInequalityNonEntState}) for
non-entangled states in which the $\widehat{\rho }_{R}^{A}$, $\widehat{\rho }%
_{R}^{B}$ are physical sub-system states. As there are entangled states that
violate the inequality (\ref{Eq.DuanInequalityNonEntState}), this inequality
is valid for determining whether a state is entangled.

\subsection{He et al 2012}

\label{SubSection - He 2012}

In two papers dealing with EPR entanglement He et al \cite{He11a}, \cite%
{He12a} a \emph{four mode} system associated with a double well potential is
considered. In the left well $A$ there are two localised modes with
annihilation operators $\widehat{a}_{1}$, $\widehat{a}_{2}$ and in the right
well $B$ there are two localised modes with annihilation operators $\widehat{%
b}_{1}$, $\widehat{b}_{2}$. The modes in each well could be associated with
different internal states or they could be associated with different spatial
modes of the same internal state. This four mode system provides for the
possibility of entanglement of \emph{two sub-systems} each consisting of 
\emph{pairs of modes}. With four modes there are three different choices of
such sub-systems but perhaps the most interesting from the point of view of
entanglement of spatially separated modes - and hence implications for EPR
entanglement - would be to have the two \emph{left well modes} $\widehat{a}%
_{1}$, $\widehat{a}_{2}$ as sub-system $A$ and the two \emph{right well modes%
} $\widehat{b}_{1}$, $\widehat{b}_{2}$ as sub-system $B$. Consistent with
the requirement that the sub-system density operators $\widehat{\rho }%
_{R}^{A}$, $\widehat{\rho }_{R}^{B}$ conform to the symmetrisation principle
and the super-selection rule, these density operators may now be of the form
given in Eq. (\ref{Eq.PhysStateSubSystPairA}). Hence as discussed in
Sub-System \ref{SubSystem - Two SubSystems of Pairs}, when considering \emph{%
non-entangled} states for the sub-systems $A$ and $B$ we now have as in Eq. (%
\ref{Eq.CondNonEntStatePairs}) 
\begin{eqnarray}
\left\langle (\widehat{a}_{i})^{n}\right\rangle _{A} &=&Tr(\widehat{\rho }%
_{R}^{A}(\widehat{a}_{i})^{n})\neq 0\qquad \left\langle (\widehat{a}%
_{i}^{\dag })^{n}\right\rangle _{A}=Tr(\widehat{\rho }_{R}^{A}(\widehat{a}%
_{i}^{\dag })^{n})\neq 0  \nonumber \\
\left\langle (\widehat{b}_{j})^{m}\right\rangle _{B} &=&Tr(\widehat{\rho }%
_{R}^{B}(\widehat{b}_{j})^{m})\neq 0\qquad \left\langle (\widehat{b}%
_{j}^{\dag })^{m}\right\rangle _{B}=Tr(\widehat{\rho }_{R}^{B}(\widehat{b}%
_{j}^{\dag })^{m})\neq 0  \nonumber \\
&&  \label{Eq.MeanAnnihCreatOprsModePairs}
\end{eqnarray}%
in general. Hence in this case where the sub-systems are \emph{pairs} of
modes the entanglement test in Eq. (\ref{Eq.EntangTest}) for sub-systems
consisting of \emph{single} modes cannot be applied.

\subsubsection{Correlation Tests for Entanglement}

However, the inequalities derived by Hillery et al \cite{Hillery09a} (see
SubSection \ref{SubSection - Hillery 2009}) 
\begin{equation}
|\left\langle (\widehat{a}_{i})^{m}\,(\widehat{b}_{j}^{\dag
})^{n}\right\rangle |^{2}\leq \left\langle (\widehat{a}_{i}^{\dag })^{m}(%
\widehat{a}_{i})^{m}\,(\widehat{b}_{j}^{\dag })^{n}(\widehat{b}%
_{j})^{n}\right\rangle  \label{Eq.HilleryIneqNonEntModePairs}
\end{equation}%
that apply for two non-entangled sub-systems $A$ and $B$ can now be usefully
applied, since in this case the quantities $\left\langle (\widehat{a}%
_{i})^{m}\,(\widehat{b}_{j}^{\dag })^{n}\right\rangle $ are in general no
longer zero. Thus there is an \emph{entanglement test }for two sub-systems
consisting of \emph{pairs of modes}. If 
\begin{eqnarray}
|\left\langle (\widehat{a}_{i})^{m}\,(\widehat{b}_{j}^{\dag
})^{n}\right\rangle |^{2} &>&\left\langle (\widehat{a}_{i}^{\dag })^{m}(%
\widehat{a}_{i})^{m}\,(\widehat{b}_{j}^{\dag })^{n}(\widehat{b}%
_{j})^{n}\right\rangle \qquad  \nonumber \\
for\;any\;of\;i,j &=&1,2  \label{Eq.HilleryEntTestPairsModes}
\end{eqnarray}%
then the quantum state for two sub-systems $A$ and $B$ - \emph{each}
consisting of \emph{two modes} localised in each well - is entangled.

\subsubsection{Spin Squeezing Tests for Entanglement}

There are numerous choices for defining spin operators but the most useful
would be the \emph{local spin operators} for each well \cite{He12a} defined
by 
\begin{eqnarray}
\widehat{S}_{x}^{A} &=&(\widehat{a}_{2}^{\dag }\widehat{a}_{1}+\widehat{a}%
_{1}^{\dag }\widehat{a}_{2})/2\qquad \widehat{S}_{y}^{A}=(\widehat{a}%
_{2}^{\dag }\widehat{a}_{1}-\widehat{a}_{1}^{\dag }\widehat{a}_{2})/2i\qquad 
\widehat{S}_{z}^{A}=(\widehat{a}_{2}^{\dag }\widehat{a}_{2}-\widehat{a}%
_{1}^{\dag }\widehat{a}_{1})/2  \nonumber \\
\widehat{S}_{x}^{B} &=&(\widehat{b}_{2}^{\dag }\widehat{b}_{1}+\widehat{b}%
_{1}^{\dag }\widehat{b}_{2})/2\qquad \widehat{S}_{y}^{B}=(\widehat{b}%
_{2}^{\dag }\widehat{b}_{1}-\widehat{b}_{1}^{\dag }\widehat{b}_{2})/2i\qquad 
\widehat{S}_{z}^{B}=(\widehat{b}_{2}^{\dag }\widehat{b}_{2}-\widehat{b}%
_{1}^{\dag }\widehat{b}_{1})/2  \nonumber \\
&&  \label{Eq.LocalSpinOprs}
\end{eqnarray}%
These satisfy the usual angular momentum commutation rules and those or the
different wells commute. The squares of the local vector spin operators are
related to the total number operators $\widehat{N}_{A}=\widehat{a}_{2}^{\dag
}\widehat{a}_{2}+\widehat{a}_{1}^{\dag }\widehat{a}_{1}$ and $\widehat{N}%
_{B}=\widehat{b}_{2}^{\dag }\widehat{b}_{2}+\widehat{b}_{1}^{\dag }\widehat{b%
}_{1}$ as $\tsum\limits_{\alpha }(\widehat{S}_{\alpha }^{A})^{2}=($ $%
\widehat{N}_{A}/2)(\widehat{N}_{A}/2+1)$ and $\tsum\limits_{\alpha }(%
\widehat{S}_{\alpha }^{B})^{2}=($ $\widehat{N}_{B}/2)(\widehat{N}_{B}/2+1)$.

For the local spin operators we have in general 
\begin{equation}
\left\langle \widehat{S}_{\alpha }^{A}\right\rangle _{A}=Tr(\widehat{\rho }%
_{R}^{A}\,\widehat{S}_{\alpha }^{A})\neq 0\qquad \left\langle \widehat{S}%
_{\alpha }^{B}\right\rangle _{B}=Tr(\widehat{\rho }_{R}^{B}\,\widehat{S}%
_{\alpha }^{B})\neq 0  \label{Eq.MeanSpinOprsModePairs}
\end{equation}%
since the pair of modes $\widehat{a}_{1}$, $\widehat{a}_{2}$ and/or $%
\widehat{b}_{1}$, $\widehat{b}_{2}$ may now be of the form given in Eq. (\ref%
{Eq.PhysStateSubSystPairA}).

In SubSection \ref{SubSection - Non-local Correlations} it was shown that $%
|\left\langle \widehat{\Omega }_{A}^{\dag }\widehat{\Omega }%
_{B}\right\rangle |^{2}\leq \left\langle \widehat{\Omega }_{A}^{\dag }\,%
\widehat{\Omega }_{A}\,\widehat{\Omega }_{B}^{\dag }\,\widehat{\Omega }%
_{B}\right\rangle $ for a non-entangled state, so with $\widehat{\Omega }%
_{A}=\widehat{S}_{-}^{A}=\widehat{S}_{x}^{A}-i\widehat{S}_{y}^{A}$ and $%
\widehat{\Omega }_{B}=\widehat{S}_{-}^{B}=\widehat{S}_{x}^{B}-i\widehat{S}%
_{y}^{B}=(\widehat{S}_{+}^{B})^{\dag }$ to give 
\begin{equation}
|\left\langle \widehat{S}_{+}^{A}\,\widehat{S}_{-}^{B}\right\rangle
|^{2}\leq \left\langle \widehat{S}_{+}^{A}\,\widehat{S}_{-}^{A}\,\widehat{S}%
_{+}^{B}\,\widehat{S}_{-}^{B}\,\right\rangle
\label{Eq.InequalitySpinOprsNonEntStatesModePairs}
\end{equation}%
for a non-entangled state of sub-systems $A$ and $B$. For the non-entangled
state of these two sub-systems we have 
\begin{equation}
\left\langle \widehat{S}_{+}^{A}\,\widehat{S}_{-}^{B}\right\rangle
=\dsum\limits_{R}P_{R}\,\left\langle \widehat{S}_{+}^{A}\right\rangle
_{A}^{R}\left\langle \widehat{S}_{-}^{B}\right\rangle _{B}^{R}
\end{equation}%
which in general is non-zero from Eq.(\ref{Eq.MeanSpinOprsModePairs}).

Hence a valid \emph{entanglement test} involving \emph{spin operators} for
sub-systems $A$ and $B$ - \emph{each} consisting of \emph{two modes}
localised in each well exists, and is if 
\begin{equation}
|\left\langle \widehat{S}_{+}^{A}\,\widehat{S}_{-}^{B}\right\rangle
|^{2}>\left\langle \widehat{S}_{+}^{A}\,\widehat{S}_{-}^{A}\,\widehat{S}%
_{+}^{B}\,\widehat{S}_{-}^{B}\,\right\rangle
\label{Eq.SpinEntTestPairsModes}
\end{equation}%
then the two sub-systems are entangled. A similar conclusion is stated in 
\cite{He12a}. This test for entanglement involves the local spin operators,
though it is not then the same as spin squeezing criteria. It is referred to
as \emph{spin entanglement}.

\pagebreak

\section{Experiments on Spin Squeezing}

\label{Section - Experiments}

There are several papers \cite{Esteve08a}, \cite{Gross10a}, \cite{Riedel10a}
which contain the results of measuring the spin squeezing parameter
analogous to the expression in Eq. (\ref{Eq.SpinSqueezingMeasure}) and
showing that spin squeezing occurs. The presence of entanglement is then
inferred by reference to theoretical papers such as \cite{Sorensen01a} that
show that spin squeezing only occurs for entangled states - it is an
entanglement \emph{witness}. As no independent \emph{measures} of
entanglement (however defined) are presented, nor are other independent
tests for entanglement carried out, it cannot be said that these paper shows 
\emph{experimentally} that spin squeezing requires entanglement. In \cite%
{Riedel10a} the emphasis is on showing how the spin squeezing can be
generated via the non-linear terms in the Josephson Hamiltonian.

\pagebreak

\section{Discussion and Summary of Key Results}

\label{Section - Discussion & Summary of Key Results}

This paper is mainly concerned with two mode entanglement for systems of
identical massive bosons and the relationship to spin squeezing. These
bosons may be atoms or molecules as in cold quantum gases.

A careful analysis is first given regarding the proper definition of a
non-entangled state for systems of identical particles, and hence by
implication the proper definition of an entangled state. Noting that
entanglement is meaningless until the subsystems being entangled are
specified, it is pointed out that whereas it is not possible to distinguish
identical particles and hence the individual particles are not legitimate
sub-systems, the same is not the case for the single particle states or
modes, so the \emph{modes} are then the the rightful \emph{sub-systems} to
be considered as being entangled or not. In this approach where the
sub-systems are modes, situations where there are differing numbers of
identical particles are treated as different physical states, not as
differing physical systems, and the \emph{symmetrisation principle} required
of physical states for identical particle systems will be satisfied by using
Fock states to describe the states.

Furthermore, it is argued that the overall physical states should conform to
the \emph{superselection rule} that excludes quantum superposition states of
the form (\ref{Eq.ForbiddenStates}) as physical states for systems of
identical particles - massive or otherwise. Although the justication of the
SSR in terms of observers and their \emph{reference frames} formulated by
other authors has also been presented for completeness, a number of fairly 
\emph{straightforward reasons} were given for why it is appropriate to apply
this superselection rule, which may be summarised as: 1. No way is known for
creating such states. 2 \ No way is known for measuring all the properties
of such states, even if they existed. 3. There is no need to invoke the
existence of such states in order to understand coherence and interference
effects. 4. The stability of such states against decoherence processes may
not be great, so even if they could be created they could rapidly change to
other states. The last reason is of lesser importance. Invoking the physical
existence of states that as far as we know cannot be made or measured, and
for which there are no known physical effects that require their presence
seems a rather unnecessary feature to add to the non-relativistic quantum
physics of many body systems, and considerations based on the general
principle of simplicity (Occam's razor) would suggest not doing this until a
clear physical justification for including them is found. As two mode
fermionic systems are restricted to states with at most two fermions, the
focus of the paper is then on bosonic systems, where large numbers of bosons
can occupy two mode systems.

However, although there is related work involving local particle number
super-selection rules, \emph{this paper differs} from a number of others by 
\emph{extending} the \emph{super-selection rule} to also apply to the
density operators $\widehat{\rho }_{R}^{A}$, $\widehat{\rho }_{R}^{B}$, ..
for the \emph{mode sub-systems} $A$, $B$, .that occur in the definition (\ref%
{Eq.NonEntangledState}) of a \emph{general non-entangled} state for systems
of identical particles. Hence it follows that the definition of \emph{%
entangled states} will differ in this paper from that which would apply if
density operators $\widehat{\rho }_{R}^{A}$, $\widehat{\rho }_{R}^{B}$,
.allowed for coherent superpositions of number states within each mode. In
fact more states are regarded as entangled in terms of the definition in the
present paper. Indeed, if \emph{further restrictions} are placed on the
sub-system density operators - such as requiring them to specify a fixed
number of bosons - the set of entangled states is further enlarged. The 
\emph{simple justification} for our viewpoint on applying the \emph{local
particle number super-selection rule} has three aspects. Firstly, since
experimental arrangements in which only one bosonic mode is involved can be
created, the same reasons (see last paragraph) justify applying the
super-selection rule to this mode system as applied for the system as a
whole. Secondly, measurements can be carried out on the separate modes, and
the joint probability for the outcomes of these measurements determined. For
a non-entangled state the joint probability (\ref{Eq.JointProbNonEntState})
for these measurements depends on all the density operators $\widehat{\rho }%
_{R}^{A}$, $\widehat{\rho }_{R}^{B}$, .. for the mode sub-systems as well as
the probability $P_{R}$ for the product state $\widehat{\rho }%
_{R}^{A}\otimes \widehat{\rho }_{R}^{B}\otimes ..$occuring when the general
mixed non-entangled state is prepared, which can be accomplished by local
preparations and classical communication. For the non-entangled state the
form of the joint probability $P_{AB..}(i,j,..)$ for measurements on all the
sub-systems is given by the products of the individual sub-system
probabilities $P_{A}^{R}(i)=Tr(\widehat{\Pi }_{i}^{A}\,\widehat{\rho }%
_{R}^{A})$, etc that measurements on the sub-systems $A,B,..$yield the
outcomes $\lambda _{i}^{A}$ etc when the sub-systems are in states $\widehat{%
\rho }_{R}^{A}$, $\widehat{\rho }_{R}^{B}$, ., the overall products being
weighted by the probability $P_{R}$ that a particular product state is
prepared. If $\widehat{\rho }_{R}^{A}$, $\widehat{\rho }_{R}^{B}$, did not
represent physical states then the interpretation of the joint probability
as this statistical average would be unphysical \ Thirdly, attempts to allow
the density operators $\widehat{\rho }_{R}^{A}$, $\widehat{\rho }_{R}^{B}$,
.. for the mode sub-systems to violate the super-selection rule provided
that the reduced density operators $\widehat{\rho }_{A}$, $\widehat{\rho }%
_{B}$ for the separate modes are consistent with it are shown not to be
possible in general.

As well as the above justifications for applying the super-selection rule to
both the overall multi-mode state for systems of identical particles and the
separate sub-system states in the definition of non-entangled states, a more
sophisticated justification based on considering SSR to be the consequence
of describing the quantum state by an observer (Charlie) whose phase
reference is unknown has also been presented in detail in Appendix \ref%
{Appendix - Reference Frames and SSR} for completeness. For the sub-systems 
\emph{local reference frames} are involved. The SSR is seen as a special
case of a general SSR which forbids quantum states from exhibiting
coherences between states associated with \emph{irreducible representations}
of the transformation group that relates reference frames, and which may be
the \emph{symmetry group} for the system.

The present paper then defines spin squeezing for two mode systems and
discusses the desirability of defining \emph{spin squeezing} in terms of the 
\emph{principal spin operators} $\widehat{J}_{x},\widehat{J}_{y},\widehat{J}%
_{z}$ for which the \emph{covariance matrix} is diagonal, rather than via
the original spin operators $\widehat{S}_{x},\widehat{S}_{y},\widehat{S}_{z}$
defined in terms of the original mode annihilation and creation operators $%
\widehat{a},\widehat{b}$ and $\widehat{a}^{\dag },\widehat{b}^{\dag }$ and
for which the covariance matrix is non-diagonal in general. It is seen that
the two sets of spin operators are related via a rotation operator and the
principal spin operators are given in terms of \emph{new mode operators} $%
\widehat{c},\widehat{d}$ and $\widehat{c}^{\dag },\widehat{d}^{\dag }$, with 
$\widehat{c},\widehat{d}$ obtained as linear combinations of the original
mode operators $\widehat{a},\widehat{b}$ and hence defining two new modes.

The immediate consequence for the case of two mode systems of identical
bosons of the present approach to defining entangled states is that spin
squeezing in \emph{any} of the principle spin operators $\widehat{J}_{x}$, $%
\widehat{J}_{y}$ or $\widehat{J}_{z}$ \emph{requires} entanglement of the
new modes $\widehat{c},\widehat{d}$. Similarly, spin squeezing in \emph{any}
of the original spin operators $\widehat{S}_{x}$, $\widehat{S}_{y}$ or $%
\widehat{S}_{z}$ requires entanglement of the original modes $\widehat{a},%
\widehat{b}$. A typical test for entanglement is $\left\langle \Delta 
\widehat{S}\,_{x}^{2}\right\rangle <|\left\langle \widehat{S}%
_{z}\right\rangle |/2$ or $\left\langle \Delta \widehat{S}%
\,_{y}^{2}\right\rangle <|\left\langle \widehat{S}_{z}\right\rangle |/2$. It
is noted that though spin squeezing requires entanglement, the opposite is
not the case and the $\emph{NOON}$ state provides an example of an entangled
physical state that is not spin squeezed. Also, the \emph{binomial state}
provides an example of a state that is entangled and spin squeezed for one
choice of mode sub-systems may be non-entangled and not spin squeezed for
another choice. The \emph{relative phase state} provides an example that is
entangled for new modes $\widehat{c},\widehat{d}$ and is highly spin
squeezed in $\widehat{J}_{y}$ and very unsqueezed in $\widehat{J}_{x}$. The
connection between spin squeezing and entanglement is regarded as
well-known, but up to now only proofs based on non-entangled states that
either disregard the symmetrization principle or the sub-system
super-selection rules exist, placing the connection between spin squeezing
and entanglement on a somewhat shaky basis. On the other hand, the proof
given here is based on a definition of non-entangled (and hence entangled)
states that is compatible with both these requirements.

There are several papers that obtain \emph{different tests} for whether a
state is entangled from those involving spin squeezing that are obtained in
this paper, the proofs often being based on a definition of non-entangled
states that ignores symmetrization or SSR. Hillery et al \cite{Hillery06a}
obtain criteria of this type, such as the entanglement test $\left\langle
\Delta \widehat{S}\,_{x}^{2}\right\rangle +\left\langle \Delta \widehat{S}%
\,_{y}^{2}\right\rangle $ $<\frac{1}{2}\left\langle \widehat{N}\right\rangle 
$. This test is also valid if the non-entangled state definition is
consistent with the SSR, but is different to the test $\left\langle \Delta 
\widehat{S}\,_{x}^{2}\right\rangle +\left\langle \Delta \widehat{S}%
\,_{y}^{2}\right\rangle $ $<|\left\langle \widehat{S}_{z}\right\rangle |$
suggested by the requirement that $\left\langle \Delta \widehat{S}%
\,_{x}^{2}\right\rangle +\left\langle \Delta \widehat{S}\,_{y}^{2}\right%
\rangle \geq |\left\langle \widehat{S}_{z}\right\rangle |$ for non-entangled
states - since both $\left\langle \Delta \widehat{S}\,_{x}^{2}\right\rangle
\geq |\left\langle \widehat{S}_{z}\right\rangle |/2$ and $\left\langle
\Delta \widehat{S}\,_{y}^{2}\right\rangle \geq |\left\langle \widehat{S}%
_{z}\right\rangle |/2$. The latter inequality is of no use since $%
\left\langle \Delta \widehat{S}\,_{x}^{2}\right\rangle +\left\langle \Delta 
\widehat{S}\,_{y}^{2}\right\rangle $ $\geq |\left\langle \widehat{S}%
_{z}\right\rangle |$ for all states. However as previously noted, showing
that either $\left\langle \Delta \widehat{S}\,_{x}^{2}\right\rangle
<|\left\langle \widehat{S}_{z}\right\rangle |/2$ or $\left\langle \Delta 
\widehat{S}\,_{y}^{2}\right\rangle <|\left\langle \widehat{S}%
_{z}\right\rangle |/2$ - or the analogous tests for other pairs of spin
operators - already provides a test for the entanglement of the original
modes $\widehat{a},\widehat{b}$. This test is a different test for
entanglement than that of Hillery et al \cite{Hillery06a}. The case of the 
\emph{relative phase eigenstate} is an example of an entangled state in
which the spin squeezing test for entanglement \emph{succeeds} whereas that
of Hillery et al \cite{Hillery06a} \emph{fails}. Other inequalities found by
Hillery et al \cite{Hillery09a} for non-entangled states which also do not
depend on whether non-entangled states satisfy the super-selection rule
include $|\left\langle (\widehat{a})^{m}\,(\widehat{b}^{\dag
})^{n}\right\rangle |^{2}\leq \left\langle (\widehat{a}^{\dag })^{m}(%
\widehat{a})^{m}\,(\widehat{b}^{\dag })^{n}(\widehat{b})^{n}\right\rangle $,
giving another valid test $|\left\langle (\widehat{a})^{m}\,(\widehat{b}%
^{\dag })^{n}\right\rangle |^{2}>\left\langle (\widehat{a}^{\dag })^{m}(%
\widehat{a})^{m}\,(\widehat{b}^{\dag })^{n}(\widehat{b})^{n}\right\rangle $
for an entangled state. However, with entanglement defined as in the present
paper we have $|\left\langle (\widehat{a})^{m}\,(\widehat{b}^{\dag
})^{n}\right\rangle |^{2}=0$ for a non-entangled state, so an entanglement
test in the form $|\left\langle (\widehat{a})^{m}\,(\widehat{b}^{\dag
})^{n}\right\rangle |^{2}>0$ immediately follows. This test is less
stringent than that of Hillery et al \cite{Hillery09a}., as $|\left\langle (%
\widehat{a})^{m}\,(\widehat{b}^{\dag })^{n}\right\rangle |^{2}$ is then
required to be larger. Sorensen et al \cite{Sorensen01a} show that spin
squeezing is a test for a state being entangled, but define non-entangled
states for identical particle systems (such as BECs) in a form that is \emph{%
inconsistent} with the symmetrisation principle - the sub-systems being
regarded as individual identical particles. However, the treatment of
Sorensen et al \cite{Sorensen01a} can be modified to apply to a system of
identical bosons if the particle index $i$ is \emph{re-interpreted} as
specifying diffferent modes, for example modes localised on optical lattice
sites $i=1,2,..,N$. With two single particle states $\left\vert \phi
_{ai}\right\rangle $ ,$\left\vert \phi _{bi}\right\rangle $ with
annihilation operators $a_{i},b_{i}$ available on each site, there would
then be $2N$ modes involved, but spin operators can still be defined. If the
definitions of non--entangled and entangled states in the present paper are
applied, it can be shown that spin squeezing in either of the spin operators 
$\widehat{S}_{x}$ or $\widehat{S}_{y}$ requires entanglement of \emph{all}
the original modes $\widehat{a}_{i},\widehat{b}_{i}$. Alternatively, if the
sub-systems are \emph{pairs} of modes $\widehat{a}_{i},\widehat{b}_{i}$ 
\emph{and} the sub-system density operators $\widehat{\rho }_{R}^{i}$ were
restricted to states with exactly \emph{one boson}, then it can be shown
that spin squeezing in $\widehat{S}_{z}$ requires entanglement of all the
pairs of modes. With this restriction the pair of modes $\widehat{a}_{i}$, $%
\widehat{b}_{i}$ behave like \emph{distinguishable} two state particles,
which was essentially the case that Sorensen et al \cite{Sorensen01a}
implicitly considered. This type of entanglement is a multi-mode
entanglement of a special type - since the modes $\widehat{a}_{i}$, $%
\widehat{b}_{i}$ may themselves be entangled there is an "entanglement of
entanglement". So with either of these key revisions, the work of Sorensen
et al \cite{Sorensen01a} could be said to show that spin squeezing requires
entanglement. Sorensen and Molmer \cite{Sorensen01b} have deduced an
inequality involving $\left\langle \Delta \widehat{J}\,_{x}^{2}\right\rangle 
$ and $|\left\langle \widehat{J}\,_{z}\right\rangle |$ for states where $%
\left\langle \widehat{J}\,_{x}\right\rangle =\left\langle \widehat{J}%
\,_{y}\right\rangle =0$ based on just the Heisenberg uncertainty principle.
This is useful in terms of confirming that states do exist that are spin
squeezed still conform to this principle. Duan et al \cite{Duan00a}, Toth et
al \cite{Toth03a} devise a test for entanglement based on the sum of the
quadrature variances $\left\langle \Delta (\widehat{x}_{A}+\widehat{x}%
_{B})^{2}\right\rangle +\left\langle \Delta (\widehat{p}_{A}-\widehat{p}%
_{B})^{2}\right\rangle $, which involve quadrature components $\widehat{x}%
_{A},\widehat{p}_{A},\widehat{x}_{B},\widehat{p}_{B}$ constructed from the
original mode annihilation, creation operators for modes $A$, $B$. Their
conclusion that if the sum is less than $2$ then the state is entangled is
valid both for the present definition of entanglement and for that in which
the application of the super-selection rule is ignored. He et al \cite{He11a}%
, \cite{He12a} consider a \emph{four mode} system, with two modes localised
in each well of a double well potential. If the two sub-systems $A$ and $B$
each consist of two modes - with $\widehat{a}_{1}$, $\widehat{a}_{2}$ as
sub-system $A$ and $\widehat{b}_{1}$, $\widehat{b}_{2}$ as sub-system $B$,
then tests of entanglement of the two sub-systems of the Hillery \cite%
{Hillery09a} type $|\left\langle (\widehat{a}_{i})^{m}\,(\widehat{b}%
_{j}^{\dag })^{n}\right\rangle |^{2}>\left\langle (\widehat{a}_{i}^{\dag
})^{m}(\widehat{a}_{i})^{m}\,(\widehat{b}_{j}^{\dag })^{n}(\widehat{b}%
_{j})^{n}\right\rangle $ for any $i$, $j=1$, $2$ or involving local spin
operators $|\left\langle \widehat{S}_{+}^{A}\,\widehat{S}_{-}^{B}\right%
\rangle |^{2}>\left\langle \widehat{S}_{+}^{A}\,\widehat{S}_{-}^{A}\,%
\widehat{S}_{+}^{B}\,\widehat{S}_{-}^{B}\,\right\rangle $ apply.

Overall then, all of the \emph{entanglement tests} (spin squeezing and
other) in the other papers discussed here are \emph{still valid} when
reconsidered in accord with the definition of entanglement based on the
symmetrisation and super-selection rules, though in one case Sorensen et al 
\cite{Sorensen01a} a re-definition of the sub-systems is required to satisfy
the symmetrization principle. However, \emph{further} tests for entanglement
are obtained in the present paper based on non-entangled states that are
consistent with the symmetrizaton and super-selection rules. In some cases
they are less stringent - the correlation test in Eq.(\ref{Eq.EntangTest})
being easier to satisfy than that of Hillery et al \cite{Hillery09a} in Eq. (%
\ref{Eq.HilleryEntangTest}). They are certainly \emph{different} to others
previously discovered.

At present, experiments demonstrating spin squeezing\ do not show
experimentally whether spin squeezing requires entanglement, however
defined, since no results for entanglement measures are presented, nor are
other independent tests for entanglement carried out. \pagebreak

\bigskip

\section{Acknowledgements}

The authors thank S. M. Barnett, J. F. Corney, P. D. Drummond, M. Hall, J.
Jeffers, K. Molmer, D. Oi, M. D. Reid, K. Rzazewski, T. Rudolph, J. A.
Vaccaro and V. Vedral for helpful discussions. BJD thanks the Science
Foundation of Ireland for funding this research via an E\ T S Walton
Visiting Fellowship. \pagebreak 

\section{Appendix 1 - Projective Measurements and Conditional Probabilities}

\label{Appendix - Projective Measurements}

\subsubsection{Projective Measurements}

For simplicity, we will only consider \emph{projective} (or von Neumann)
measurements rather than more general measurements involving \emph{positive
operator measurements} (POM). If $\widehat{\Omega }$ is a physical quantity
associated with the system, with eigenvalues $\lambda _{i}$ and with $%
\widehat{\Pi }_{i}$ the projector onto the subspace with eigenvalue $\lambda
_{i}$ then the probability $P(i)$ that measurement of $\widehat{\Omega }$
leads to the value $\lambda _{i}$ is given by \cite{Isham95a} 
\begin{equation}
P(i)=Tr(\widehat{\Pi }_{i}\widehat{\rho })  \label{Eq.ProbMeast}
\end{equation}%
For projective measurements $\widehat{\Pi }_{i}=\widehat{\Pi }_{i}^{2}=%
\widehat{\Pi }_{i}^{\dag }$ and $\dsum\limits_{i}\widehat{\Pi }_{i}=1$,
together with $\widehat{\Omega }\widehat{\Pi }_{i}=\widehat{\Pi }_{i}%
\widehat{\Omega }=\lambda _{i}\widehat{\Pi }_{i}$.

Following the measurement which leads to the value $\lambda _{i}$ the
density operator is different and given by%
\begin{equation}
\widehat{\rho }_{cond}(\widehat{\Omega },i)=(\widehat{\Pi }_{i}\widehat{\rho 
}\widehat{\Pi }_{i})/P(i)  \label{Eq.DensOprAfterMeast}
\end{equation}%
This is known as the \emph{reduction of the wave function}, and can be
viewed in two ways. From an ontological point of view a quantum projective
measurement \emph{changes} the quantum state significantly because the
interaction with the measurement system is not just a small perturbation, as
it can be in classical physics. From the epistomological point of view we
know what value the physical quantity $\widehat{\Omega }$ now has, so if
measurement of $\widehat{\Omega }$ were to be \emph{repeated} immediately it
would be expected -- with a probability of unity - that the value would be $%
\lambda _{i}$. The new density operator $\widehat{\rho }_{cond}(\widehat{%
\Omega },i)$ satisfies this requirement. It also satisfies the standard
requirements of Hermitiancy, unit trace, positivity - as is easily shown.

To show this formally we have for the \emph{mean value} for $\widehat{\Omega 
}$ following the measurement%
\begin{eqnarray}
\left\langle \widehat{\Omega }\right\rangle _{i} &=&Tr(\widehat{\Omega }\,%
\widehat{\rho }_{cond}(\widehat{\Omega },i))  \nonumber \\
&=&Tr(\widehat{\Omega }\,(\widehat{\Pi }_{i}\widehat{\rho }\widehat{\Pi }%
_{i}))/P(i)  \nonumber \\
&=&\lambda _{i}Tr(\widehat{\Pi }_{i}\widehat{\rho })/P(i)  \nonumber \\
&=&\lambda _{i}  \label{Eq.MeanAfterMeast}
\end{eqnarray}%
whilst for the \emph{variance}%
\begin{eqnarray}
\left\langle \Delta \widehat{\Omega }^{2}\right\rangle _{i} &=&Tr((\widehat{%
\Omega }-\left\langle \widehat{\Omega }\right\rangle _{i})^{2}\,\widehat{%
\rho }_{cond}(\widehat{\Omega },i))  \nonumber \\
&=&Tr(\widehat{\Omega }^{2}\,\widehat{\rho }_{red}(i))-\left\langle \widehat{%
\Omega }\right\rangle _{i}^{2}  \nonumber \\
&=&\lambda _{i}^{2}-\lambda _{i}^{2}  \nonumber \\
&=&0  \label{Eq.VarianceAfterMeast}
\end{eqnarray}%
which is zero as expected.

If following the measurement of $\widehat{\Omega }$ the results of the
measurement were discarded then the density operator after the measurement
is 
\begin{equation}
\widehat{\rho }_{cond}(\widehat{\Omega })=\dsum\limits_{i}P(i)\,\widehat{%
\rho }_{cond}(\widehat{\Omega },i)=\dsum\limits_{i}\widehat{\Pi }_{i}%
\widehat{\rho }\widehat{\Pi }_{i}  \label{Eq.DensOprAfterMeastDiscarded}
\end{equation}%
which is the sum of the $\widehat{\rho }_{cond}(\widehat{\Omega },i)$ each
weighted by the probability $P(i)$ of the result $\lambda _{i}$ occuring.
Note that the expression for $\widehat{\rho }_{cond}(\widehat{\Omega })$ is
not the same as the original density operator $\widehat{\rho }$. This is to
be expected from both the epistimological and ontological points of view,
since although we do \emph{not} know \emph{what} value $\lambda _{i}$ has
occurred, it is known that \emph{a} definite value for $\widehat{\Omega }$
has been found, or that measurement process has destroyed any \emph{%
coherences} that previously existed between different eigenstates of $%
\widehat{\Omega }$. We note that $\widehat{\rho }_{cond}(\widehat{\Omega })$
also satisfies the standard requirements of Hermitiancy, unit trace,
positivity - as is easily shown.

\subsubsection{Conditional Probabilities}

Suppose we follow the measurement of $\widehat{\Omega }$ resulting in
eigenvalue $\lambda _{i}$ with a measurement of $\widehat{\Lambda }$
resulting in eigenvalue $\mu _{j}$ where the projector associated with the
latter measurement is $\widehat{\Xi }_{j}$. Then the \emph{conditional
probabiltity} of measuring $\widehat{\Lambda }$ resulting in eigenvalue $\mu
_{j}$ following the measurement of $\widehat{\Omega }$ that resulted in
eigenvalue $\lambda _{i}$ would be 
\begin{eqnarray}
P(j|i) &=&Tr(\widehat{\Xi }_{j}\widehat{\rho }_{cond}(\widehat{\Omega },i)) 
\nonumber \\
&=&Tr(\widehat{\Xi }_{j}(\widehat{\Pi }_{i}\widehat{\rho }\widehat{\Pi }%
_{i}))/P(i)  \nonumber \\
&=&Tr((\widehat{\Xi }_{j}\widehat{\Pi }_{i})\,\widehat{\rho }\,(\widehat{\Pi 
}_{i}\widehat{\Xi }_{j}))/P(i)  \label{Eq.CondProbMeastJAfterI}
\end{eqnarray}%
where the cyclic properties of the trace and the idempotent property of the
projector have been used. If the measurements had taken place in the reverse
order the conditional probabiltity of measuring $\widehat{\Omega }$
resulting in eigenvalue $\lambda _{i}$ following the measurement of $%
\widehat{\Lambda }$ that resulted in eigenvalue $\mu _{j}$ would be%
\begin{equation}
P(i|j)=Tr((\widehat{\Pi }_{i}\widehat{\Xi }_{j})\,\widehat{\rho }\,(\widehat{%
\Xi }_{j}\widehat{\Pi }_{i}))/P(j)  \label{Eq.CondProbMeastIAfterJ}
\end{equation}

We note that the actual probability of measuring $\lambda _{i}$ then $\mu
_{j}$ would be the \emph{joint probability} 
\begin{equation}
P(j\;after\;i)=P(j|i)\,P(i)=Tr((\widehat{\Xi }_{j}\widehat{\Pi }_{i})\,%
\widehat{\rho }\,(\widehat{\Pi }_{i}\widehat{\Xi }_{j}))
\label{Eq.JointProbJAfterI}
\end{equation}%
whilst the actual probability of measuring $\mu _{j}$ then $\lambda _{i}$
would be the joint probability%
\begin{equation}
P(i\;after\;j)=P(i|j)\,P(j)=Tr((\widehat{\Pi }_{i}\widehat{\Xi }_{j})\,%
\widehat{\rho }\,(\widehat{\Xi }_{j}\widehat{\Pi }_{i}))
\label{Eq. JointProbIAfterJ}
\end{equation}%
and we note that in general these two joint probabilities are different.

If however, the two physical quantities \emph{commute}, then there are a
complete set of simultaneous eigenvectors $\left\vert \lambda _{i},\mu
_{j}\right\rangle $ for $\widehat{\Omega }$ and $\widehat{\Lambda }$. It is
then straightforward to show that $\widehat{\Pi }_{i}\widehat{\Xi }_{j}=%
\widehat{\Xi }_{j}\widehat{\Pi }_{i}$, in which case $P(j\;after\;i)=P(i%
\;after\;j)=P(i,j)$, so it does not matter which order the measurements are
carried out. The overall result 
\begin{eqnarray}
P(i,j) &=&P(j|i)\,P(i)=P(i|j)\,P(j)  \nonumber \\
&=&Tr(\widehat{\Pi }_{i}\widehat{\Xi }_{j}\,\widehat{\rho }\,\widehat{\Xi }%
_{j}\widehat{\Pi }_{i})  \nonumber \\
&=&Tr(\widehat{\Pi }_{i}\widehat{\Xi }_{j}\,\widehat{\rho })
\label{Eq.BayesThm}
\end{eqnarray}%
is an expression of \emph{Bayes theorem}.

A case of particular importance where this occurs is in situations involving
two or more distinct sub-systems, in which the operators $\widehat{\Omega }$
and $\widehat{\Lambda }$ are associated with different sub-systems. For two
sub-systems $A$ and $B$ the operators $\widehat{\Omega }$ and $\widehat{%
\Lambda }$ are of the form $\widehat{\Omega }_{A}$ and $\widehat{\Omega }%
_{B} $, or more strictly $\widehat{\Omega }_{A}\otimes \widehat{1}_{B}$ and $%
\widehat{1}_{A}\otimes \widehat{\Omega }_{B}$. It is easy to see that $(%
\widehat{\Omega }_{A}\otimes \widehat{1}_{B})(\widehat{1}_{A}\otimes 
\widehat{\Omega }_{B})=\widehat{\Omega }_{A}\otimes \widehat{\Omega }_{B}=(%
\widehat{1}_{A}\otimes \widehat{\Omega }_{B})(\widehat{\Omega }_{A}\otimes 
\widehat{1}_{B})$, so the operators commute and results such as in Bayes
theorem (\ref{Eq.BayesThm}) apply.

\subsubsection{Conditional Mean and Variance}

To determine the \emph{conditioned mean value} of $\widehat{\Lambda }$ after
measurement of $\widehat{\Omega }$ has led to the eigenvalue $\lambda _{i}$
we use $\widehat{\rho }_{cond}(\widehat{\Omega },i)$ rather than $\,\widehat{%
\rho }$ in the mean formula $\left\langle \widehat{\Lambda }\right\rangle
=Tr(\widehat{\Lambda }\widehat{\rho })$. Hence%
\begin{eqnarray}
\left\langle \widehat{\Lambda }\right\rangle _{i} &=&Tr(\widehat{\Lambda }%
\widehat{\rho }_{cond}(\widehat{\Omega },i))  \nonumber \\
&=&Tr(\widehat{\Lambda }\,(\widehat{\Pi }_{i}\widehat{\rho }\widehat{\Pi }%
_{i}))/P(i)  \label{Eq.CondMean}
\end{eqnarray}%
Now 
\begin{equation}
\widehat{\Lambda }=\dsum\limits_{j}\mu _{j}\widehat{\Xi }_{j}
\label{Eq.OprProjectors}
\end{equation}%
so that 
\begin{eqnarray}
\left\langle \widehat{\Lambda }\right\rangle _{i} &=&\dsum\limits_{j}\mu
_{j}\,Tr(\widehat{\Xi }_{j}\,\widehat{\Pi }_{i}\widehat{\rho }\widehat{\Pi }%
_{i})/P(i)  \nonumber \\
&=&\dsum\limits_{j}\mu _{j}\,Tr(\widehat{\Xi }_{j}\widehat{\Pi }_{i}\widehat{%
\rho }\widehat{\Pi }_{i}\widehat{\Xi }_{j})/P(i)  \nonumber \\
&=&\dsum\limits_{j}\mu _{j}\,P(j|i)  \label{Eq.CondMeanResult}
\end{eqnarray}%
using $\widehat{\Xi }_{j}=\widehat{\Xi }_{j}^{2}$, the cyclic trace
properties and Eq.(\ref{Eq.CondProbMeastJAfterI}). Hence the conditional
mean value is as expected, with the conditional probability $P(j|i)$
replacing $P(j)$ in the averaging process.

For the \emph{conditioned variance} of $\widehat{\Lambda }$ after
measurement of $\widehat{\Omega }$ has led to the eigenvalue $\lambda _{i}$
we use $\widehat{\rho }_{cond}(\widehat{\Omega },i)$ rather than $\,\widehat{%
\rho }$ and the conditioned mean $\left\langle \widehat{\Lambda }%
\right\rangle _{i}$ rather than $\left\langle \widehat{\Lambda }%
\right\rangle $ in the variance formula $\left\langle \Delta \widehat{%
\Lambda }^{2}\right\rangle =Tr((\widehat{\Lambda }-\left\langle \widehat{%
\Lambda }\right\rangle )^{2}\widehat{\rho })$. Hence%
\begin{eqnarray}
\left\langle \Delta \widehat{\Lambda }^{2}\right\rangle _{i} &=&Tr((\widehat{%
\Lambda }-\left\langle \widehat{\Lambda }\right\rangle _{i})^{2}\widehat{%
\rho }_{cond}(\widehat{\Omega },i))  \nonumber \\
&=&Tr((\widehat{\Lambda }-\left\langle \widehat{\Lambda }\right\rangle
_{i})^{2}(\widehat{\Pi }_{i}\widehat{\rho }\widehat{\Pi }_{i}))/P(i)
\label{Eq.CondVariance}
\end{eqnarray}%
Now 
\begin{equation}
(\widehat{\Lambda }-\left\langle \widehat{\Lambda }\right\rangle
_{i})^{2}=\dsum\limits_{j}(\mu _{j}-\left\langle \widehat{\Lambda }%
\right\rangle _{i})^{2}\widehat{\Xi }_{j}  \label{Eq.FluctnProjectors}
\end{equation}%
so that 
\begin{eqnarray}
\left\langle \Delta \widehat{\Lambda }^{2}\right\rangle _{i}
&=&\dsum\limits_{j}(\mu _{j}-\left\langle \widehat{\Lambda }\right\rangle
_{i})^{2}\,Tr(\widehat{\Xi }_{j}\,\widehat{\Pi }_{i}\widehat{\rho }\widehat{%
\Pi }_{i})/P(i)  \nonumber \\
&=&\dsum\limits_{j}(\mu _{j}-\left\langle \widehat{\Lambda }\right\rangle
_{i})^{2}\,P(j|i)  \label{Eq.CondVarianceResult}
\end{eqnarray}%
using the same steps as for the conditioned mean. Hence the conditional
variance is as expected, with the conditional probability $P(j|i)$ replacing 
$P(j)$ in the averaging process.

\pagebreak

\section{Appendix 2 - Inequalities}

\label{Appendix - Inequalities}

These inequalities are examples of Schwarz inequalities.

\subsection{Integral Inequality}

If $C(\lambda ),D(\lambda )$ are real, positive functions of $\lambda $ and $%
P(\lambda )$ is another real, positive function then we can show that%
\begin{equation}
\tint d\lambda \,P(\lambda )C(\lambda ).\tint d\lambda \,P(\lambda
)D(\lambda )\geq \left( \tint d\lambda \,P(\lambda )\sqrt{C(\lambda
)D(\lambda )}\right) ^{2}  \label{Eq.IntegralInequality0}
\end{equation}

To show this write $x=\tint d\lambda \,P(\lambda )C(\lambda )$ and $y=\tint
d\lambda \,P(\lambda )D(\lambda )$. Then 
\begin{eqnarray}
xy &=&\tint d\lambda \,P(\lambda )C(\lambda )\tint d\mu \,P(\mu )D(\mu ) 
\nonumber \\
&=&\tint \tint d\lambda \,d\mu \,P(\lambda )P(\mu )C(\lambda )D(\mu ) 
\nonumber \\
&=&\tint d\lambda \,P(\lambda )^{2}C(\lambda )D(\lambda )+\tint \tint
d\lambda \,d\mu \,(1-\delta (\lambda -\mu ))\,P(\lambda )P(\mu )C(\lambda
)D(\mu )  \nonumber \\
&&  \label{Eq.R1}
\end{eqnarray}%
Also, write $z=\left( \tint d\lambda \,P(\lambda )\sqrt{C(\lambda )D(\lambda
)}\right) ^{2}$. Then%
\begin{eqnarray}
z &=&\tint d\lambda \,P(\lambda )\sqrt{C(\lambda )D(\lambda )}\,\tint d\mu
\,P(\mu )\sqrt{C(\mu )D(\mu )}\,  \nonumber \\
&=&\tint \tint d\lambda \,d\mu \,P(\lambda )P(\mu )\,\sqrt{C(\lambda
)D(\lambda )}\sqrt{C(\mu )D(\mu )}  \nonumber \\
&=&\tint d\lambda \,P(\lambda )^{2}C(\lambda )D(\lambda )+\tint \tint
d\lambda \,d\mu \,(1-\delta (\lambda -\mu ))\,P(\lambda )P(\mu )\,\sqrt{%
C(\lambda )D(\lambda )}\sqrt{C(\mu )D(\mu )}  \nonumber \\
&&  \label{Eq.R2}
\end{eqnarray}%
so that 
\begin{eqnarray}
xy-z &=&\tint \tint d\lambda \,d\mu \,(1-\delta (\lambda -\mu ))\,P(\lambda
)P(\mu )\,\left( C(\lambda )D(\mu )-\sqrt{C(\lambda )D(\lambda )}\sqrt{C(\mu
)D(\mu )}\right)  \nonumber \\
&=&\frac{1}{2}\tint \tint d\lambda \,d\mu \,(1-\delta (\lambda -\mu
))\,P(\lambda )P(\mu )\,\left( C(\lambda )D(\mu )+C(\mu )D(\lambda )-2\sqrt{%
C(\lambda )D(\mu )}\sqrt{C(\mu )D(\lambda )}\right)  \nonumber \\
&=&\frac{1}{2}\tint \tint d\lambda \,d\mu \,(1-\delta (\lambda -\mu
))\,P(\lambda )P(\mu )\,\left( \sqrt{C(\lambda )D(\mu )}-\sqrt{C(\mu
)D(\lambda )}\right) ^{2}  \nonumber \\
&\geq &0  \label{Eq.R3}
\end{eqnarray}%
which proves the result.

For the special case where $D(\lambda )=1$ and where $\tint d\lambda
\,P(\lambda )=1$ we get the simpler result%
\begin{equation}
\tint d\lambda \,P(\lambda )C(\lambda )\geq \left( \tint d\lambda
\,P(\lambda )\sqrt{C(\lambda )}\right) ^{2}  \label{Eq.IntegralInequality}
\end{equation}

\subsection{Sum Inequality}

If $C_{R}$ and $D_{R}$ are real, positive quantities for various $R$ and $%
P_{R}$ is another real, positive quantity then we can show that%
\begin{equation}
\tsum\limits_{R}\,P_{R}\,C_{R}\,\tsum\limits_{R}\,P_{R}\,D_{R}\geq \left(
\tsum\limits_{R}\,P_{R}\,\sqrt{C_{R}D_{R}}\right) ^{2}
\label{Eq.SumInequality0}
\end{equation}%
To prove this write $x=\tsum\limits_{R}\,P_{R}\,C_{R}$ and $%
y=\tsum\limits_{R}\,P_{R}\,D_{R}$ Then 
\begin{eqnarray}
xy &=&\tsum\limits_{R}\,P_{R}\,C_{R}\,\tsum\limits_{S}\,P_{S}\,D_{S} 
\nonumber \\
&=&\tsum\limits_{R}\tsum\limits_{S}\,P_{R}\,P_{S}\,C_{R}D_{S}  \nonumber \\
&=&\tsum\limits_{R}\,P_{R}^{2}\,C_{R}D_{R}+\tsum\limits_{R}\tsum\limits_{S}%
\,(1-\delta _{RS})\,P_{R}\,P_{S}\,C_{R}D_{S}  \label{Eq.R4}
\end{eqnarray}%
Also, write $z=\left( \tsum\limits_{R}\,P_{R}\,\sqrt{C_{R}D_{R}}\right) ^{2}$%
. Then%
\begin{eqnarray}
z &=&\left( \tsum\limits_{R}\,P_{R}\,\sqrt{C_{R}D_{R}}\right) \left(
\tsum\limits_{S}\,P_{S}\,\sqrt{C_{S}D_{S}}\right)  \nonumber \\
&=&\tsum\limits_{R}\tsum\limits_{S}\,P_{R}\,P_{S}\,\sqrt{C_{R}D_{R}}\sqrt{%
C_{S}D_{S}}  \nonumber \\
&=&\tsum\limits_{R}\,P_{R}^{2}\,C_{R}D_{R}+\tsum\limits_{R}\tsum\limits_{S}%
\,(1-\delta _{RS})\,P_{R}\,P_{S}\,\sqrt{C_{R}D_{R}}\sqrt{C_{S}D_{S}}
\label{Eq.R5}
\end{eqnarray}%
so that 
\begin{eqnarray}
xy-z &=&\tsum\limits_{R}\tsum\limits_{S}\,\,P_{R}\,P_{S}\,(1-\delta
_{RS})\,\left( C_{R}D_{S}-\sqrt{C_{R}D_{R}}\sqrt{C_{S}D_{S}}\right) 
\nonumber \\
&=&\frac{1}{2}\tsum\limits_{R}\tsum\limits_{S}\,\,P_{R}\,P_{S}\,(1-\delta
_{RS})\,\left( C_{R}D_{S}+C_{S}D_{R}-2\sqrt{C_{R}D_{S}}\sqrt{C_{S}D_{R}}%
\right)  \nonumber \\
&=&\frac{1}{2}\tsum\limits_{R}\tsum\limits_{S}\,\,P_{S}\,P_{R}\,(1-\delta
_{RS})\,\,\left( \sqrt{C_{R}D_{S}}-\sqrt{C_{S}D_{R}}\right) ^{2}  \nonumber
\\
&\geq &0  \label{Eq.R6}
\end{eqnarray}%
which proves the result.

For the special case where $D_{R}=1$ and where $\tsum\limits_{R}\,P_{R}=1$
we get the simpler result%
\begin{equation}
\tsum\limits_{R}\,P_{R}\,C_{R}\,\geq \left( \tsum\limits_{R}\,P_{R}\,\sqrt{%
C_{R}}\right) ^{2}  \label{Eq.SumInequality}
\end{equation}%
This inequality is used in \cite{Hillery06a}.

\pagebreak

\section{Appendix 3 - Particle and Mode Entanglement}

\label{Appendix - Particle and Mode Entanglement}

It is useful to contrast the two meanings of entanglement - mode and
particle - in terms of three examples. The first is from the textbook by
Peres (\cite{Peres93a}, see pp126-128). A system with $N=2$ identical
particles has one particle in a single particle state (mode) $\left\vert
u\right\rangle $, the other in an orthogonal single particle state $%
\left\vert v\right\rangle $. In first quantization the synmmetrized quantum
pure states for identical bosons or for identical fermions are (my notation)
are%
\begin{eqnarray}
\left\vert \Psi \right\rangle _{boson} &=&\frac{1}{\sqrt{2}}(\left\vert
u(1)\right\rangle \otimes \left\vert v(2)\right\rangle +\left\vert
u(2)\right\rangle \otimes \left\vert v(1)\right\rangle )  \nonumber \\
\left\vert \Psi \right\rangle _{fermion} &=&\frac{1}{\sqrt{2}}(\left\vert
u(1)\right\rangle \otimes \left\vert v(2)\right\rangle -\left\vert
u(2)\right\rangle \otimes \left\vert v(1)\right\rangle )
\label{Eq.SymmetrizedStates}
\end{eqnarray}%
which consequently means that "two particles of the same type are always
entangled". Peres obviously considers such entanglement is a result of \emph{%
symmetrization}. In second quantization the state in both the fermion and
boson cases is $\left\vert 1\right\rangle _{u}\otimes $ $\left\vert
1\right\rangle _{v}$ which is a separable state for modes $u,v$, and not a
(mode) entangled state.

The second example is taken from the paper of Hyllus et al \cite{Hyllus12a},
specifically a case illustrated in Fig 1(b) which shows a state with $N=5$
identical bosons. The bosons may occupy differing spatial states (eg
harmonic oscillator states) - referred to by Hyllus et al as \emph{external}
degrees of freedom - and each bosonic particle has two distinct internal
states (eg hyperfine states) -\emph{\ internal} degrees of freedom. Fig 1(b)
shows two spatial states and two internal states ($u,d$ say) , with only the
lower spatial state ($\phi _{0}$ say) being occupied by $N=5$ bosons. From
the Hyllus et al viewpoint (see last para on p 012337-4) "For
indistinguishable particles, only two possibilities are allowed in this
case: either ALL the \emph{particles} are in a separable state (that is,
product $\left\vert \phi \right\rangle ^{\otimes N}$) state, or all \emph{%
particles} are entangled due to the \emph{symmetrization}." Hyllus et al
describe the states in terms of first quantization but for purposes of
comparison we will also describe them via second quantization. What they
mean by the \emph{separable} state is in full 
\begin{equation}
\left\vert \phi \right\rangle ^{\otimes N}=\left\vert \phi _{1}\right\rangle
\left\vert \phi _{2}\right\rangle \left\vert \phi _{3}\right\rangle
\left\vert \phi _{4}\right\rangle \left\vert \phi _{5}\right\rangle
\label{Eq.HyllusSepState}
\end{equation}%
where for the $i$th particle the single particle space-spin state would of
the form 
\begin{equation}
\left\vert \phi _{i}\right\rangle =(\cos \theta \,\left\vert
u_{i}\right\rangle +\sin \theta \,\exp i\chi \,\left\vert d_{i}\right\rangle
)\otimes \left\vert \phi _{0i}\right\rangle  \label{Eq.SingleParticleState}
\end{equation}%
in which a particular internal state is chosen \ The separable state in Eq.(%
\ref{Eq.HyllusSepState}) is just a tensor product of single particle states
for the five bosons. It is symmetric, so the symmetrization principle is
satisfied. There is of course one other orthogonal separable state $%
\left\vert \xi \right\rangle ^{\otimes N}=\left\vert \xi _{1}\right\rangle
\left\vert \xi _{2}\right\rangle \left\vert \xi _{3}\right\rangle \left\vert
\xi _{4}\right\rangle \left\vert \xi _{5}\right\rangle $ with an orthogonal
single particle space-spin state $\left\vert \xi _{i}\right\rangle =(-\sin
\theta \,\left\vert u_{i}\right\rangle +\cos \theta \,\exp i\chi
\,\left\vert d_{i}\right\rangle )\otimes \left\vert \phi _{0i}\right\rangle $
in which the internal state is orthogonal to the previous one. If one of the
bosons is taken from a state $\left\vert \phi \right\rangle $ and placed in
the orthogonal state $\left\vert \xi \right\rangle $, then representing it
in the form of a single tensor product such as $\left\vert \phi
_{1}\right\rangle \left\vert \phi _{2}\right\rangle \left\vert \phi
_{3}\right\rangle \left\vert \phi _{4}\right\rangle \left\vert \xi
_{5}\right\rangle $ would not satisfy the symmetrization principle. If one
such product as $\left\vert \phi _{1}\right\rangle \left\vert \phi
_{2}\right\rangle \left\vert \phi _{3}\right\rangle \left\vert \phi
_{4}\right\rangle \left\vert \xi _{5}\right\rangle $ is subjected to an
operator which is the sum of all permutation operators $\widehat{P}\,$, then
apart from normalising factor the result will represent the situation where
one of the five bosons is in the state $\left\vert \xi \right\rangle $
rather than $\left\vert \phi \right\rangle $. Hence such a state is given by 
\begin{eqnarray}
\left\vert \Psi _{4,1}\right\rangle &=&\mathcal{N}\dsum\limits_{P}\widehat{P}%
\,\left( \left\vert \phi _{1}\right\rangle \left\vert \phi _{2}\right\rangle
\left\vert \phi _{3}\right\rangle \left\vert \phi _{4}\right\rangle
\left\vert \xi _{5}\right\rangle \right)  \nonumber \\
&=&\mathcal{N}\,^{\#}\left( 
\begin{array}{c}
\left\vert \phi _{1}\right\rangle \left\vert \phi _{2}\right\rangle
\left\vert \phi _{3}\right\rangle \left\vert \phi _{4}\right\rangle
\left\vert \xi _{5}\right\rangle +\left\vert \phi _{1}\right\rangle
\left\vert \phi _{2}\right\rangle \left\vert \phi _{3}\right\rangle
\left\vert \phi _{5}\right\rangle \left\vert \xi _{4}\right\rangle
+\left\vert \phi _{1}\right\rangle \left\vert \phi _{2}\right\rangle
\left\vert \phi _{5}\right\rangle \left\vert \phi _{4}\right\rangle
\left\vert \xi _{3}\right\rangle \\ 
+\left\vert \phi _{1}\right\rangle \left\vert \phi _{5}\right\rangle
\left\vert \phi _{3}\right\rangle \left\vert \phi _{4}\right\rangle
\left\vert \xi _{2}\right\rangle +\left\vert \phi _{5}\right\rangle
\left\vert \phi _{2}\right\rangle \left\vert \phi _{3}\right\rangle
\left\vert \phi _{4}\right\rangle \left\vert \xi _{1}\right\rangle%
\end{array}%
\right)  \nonumber \\
&=&\mathcal{N}^{\#}\,\left( 
\begin{array}{c}
\left\vert \phi _{1}\right\rangle \left\vert \phi _{2}\right\rangle
\left\vert \phi _{3}\right\rangle \left\vert \phi _{4}\right\rangle
\left\vert \xi _{5}\right\rangle +\left\vert \phi _{1}\right\rangle
\left\vert \phi _{2}\right\rangle \left\vert \phi _{3}\right\rangle
\left\vert \xi _{4}\right\rangle \left\vert \phi _{5}\right\rangle
+\left\vert \phi _{1}\right\rangle \left\vert \phi _{2}\right\rangle
\left\vert \xi _{3}\right\rangle \left\vert \phi _{4}\right\rangle
\left\vert \phi _{5}\right\rangle \\ 
+\left\vert \phi _{1}\right\rangle \left\vert \xi _{2}\right\rangle
\left\vert \phi _{3}\right\rangle \left\vert \phi _{4}\right\rangle
\left\vert \phi _{5}\right\rangle +\left\vert \xi _{1}\right\rangle
\left\vert \phi _{2}\right\rangle \left\vert \phi _{3}\right\rangle
\left\vert \phi _{4}\right\rangle \left\vert \phi _{5}\right\rangle%
\end{array}%
\right)  \nonumber \\
&&  \label{Eq.HyllusEntState}
\end{eqnarray}%
which are where the sum is over the $5!$ permuation operators and the $%
\mathcal{N}^{\prime }s$ are normalising factors. However, Hyllus et al refer
to this as \emph{entanglement by symmetrization} and regard this state as
being entangled. From this point of view it is symmetrization via $%
\dsum\limits_{P}\widehat{P}$ that is responsible for entanglement in that it
creates contributions to the state vector which becomes no longer just a
simple product. There is a term $\left\vert \phi _{1}\right\rangle
\left\vert \phi _{2}\right\rangle \left\vert \phi _{3}\right\rangle
\left\vert \phi _{4}\right\rangle \left\vert \xi _{5}\right\rangle $
followed by $\left\vert \phi _{1}\right\rangle \left\vert \phi
_{2}\right\rangle \left\vert \phi _{3}\right\rangle \left\vert \xi
_{4}\right\rangle \left\vert \phi _{5}\right\rangle $ in which particles $4$
and $5$ are in different single particle states.

However, from the opposing point of view in which it is modes, not particles
that are entangled, and the state just described would \emph{not} be
regarded as being entangled. The Fig 1(b) case would be seen as a \emph{two
mode} situation in which the two modes are $\left\vert U\right\rangle
=\left\vert u\right\rangle \otimes \left\vert \phi _{0}\right\rangle $ and $%
\left\vert D\right\rangle =\left\vert d\right\rangle \otimes \left\vert \phi
_{0}\right\rangle $. In second quantization the \emph{Fock states} are $%
\left\vert n_{U}\,,n_{D}\right\rangle =\left\vert n_{U}\right\rangle \otimes 
$ $\left\vert n_{D}\right\rangle $ with $n_{U}\,,n_{D}$ being the mode
occupancies. It is these two modes that may or may not be entangled, and
there are \emph{six} separable pure states (\emph{not} two) with a total of $%
N=5$ bosons, namely $\left\vert 5\,,0\right\rangle $, $\left\vert
4\,,1\right\rangle $, $\left\vert 3\,,2\right\rangle $, $\left\vert
2\,,3\right\rangle $, $\left\vert 1\,,4\right\rangle $, and $\left\vert
0\,,5\right\rangle $. The states $\left\vert 5\,,0\right\rangle $ and $%
\left\vert 0\,,5\right\rangle $ are of course equivalent in first
quantization to\ $\left\vert \phi _{1}\right\rangle \left\vert \phi
_{2}\right\rangle \left\vert \phi _{3}\right\rangle \left\vert \phi
_{4}\right\rangle \left\vert \phi _{5}\right\rangle $ and $\left\vert \xi
_{1}\right\rangle \left\vert \xi _{2}\right\rangle \left\vert \xi
_{3}\right\rangle \left\vert \xi _{4}\right\rangle \left\vert \xi
_{5}\right\rangle $, whilst the state in the last equation is just the \emph{%
separable} state $\left\vert 4\,,1\right\rangle $. \ The general \emph{mode
entangled} pure state with $N=5$ bosons is given by%
\begin{eqnarray*}
\left\vert \Psi \right\rangle &=&\mathcal{D}_{5,0}\left\vert
5\,,0\right\rangle +\mathcal{D}_{4,1}\left\vert 4\,,1\right\rangle \\
&&+\mathcal{D}_{3,2}\left\vert 3\,,2\right\rangle +\mathcal{D}%
_{2,3}\left\vert 2\,,3\right\rangle \\
&&+\mathcal{D}_{1,4}\left\vert 1\,,4\right\rangle +\mathcal{D}%
_{0,5}\left\vert 0\,,5\right\rangle
\end{eqnarray*}%
where the $\mathcal{D}$ are expansion coefficients, which is of course
equivalent to various first quantization expressions. But now we would say
it is the two modes $\left\vert U\right\rangle $ and $\left\vert
D\right\rangle $ that are entangled, not the $5$ bosons! Entanglement for $%
N=5$ boson pure states is associated with there being \emph{six} distinct
Fock states that occur for \emph{five} bosons being split between \emph{two}
modes. If there were four modes then for $N=5$ boson pure states there would
be many more distinct Fock states available depending on how the bosons are
divided amongst the modes. It is more a question of \emph{combinatorics}
rather than \emph{symmetrization} which is relevant in determining the \emph{%
dimension} of the space of entangled states. A quite different picture of
what is meant by an entangled state occurs when entanglement refers to modes
rather than particles.

\pagebreak

\section{Appendix 4 - Reference Frames and Super-Selection Rules}

\label{Appendix - Reference Frames and SSR}

Several papers such as \cite{Kitaev04a}, \cite{Bartlett06a}, \cite%
{Bartlett07a}, \cite{Vaccaro08a}, \cite{Tichy11a}, \cite{White09a}, \cite%
{Paterek11a} explain the link between \emph{reference frames} and \emph{%
super-selection rules} (SSR). In this Appendix we present the key ideas
involved.

\subsection{Two Observers with Different Reference Frames}

The first point to appreciate is that there are \emph{two observers} - Alice
and Charlie - who are involved in describing the \emph{same state} of a
particular \emph{quantum system}. Charlie is the \emph{external} observer,
Alice the \emph{internal} observer - perhaps closely linked to the system.
It is important to realise that it is \emph{Charlie's description} of the
quantum state which is of \emph{most interest}, in particular how this
description may differ from what Alice may regard as the system state. The
system could be a \emph{multi-mode} system involving identical particles, it
could just be a \emph{single mode} system or it could even be a \emph{single
particle} with or without spin. Alice and Charlie each describe quantum
states in terms of their own \emph{reference frames}, which might be a set
of \emph{coordinate axes} for the case of the spin or position states for
the single particle system, or it could be a \emph{large quantum system}
with a well-defined reference \emph{phase} in the case of multi-mode or
single mode systems involving identical particles. Alice and Charlie may
each choose from a set of possible reference frames - for the single
particle case there are an infinite number of difference choices of
coordinate axes for example, related to each other via \emph{rotations}
and/or \emph{translations}. In \emph{Situation A} - which \emph{is} \emph{not%
} associated with \emph{SSR} - Alice and Charlie \emph{do know} the
relationship between their two reference frames (and can communicate this
relationship via \emph{classical communications}) - such as in the case of
the single particle system when the relative orientation of their two
different coordinate axes are known. In \emph{Situation B} - which \emph{is}
associated with \emph{SSR} - Alice and Charlie \emph{do not know} the
relationship between their two reference frames - such as in the multi-mode
or single mode system involving identical particles when the relative phase
between their two large quantum phase reference systems is not known. Alice
and Charlie describe the same state via density operators $\widehat{\sigma }$
and $\widehat{\rho }$, and the key question is the \emph{relationship}
between these two operators in situations A and B and for various types of
reference frames. In terms of the notation in \cite{Bartlett07a} $\rho
\rightarrow $ $\widehat{\sigma }$ and $\widetilde{\rho }\rightarrow \widehat{%
\rho }$.

\subsection{Symmetry Groups}

A particular relationship going from Alice's to Charlie's reference frame is
specified by the \emph{parameter} $g$, which in turn defines a \emph{unitary
transformation operator} $\widehat{T}(g)$ that acts in the system space.
Particular examples will be listed below. If there was a third observer -
Donald - and the relationship going from Charlie's to Donald's reference
frame is specified by the parameter $h$, which in turn defines a unitary
operator $\widehat{T}(h)$, then if we symbolise the relationship going from
Alice's to Donald's reference frame by the parameter $hg$, it follows that $%
\widehat{T}(hg)=\widehat{T}(h)\widehat{T}(g)$. This shows that the unitary
operators satisfy one of the requirements to constitute a \emph{group},
referred to generally as the \emph{transformation} group. The other
requirements are easily confirmed. The unitary operator $\widehat{T}(0)=%
\widehat{1}$ corresponding to the case where no change of reference frame
occurs (specified by the parameter $0$) exists, and satisfies the
requirement that $\widehat{T}(0g)=\widehat{T}(0)\widehat{T}(g)=\widehat{T}%
(g0)=\widehat{T}(g)\widehat{T}(0)$. The unitary operator $\widehat{T}%
(g^{-1})=\widehat{T}(g)^{\dag }$ corresponding to the relationship specified
as $g^{-1}$ that converts Charlie's reference frame back to that of Alice
exists, and satisfies the requirement that $\widehat{T}(0)=\widehat{T}%
(g^{-1})\widehat{T}(g)=\widehat{T}(g)\widehat{T}(g^{-1})$. Hence all the
group properties are satisfied.

A few examples are as follows:

1. \emph{Translation group} - single spinless particle system, with $%
\widehat{p}$, $\widehat{x}$.the momentum, position vector operators. Here $%
\underrightarrow{a}$ is a vector giving the translation of Charlie's
cartesian axes reference frame from that of Alice, thus $g\equiv $.$%
\underrightarrow{a}$. The unitary translation operator is $\widehat{T}(%
\underrightarrow{a})=\exp (i\widehat{p}\cdot \underrightarrow{a}/\hbar )$.

2. \emph{Rotation group} - single particle system, with $\widehat{J}$ the
angular momentum vector operators. Here $\underrightarrow{u}$ is a unit
vector giving the axis and rotation angle $\phi $ for rotating Alice's
cartesian axes reference frame into that of Charlie, thus $g\equiv $.$%
\underrightarrow{u}$.$,\phi $. The unitary rotation operator is $\widehat{T}(%
\underrightarrow{u},\phi )=\exp (i\phi \widehat{J}\cdot \underrightarrow{u}%
/\hbar )$.

3. \emph{Particle number U(1) group} - single mode bosonic system, with $%
\widehat{a}$ the mode annihilation operator and $\widehat{N}_{a}=\widehat{a}%
^{\dag }\widehat{a}$ the mode number operator. Here $\theta _{a}$ is the
phase change Alice's to Charlie's reference frame. The unitary operator is $%
\widehat{T}(\theta _{a})=\exp (i\widehat{N}_{a}\theta _{a})$.

4. \emph{Particle number U(1) group} - multi-mode bosonic system, with $%
\widehat{a}$ as a typical mode annihilation operator and $\widehat{N}%
=\dsum\limits_{a}\widehat{a}^{\dag }\widehat{a}$ the total number operator.
Here $\theta $ is the phase change from Alice's to Charlie's reference
frame. The unitary operator is $\widehat{T}(\theta )=\exp (i\widehat{N}%
\theta )$.

In these examples the system operators $\widehat{p}$, $\widehat{J}$, $%
\widehat{N}_{a}$, $\widehat{N}$ etc are the \emph{generators} of the
respective groups. In many situations the generators commute with the
Hamiltonian for the system (or more generally with the evolution operator
that describes time evolution of the quantum state), in which case the group
of unitary operators $\widehat{T}(g)$ is the \emph{symmetry group}, and the
generators are \emph{conserved} physical quantities.

\subsection{Relationships - Situation A}

In \emph{Situation A}, where the relationship between the reference frames
for Alice and Charlie is \emph{known} and specified by a \emph{single}
parameter $g$, Alice's description of the state $\widehat{\sigma }$ is
related to Charlie's description $\widehat{\rho }$ for the same state via
the unitary transformation 
\begin{equation}
\widehat{\rho }=\widehat{T}(g)\,\widehat{\sigma }\,\widehat{T}(g)^{-1}
\label{Eq.AliceCharlieStatesSitnA}
\end{equation}%
Note that this is a \emph{passive} transformation - no change of state is
involved, just the same state being described by two different observers.

As an example, consider the \emph{spinless particle} and the \emph{%
translation} group. If $\left\vert \underrightarrow{x}\right\rangle $ is a
position eigenstate then $\widehat{T}(\underrightarrow{a})\left\vert 
\underrightarrow{x}\right\rangle =\left\vert \underrightarrow{x}-%
\underrightarrow{a}\right\rangle $. A pure quantum position eigenstate
described by Alice as $\,\widehat{\sigma }=\left\vert \Phi \right\rangle
\left\langle \Phi \right\vert $ with state vector $\left\vert \Phi
\right\rangle =\left\vert \underrightarrow{x}\right\rangle $ would be
described by Charlie as $\widehat{\rho }=\left\vert \Psi \right\rangle
\left\langle \Psi \right\vert $ but now with $\left\vert \Psi \right\rangle
=\left\vert \underrightarrow{x}-\underrightarrow{a}\right\rangle $, which is
also a pure quantum position eigenstate but with eigenvalue $%
\underrightarrow{x}-\underrightarrow{a}$. This is as expected since Alices's
cartesian axes have been translated by $\underrightarrow{a}$ to the origin
of Charlie's axes without change of orientation. In the case of momentum
eigenstates $\left\vert \underrightarrow{p}\right\rangle $ we have $\widehat{%
T}(\underrightarrow{a})\left\vert \underrightarrow{p}\right\rangle =\exp (i%
\underrightarrow{p}\cdot \underrightarrow{a}/\hbar )\left\vert 
\underrightarrow{p}\right\rangle $, so a pure quantum momentum eigenstate
described by Alice with $\left\vert \Phi \right\rangle =\left\vert 
\underrightarrow{p}\right\rangle $ would be described by Charlie with $%
\left\vert \Psi \right\rangle =\exp (i\underrightarrow{p}\cdot 
\underrightarrow{a}/\hbar \,\left\vert \underrightarrow{p}\right\rangle $,
which is also a pure momentum eigenstate with the same eigenvalue $%
\underrightarrow{p}$. Alice and Charlie describe the pure momentum
eigernstate with the same density operator $\widehat{\rho }=\widehat{\sigma }
$, the phase factor cancels.

For more general pure states, consider a quantum state described by Alice as 
$\,\widehat{\sigma }=\left\vert \Phi \right\rangle \left\langle \Phi
\right\vert $ with state vector $\left\vert \Phi \right\rangle =\dint d%
\underrightarrow{x}\,\phi (\underrightarrow{x})\,\left\vert \underrightarrow{%
x}\right\rangle $. States of this form can represent \emph{localised} states
when $\phi (\underrightarrow{x})$ is only significant in confined spatial
regions, or they can represent \emph{delocalised} states such as momentum
eigenstates $\left\vert \underrightarrow{p}\right\rangle $ when $\phi (%
\underrightarrow{x})=(2\pi \hbar )^{-3/2}\exp (i\underrightarrow{p}\cdot 
\underrightarrow{x}/\hbar )$. We see that Charlie also describes a pure
quantum state $\widehat{\rho }=\left\vert \Psi \right\rangle \left\langle
\Psi \right\vert $ but now with $\left\vert \Psi \right\rangle =\widehat{T}(%
\underrightarrow{a})\,\left\vert \Phi \right\rangle =\dint d\underrightarrow{%
x}\,\phi (\underrightarrow{x}+\underrightarrow{a})\,\left\vert 
\underrightarrow{x}\right\rangle =\dint d\underrightarrow{x}\,\psi (%
\underrightarrow{x})\,\left\vert \underrightarrow{x}\right\rangle $, so the
wavefunction is now $\psi (\underrightarrow{x})=\phi (\underrightarrow{x}+%
\underrightarrow{a})$.

Note that if Alice's state vector was written in terms of momentum
eigenstates $\left\vert \Phi \right\rangle =\dint d\underrightarrow{p}\,%
\widetilde{\phi }(\underrightarrow{p})\,\left\vert \underrightarrow{p}%
\right\rangle $, then Charlie's state vector $\left\vert \Psi \right\rangle
=\dint d\underrightarrow{p}\,\widetilde{\psi }(\underrightarrow{p}%
)\,\left\vert \underrightarrow{p}\right\rangle $ has a momentum wave
function $\widetilde{\psi }(\underrightarrow{p})=\exp (i\underrightarrow{p}%
\cdot \underrightarrow{a}/\hbar )\,\widetilde{\phi }(\underrightarrow{p})$
related to that of Alice by a phase factor. Note that a state which is a
quantum superposition of momentum eigenstates as described by Alice is also
described as a quantum superposition of momentum eigenstates by Charlie. A
similar feature applies in all situation A cases, and is related to SSR 
\emph{not} applying in situation A.

The case of the \emph{particle} with \emph{spin} and the \emph{rotation}
group is outlined in Ref. \cite{Bartlett06a}.

\subsection{Relationships - Situation B}

\label{AppendixSubSection - Situation B}

In \emph{Situation B}, where on the other hand the relationship between
frames is completely \emph{unknown}, all possible transformations $g$ must
be given \emph{equal weight}, and hence the relationship between Alice's and
Charlie's description of the same state becomes 
\begin{eqnarray}
\widehat{\rho } &=&\dint w(g)dg\,\widehat{T}(g)\,\widehat{\sigma }\,\widehat{%
T}(g)^{-1}  \nonumber \\
&=&\mathcal{G\,[}\widehat{\sigma }]  \label{Eq.AliceCharlieStatesSitnB}
\end{eqnarray}%
where $\dint w(g)dg$ is a symbolic integral over the parameter $g$, which
includes a weight factor $w(g)$ so that $\dint w(g)dg=1$. This linear
process connecting $\widehat{\sigma }$ to $\widehat{\rho }$ is the "$%
\mathcal{G}$- \emph{twirling}" operation. Again, this is a paassive
transformation.

It is straightforward to show that for any fixed parameter $h$ that 
\begin{equation}
\widehat{T}(h)\,\widehat{\rho }\,\widehat{T}(h)^{-1}=\widehat{\rho }
\label{Eq.CharlieStateInvarSitnB}
\end{equation}%
showing that Charlie's density operator is $\mathcal{G}$ invariant under the
transformation group - unlike the case for Situation A.

As an example, consider the \emph{single mode} bosonic system and the \emph{%
U(1)} group. If $\left\vert n_{a}\right\rangle $ is a Fock state then $%
\widehat{T}(\theta _{a})\,\left\vert n_{a}\right\rangle =\exp (in_{a}\theta
_{a})\,\left\vert n_{a}\right\rangle $. Consider a pure quantum state
described by Alice as the \emph{Glauber coherent state} $\widehat{\sigma }%
=\left\vert \Phi \right\rangle \left\langle \Phi \right\vert $ with state
vector $\left\vert \Phi (\beta )\right\rangle
=\dsum\limits_{n_{a}}\,C(n_{a},\beta )\,\left\vert n_{a}\right\rangle $,
where $C(n_{a},\beta )=\exp (-|\beta |^{2}/2)\,\beta ^{n_{a}}\,/\sqrt{%
(n_{a})!}$. It is straightforward to show that 
\begin{equation}
\widehat{T}(\theta _{a})\,\left\vert \Phi (\beta )\right\rangle =\left\vert
\Phi (\beta \exp (i\theta _{a}))\right\rangle  \label{Eq.GlauberStateTransfn}
\end{equation}%
so that the Glauber coherent state is transformed into another Glauber
coherent state, but with $\beta $ changed via a phase factor to $\beta \exp
(i\theta _{a})$. The quantum state described by Charlie is given by%
\begin{eqnarray}
\widehat{\rho } &=&\dint \frac{d\theta _{a}}{2\pi }\left\vert \Phi (\beta
\exp (i\theta _{a}))\right\rangle \left\langle \Phi (\beta \exp (i\theta
_{a}))\right\vert  \label{Eq.CharlieStateGlauberForm1} \\
&=&\dint \frac{d\theta _{a}}{2\pi }\dsum\limits_{n_{a}}\dsum\limits_{m_{a}}%
\,C(n_{a},\beta )\,C(m_{a},\beta )^{\ast }\,\widehat{T}(\theta
_{a})\,\left\vert n_{a}\right\rangle \left\langle m_{a}\right\vert \,%
\widehat{T}(\theta _{a})^{\dag }  \nonumber \\
&=&\dsum\limits_{n_{a}}\dsum\limits_{m_{a}}\,C(n_{a},\beta )\,C(m_{a},\beta
)^{\ast }\,\left\vert n_{a}\right\rangle \left\langle m_{a}\right\vert
\,\dint \frac{d\theta _{a}}{2\pi }\exp (i[n_{a}-m_{a}]\theta _{a})  \nonumber
\\
&=&\dsum\limits_{n_{a}}\,|C(n_{a},\beta )|^{2}\,\left\vert
n_{a}\right\rangle \left\langle n_{a}\right\vert  \nonumber \\
&=&\dsum\limits_{n_{a}}\,\exp (-|\beta |^{2})\,\frac{(|\beta |^{2})^{n_{a}}}{%
(n_{a})!}\,\left\vert n_{a}\right\rangle \left\langle n_{a}\right\vert
\label{Eq.CharlieStateGlauberForm2}
\end{eqnarray}%
which is a \emph{mixed state} consisting of a \emph{Poisson distribution} of 
\emph{Fock states} with mean occupation number $\overline{n}_{a}=|\beta
|^{2} $. In view of the first expression for $\widehat{\rho }$ it can also
be thought of as a mixed state consisting of Glauber coherent states each
with the same amplitude $|\beta |\,=\,\sqrt{\overline{n}_{a}}$, but with all
phases $(\arg \beta +\theta _{a})$ equally probable. Thus, whereas Alice
describes the state as a pure state that is a quantum superposition of Fock
states with differing occupancy numbers, Charlie describes the same state as
a mixed state involving a statistical mixture of number states. The former
violates the SSR whereas the latter does not. A similar feature applies in
all situation B cases, and is related to SSR applying in Situation B.
Whether Alice could ever prepare such a state in the first place is
controversial - see the discussion presented above in SubSections \ref%
{SubSection - Super-Selection Rule} and \ref{SubSection - SSR Separate Modes}%
. However, \emph{assuming} she could, the quantum state as described by
Charlie is a mixed state.

The situation just studied relates of course to the debate \cite{Molmer97a}
regarding whether the quantum state for a \emph{single mode laser} operating
well above threshold should be described by a Glauber coherent state or as a
Poisson statistical mixture of photon number states. The first viewpoint
(Alice) describes the state from the point of view of an internal observer
with a reference frame, the second (Charlie) describes the same state from
the point of view of an external observer for whose reference frame
relationship to that of the internal observer is unknown. The debate is
regarded by \cite{Bartlett06a} as settled on the basis that both viewpoints
are valid, they are just at cross purposes because they refer to
descriptions of the same quantum state by two different observers.

It should not be thought however that the quantum state would always be
described in such a fundamentally different manner for all Situation B
cases. As an example, consider the \emph{multi-mode bosonic} system and the 
\emph{U(1)} group. Consider the pure quantum state described by Alice as the
multi-mode $N$ boson \emph{Fock state} $\widehat{\sigma }=\left\vert \Phi
\right\rangle \left\langle \Phi \right\vert $ with state vector $\left\vert
\Phi (N)\right\rangle =\left\vert n_{1}n_{2}...n_{a}...;N\right\rangle
=\dprod\limits_{a}\left\vert n_{1}\right\rangle \left\vert
n_{2}\right\rangle ..\left\vert n_{a}\right\rangle ...$, where $%
N=\dsum\limits_{a}n_{a}$. We have $\widehat{T}(\theta )\left\vert
n_{1}n_{2}...n_{a}...;N\right\rangle =\exp (iN\theta )\,\left\vert
n_{1}n_{2}...n_{a}...;N\right\rangle $, so that the same state would be
described by Charlie as $\widehat{\rho }=\left\vert \Psi \right\rangle
\left\langle \Psi \right\vert $ and with $\left\vert \Psi \right\rangle
=\left\vert n_{1}n_{2}...n_{a}...;N\right\rangle $. This is also a
multi-mode $N$ boson Fock state with exactly the same occupancies. The
product $\exp (iN\theta )\,\exp (-iN\theta )$ of phase factors averages out
to unity and here $\widehat{\rho }=\widehat{\sigma }$, so Alice and Charlie
both describe the multi-mode Fock states in the same way. Another example
for \emph{two mode bosonic} systems and the \emph{U(1)} group\ is provided
by the one boson \emph{Bell states} (the BS\ notation used here is
non-conventional). These are entangled two mode states that Alice would
describe via the state vectors $\left\vert \Phi ^{\pm }\right\rangle
=(\left\vert 10\right\rangle \pm \left\vert 01\right\rangle )/\sqrt{2}$. We
have $\widehat{T}(\theta )\,\left\vert \Phi ^{\pm }\right\rangle =\exp
(i\theta )\,\left\vert \Phi ^{\pm }\right\rangle $, so that the same state
would be described by Charlie with $\left\vert \Psi ^{\pm }\right\rangle
=(\left\vert 10\right\rangle \pm \left\vert 01\right\rangle )/\sqrt{2}$.
Again the product of phase factors averages to unity and $\widehat{\rho }=%
\widehat{\sigma }$, so Alice and Charlie both describe the quantum states as
Bell states, and in the same form.

\subsection{Dynamical and Measurement Considerations}

Discussions of the relationship between equations governing the dynamical
behaviour of Alice's and Charlie's density operators depend on whether the
evolution is just governed by a Hamiltonian or whether master equations
describing evolution affected by interactions with an external environment
are involved. Such matters will not be treated in detail here, nor will the
issue of relating Alice's and Charlie's measurements. The latter issue is
dealt with in \cite{Bartlett07a}.

However, in the case where Alice describes the \emph{Hamiltonian evolution}
of her density operator via the Liouville - von-Neumann equation 
\begin{equation}
i\hbar \frac{\partial }{\partial t}\widehat{\sigma }=[\widehat{H},\widehat{%
\sigma }]  \label{Eq.LVNAlice}
\end{equation}%
where in Alice's frame the Hamiltonian is $\widehat{H}$, and where in
addition the transformation group is also the \emph{symmetry group} so that $%
\widehat{T}(g)\widehat{H}\widehat{T}(g)^{-1}=\widehat{H}$ for all $g$, it is
easy to see that for both Situations A and B, Charlie's density operator
will evolve via the same LVN equation%
\begin{equation}
i\hbar \frac{\partial }{\partial t}\widehat{\rho }=[\widehat{H},\widehat{%
\rho }]  \label{Eq.LVNCharlie}
\end{equation}%
Thus both Alice and Charlie will describe the same dynamical evolution,
though of course the initial (and hence evolved) states may differ in the
two cases.

\subsection{Nature of Reference Frames}

Reference frames of differing types are involved for the various
transformation groups. The common feature is that they are thought of as 
\emph{actual physical systems} themselves which are either macroscopic \emph{%
classical} systems or macroscopic \emph{quantum} systems in states
associated with the \emph{classical limit}. They are intended to be \emph{%
essentially unaffected} by the presence of the systems for which they are
acting as reference frames. In some cases relatively uncontroversial
examples exist, such as for the \emph{cartesian axes} associated with the 
\emph{translation} and \emph{rotation} groups associated with the single
particle system. The physical reference system may be a large \emph{magnet}
whose magnetic field points in a well defined direction and defines a $z$
axis, combined with an \emph{electrostatic generator} whose electric field
is in another well defined direction at right angles that defines an $x$
axis. In other cases the existence of suitable reference frames is less
clear.

In this SubSection we will describe possible phase reference frames as if
they are entirely separated (or uncorrelated) with the system of interest.
In terms of the treatment by Bartlett et al \cite{Bartlett06a}, \cite%
{Bartlett07a} these are \emph{non-implicated} reference frames. In the next
SubSection and in the next Appendix phase reference frames that are
correlated with the system of interest will be described - these are the
so-called \emph{implicated} reference frames of Bartlett et al.

For the large quantum system with a well-defined reference \emph{phase}
associated with the \emph{U(1)} group in the case of multi-mode or single
mode systems involving identical particles, the usual choice is a single
mode bosonic system such as a single mode \emph{BEC} or a \emph{laser} with
a large mean occupancy, and which is thought of as being prepared in a
Glauber coherent state $\left\vert \Phi (\alpha )\right\rangle $ in order to
provide the \emph{phase reference frame}, the reference phase being $\arg
\alpha $. Whether such a reference frame really exists is controversial. The
discussion presented above in SubSections \ref{SubSection - Super-Selection
Rule} and \ref{SubSection - SSR Separate Modes} raises the question of
whether such a phase reference state could ever be prepared, so this choice
of a physical phase reference is rather unsatisfactory. However, from the
point of view of this presentation we \emph{assume} it does, so that - as in
the previous example - Alice can describe the reference state as another
coherent state. Again, whether Alice could ever prepare such a state is
questionable.

Another possibility for a physical phase reference is a \emph{macroscopic }%
low frequency\emph{\ harmonic oscillator}, whose quantum energy eigenstates $%
\left\vert n\right\rangle $ - with $n=0,1,..,n_{\max }$ and energies $%
n\,\hbar \omega $ can be used to construct phase eigenstates $\left\vert
\theta _{p}\right\rangle $ with $p=0,1,..,n_{\max }$ and $\theta
_{p}=p\times 2\pi /(n_{\max }+1)$, and which are defined by \cite{Barnett89a}%
\begin{equation}
\left\vert \theta _{p}\right\rangle =\frac{1}{\sqrt{n_{\max }+1}}%
\dsum\limits_{n=0}^{n_{\max }}\,\exp (in\theta _{p})\,\left\vert
n\right\rangle  \label{Eq.PhaseState}
\end{equation}%
These states are orthonormal. The separation between the equally spaced
phase angles $\Delta \theta =2\pi /(n_{\max }+1)$ can be made very small if $%
n_{\max }$ is large enough. Under the effect of the harmonic oscillator
Hamiltonian $\widehat{H}=\hbar \omega \widehat{N}$, where $\widehat{N}$ is
the number operator, the phase state $\left\vert \theta _{p}\right\rangle $
evolves into $\left\vert \theta _{p}-\omega \Delta t\right\rangle $ during a
time interval $\Delta t$, so if the time intervals are chosen so that $%
\omega \Delta t=2\pi /(n_{\max }+1)$, the phase angle $\theta _{p}$ changes
into $\theta _{p-1}$. Thus the system behaves like a backwards running \emph{%
clock }\cite{Pegg91a}, the phase angles $\theta _{p}$ defining the positions
of the hands. If the clock initially has phase $\theta _{p}$ the probability
of finding the clock to have phase $\theta _{q}$ after a time interval $%
\Delta t$ is given by 
\begin{equation}
P(\theta _{q},\theta _{p},\Delta t)=\frac{1}{(n_{\max }+1)^{2}}\frac{\sin
^{2}((n_{\max }+1)\Delta /2)}{\sin ^{2}(\Delta /2)}  \label{Eq.PhaseProb}
\end{equation}%
where $\Delta =\theta _{p}-\theta _{q}-\omega \Delta t$. For times $\Delta t$
such that $\omega \Delta t\ll 2\pi /(n_{\max }+1)$ the probability of the
phase remaining as $\theta _{p}$ is close to unity. Thus if the phase state $%
\left\vert \theta _{p}\right\rangle $ is used as a phase reference, it will
remain stable for a time $\Delta t$ satisfying the last inequality. For $%
\Delta t\sim 100\mu s$ and $n_{\max }\sim 10^{4\text{ }}$so that phase is
defined to $\sim 10^{-3}$ radians, an oscillator frequency $\omega \sim
10^{0}$ s$^{-1}$ would suffice for this phase reference standard. Such
macroscopic oscillators do exist, though the process to prepare them in the
phase reference \emph{quantum} state $\left\vert \theta _{p}\right\rangle $
would be technically difficult. Whether such a system would be useful as a
phase reference for optical fields or a BEC is another issue

\subsection{Relational Description of Phase References}

In this SubSection phase reference frames that are correlated with the
system of interest will be described - these are the so-called \emph{%
implicated} reference frames of Bartlett et al \cite{Bartlett06a}, \cite%
{Bartlett07a}.

One such approach to describing phase references in the \emph{U(1)} group
case is via the concept of \emph{maps}. For simplicity consider a one mode
system \emph{S}, the basis vectors for which are Fock states $\left\vert
m\right\rangle _{S}$, where it is sufficient to restrict $m=0,1,..,m_{\max }$%
. The reference system $R$, will also be a one mode system with Fock states $%
\left\vert n\right\rangle _{R}$, where $n$ is large. Product states $%
\left\vert m\right\rangle _{S}\otimes \left\vert n\right\rangle _{R}$ for
the combined modes exist in the Hilbert space $H_{S}\otimes H_{R}$ and are
eigenstates of the various number operators, including the total number
operator $\widehat{N}_{T}=\widehat{N}_{S}+\widehat{N}_{R}$ - where the
eigenvalue is $l=m+n$. The product states may be listed via $%
m=0,1,..,m_{\max }$ and $n=0,1,..$or $m=0,1,..,m_{\max }$ and $l=m,m+1,...$.
Here we will describe how a \emph{coherent superpostion} of \emph{number
states}, such as a Glauber coherent state can be represented.

In the so-called \emph{internalisation} or \emph{quantisation} of the
reference frame the product state $\left\vert m\right\rangle _{S}\otimes
\left\vert n\right\rangle _{R}$ is mapped onto the product state $\left\vert
m\right\rangle _{S}\otimes \left\vert n-m\right\rangle _{R}$ where $n\geq
m_{\max }$. Thus 
\begin{equation}
\left\vert m\right\rangle _{S}\otimes \left\vert n\right\rangle
_{R}\rightarrow \left\vert m\right\rangle _{S}\otimes \left\vert
n-m\right\rangle _{R}  \label{Eq.IntMap}
\end{equation}%
Hence for a linear combination of system states given by 
\begin{equation}
\left\vert \Phi \right\rangle _{S}=\dsum\limits_{m=0}^{m_{\max
}}\,C_{m}\left\vert m\right\rangle _{S}  \label{Eq.SysState}
\end{equation}%
we have for the state $\left\vert \Phi \right\rangle _{S}\otimes \left\vert
n\right\rangle _{R}$ in $H_{S}\otimes H_{R}$ 
\begin{equation}
\left\vert \Phi \right\rangle _{S}\otimes \left\vert n\right\rangle
_{R}=\dsum\limits_{m=0}^{m_{\max }}\,C_{m}\left\vert m\right\rangle
_{S}\otimes \left\vert n\right\rangle _{R}\rightarrow
\dsum\limits_{m=0}^{m_{\max }}\,C_{m}\left\vert m\right\rangle _{S}\otimes
\left\vert n-m\right\rangle _{R}=\left\vert \Psi _{n}\right\rangle _{RS}
\label{Eq.MapCoherSuper}
\end{equation}%
The mapping results in an entangled state where there are $n$ bosons
distributed betweeen the two modes. This state $\left\vert \Psi
_{n}\right\rangle _{RS}$ is a pure state which is compatible with the SSR
and is in one-one correspondence with the original system state $\left\vert
\Phi \right\rangle _{S}$. Note that to create this state the reference state 
$\left\vert n\right\rangle _{R}$ must have more bosons in it than $m_{\max }$%
. The density operator for the original pure system $S$ state would be $%
\widehat{\sigma }_{S}=\left\vert \Phi \right\rangle _{S}\left\langle \Phi
\right\vert _{S}$, and we note that this state violates the SSR. The state $%
\left\vert \Phi \right\rangle _{S}$ would be essentially a Glauber coherent
state if $C_{m}=\exp (-|\alpha |^{2}/2)\alpha ^{m}/(\sqrt{m!})$, with $%
m_{\max }\gg |\alpha |^{2}$. However, for the mapped state $\left\vert \Psi
_{n}\right\rangle _{RS}$ the reduced density operator $\widehat{\rho }_{S}$
is given by 
\begin{eqnarray}
\widehat{\rho }_{S} &=&Tr_{R}(\left\vert \Psi _{n}\right\rangle
_{RS}\left\langle \Psi \right\vert _{RS})  \nonumber \\
&=&\dsum\limits_{m=0}^{m_{\max }}\,|C_{m}|^{2}\,\left\vert m\right\rangle
_{S}\left\langle m\right\vert _{S}  \label{Eq.RDOSystem}
\end{eqnarray}%
This is a mixed state and is compatible with the SSR. For the Glauber
coherent state $\left\vert \Phi \right\rangle _{S}$ this is the Poisson
distribution of number states. Hence the original SSR violating
superposition of number states for system $S$ is mapped onto a state in the
combined system for which the reduced density operator is a statistical
mixture and is consistent with the SSR. $\widehat{\sigma }_{S}$ would
correspond to Alice's description of the state, $\widehat{\rho }_{S}$ to
Charlie's.

In the alternative so-called \emph{externalisation} of the reference frame
the mapping is between product states, and is the reverse of the previous
mapping. The product state $\left\vert m\right\rangle _{S}\otimes \left\vert
n\right\rangle _{R}$ is mapped onto the product state $\left\vert
m\right\rangle _{S}\otimes \left\vert m+n\right\rangle _{R}$ in the Hilbert
space $H_{S}\otimes H_{R}$ where the former is spanned by vectors $%
\left\vert m\right\rangle _{S}$ and the latter by vectors $\left\vert
m+n\right\rangle _{R}$, and where $n\geq m_{\max }$. Thus 
\begin{equation}
\left\vert m\right\rangle _{S}\otimes \left\vert n\right\rangle
_{R}\rightarrow \left\vert m\right\rangle _{S}\otimes \left\vert
m+n\right\rangle _{R}  \label{Eq.ExtMap}
\end{equation}%
The mapping of the $H_{S}\otimes H_{R}$ state $\left\vert \Psi
_{n}\right\rangle _{RS}$ then is 
\begin{eqnarray}
\left\vert \Psi _{n}\right\rangle _{RS} &=&\dsum\limits_{m=0}^{m_{\max
}}\,C_{m}\left\vert m\right\rangle _{S}\otimes \left\vert n-m\right\rangle
_{R}  \nonumber \\
&\rightarrow &\dsum\limits_{m=0}^{m_{\max }}\,C_{m}\left\vert m\right\rangle
_{S}\otimes \left\vert n\right\rangle _{R}=\left(
\dsum\limits_{m=0}^{m_{\max }}\,C_{m}\left\vert m\right\rangle _{S}\right)
\otimes \left\vert n\right\rangle _{R}=\left\vert \Xi _{n}\right\rangle _{RS}
\nonumber \\
&&  \label{Eq.MapMapCoherSuper}
\end{eqnarray}%
The mapping results in a non-entangled state which is incompatible with the
SSR. The state in the subspace $H_{S}$ is a coherent superposition of number
states, whilst that in $H_{R}$ is a Fock state. The reduced density operator
in $H_{S}$ is $\widehat{\sigma }_{S}^{\#}$ given by 
\begin{eqnarray}
\widehat{\sigma }_{S}^{\#} &=&Tr_{R}(\left\vert \Xi _{n}\right\rangle
_{RS}\left\langle \Xi _{n}\right\vert _{RS})  \nonumber \\
&=&\dsum\limits_{m=0}^{m_{\max }}\dsum\limits_{k=0}^{m_{\max
}}\,C_{m}C_{k}^{\ast }\,\left\vert m\right\rangle _{S}\left\langle
k\right\vert _{S}  \label{Eq.RDORel}
\end{eqnarray}%
which is the same as $\widehat{\sigma }_{S}=\left\vert \Phi \right\rangle
_{S}\left\langle \Phi \right\vert _{S}$ and involves coherences between
different number states in contradiction to the SSR. Clearly this second
mapping just reverses the first one.

Of these two treatments of phase reference frames, the internalisation
version has a closer link to physics in that the pure state $\left\vert \Psi
_{n}\right\rangle _{RS}$ can in principle be created and does lead to a way
of creating a state that is in one-one correspondence with any SSR violating
pure state $\left\vert \Phi \right\rangle _{S}$, though it is in the form of
an entangled state of the $S$, $R$ sub-systems rather than just $S$ alone.
This is an important point to note - the original SSR violating state does
not exist as a state of a separate system, all that exists is an SSR
compatible entangled state that is in one-one correspondence with it.
However, the general process for creating a state such as $\left\vert \Psi
_{n}\right\rangle _{RS}$ is not explained. For simple cases such as $%
\left\vert \Phi \right\rangle _{S}=(\left\vert 0\right\rangle
_{S}+\left\vert 1\right\rangle _{S})/\sqrt{2}$ the creation of the required
state $\left\vert \Psi _{n}\right\rangle _{RS}=(\left\vert 0\right\rangle
_{S}\otimes \left\vert n\right\rangle _{R}+\left\vert 1\right\rangle
_{S}\otimes \left\vert n-1\right\rangle _{R})/\sqrt{2}$, where $n\geq 1$
would seem feasible via the ejection of one boson from a BEC in a Fock state 
$\left\vert n\right\rangle _{R}$ into a previously unoccupied mode. .

\subsection{Irreducible Matrix Representations and Super-selection Rules}

If $\left\vert i\right\rangle $ $(i=1,2,..)$ are a set of orthonormal basis
vectors in the system state space, then the group of unitary operators $%
\widehat{T}(g)$ is represented by a group of \emph{unitary matrices} $D(g)$%
\begin{equation}
\widehat{T}(g)\,\left\vert i\right\rangle
=\dsum\limits_{j}D_{ji}(g)\,\left\vert j\right\rangle
\label{Eq.MatrixRepnTransfnGrp}
\end{equation}%
with elements $D_{ji}(g)$, and such that $D(hg)=D(h)D(g)$ etc corresponding
to the group properties of the operators. This is a \emph{matrix
representation} of the transformation group.

The theory of such group representations and their application to quantum
systems is well established, following the pioneering work of Wigner in the
1930s. We can just use the results here. A key concept is that of \emph{%
irreducible} representations. Within the system state space we can in
general choose so-called irreducible sub-spaces, denoted as $\Gamma _{\alpha
}$ of dimension $d_{\alpha }$ and spanned by new orthonormal basis vectors $%
\left\vert \Gamma _{\alpha }\lambda \right\rangle $ $(\lambda
=1,2,..,d_{\alpha })$ such that 
\begin{equation}
\widehat{T}(g)\,\left\vert \Gamma _{\alpha }\lambda \right\rangle
=\dsum\limits_{\mu =1}^{d_{\alpha }}D_{\mu \lambda }^{\alpha
}(g)\,\left\vert \Gamma _{\alpha }\mu \right\rangle
\label{Eq.IrreducMatrixRepnTransfnGrp}
\end{equation}
For each irreducible sub-space $\Gamma _{\alpha }$ there is \emph{no}
smaller sub-space for which the operation of all $\widehat{T}(g)$ just leads
to linear combinations of vectors within that sub-space. The $d_{\alpha
}\times d_{\alpha }$ matrices $D^{\alpha }(g)$ then form an irreducible
matrix representation for the transformation group. For different $\alpha $
the representations are said to be \emph{inequivalent}.

The irreducible matrices satisfy the so-called \emph{great orthogonality
theorem} \cite{Tinkham64a}%
\begin{equation}
\dint w(g)dg\,D_{\mu \lambda }^{\alpha }(g)D_{\xi \tau }^{\beta }(g)^{\ast }=%
\frac{1}{d_{\alpha }}\delta _{\alpha \beta }\delta _{\mu \xi }\delta
_{\lambda \tau }  \label{Eq.GreatOrthogThm}
\end{equation}%
The proof of this result is based on Schur's lemma.

The importance of the irreducible representations and the consequent
orthogonality theorem lies in its application to Situation B cases, where we
have seen that Charlie's density operator $\widehat{\rho }$ is invariant
under any of the transformations $\widehat{T}(h)\,\widehat{\rho }\,\widehat{T%
}(h)^{-1}=\widehat{\rho }$. Suppose we represent $\widehat{\rho }$ in terms
of the basis vectors $\left\vert \Gamma _{\alpha }\lambda \right\rangle $
associated with the irreducible representations%
\begin{equation}
\widehat{\rho }=\dsum\limits_{\alpha \lambda }\dsum\limits_{\beta \tau
}\,R_{\lambda \tau }^{\alpha \beta }\,\left\vert \Gamma _{\alpha }\lambda
\right\rangle \left\langle \Gamma _{\beta }\tau \right\vert
\label{Eq.CharlieDensOprIrredBasis}
\end{equation}%
where $R$ will be a Hermitian, positive definite matrix with unit trace
since it represents a density operator. Applying the transformation gives%
\begin{eqnarray}
\widehat{T}(h)\,\widehat{\rho }\,\widehat{T}(h)^{-1} &=&\dsum\limits_{\alpha
\lambda \mu }\dsum\limits_{\beta \tau \xi }\,R_{\lambda \tau }^{\alpha \beta
}\,D_{\mu \lambda }^{\alpha }(h)\,\left\vert \Gamma _{\alpha }\mu
\right\rangle \left\langle \Gamma _{\beta }\xi \right\vert \,D_{\xi \tau
}^{\beta }(h)^{\ast }  \nonumber \\
&=&\widehat{\rho }
\end{eqnarray}%
Averaging over $h$ and using the great orthogonality theorem gives%
\begin{equation}
\widehat{\rho }=\dsum\limits_{\alpha }\dsum\limits_{\mu }\,\left(
\dsum\limits_{\lambda }\frac{1}{d_{\alpha }}R_{\lambda \lambda }^{\alpha
\alpha }\,\right) \,\left\vert \Gamma _{\alpha }\mu \right\rangle
\left\langle \Gamma _{\alpha }\mu \right\vert
\end{equation}%
This is in the form of a mixed state involving irreducible state vectors $%
\left\vert \Gamma _{\alpha }\mu \right\rangle $ each occuring with a
probability $P_{\mu }^{\alpha }$ given by 
\begin{equation}
P_{\mu }^{\alpha }=\dsum\limits_{\lambda }\frac{1}{d_{\alpha }}R_{\lambda
\lambda }^{\alpha \alpha }\,=P^{\alpha }  \label{Eq.ProbCharlieDensOpr}
\end{equation}%
which is the same for all $\mu $ associated with a given irreducible
representation $\Gamma _{\alpha }$. This is clearly a positive real quantity
and since 
\begin{eqnarray}
\dsum\limits_{\alpha }\dsum\limits_{\mu }P_{\mu }^{\alpha }
&=&\dsum\limits_{\alpha }\dsum\limits_{\mu }\dsum\limits_{\lambda }\frac{1}{%
d_{\alpha }}R_{\lambda \lambda }^{\alpha \alpha }\,=\dsum\limits_{\alpha
}\dsum\limits_{\lambda }R_{\lambda \lambda }^{\alpha \alpha }  \nonumber \\
&=&Tr\,\widehat{\rho }=1  \label{Eq.ProbSumUnity}
\end{eqnarray}%
the probabilities sum to unity as required.

The final result for Charlie's density operator

\begin{equation}
\widehat{\rho }=\dsum\limits_{\alpha }\dsum\limits_{\mu }\,P^{\alpha
}\,\left\vert \Gamma _{\alpha }\mu \right\rangle \left\langle \Gamma
_{\alpha }\mu \right\vert  \label{Eq.CharlieDensOprSSRForm}
\end{equation}%
demonstrates the presence of a \emph{super-selection rule}. In Charlie's
description of the quantum state there are \emph{no coherences} between
states $\left\vert \Gamma _{\alpha }\mu \right\rangle $ associated with
differing irreducible representations of the transformation group. This
represents the general form of the SSR for all transformation groups in
Situation B cases.

As an example, consider the \emph{U(1) group} and the \emph{single mode}
bosonic system. Since the Fock states satisfy $\widehat{T}(\theta
_{a})\,\left\vert n_{a}\right\rangle =\exp (in_{a}\theta _{a})\,\left\vert
n_{a}\right\rangle $ they form the basis for the irreducible representations
of the U(1) group, the occupation number $n_{a}$ specifying the irreducible
representation and the $1\times 1$ matrices $\exp (in_{a}\theta _{a})$ being
the unitary matrices. Hence Charlie will describe the quantum state as 
\begin{equation}
\widehat{\rho }=\dsum\limits_{n_{a}}P(n_{a})\,\left\vert n_{a}\right\rangle
\left\langle n_{a}\right\vert  \label{Eq.CharlieSingleModeSSRForm}
\end{equation}%
which is a statistical mixture of Fock states with no coherences between
different Fock states. This result is of the same form as in Eq.(\ref%
{Eq.PhysicalStatesSubSys}) and is in accord with the SSR on boson number.

As another example, consider the \emph{U(1) group} and the \emph{multi-mode}
bosonic system. Here sums of products of Fock states 
\begin{equation}
\left\vert n_{1}n_{2}...n_{a}...;N\right\rangle =\dprod\limits_{a}\left\vert
n_{1}\right\rangle \left\vert n_{2}\right\rangle ..\left\vert
n_{a}\right\rangle ...\qquad N=\dsum\limits_{a}n_{a}
\label{Eq.FockStatesMultiMode}
\end{equation}%
such that the total occupancy is $N=\dsum\limits_{a}n_{a}$ can be used to
form irreducible representations for the transformation group in terms of
linear combinations of the products with the same $N$. Writing these linear
combinations as 
\begin{equation}
\left\vert \Psi _{N}^{\mu }\right\rangle
=\dsum\limits_{\{n_{1}n_{2}...n_{a}...\}}C_{\{n_{1}n_{2}...n_{a}...\}}^{N\mu
}\left\vert n_{1}n_{2}...n_{a}...;N\right\rangle
\label{Eq.MultiModeStatesSameN}
\end{equation}%
we have since $\widehat{T}(\theta )\left\vert
n_{1}n_{2}...n_{a}...;N\right\rangle =\exp (iN\theta )\,\left\vert
n_{1}n_{2}...n_{a}...;N\right\rangle $ we see that $\widehat{T}(\theta
)\,\left\vert \Psi _{N}^{\mu }\right\rangle =\exp (iN\theta )\,\left\vert
\Psi _{N}^{\mu }\right\rangle $ also, so the $\left\vert \Psi _{N}^{\mu
}\right\rangle $ define the irreducible basis states. The total occupancy $N$
specifies the irreducible representation, but here there are many
irreducible representations with the same $N$ depending on the various $\mu $%
. In this case Charlie will describe the state as 
\begin{equation}
\widehat{\rho }=\dsum\limits_{N}\dsum\limits_{\mu }P_{\mu }^{N}\,\left\vert
\Psi _{N}^{\mu }\right\rangle \left\langle \Psi _{N}^{\mu }\right\vert
\label{Eq.CharlieMultModeSSRForm}
\end{equation}%
which is a statistical mixture of multi-mode states $\left\vert \Psi
_{N}^{\mu }\right\rangle $ all with the same total occupancy $N$. Although
there are coherence terms between individual modal Fock states, there are no
coherences between states with different total occupancy. This result is of
the same form as in Eq.(\ref{Eq.PhysicalState}) and again is an example of a
super-selection rule operating in terms of Charlie's description of the
quantum state.

Finally, we note that in situation A where the relationship between the
frames is known and there is no invariance for Charlie's density operator,
we do not have SSR applying. For the \emph{single particle }case and the 
\emph{translation group} the momentum states $\left\vert \underrightarrow{p}%
\right\rangle $ define the irreducible representations, each specified by $%
\underrightarrow{p}$, and as we saw Charlie's description of the quantum
state involved linear combinations of these irreducible basis vectors, in
contradiction to the SSR.

\subsection{Non-Entangled States}

\label{Appendix SubSection - Non Ent States}

The essential feature of an \emph{non-entangled} or \emph{separable} state
is that the sub-systems are considered to be \emph{unrelated} to each other.
Hence, both for Alice and Charlie there will be \emph{separate reference
frames} for each sub-system, with transformation groups - $\widehat{T}%
_{A}(g_{a})$ for sub-system $A$, $\widehat{T}_{B}(g_{b})$ for sub-system $B$%
, etc which relate the reference systems of Alice to those of Charlie. The
transformations $g_{a}$, $g_{b}$, $..$ are different. The \emph{overall}
transformation operator would be of the form $\widehat{T}(g_{a},g_{b},..)=%
\widehat{T}_{A}(g_{a})\otimes \widehat{T}_{B}(g_{b})\otimes ..$. Alice would
describe a general non-entangled state as having a density operator 
\begin{equation}
\widehat{\sigma }=\sum_{R}P_{R}\,\widehat{\sigma }_{R}^{A}\otimes \widehat{%
\sigma }_{R}^{B}\otimes \widehat{\sigma }_{R}^{C}\otimes ...
\label{Eq.AliceDensOprNonEntState}
\end{equation}%
It then follows for Situation B where the reference frames for Alice and
Charlie are unrelated, that Charlie would describe the same state via the
density operator 
\begin{equation}
\widehat{\rho }=\sum_{R}P_{R}\,\widehat{\rho }_{R}^{A}\otimes \widehat{\rho }%
_{R}^{B}\otimes \widehat{\rho }_{R}^{C}\otimes ...
\label{Eq.CharlieDensOprNonEntState}
\end{equation}%
where

\begin{equation}
\widehat{\rho }_{R}^{C}=\dint w(g_{c})dg_{c}\,\widehat{T}_{C}(g_{c})\,%
\widehat{\sigma }_{R}^{C}\,\widehat{T}_{C}(g_{c})^{-1}\qquad C=A,B,..
\label{Eq,SubSysDensityOprSitnB}
\end{equation}%
Note that \emph{separate} twirl operations are applied to the different
sub-systems, as explicitly shown in the papers by Vaccaro et al \cite%
{Vaccaro08a} (see Section IIIA, Eqn. 3.3 therein) and Paterek et al \cite%
{Paterek11a} (see Section 6). This leads for general transformation groups
to the \emph{local group super-selection rule}, where the $\widehat{\rho }%
_{R}^{C}$ involve \emph{no coherences} between states associated with
differing irreducible representations of the transformation group. We see
that Charlie also describes a non-entangled state and with the same mixture
probability $P_{R}$ as for Alice. Thus non-entanglement or separability is a
feature that is the $\emph{same}$ for both Alice and Charlie, as ought to be
the case.

In the context of sub-systems consisting of \emph{modes} (or sets of modes)
occupied by \emph{identical bosons}, the case of interest is Situation B,
with each transformation group being U(1). Here the relationship between
Charlie's and Alice's \emph{phase reference} frames are unknown. Hence
irrespective of Alice's description of the sub-system states $\widehat{%
\sigma }_{R}^{A}$, $\widehat{\sigma }_{R}^{B}$,... we see from the previous
section that Charlie will describe the separate sub-system states $\widehat{%
\rho }_{R}^{A}$, $\widehat{\rho }_{R}^{B}$, as statistical mixtures of
number states for the separate modes (or total number states for the sets of
modes in each sub-system). Thus from Charlie's point of view the separate
mode density operators will satisfy the SSR. Thus we see that the
introduction of reference frames and two observers - Charlie being the
external one whose description of the quantum states is of primary interest
- leads to the \emph{same SSR outcome} as the simpler considerations set out
in SubSections \ref{SubSection - Super-Selection Rule} and \ref{SubSection -
SSR Separate Modes}. Essentially the same considerations have been used in 
\cite{Bartlett06b}, \cite{Vaccaro08a} and the other papers to justify the 
\emph{local photon number superselection rule}. \pagebreak

\section{Appendix 5 - Super-Selection Rule Violations ?}

\label{Appendix - Super-Selection Rule Violations ?}

\subsection{Preparation of Coherent Superposition of an Atom and a Molecule ?%
}

A key paper dealing with the coherent superposition of an atom and a
molecule is that by Dowling et al \cite{Dowling06a}, entitled
\textquotedblleft Observing a coherent superposition of an atom and a
molecule\textquotedblright . Essentially the process involves one atom A
interacting with a BEC of different atoms B leading to the creation of one
molecule AB, with the BEC being depleted by one B atom.

\subsubsection{Hamiltonian}

The Hamiltonian is given by 
\begin{equation}
\widehat{H}=\hbar \omega _{A}\widehat{b}_{A}^{\dag }\widehat{b}_{A}+\hbar
\omega _{M}\widehat{b}_{M}^{\dag }\widehat{b}_{M}+\hbar \omega _{2}\widehat{b%
}_{2}^{\dag }\widehat{b}_{2}+\frac{\hbar \kappa }{2}(\widehat{b}_{M}^{\dag }%
\widehat{b}_{A}\widehat{b}_{2}+\widehat{b}_{M}\widehat{b}_{A}^{\dag }%
\widehat{b}_{2}^{\dag })  \label{Eq.HamiltonianDowling}
\end{equation}%
where $\widehat{b}_{A},\widehat{b}_{M}$ and $\widehat{b}_{2}$ are standard
bosonic annihilation operators for the atom, molecule and BEC modes
respectively, $\omega _{A},\omega _{M}$ and $\omega _{2}$ are the
corresponding mode frequencies and $\kappa $ defines the interaction
strength for the process where a molecule is created or destroyed from/to an
atom A and a BEC\ atom B. $\Delta $ is the frequency difference between the
molecular state AB and the two separate states for atoms A and B -- this is
zero on Feshbach resonance - and is given by 
\begin{equation}
\Delta =\omega _{M}-\omega _{A}-\omega _{2}  \label{Eq.FeshbachDetuning}
\end{equation}%
The Hamiltonian commutes with the total number operator $\widehat{N}_{tot}$,
where 
\begin{equation}
\widehat{N}_{tot}=2\,\widehat{b}_{M}^{\dag }\widehat{b}_{M}+\widehat{b}%
_{A}^{\dag }\widehat{b}_{A}+\widehat{b}_{2}^{\dag }\widehat{b}_{2}
\label{Eq.TotalNumberOpr}
\end{equation}%
where the molecule number operator is multipled by two.

\subsubsection{Initial State}

Initially the state of the system is given by the density operator Eqs (10)
and (11) in the paper 
\begin{eqnarray}
\widehat{W}_{0L} &=&\dint \frac{d\theta }{2\pi }\exp (-i\widehat{N}%
_{tot}\theta )\left\vert \Psi \right\rangle _{0L}\left\langle \Psi
\right\vert _{0L}\exp (+i\widehat{N}_{tot}\theta )
\label{Eq.InitialdensityOpr} \\
\left\vert \Psi \right\rangle _{0L} &=&\left\vert A\right\rangle \left\vert
\beta \right\rangle  \label{Eq.BasicInitialStateVector}
\end{eqnarray}%
where $\left\vert A\right\rangle $ is a state with one atom A and $%
\left\vert \beta \right\rangle $ is a Glauber coherent state for the BEC of
atoms B.The super-operator acting on the pure state $\left\vert \Psi
\right\rangle _{0L}\left\langle \Psi \right\vert _{0L}$ is called the \emph{%
twirling operator}, the group of unitary operators $\exp (-i\widehat{N}%
_{tot}\theta )$ depend on a \emph{phase} variable $\theta $ and are a
unitary representation of $U(1)$, the \emph{generator} being $\widehat{N}%
_{tot}$. These operators act as a \emph{symmetry group} for the system and
leave the Hamiltonian invariant. The \emph{initial state} is also given by 
\begin{eqnarray}
\widehat{W}_{0L} &=&\widehat{\rho }_{A-M}(0)\otimes \widehat{\rho }_{2}(0)
\label{Eq.InitialDensityOpr2} \\
\widehat{\rho }_{A-M}(0) &=&\left\vert A\right\rangle \left\langle
A\right\vert  \label{Eq.AtomMolDensityOpr} \\
\widehat{\rho }_{2}(0) &=&\dint \frac{d\theta }{2\pi }\exp (-i\widehat{n}%
_{2}\theta )\left\vert \beta \right\rangle \left\langle \beta \right\vert
\exp (+i\widehat{n}_{2}\theta )  \label{Eq.BECDensityOpr1} \\
&=&\dsum\limits_{n}p_{n}(<n>)\left\vert n\right\rangle \left\langle
n\right\vert  \label{Eq.BECDensityOpr2} \\
&=&\dint \frac{d\theta }{2\pi }\left\vert \beta \exp (-i\theta
)\right\rangle \left\langle \beta \exp (-i\theta )\right\vert
\label{Eq.BECDensityOpr3}
\end{eqnarray}%
where $\widehat{n}_{2}=\widehat{b}_{2}^{\dag }\widehat{b}_{2}$ is the number
uperator for the BEC mode and $p_{n}(<n>)=\{\exp (-<n>)\,<n>^{n}/n!\}$ is a
Poisson distribution, whose mean is $<n>=|\beta |^{2}$. Initially then there
is one atom A and the BEC is in a statistical mixture of number states with
a Poisson distribution, which is mathematically equivalent to a statistical
mixture of Glauber coherent states $\left\vert \beta \exp (-i\theta
)\right\rangle $ with the same amplitude $\sqrt{<n>}$ but with all phases $%
(\arg \beta +\theta )$ being equally weighted.

\subsubsection{Implicated Reference Frame}

In the paper by Dowling et al \cite{Dowling06a} the BEC is acting as an 
\emph{implicated phase reference frame} (see \cite{Bartlett06a}, \cite%
{Bartlett07a}). The state of the reference frame as described by Charlie is
given by 
\begin{equation}
\widehat{\rho }_{REF}=\,\widehat{\rho }_{2}(0)=\dint \frac{d\theta }{2\pi }%
\exp (-i\widehat{n}_{2}\theta )\left\vert \beta \right\rangle \left\langle
\beta \right\vert \exp (+i\widehat{n}_{2}\theta )  \label{Eq.RefFrame}
\end{equation}%
and from Eq. (\ref{Eq.HamiltonianDowling}), there is an interaction between
the reference BEC and the separate atom A and molecule M systems. However,
because $<n>=|\beta |^{2}$ is very large, the BEC is essentially unchanged
during the process, as reflected in the use of approximations in eqs (27),
(28) of the paper. Another implicated phase reference frame situation, but
involving a two mode reference frame is discussed in the paper by Paterek et
al \cite{Paterek11a}

Overall, in terms of the discussion in Appendix \ref{Appendix - Reference
Frames and SSR} $\widehat{W}_{0L}$ would be \emph{Charlie}'s description of
the initial state, whereas \emph{Alice} would describe it as $\left\vert
\Psi \right\rangle _{0L}\left\langle \Psi \right\vert _{0L}$. Presumably in
the paper by Dowling et al \cite{Dowling06a} what is referred to as the
"state of the laboratory" be Charlie's reference frame, and what they refer
to as the "internal reference frame" would refer to that of Alice. However,
whether Alice could actually prepare such a state as $\left\vert \Psi
\right\rangle _{0L}\left\langle \Psi \right\vert _{0L}$ is controversial -
see SubSections \ref{SubSection - Super-Selection Rule} and \ref{SubSection
- SSR Separate Modes}, though here this is assumed to be possible.

\subsubsection{Process - Alice and Charlie Descriptions}

There are three stages in the process, the first being with the interaction
that turns separate atoms A and B into the molecule AB turned on at Feshbach
resonance for a time $t=\pi /(2\kappa <n>)$, the second being free evolution
at large Feshbach detuning $\Delta $ for a time $\tau $ leading to a phase
factor $\phi =\Delta \tau $, the third being again with the interaction
turned on at Feshbach resonance for a further time $t=\pi /(2\kappa <n>)$.
The typical initial state $\left\vert \Psi \right\rangle _{0L}$ given by $%
\left\vert A\right\rangle \left\vert \beta \right\rangle $ (eq (11)) evolves
into $\left\vert \Psi \right\rangle _{3L}$ given by (see eq. (32) of paper) 
\begin{equation}
\left\vert \Psi \right\rangle _{3L}=\left( \sin (\frac{\phi }{2}%
)\,\left\vert A\right\rangle -\exp (i\arg \beta )\,\cos (\frac{\phi }{2}%
)\,\left\vert M\right\rangle \right) \left\vert \beta \right\rangle
\label{Eq.BasicFinalStateVector}
\end{equation}%
using approximations set out in eqs (27), (28) of the paper that depend on $%
<n>$ being large. Here $\left\vert M\right\rangle $ is a state with one
molecule AB. Thus it looks like a coherent superposition of an atom state $%
\left\vert A\right\rangle $ and a molecule state $\left\vert M\right\rangle $
has been prepared, the atom plus molecule system being disentangled from the
BEC. \emph{Alice} would describe the final state of the system as $%
\left\vert \Psi \right\rangle _{3L}\left\langle \Psi \right\vert _{3L}$, so
from her point of view a coherent superposition of an atom and a molecule
has been prepared.

However, for \emph{Charlie} the \emph{final state} of the system is
described by a density operator $\widehat{W}_{3L}$ which is reconstructed by
applying the twirling operator to $\left\vert \Psi \right\rangle
_{3L}\left\langle \Psi \right\vert _{3L}$ . Noting that%
\begin{equation}
\exp (-i\widehat{N}_{tot}\theta )\left\vert \Psi \right\rangle _{3L}=\left(
\exp (-i\theta )\,\sin (\frac{\phi }{2})\,\left\vert A\right\rangle -\exp
(-2i\theta )\,\exp (i\arg \beta )\,\cos (\frac{\phi }{2})\,\left\vert
M\right\rangle \right) \left\vert \beta \exp (-i\theta )\right\rangle
\label{Eq.TwirlingEffect}
\end{equation}%
and using 
\begin{equation}
Tr_{2}(\left\vert \beta \exp (-i\theta )\right\rangle \left\langle \beta
\exp (-i\theta )\right\vert )=\left\langle \beta \exp (-i\theta )|\beta \exp
(-i\theta )\right\rangle =1  \label{Eq.BECTrace}
\end{equation}%
we see that Charlie's final reduced density operator for the \emph{%
atom-molecule system} is 
\begin{eqnarray}
\widehat{\rho }_{A-M}(3) &=&Tr_{2}\widehat{W}_{3L}  \nonumber \\
&=&Tr_{2}\dint \frac{d\theta }{2\pi }\exp (-i\widehat{N}_{tot}\theta
)\left\vert \Psi \right\rangle _{3L}\left\langle \Psi \right\vert _{3L}\exp
(+i\widehat{N}_{tot}\theta )  \nonumber \\
&=&\dint \frac{d\theta }{2\pi }\left( \exp (-i\theta )\,\sin (\frac{\phi }{2}%
)\,\left\vert A\right\rangle -\exp (-2i\theta )\,\exp (i\arg \beta )\,\cos (%
\frac{\phi }{2})\,\left\vert M\right\rangle \right)  \nonumber \\
&&\times \left( \exp (+i\theta )\,\sin (\frac{\phi }{2})\,\left\langle
A\right\vert -\exp (+2i\theta )\,\exp (-i\arg \beta )\,\cos (\frac{\phi }{2}%
)\,\left\langle M\right\vert \right)  \nonumber \\
&=&\sin ^{2}(\frac{\phi }{2})\,\left\vert A\right\rangle \left\langle
A\right\vert +\cos ^{2}(\frac{\phi }{2})\,\left\vert M\right\rangle
\left\langle M\right\vert  \label{Eq.FinalRDOAtomMol}
\end{eqnarray}%
Thus the coherence terms like $\left\vert A\right\rangle \left\langle
M\right\vert $ and $\left\vert M\right\rangle \left\langle A\right\vert $ do
not appear in the final density operator when the average over $\theta $
(not $\beta $) is carried out.

For Charlie the density operator for the atom and molecule is of course a
statistical mixture of a state with one atom and no molecule and a state
with no atom and one molecule. The authors of \cite{Dowling06a} actually
point this out in the paragraph after eq (35) where (presumably for the case 
$\phi =\pi /4$) it is stated \textquotedblleft the state is found to be
\ldots\ an incoherent mixture of an atom and a molecule.\textquotedblright .
The probabilities for detecting an atom A or a molecule AB are as in eq (33)
of the paper. In terms of Charlie's description, the density operator at the
end of the preparation process does \emph{not} signify the existence of a
coherent superposition of an atom and a molecule, as the title to the paper
might be taken to imply. The existence of such a coherent superposition
would of course be present in Alice's description, but it is Charlie's
(laboratory) description that is more relevant.

\subsubsection{Interference Effects Without SSR Violation}

Note that \emph{interference effects} are still present since the atom or
molecule detection probabilities depend on the phase $\phi $ associated with
the free evolution stage of the process. However, as in many other
instances, the presence of coherence effects does not require the existence
of \emph{coherent superposition states} that violate the super-selection
rule. The authors actually point this out in the paragraph after eq (35),
where it is stated \textquotedblleft we have clearly predicted the standard
operational signature of coherence, namely Ramsey type fringes, but the
coherence is not present in our mathematical description of the
system.\textquotedblright\ What they are referring to is Charlie's
description of the final state - which indeed shows no such coherence, but
the belief that coherent superposition states are needed to predict
coherence effects is mistaken.

To drive this point home, the process can be treated with the initial state
for the BEC being given as a Fock state $\left\vert N\right\rangle $. With
the interaction being given as in Eq.(\ref{Eq.HamiltonianDowling}) (eq (14)
in the paper) the state vector is a simple linear combination of two terms%
\begin{equation}
\left\vert \Psi (t)\right\rangle =A(t)\left\vert A\right\rangle \,\left\vert
N\right\rangle +B(t)\left\vert M\right\rangle \,\left\vert N-1\right\rangle
\label{Eq.StateVectorFockMethod}
\end{equation}%
This is of course an entangled state. Coupled equations for the two
amplitudes $A(t)$ and $B(t)$ can easily be obtained and simple solutions
obtained for stages where the Feshbach detuning is either zero or large. The
state vector is continuous from one stage to the next , and the reduced
density operator at the end of the three stage process for the atom plus
molecule sub-system can be obtained. It is of the form%
\begin{eqnarray}
\widehat{\rho }_{A-M}(3) &=&Tr_{2}(\left\vert \Psi (3)\right\rangle
\left\langle \Psi (3)\right\vert )  \nonumber \\
&=&\sin ^{2}(\frac{\phi }{2})\,\left\vert A\right\rangle \left\langle
A\right\vert +\cos ^{2}(\frac{\phi }{2})\,\left\vert M\right\rangle
\left\langle M\right\vert  \label{Eq.FinalRDOAtomMolFockMethod}
\end{eqnarray}%
which is of course a statistical mixture of a state with one atom and no
molecule and a state with no atom and one molecule - and is exactly the same
result as obtained in the paper by Dowling et al.\cite{Dowling06a}. Note
that coherence effects in regard to the interferometric dependence on $\phi $
for measurements on the final state has been found without invoking either
the description of the BEC via Glauber coherent states or the presence of a
coherent superposition of an atomic and a molecular state. The result can
easily be extended for the case where the BEC is initially in a statistical
mixture of Fock states with differing $N$ occuring with a probability $P_{N}$%
. Each initial state $\left\vert A\right\rangle \,\left\vert N\right\rangle $
evolves as in Eq. (\ref{Eq.StateVectorFockMethod}). We then would have 
\begin{eqnarray}
\widehat{\rho }_{A-M}(3) &=&Tr_{2}(\tsum\limits_{N}P_{N}\,\left\vert \Psi
_{N}(3)\right\rangle \left\langle \Psi _{N}(3)\right\vert )  \nonumber \\
&=&\tsum\limits_{N}P_{N}\left( \sin ^{2}(\frac{\phi }{2})\,\left\vert
A\right\rangle \left\langle A\right\vert +\cos ^{2}(\frac{\phi }{2}%
)\,\left\vert M\right\rangle \left\langle M\right\vert \right)  \nonumber \\
&=&\sin ^{2}(\frac{\phi }{2})\,\left\vert A\right\rangle \left\langle
A\right\vert +\cos ^{2}(\frac{\phi }{2})\,\left\vert M\right\rangle
\left\langle M\right\vert  \label{Eq.FinalRDOStatMixtFocks}
\end{eqnarray}%
which is the same as before. Allowing for a statistical mixture of Fock
states makes no difference to the interferometric result.

\subsubsection{Conclusion}

Dowling et al \cite{Dowling06a} state in their abstract that
\textquotedblleft we demonstrate that it is possible to perform a
Ramsey-type interference experiment to exhibit a coherent superposition of a
single atom and a diatomic molecule\textquotedblright\ . However the
interferometric effects (involving the dependence on $\phi $) cannot be said
to exhibit the \emph{existence} of such a coherent superposition, since the
same interferometric results can be obtained \emph{without} ever introducing
such a quantum state. There is \emph{not} a convincing case that quantum
states that violate the super-selection rule forbidding the creation of
coherent superpositions of Fock states with differing particle numbers can
be \emph{created}, even in Alice's reference system. The fact that an SSR
violating state $\left\vert \Psi \right\rangle _{3L}\left\langle \Psi
\right\vert _{3L}$ is created in Alice's reference system is not surprising,
because in the process considered the initial state $\left\vert \beta
\right\rangle $ for the BEC was assumed as a factor in Alice's initial
state, and this was itself inconsistent with the SSR. Furthermore, such SSR
violating states are not \emph{needed} to describe coherence and
interference effects, so that justification for their physical existence
also fails.

\subsection{Detection of Coherent Superposition of a Vacuum and a One-Boson
State ?}

Whether such super-selection rule violating states can be detected has also
not been justified. For example, consider the state given by a superposition
of a one boson state and the vacuum state (as discussed in \cite%
{Dunningham11a}). Consider an interferometric process in which one mode $A$
for a two mode BEC interferometer is initially in the state $\alpha
\left\vert 0\right\rangle +\beta \left\vert 1\right\rangle $, and the other
mode $B$ is initially in the state $\left\vert 0\right\rangle $ - thus $%
\left\vert \Psi (i)\right\rangle =(\alpha \left\vert 0\right\rangle +\beta
\left\vert 1\right\rangle )_{A}\otimes \left\vert 0\right\rangle _{B}$ in
the usual occupancy number notation, where $|\alpha |^{2}+|\beta |^{2}=1$.
Modes $A$, $B$ could refer to two different hyperfine states of a bosonic
atom with non-relativistic energies $\hbar \omega _{A}$.and $\hbar \omega
_{B}$, mode annihilation operators $\widehat{a}$, $\widehat{b}$. The modes
are first coupled by a \emph{beam splitter}, which could be a resonant
microwave pulse that causes transitions between the two hyperfine states and
which can be described via a unitary operator $\widehat{U}_{BS}$ such that%
\begin{eqnarray}
\widehat{U}_{BS}(\left\vert 1\right\rangle _{A}\otimes \left\vert
0\right\rangle _{B}) &=&(\left\vert 1\right\rangle _{A}\otimes \left\vert
0\right\rangle _{B}-i\left\vert 0\right\rangle _{A}\otimes \left\vert
1\right\rangle _{B})/\sqrt{2}  \nonumber \\
\widehat{U}_{BS}(\left\vert 0\right\rangle _{A}\otimes \left\vert
1\right\rangle _{B}) &=&(-i\left\vert 1\right\rangle _{A}\otimes \left\vert
0\right\rangle _{B}+\left\vert 0\right\rangle _{A}\otimes \left\vert
1\right\rangle _{B})/\sqrt{2}  \nonumber \\
\widehat{U}_{BS}(\left\vert 0\right\rangle _{A}\otimes \left\vert
0\right\rangle _{B}) &=&(\left\vert 0\right\rangle _{A}\otimes \left\vert
0\right\rangle _{B}).  \label{Eq.BeamSplitter}
\end{eqnarray}%
After passing through the beam splitter the system is allowed to evolve
freely for a time $\tau $, the Hamiltonian being $\widehat{H}%
_{free}=(mc^{2}+\hbar \omega _{A})\widehat{a}^{\dag }\widehat{a}%
+(mc^{2}+\hbar \omega _{B})\widehat{b}^{\dag }\widehat{b}$ - where
collisional effects have been ignored and the rest mass energy included for
completeness. Following the free evolution stage, the modes are then coupled
again via a beam splitter, and the probability of an atom being found in
modes $A$, $B$ then being measured. A straightforward treatment of the
evolution shows that the final state is given by 
\begin{eqnarray}
\left\vert \Psi (f)\right\rangle &=&\alpha (\left\vert 0\right\rangle
_{A}\otimes \left\vert 0\right\rangle _{B})  \nonumber \\
&&+\beta \exp (-i\{mc^{2}/\hbar +\omega _{A}\}\tau )  \nonumber \\
&&\times \left( \frac{1-\exp (-i\Delta \tau )}{2}(\left\vert 1\right\rangle
_{A}\otimes \left\vert 0\right\rangle _{B})-i\frac{1+\exp (-i\Delta \tau )}{2%
}(\left\vert 0\right\rangle _{A}\otimes \left\vert 1\right\rangle
_{B})\right)  \nonumber \\
&&  \label{Eq.FinalState}
\end{eqnarray}%
where $\Delta =\omega _{B}-\omega _{A}$ is the detuning. The probabilities
of finding one atom in modes $A$, $B$ respectively are 
\begin{equation}
P_{10}=|\beta |^{2}\sin ^{2}(\Delta \tau /2)\qquad P_{01}=|\beta |^{2}\cos
^{2}(\Delta \tau /2)  \label{Eq.AtomDetnProb}
\end{equation}%
Thus whilst coherence effects occur depending on the phase difference $\phi
=\Delta \tau $ associated with the interferometric process, the overall
detection probabilities only depend on the initial state via $|\beta |^{2}$.
There is no dependence on the \emph{relative phase} between $\alpha $ and $%
\beta $, as would be required if the superposition state $\alpha \left\vert
0\right\rangle +\beta \left\vert 1\right\rangle $ is to be specified from
the measurement results. Exactly the same detection probabilities are
obtained if the initial state is the mixed state $\widehat{\rho }(i)=|\alpha
|^{2}(\left\vert 0\right\rangle _{A}\left\langle 0\right\vert _{A}\otimes
\left\vert 0\right\rangle _{B}\left\langle 0\right\vert _{B})+|\beta
|^{2}(\left\vert 1\right\rangle _{A}\left\langle 1\right\vert _{A}\otimes
\left\vert 0\right\rangle _{B}\left\langle 0\right\vert _{B})$, in which the
vacuum state for mode $A$ occurs with a probability $|\alpha |^{2}$ and the
one boson state for mode $A$ occurs with a probability $|\beta |^{2}$. In
this example the coherent superposition associated with the super-selection
rule violating state would not be detected in the interferometric process.
The paper by Dunningham et al \cite{Dunningham11a} considers first a
detection process that involves using a Glauber coherent state as one of the
input states. Similar interference effects as in Eq. (\ref{Eq.AtomDetnProb})
are obtained. A second detection process in which the single term Glauber
coherent state is replaced by a statistical mixture with all phases equally
weighted in considered next, leading to the same interference effects. This
again confirms that it is not necessary to invoke the existence of coherent
superpositions of number states in order to demonstate interference effects.
\pagebreak

\section{Appendix 6 - Non-Physical Two Mode States}

\label{Appendix - Non-Physical Two Mode States}

We now consider some possible states for the second mode $B$ - to be
combined with $\widehat{\rho }_{1}^{A}$, $\widehat{\rho }_{2}^{A}$ and $%
P_{1} $, $P_{2}$ as in Eq. (\ref{Eq.NonPhysicalStates}). These states are
two general pure orthogonal states of the form $\alpha \left\vert
0\right\rangle _{B}+\beta \left\vert 1\right\rangle _{B}$ and $-\beta ^{\ast
}\left\vert 0\right\rangle _{B}+\alpha ^{\ast }\left\vert 1\right\rangle
_{B} $ with $(|\alpha |^{2}+|\beta |^{2})=1$. We have

\begin{eqnarray}
\widehat{\rho }_{1}^{B} &=&\left( (\alpha \left\vert 0\right\rangle
_{B}+\beta \left\vert 1\right\rangle _{B})\right) \left( (\alpha ^{\ast
}\left\langle 0\right\vert _{B}+\beta ^{\ast }\left\langle 1\right\vert
_{B})\right)  \nonumber \\
\widehat{\rho }_{2}^{B} &=&\left( (-\beta ^{\ast }\left\vert 0\right\rangle
_{B}+\alpha ^{\ast }\left\vert 1\right\rangle _{B})\right) \left( (-\beta
\left\langle 0\right\vert _{B}+\alpha \left\langle 1\right\vert _{B})\right)
\end{eqnarray}%
This gives the reduced density operator 
\begin{equation}
\widehat{\rho }_{B}=\frac{1}{2}\left( \left\vert 0\right\rangle
_{B}\left\langle 0\right\vert _{B}\right) +\frac{1}{2}\left( \left\vert
1\right\rangle _{B}\left\langle 1\right\vert _{B}\right)
\end{equation}

A straightforward calculation gives for the overall density operator for the
two mode non-entangled state as in Eq. (\ref{Eq.NonEntangledState})%
\[
\widehat{\rho }=\widehat{\rho }_{1}+\widehat{\rho }_{2} 
\]%
where the $\widehat{\rho }_{1}$, $\widehat{\rho }_{2}$ are contributions
that are consistent with or inconsistent with the super-selection rule. We
have%
\begin{eqnarray}
\widehat{\rho }_{1} &=&\frac{1}{4}\left\vert 0\right\rangle _{A}\left\langle
0\right\vert _{A}\otimes \left\vert 0\right\rangle _{B}\left\langle
0\right\vert _{B}+\frac{1}{4}\left\vert 1\right\rangle _{A}\left\langle
1\right\vert _{A}\otimes \left\vert 0\right\rangle _{B}\left\langle
0\right\vert _{B}  \nonumber \\
&&+\frac{1}{4}\left\vert 0\right\rangle _{A}\left\langle 0\right\vert
_{A}\otimes \left\vert 1\right\rangle _{B}\left\langle 1\right\vert _{B}+%
\frac{1}{4}\left\vert 1\right\rangle _{A}\left\langle 1\right\vert
_{A}\otimes \left\vert 1\right\rangle _{B}\left\langle 1\right\vert _{B} 
\nonumber \\
&&+\frac{1}{2}\alpha ^{\ast }\beta \,\left\vert 0\right\rangle
_{A}\left\langle 1\right\vert _{A}\otimes \left\vert 1\right\rangle
_{B}\left\langle 0\right\vert _{B}+\frac{1}{2}\alpha \beta ^{\ast
}\,\left\vert 1\right\rangle _{A}\left\langle 0\right\vert _{A}\otimes
\left\vert 0\right\rangle _{B}\left\langle 1\right\vert _{B}
\end{eqnarray}%
and 
\begin{eqnarray}
\widehat{\rho }_{2} &=&\frac{1}{4}(|\alpha |^{2}-|\beta |^{2})\left\vert
0\right\rangle _{A}\left\langle 1\right\vert _{A}\otimes \left\vert
0\right\rangle _{B}\left\langle 0\right\vert _{B}+\frac{1}{4}(|\alpha
|^{2}-|\beta |^{2})\left\vert 1\right\rangle _{A}\left\langle 0\right\vert
_{A}\otimes \left\vert 0\right\rangle _{B}\left\langle 0\right\vert _{B} 
\nonumber \\
&&+\frac{1}{4}(|\beta |^{2}-|\alpha |^{2})\left\vert 0\right\rangle
_{A}\left\langle 1\right\vert _{A}\otimes \left\vert 1\right\rangle
_{B}\left\langle 1\right\vert _{B}+\frac{1}{4}(|\beta |^{2}-|\alpha
|^{2})\left\vert 1\right\rangle _{A}\left\langle 0\right\vert _{A}\otimes
\left\vert 1\right\rangle _{B}\left\langle 1\right\vert _{B}  \nonumber \\
&&+\frac{1}{2}\alpha ^{\ast }\beta \,\left\vert 1\right\rangle
_{A}\left\langle 0\right\vert _{A}\otimes \left\vert 1\right\rangle
_{B}\left\langle 0\right\vert _{B}+\frac{1}{2}\alpha \beta ^{\ast
}\,\left\vert 0\right\rangle _{A}\left\langle 1\right\vert _{A}\otimes
\left\vert 0\right\rangle _{B}\left\langle 1\right\vert _{B}
\end{eqnarray}

Now to make $\widehat{\rho }_{2}=0$ requires $|\alpha |^{2}=|\beta |^{2}$so
that the first four terms in $\widehat{\rho }_{2}$ are zero. This in
combination with $|\alpha |^{2}+|\beta |^{2})=1$ leads to $|\alpha |=|\beta
|=\frac{1}{\sqrt{2}}$. However this results in the remaining two terms in $%
\widehat{\rho }_{2}$ - which are coherences between $N=0$ and $N=2$ states -
always being non-zero. Overall then, no choice of $\alpha $, $\beta $ will
lead to a overall density operator which is physical. Adding further states $%
\left\vert 2\right\rangle _{B}$, $\left\vert 3\right\rangle _{B}$, ...does
not rectify the problem. \pagebreak

\section{Appendix 7 - Classical Entanglement}

\label{Appendix - Classical Entanglement}

In addition to quantum entanglement there is a body of work (see \cite%
{Gisin91a}, \cite{Spreeuw98a}, \cite{Borges10a})\ dealing with so-called%
\textbf{\ }\emph{classical entanglement}\textbf{. }Here the\textbf{\ }\emph{%
states}\textbf{\ }of classical systems are represented via formalisms
involving\textbf{\ }\emph{linear vector spaces}\textbf{\ }and classical
entanglement is defined mathematically. The case of the\textbf{\ }\emph{%
electromagnetic (EM) field}\textbf{\ }provides such an example, here the
real vector electric field\textbf{\ }$\underrightarrow{E}(\mathbf{r},t)$ for
the simple non-stochastic situation may be written as\textbf{\ }%
\begin{equation}
\underrightarrow{E}(\mathbf{r},t)=\widehat{x}\,E_{x}(\mathbf{r},t)+\widehat{y%
}\,E_{y}(\mathbf{r},t)+\widehat{z}\,E_{z}(\mathbf{r},t)
\label{Eq.ClassicalEField}
\end{equation}%
where $\widehat{x},\widehat{y},\widehat{z}$ are orthogonal unit\textbf{\ }%
\emph{polarization}\textbf{\ }vectors and $E_{x},E_{y},E_{z}$ are field
functions for the corresponding\textbf{\ }\emph{components}\textbf{. }Both
the unit vectors and components may be complex if circular rather than
linear polarization vectors are used, and\textbf{\ }$\underrightarrow{E}(%
\mathbf{r},t)$ can be expanded in terms of different choices of polarization
vectors and their associated components. In classical physics\textbf{\ }$%
\underrightarrow{E}(\mathbf{r},t)$ specifies the\textbf{\ }\emph{state}%
\textbf{\ }of the field, so in effect the six quantities $\widehat{x},%
\widehat{y},\widehat{z}$ and $E_{x},E_{y},E_{z}$ - all of which can be
simultaneously specified and which determine the $\underrightarrow{E}$ field
state. \ Strictly speaking, the\textbf{\ }\emph{magnetic field }$%
\underrightarrow{B}(\mathbf{r},t)$ should also be included, but for
simplicity we ignore it. Also, the field may require an ensemble of\textbf{\ 
}\emph{stochastic}\textbf{\ }fields\textbf{\ }$\{\underrightarrow{E}(\mathbf{%
r},t)\}$to specify it, described via a distribution function, but here we
just consider the simple non-stochastic case. In a different specification
of the same classical field state in the form\textbf{\ }$\underrightarrow{E}(%
\mathbf{r},t)=\widehat{u}\,E_{u}(\mathbf{r},t)+\widehat{v}\,E_{v}(\mathbf{r}%
,t)+\widehat{w}\,E_{w}(\mathbf{r},t)$ the new polarizations $\widehat{u},%
\widehat{v},\widehat{w}$ and components $E_{u},E_{v},E_{w}$ also specify the
state, and the new polarizations and components are just linear combinations
of the old.

\subsection{Classical Ket Vector Formalism and Entangled States}

Because of this linearity the polarizations and components can be described
formally as vectors in two\textbf{\ }\emph{linear vector spaces}, one for
polarizations the other for components. Thus the $\underrightarrow{E}$ field
state is now written as%
\begin{equation}
\left\vert \underrightarrow{E}(\mathbf{r},t)\right\rangle =\left\vert 
\widehat{x}\right\rangle \,\left\vert E_{x}(\mathbf{r},t)\right\rangle
+\left\vert \widehat{y}\right\rangle \,\left\vert E_{y}(\mathbf{r}%
,t)\right\rangle +\left\vert \widehat{z}\right\rangle \,\left\vert E_{z}(%
\mathbf{r},t)\right\rangle  \label{Eq.ClassEFieldVector}
\end{equation}%
in a form simular to $\left\vert \Phi \right\rangle =\tsum\limits_{\alpha
\beta ..}C_{\alpha \beta }\,\left\vert \Phi _{A}^{\alpha }\right\rangle
\otimes \left\vert \Phi _{B}^{\beta }\right\rangle $\ for a pure state of a
bipartite quantum system. The quantities $\left\vert \widehat{x}%
\right\rangle ,\left\vert \widehat{y}\right\rangle ,\left\vert \widehat{z}%
\right\rangle $\ would then specify particular polarization states and the
quantities\textbf{\ }$\,\left\vert E_{x}(\mathbf{r},t)\right\rangle
,\,\left\vert E_{y}(\mathbf{r},t)\right\rangle ,\,\left\vert E_{z}(\mathbf{r}%
,t)\right\rangle $ would then specify the states of their\textbf{\ }\emph{%
related}\textbf{\ }components. One could even introduce\textbf{\ }\emph{%
scalar products}\textbf{\ }in these two state spaces, thus\textbf{\ }%
\begin{equation}
\left\langle \widehat{a}\,|\widehat{b}\right\rangle \equiv \widehat{a}^{\ast
}\,\cdot \widehat{b}\qquad \left\langle E_{a}(\mathbf{r},t)\,|E_{b}(\mathbf{r%
},t)\right\rangle \equiv \tint d\mathbf{r\,}E_{a}(\mathbf{r},t)^{\ast }E_{b}(%
\mathbf{r},t)  \label{Eq.ScalarProd}
\end{equation}%
which satisfy the standard rules for scalar products. In this formalism the
classical state\textbf{\ }$\left\vert \underrightarrow{E}(\mathbf{r}%
,t)\right\rangle $ is a state of a composite system in which polarization
and components are thought of as being the two\textbf{\ }\emph{sub-systems}%
\textbf{. \ }The state\textbf{\ }$\left\vert \underrightarrow{E}(\mathbf{r}%
,t)\right\rangle $ is then defined as a \emph{classically entangled} state
for these two sub-systems just as $\tsum\limits_{\alpha \beta ..}C_{\alpha
\beta }\,\left\vert \Phi _{A}^{\alpha }\right\rangle \otimes \left\vert \Phi
_{B}^{\beta }\right\rangle $ is defined as a quantum entangled state of the
two sub-systems $A$ and $B$. A state of the form\textbf{\ }$\left\vert 
\underrightarrow{E}(\mathbf{r},t)\right\rangle =\left\vert \widehat{u}%
\right\rangle \,\left\vert E_{u}(\mathbf{r},t)\right\rangle $ would then be
a non-entangled or\textbf{\ }\emph{separable}\textbf{\ }state, and this
corresponds to a classical EM field state\textbf{\ }$\underrightarrow{E}(%
\mathbf{r},t)=\widehat{u}\,E_{u}(\mathbf{r},t)$\textbf{\ }with a definite
polarization. Polarization and components are like two sub-systems in that
they are clearly different to each other.

\subsection{Circular Polarization States}

Of course the states of these two distinct sub-systems are such that they
cannot be changed independently of each other when describing the same
classical state - as is also the case for quantum sub-systems. If the
polarizations states are changed as in 
\begin{equation}
\left\vert \widehat{u}\right\rangle =\frac{1}{\sqrt{2}}(\left\vert \widehat{x%
}\right\rangle +i\left\vert \widehat{y}\right\rangle )\quad \left\vert 
\widehat{v}\right\rangle =\frac{1}{\sqrt{2}}(\left\vert \widehat{x}%
\right\rangle -i\left\vert \widehat{y}\right\rangle )\quad \left\vert 
\widehat{w}\right\rangle =\left\vert \widehat{z}\right\rangle
\label{Eq.PolnChange}
\end{equation}%
corresponding to changing from linear to circular polarization, then the
components are changed as\textbf{\ }%
\begin{equation}
\left\vert E_{u}(\mathbf{r},t)\right\rangle =\frac{1}{\sqrt{2}}(\left\vert
E_{x}(\mathbf{r},t)\right\rangle -i\left\vert E_{y}(\mathbf{r}%
,t)\right\rangle )\quad \left\vert E_{v}(\mathbf{r},t)\widehat{v}%
\right\rangle =\frac{1}{\sqrt{2}}(\left\vert E_{x}(\mathbf{r}%
,t)\right\rangle +i\left\vert E_{y}(\mathbf{r},t)\right\rangle )\quad
\left\vert E_{w}(\mathbf{r},t)\right\rangle =\left\vert E_{z}(\mathbf{r}%
,t)\right\rangle  \label{Eq.ComptChange}
\end{equation}%
\textbf{\ }Independent changes to the polarization and component sub-systems
would describe differing classical states.

\subsection{Quantum Treatment}

The classical ket vector formalism may be contrasted to the quantum
treatment of the EM\ field. \ Expressions such as (\ref{Eq.ClassicalEField})
apply to the electric field operator%
\begin{equation}
\widehat{\underrightarrow{E}}(\mathbf{r},t)=\widehat{x}\,\widehat{E}_{x}(%
\mathbf{r},t)+\widehat{y}\,\widehat{E}_{y}(\mathbf{r},t)+\widehat{z}\,%
\widehat{E}_{z}(\mathbf{r},t)
\end{equation}%
where the components now become field operators. A pure quantum state for
the EM field is represented by a ket vector $\left\vert \Phi \right\rangle $
and the equivalent classical state for the field is specified by the mean
value of the vector field operator\textbf{\ }%
\begin{equation}
\left\langle \widehat{\underrightarrow{E}}(\mathbf{r},t)\right\rangle =%
\widehat{x}\,\left\langle \widehat{E}_{x}(\mathbf{r},t)\right\rangle +%
\widehat{y}\,\left\langle \widehat{E}_{y}(\mathbf{r},t)\right\rangle +%
\widehat{z}\,\left\langle \widehat{E}_{z}(\mathbf{r},t)\right\rangle
\end{equation}%
This expression is the true equivalent of (\ref{Eq.ClassicalEField}), here
components are c-number mean values and polarization vectors remain ordinary
vectors - not turned into ket vectors \ We note that the quantum expression
for a general pure state of the EM\ field would be an entangled state of the
form $\tsum\limits_{\{n_{a},n_{b,..\}}}C(n_{a},n_{b},...)\,\left\vert
n_{a}\right\rangle \left\vert n_{b}\right\rangle ...$, where the $\left\vert
n_{k}\right\rangle $ are Fock states with $n_{k}$ photons in mode $k$. In
the quantum approach the sub-systems are\textbf{\ }\emph{modes}, each
specified by a\textbf{\ }\emph{wave-vector}\textbf{\ }and a\textbf{\ }\emph{%
polarization}. \ In the classical case the\textbf{\ }$E_{a}(\mathbf{r},t)$
may be expanded in terms of plane-wave modes with differing wave-vectors, so
both wave-vectors and polarizations are involved there also. However, in the
quantum approach the sub-systems are entities that are specified jointly by
wave-vectors and polarizations, unlike in the classical ket vector formalism
where they are treated separately.

\subsection{Several Light Beams}

The classical ket vector formalism has been extended to treat EM fields
involving several\textbf{\ }\emph{spatially separate}\textbf{\ }light beams.
In the case of two separate light beams with classical field states\textbf{\ 
}$\underrightarrow{E}(\mathbf{r},t)_{1}$ and\textbf{\ }$\underrightarrow{E}(%
\mathbf{r},t)_{2}$ one would normally write the total classical field as%
\textbf{\ }%
\begin{equation}
\underrightarrow{E}(\mathbf{r},t)=\underrightarrow{E}(\mathbf{r},t)_{1}+%
\underrightarrow{E}(\mathbf{r},t)_{2}  \label{Eq.TwoBeamField}
\end{equation}%
However in the classical ket vector formalism the classical state is
represented by a tensor product \cite{Spreeuw98a}%
\begin{equation}
\left\vert \underrightarrow{E}(\mathbf{r},t)\right\rangle =\left\vert 
\underrightarrow{E}(\mathbf{r},t)\right\rangle _{1}\otimes \left\vert 
\underrightarrow{E}(\mathbf{r},t)\right\rangle _{2}
\label{Eq.TwoBeamKetState}
\end{equation}%
and if there were more than one state for each of the separate light beams%
\textbf{\ }$\underrightarrow{E}(\mathbf{r},t)_{1a}$, $\underrightarrow{E}(%
\mathbf{r},t)_{1b}$ etc, then entangled states of the form\textbf{\ }%
\begin{equation}
\left\vert \underrightarrow{E}(\mathbf{r},t)\right\rangle =A\left\vert 
\underrightarrow{E}(\mathbf{r},t)\right\rangle _{1a}\otimes \left\vert 
\underrightarrow{E}(\mathbf{r},t)\right\rangle _{2a}+B\left\vert 
\underrightarrow{E}(\mathbf{r},t)\right\rangle _{1b}\otimes \left\vert 
\underrightarrow{E}(\mathbf{r},t)\right\rangle _{2b}
\label{Eq.TwoBeamClassicalEntState}
\end{equation}%
can exist, where $A$, $B$ can be any c-numbers. \ It is not obvious what the
overall classical state associated with this vector would be. \ Is it $A(%
\underrightarrow{E}(\mathbf{r},t)_{1a}+\underrightarrow{E}(\mathbf{r}%
,t)_{2a})+B(\underrightarrow{E}(\mathbf{r},t)_{1b}+\underrightarrow{E}(%
\mathbf{r},t)_{2b})$ or something else ?

Again we may contrast this with the quantum treatment. Separate beams would
be associated with differing field modes - presumably localised in different
spatial regions, so the electric field operator may be written in full
tensor notation as the sum of contributions from the two separate sets of
modes\textbf{\ }%
\begin{equation}
\underrightarrow{\widehat{E}}(\mathbf{r},t)=\underrightarrow{\widehat{E}}(%
\mathbf{r},t)_{1}\otimes \widehat{1}_{2}+\widehat{1}_{1}\otimes 
\underrightarrow{\widehat{E}}(\mathbf{r},t)_{2}  \label{Eq.TwoBeamEFieldOpr}
\end{equation}%
Separate states for the two beams would be of the form $\left\vert \Phi
\right\rangle =\left\vert \Phi \right\rangle _{1}\otimes \left\vert \Phi
\right\rangle _{2}$ and pure entangled states would be expressed as 
\begin{equation}
\left\vert \Phi \right\rangle =\alpha \left\vert \Phi \right\rangle
_{1a}\otimes \left\vert \Phi \right\rangle _{2a}+\beta \left\vert \Phi
\right\rangle _{1b}\otimes \left\vert \Phi \right\rangle _{2b}
\label{Eq.TwoBeamQuantEntState}
\end{equation}%
where we assume the states for each beam are normalised to unity. A
straight-forward calculation shows the the mean electric field is\textbf{\ }%
\begin{eqnarray}
\left\langle \widehat{\underrightarrow{E}}(\mathbf{r},t)\right\rangle
&=&|\alpha |^{2}\left\langle \Phi \right\vert _{1a}\underrightarrow{\widehat{%
E}}(\mathbf{r},t)_{1}\left\vert \Phi \right\rangle _{1a}+|\beta
|^{2}\left\langle \Phi \right\vert _{1b}\underrightarrow{\widehat{E}}(%
\mathbf{r},t)_{1}\left\vert \Phi \right\rangle _{1b}  \nonumber \\
&&+|\alpha |^{2}\left\langle \Phi \right\vert _{2a}\underrightarrow{\widehat{%
E}}(\mathbf{r},t)_{2}\left\vert \Phi \right\rangle _{2a}+|\beta
|^{2}\left\langle \Phi \right\vert _{2b}\underrightarrow{\widehat{E}}(%
\mathbf{r},t)_{2}\left\vert \Phi \right\rangle _{2b}  \nonumber \\
&&+\alpha ^{\ast }\beta \left\langle \Phi \right\vert _{1a}\underrightarrow{%
\widehat{E}}(\mathbf{r},t)_{1}\left\vert \Phi \right\rangle
_{1b}\left\langle \Phi \right\vert _{2a}\left\vert \Phi \right\rangle
_{2b}+\alpha ^{\ast }\beta \left\langle \Phi \right\vert _{1a}\left\vert
\Phi \right\rangle _{1b}\left\langle \Phi \right\vert _{2a}\underrightarrow{%
\widehat{E}}(\mathbf{r},t)_{2}\left\vert \Phi \right\rangle _{2b}  \nonumber
\\
&&+\beta ^{\ast }\alpha \left\langle \Phi \right\vert _{1b}\underrightarrow{%
\widehat{E}}(\mathbf{r},t)_{1}\left\vert \Phi \right\rangle
_{1a}\left\langle \Phi \right\vert _{2b}\left\vert \Phi \right\rangle
_{2a}+\beta ^{\ast }\alpha \left\langle \Phi \right\vert _{1b}\left\vert
\Phi \right\rangle _{1a}\left\langle \Phi \right\vert _{2b}\underrightarrow{%
\widehat{E}}(\mathbf{r},t)_{2}\left\vert \Phi \right\rangle _{2a}  \nonumber
\\
&&  \label{Eq.EquivClassFieldQuantEntState}
\end{eqnarray}%
There is no obvious similarity with possible forms for the electric field
classical state associated with (\ref{Eq.TwoBeamClassicalEntState}). Even if
we made the different states associated with the beams orthogonal $%
\left\langle \Phi \right\vert _{2a}\left\vert \Phi \right\rangle
_{2b}=\left\langle \Phi \right\vert _{1a}\left\vert \Phi \right\rangle
_{1b}=0$ to eliminate terms that are associated with both beams together,
contributions such as\textbf{\ }$(\left\langle \Phi \right\vert _{1a}%
\underrightarrow{\widehat{E}}(\mathbf{r},t)_{1}\left\vert \Phi \right\rangle
_{1a}+\left\langle \Phi \right\vert _{2a}\underrightarrow{\widehat{E}}(%
\mathbf{r},t)_{2}\left\vert \Phi \right\rangle _{2a})$ that would be
associated with the classical ket tensor product state\textbf{\ }$\left\vert 
\underrightarrow{E}(\mathbf{r},t)\right\rangle _{1a}\otimes \left\vert 
\underrightarrow{E}(\mathbf{r},t)\right\rangle _{2a}$ are weighted by $%
|\alpha |^{2}$ -- which is always real and positive, whereas the coefficient 
$A$ in the classical entangled state could be any c-number. There just does
seem to be a meaningful way to reproduce the classical field associated with
an actual quantum entangled state for two separate beams via the classical
tensor product formalism.

\subsection{Classically Entangled States - Joint Measurements on Sub-Systems}

In the case of the classically entangled state for the electric field (\ref%
{Eq.ClassEFieldVector})\ defined in SubSection \ref{SubSubSection -
Classical Entanglement} classical measurements can be performed to give%
\textbf{\ }\emph{deterministic}\textbf{\ }results for every one of the
polarization and component states. Thus for (\ref{Eq.ClassEFieldVector}) the
probability for measuring polarizations $\widehat{u},\widehat{v},\widehat{w}$
and components $E_{u},E_{v},E_{w}$ is%
\begin{eqnarray}
P(\widehat{u},\widehat{v},\widehat{w},E_{u},E_{v},E_{w}) &=&1\quad If(%
\widehat{u}=\widehat{x},\widehat{v}=\widehat{y},\widehat{w}=\widehat{z}%
,E_{u}=E_{x},E_{v}=E_{y},E_{w}=E_{z})  \nonumber \\
&=&0\quad Otherwise  \label{Eq.JointProbClassEntState}
\end{eqnarray}%
Introducing the sub-system probabilities\textbf{\ }%
\begin{eqnarray}
P_{Pol}(\widehat{u},\widehat{v},\widehat{w}) &=&1\quad If(\widehat{u}=%
\widehat{x},\widehat{v}=\widehat{y},\widehat{w}=\widehat{z})  \nonumber \\
&=&0\quad Otherwise  \label{Eq.PolnProb}
\end{eqnarray}%
and\textbf{\ }%
\begin{eqnarray}
P_{Compt}(E_{u},E_{v},E_{w}) &=&1\quad
If(E_{u}=E_{x},E_{v}=E_{y},E_{w}=E_{z})  \nonumber \\
&=&0\quad Otherwise  \label{Eq.ComptProb}
\end{eqnarray}%
we see that\textbf{\ }%
\begin{equation}
P(\widehat{u},\widehat{v},\widehat{w},E_{u},E_{v},E_{w})=P_{Pol}(\widehat{u},%
\widehat{v},\widehat{w})\times P_{Compt}(E_{u},E_{v},E_{w})
\label{Eq.ClassicalJointProb}
\end{equation}%
This is the characteristic result for a\textbf{\ }\emph{non-entangled}%
\textbf{\ }state,\textbf{\ }\emph{not}\textbf{\ }an entangled state. Hence
the classical entangled states - of which (\ref{Eq.ClassEFieldVector}) is a
typical example, fail to exhibit the characteristic key feature of quantum
entanglement. Classical and quantum entanglement are basically different in
spite of their mathematical similarities.

\subsection{Classical Entanglement and Bell Inequality}

\textbf{TO BE\ WRITTEN}

Thus, in spite of some similarities, there are key features that is not
analogous to that for quantum sub-systems. In the end, classical and quantum
entanglement are fundamentally different when the physics of the two
different types of system - one classical and deterministic, the other
quantum and probabilistic are taken into account rather than just focusing
on similarities in the mathematical formalisms. The key feature of quantum
entanglement is different to the corresponding one for classical
entanglement.\pagebreak

\section{Appendix 8 - Derivation of Sorensen et al Results}

\label{Appendix - Sorensen Results}

Sorensen et al \cite{Sorensen01a} derive a number of ineqalities from which
they deduce a further inequality for the spin squeezing parameter in the
case of a non-entangled state. From this result they conclude that spin
squeezing implies entanglement. The final inequality they obtain for a
non-entangled state is 
\begin{equation}
\left\langle \Delta \widehat{S}\,_{z}^{2}\right\rangle \geq \frac{1}{N}%
\left( \left\langle \widehat{S}\,_{x}\right\rangle ^{2}+\left\langle 
\widehat{S}\,_{y}\right\rangle ^{2}\right)
\end{equation}

Their approach is based on writing the density operator for a non-entangled
state of $N$ identical particles as in Eq. (\ref%
{Eq.NonEntStateIdenticalAtoms})

\begin{equation}
\widehat{\rho }=\sum_{R}P_{R}\,\widehat{\rho }_{R}^{1}\otimes \widehat{\rho }%
_{R}^{2}\otimes \widehat{\rho }_{R}^{3}\otimes ...=\sum_{R}P_{R}\,\widehat{%
\rho }_{R}
\end{equation}%
The spin operators are defined as $\widehat{S}_{x}=\tsum\limits_{i}\widehat{S%
}_{x}^{i}=\sum_{i}(\left\vert \phi _{b}(i)\right\rangle \left\langle \phi
_{a}(i)\right\vert +\left\vert \phi _{a}(i)\right\rangle \left\langle \phi
_{b}(i)\right\vert )/2$ ; $\widehat{S}_{y}=\tsum\limits_{i}\widehat{S}%
_{y}^{i}=\sum_{i}(\left\vert \phi _{b}(i)\right\rangle \left\langle \phi
_{a}(i)\right\vert -\left\vert \phi _{a}(i)\right\rangle \left\langle \phi
_{b}(i)\right\vert )/2i$ ; $\widehat{S}_{z}=\tsum\limits_{i}\widehat{S}%
_{z}^{i}=\sum_{i}(\left\vert \phi _{b}(i)\right\rangle \left\langle \phi
_{b}(i)\right\vert -\left\vert \phi _{a}(i)\right\rangle \left\langle \phi
_{a}(i)\right\vert )/2$ , where the sum $i$ is over the identical atoms and
each atom is associated with two states $\left\vert \phi _{a}\right\rangle $
and $\left\vert \phi _{b}\right\rangle $. Clearly, the spin operators
satisfy the standard commutation rules for agular momentum operators.

Sorensen et al \ \cite{Sorensen01a} state that the variance for $\widehat{S}%
_{z}$ satisfies the result%
\begin{equation}
\left\langle \Delta \widehat{S}\,_{z}^{2}\right\rangle =\frac{N}{4}%
-\tsum\limits_{R}P_{R}\tsum\limits_{i}\left\langle \widehat{S}%
_{z}^{i}\right\rangle _{R}^{2}+\tsum\limits_{R}P_{R}\left\langle \widehat{S}%
_{z}\right\rangle _{R}^{2}-\left\langle \widehat{S}_{z}\right\rangle ^{2}
\label{Eq.VarianceSZ}
\end{equation}%
To prove this we have 
\begin{eqnarray}
\left\langle \widehat{S}\,_{z}^{2}\right\rangle &=&\sum_{R}P_{R}\,Tr(%
\widehat{\rho }_{R}\tsum\limits_{i}\tsum\limits_{j}\widehat{S}_{z}^{i}%
\widehat{S}_{z}^{j})  \nonumber \\
&=&\sum_{R}P_{R}\,\left( \tsum\limits_{i}\left\langle \left( \widehat{S}%
_{z}^{i}\right) ^{2}\right\rangle _{R}+\tsum\limits_{i\neq j}\left\langle 
\widehat{S}_{z}^{i}\right\rangle _{R}\left\langle \widehat{S}%
_{z}^{j}\right\rangle _{R}\right)  \nonumber \\
&=&\frac{N}{4}+\sum_{R}P_{R}\,\left( \tsum\limits_{i\neq j}\left\langle 
\widehat{S}_{z}^{i}\right\rangle _{R}\left\langle \widehat{S}%
_{z}^{j}\right\rangle _{R}\right)
\end{eqnarray}%
where we have used 
\begin{eqnarray}
\left( \widehat{S}_{z}^{i}\right) ^{2} &=&\frac{1}{4}(\left\vert \phi
_{b}(i)\right\rangle \left\langle \phi _{b}(i)\right\vert -\left\vert \phi
_{a}(i)\right\rangle \left\langle \phi _{a}(i)\right\vert )^{2}  \nonumber \\
&=&\frac{1}{4}(\left\vert \phi _{b}(i)\right\rangle \left\langle \phi
_{b}(i)|\phi _{b}(i)\right\rangle \left\langle \phi _{b}(i)\right\vert
-(\left\vert \phi _{b}(i)\right\rangle \left\langle \phi _{b}(i)|\phi
_{a}(i)\right\rangle \left\langle \phi _{a}(i)\right\vert )  \nonumber \\
&&+\frac{1}{4}(-(\left\vert \phi _{a}(i)\right\rangle \left\langle \phi
_{a}(i)|\phi _{b}(i)\right\rangle \left\langle \phi _{b}(i)\right\vert
+(\left\vert \phi _{a}(i)\right\rangle \left\langle \phi _{a}(i)|\phi
_{a}(i)\right\rangle \left\langle \phi _{a}(i)\right\vert )  \nonumber \\
&=&\frac{1}{4}((\left\vert \phi _{b}(i)\right\rangle \left\langle \phi
_{b}(i)\right\vert +(\left\vert \phi _{a}(i)\right\rangle \left\langle \phi
_{a}(i)\right\vert )  \nonumber \\
&=&\frac{1}{4}\widehat{1}_{i}
\end{eqnarray}%
a result based on the orthogonality, normalisation and completeness of the
states $\left\vert \phi _{a}(i)\right\rangle ,\left\vert \phi
_{b}(i)\right\rangle $. Also 
\begin{eqnarray}
\left\langle \widehat{S}\,_{z}\right\rangle _{R} &=&Tr(\widehat{\rho }%
_{R}\tsum\limits_{i}\widehat{S}_{z}^{i})  \nonumber \\
&=&\tsum\limits_{i}\left\langle \widehat{S}_{z}^{i}\right\rangle _{R} 
\nonumber \\
\tsum\limits_{R}P_{R}\left\langle \widehat{S}_{z}\right\rangle _{R}^{2}
&=&\tsum\limits_{R}P_{R}\left( \tsum\limits_{i}\left\langle \widehat{S}%
_{z}^{i}\right\rangle _{R}^{2}+\tsum\limits_{i\neq j}\left\langle \widehat{S}%
_{z}^{i}\right\rangle _{R}\left\langle \widehat{S}_{z}^{j}\right\rangle
_{R}\right)
\end{eqnarray}%
so eliminating the term $\sum_{R}P_{R}\,\left( \tsum\limits_{i\neq
j}\left\langle \widehat{S}_{z}^{i}\right\rangle _{R}\left\langle \widehat{S}%
_{z}^{j}\right\rangle _{R}\right) $ gives the required expression for $%
\left\langle \Delta \widehat{S}\,_{z}^{2}\right\rangle =\left\langle 
\widehat{S}\,_{z}^{2}\right\rangle -\left\langle \widehat{S}%
_{z}\right\rangle ^{2}$.

Next, Sorensen et al \cite{Sorensen01a} state that 
\begin{equation}
\left\langle \widehat{S}_{x}\right\rangle ^{2}\leq
N\tsum\limits_{R}P_{R}\tsum\limits_{i}\left\langle \widehat{S}%
_{x}^{i}\right\rangle _{R}^{2}\qquad \left\langle \widehat{S}%
\,_{y}\right\rangle ^{2}\leq N\sum_{R}P_{R}\,\tsum\limits_{i}|\left\langle 
\widehat{S}_{y}^{i}\right\rangle _{R}|^{2}  \label{Eq.InequalSX2}
\end{equation}%
To prove this we have 
\begin{eqnarray}
\left\langle \widehat{S}\,_{x}\right\rangle &=&\sum_{R}P_{R}\,Tr(\widehat{%
\rho }_{R}\tsum\limits_{i}\widehat{S}_{x}^{i})  \nonumber \\
&=&\sum_{R}P_{R}\,\tsum\limits_{i}\left\langle \widehat{S}%
_{x}^{i}\right\rangle _{R}  \nonumber \\
|\left\langle \widehat{S}\,_{x}\right\rangle | &\leq
&\sum_{R}P_{R}\,\tsum\limits_{i}|\left\langle \widehat{S}_{x}^{i}\right%
\rangle _{R}|
\end{eqnarray}%
since the modulus of a sum is less than or equal to the sum of the moduli.
Now

\begin{eqnarray}
\left\langle \widehat{S}\,_{x}\right\rangle ^{2} &=&|\left\langle \widehat{S}%
\,_{x}\right\rangle |^{2}\leq \left(
\sum_{R}P_{R}\,\tsum\limits_{i}|\left\langle \widehat{S}_{x}^{i}\right%
\rangle _{R}|\right) ^{2}  \nonumber \\
&\leq &\sum_{R}P_{R}\,\left( \tsum\limits_{i}|\left\langle \widehat{S}%
_{x}^{i}\right\rangle _{R}|\right) ^{2}
\end{eqnarray}%
using the general result that $\left( \tsum\limits_{R}P_{R}\,\sqrt{C_{R}}%
\right) ^{2}\leq \tsum\limits_{R}P_{R}\,C_{R}$, where $\tsum%
\limits_{R}P_{R}=1$ with here $\sqrt{C_{R}}=\tsum\limits_{i}|\left\langle 
\widehat{S}_{x}^{i}\right\rangle _{R}|$. Next consider 
\begin{eqnarray}
y &=&N\tsum\limits_{i}|\left\langle \widehat{S}_{x}^{i}\right\rangle
_{R}|^{2}  \nonumber \\
z &=&\left( \tsum\limits_{i}|\left\langle \widehat{S}_{x}^{i}\right\rangle
_{R}|\right) ^{2}=\left( \tsum\limits_{i}|\left\langle \widehat{S}%
_{x}^{i}\right\rangle _{R}|\right) ^{2}  \nonumber \\
y-z &=&\tsum\limits_{i<j}(|\left\langle \widehat{S}_{x}^{i}\right\rangle
_{R}|-|\left\langle \widehat{S}_{x}^{j}\right\rangle _{R}|)^{2}\geq 0
\end{eqnarray}%
so that 
\begin{equation}
\left\langle \widehat{S}\,_{x}\right\rangle ^{2}\leq
N\sum_{R}P_{R}\,\tsum\limits_{i}|\left\langle \widehat{S}_{x}^{i}\right%
\rangle _{R}|^{2}\qquad \left\langle \widehat{S}\,_{y}\right\rangle ^{2}\leq
N\sum_{R}P_{R}\,\tsum\limits_{i}|\left\langle \widehat{S}_{y}^{i}\right%
\rangle _{R}|^{2}
\end{equation}%
which is the required result. The inequality for $\left\langle \widehat{S}%
\,_{y}\right\rangle ^{2}$ is proved similarly.

Another inequality is stated \cite{Sorensen01a} for $\left\langle \widehat{S}%
\,_{z}\right\rangle ^{2}$. This is 
\begin{equation}
\left\langle \widehat{S}\,_{z}\right\rangle ^{2}\leq
\sum_{R}P_{R}\,\left\langle \widehat{S}_{z}\right\rangle _{R}^{2}
\label{Eq.IneqalSZ2}
\end{equation}%
To show this we have 
\begin{eqnarray}
\left\langle \widehat{S}\,_{z}\right\rangle &=&\sum_{R}P_{R}\,Tr(\widehat{%
\rho }_{R}\tsum\limits_{i}\widehat{S}_{z}^{i})  \nonumber \\
&=&\sum_{R}P_{R}\,\tsum\limits_{i}\left\langle \widehat{S}%
_{z}^{i}\right\rangle _{R}  \nonumber \\
&=&\sum_{R}P_{R}\,\left\langle \widehat{S}_{z}\right\rangle _{R}  \nonumber
\\
|\left\langle \widehat{S}\,_{z}\right\rangle | &\leq
&\sum_{R}P_{R}\,|\left\langle \widehat{S}_{z}\right\rangle _{R}|
\end{eqnarray}%
so that 
\begin{eqnarray}
\left\langle \widehat{S}\,_{z}\right\rangle ^{2} &=&|\left\langle \widehat{S}%
\,_{z}\right\rangle |^{2}\leq \left( \sum_{R}P_{R}\,|\left\langle \widehat{S}%
_{z}\right\rangle _{R}|\right) ^{2}  \nonumber \\
&\leq &\sum_{R}P_{R}\,|\left\langle \widehat{S}_{z}\right\rangle _{R}|^{2} 
\nonumber \\
&=&\sum_{R}P_{R}\,\left\langle \widehat{S}_{z}\right\rangle _{R}^{2}
\end{eqnarray}%
using the general result that $\left( \tsum\limits_{R}P_{R}\,\sqrt{C_{R}}%
\right) ^{2}\leq \tsum\limits_{R}P_{R}\,C_{R}$, where $\tsum%
\limits_{R}P_{R}=1$ with here $\sqrt{C_{R}}=|\left\langle \widehat{S}%
_{z}\right\rangle _{R}|$.

Finally, we find that 
\begin{eqnarray}
\sum_{R}P_{R}\,\tsum\limits_{i}\left( \left\langle \widehat{S}%
_{x}^{i}\right\rangle _{R}^{2}+\left\langle \widehat{S}_{y}^{i}\right\rangle
_{R}^{2}+\left\langle \widehat{S}_{z}^{i}\right\rangle _{R}^{2}\right) &\leq
&\frac{1}{4}N  \nonumber \\
-\sum_{R}P_{R}\,\tsum\limits_{i}\left( \left\langle \widehat{S}%
_{z}^{i}\right\rangle _{R}^{2}\right) &\geq &-\frac{1}{4}N+\sum_{R}P_{R}\,%
\tsum\limits_{i}\left( \left\langle \widehat{S}_{x}^{i}\right\rangle
_{R}^{2}+\left\langle \widehat{S}_{y}^{i}\right\rangle _{R}^{2}\right) 
\nonumber \\
&&  \label{Eq.InequalBloch}
\end{eqnarray}%
To show this we use the properties of the density operator $\widehat{\rho }%
_{R}^{i}$ for the $i$th particle of Hermitiancy, positiveness, unit trace $%
Tr(\widehat{\rho }_{R}^{i})=1$ and $Tr(\widehat{\rho }_{R}^{i})^{2}\leq 1$.
In terms of matrix elements of the density operator $\widehat{\rho }_{R}^{i}$
between the two states $\left\vert \phi _{a}(i)\right\rangle $, $\left\vert
\phi _{b}(i)\right\rangle $ the quantities $\left\langle \widehat{S}%
_{x}^{i}\right\rangle _{R}$, $\left\langle \widehat{S}_{y}^{i}\right\rangle
_{R}$ and $\left\langle \widehat{S}_{z}^{i}\right\rangle _{R}$ are%
\begin{eqnarray}
\left\langle \widehat{S}_{x}^{i}\right\rangle _{R} &=&Tr(\widehat{\rho }%
_{R}^{i}\,\frac{1}{2}(\left\vert \phi _{b}(i)\right\rangle \left\langle \phi
_{a}(i)\right\vert +\left\vert \phi _{a}(i)\right\rangle \left\langle \phi
_{b}(i)\right\vert ))  \nonumber \\
&=&\frac{1}{2}\left( \rho _{ab}^{i}+\rho _{ba}^{i}\right)  \nonumber \\
\left\langle \widehat{S}_{y}^{i}\right\rangle _{R} &=&\frac{1}{2i}\left(
\rho _{ab}^{i}-\rho _{ba}^{i}\right)  \nonumber \\
\left\langle \widehat{S}_{z}^{i}\right\rangle _{R} &=&\frac{1}{2}\left( \rho
_{bb}^{i}-\rho _{aa}^{i}\right)  \label{Eq.SpinExpnValuesMatrixElements}
\end{eqnarray}%
where $\rho _{cd}^{i}=\left\langle \phi _{c}(i)\right\vert \widehat{\rho }%
_{R}^{i}$ $\left\vert \phi _{d}(i)\right\rangle $. The Hermitiancy and
positiveness of $\widehat{\rho }_{R}^{i}$ show that $\rho _{bb}^{i}$ and $%
\rho _{aa}^{i}$ are real and positive, $\rho _{ab}^{i}=(\rho
_{ba}^{i})^{\ast }$ and $\rho _{aa}^{i}\rho _{bb}^{i}-|\rho
_{ab}^{i}|^{2}\geq 0$. The condition $Tr(\widehat{\rho }_{R}^{i})=1$ leads
to $\rho _{aa}^{i}+\rho _{bb}^{i}=1$, from which $Tr(\widehat{\rho }%
_{R}^{i})^{2}\leq 1$ follows using the previous positivity results. Taken
together these conditions lead to the following useful parametrisation of
the density matrix elements 
\begin{eqnarray}
\rho _{aa}^{i} &=&\sin ^{2}\alpha _{i}\qquad \rho _{bb}^{i}=\cos ^{2}\alpha
_{i}  \nonumber \\
\rho _{ab}^{i} &=&\sqrt{\sin ^{2}\alpha _{i}\,\cos ^{2}\alpha _{i}}\,\sin
^{2}\beta _{i}\,\exp (+i\phi _{i})\qquad \rho _{ba}^{i}=\sqrt{\sin
^{2}\alpha _{i}\,\cos ^{2}\alpha _{i}}\,\sin ^{2}\beta _{i}\,\exp (-i\phi
_{i})  \nonumber \\
&&  \label{Eq.ParamMatrixElements}
\end{eqnarray}%
where $\alpha _{i}$, $\beta _{i}$ and $\phi _{i}$ are real. In terms of
these quantities we then have 
\begin{eqnarray}
\left\langle \widehat{S}_{x}^{i}\right\rangle _{R} &=&\frac{1}{2}\sin
2\alpha _{i}\,\sin ^{2}\beta _{i}\,\cos \phi _{i}  \nonumber \\
\left\langle \widehat{S}_{y}^{i}\right\rangle _{R} &=&\frac{1}{2}\sin
2\alpha _{i}\,\sin ^{2}\beta _{i}\,\sin \phi _{i}  \nonumber \\
\left\langle \widehat{S}_{z}^{i}\right\rangle _{R} &=&\frac{1}{2}\cos
2\alpha _{i}\,  \label{Eq.SpinExpnValuesParam}
\end{eqnarray}%
It is then easy to show that 
\begin{eqnarray}
\left\langle \widehat{S}_{x}^{i}\right\rangle _{R}^{2}+\left\langle \widehat{%
S}_{y}^{i}\right\rangle _{R}^{2}+\left\langle \widehat{S}_{z}^{i}\right%
\rangle _{R}^{2} &=&\frac{1}{4}-\frac{1}{4}\sin ^{2}2\alpha _{i}\,(1-\sin
^{4}\beta _{i}\,)  \nonumber \\
&\leq &\frac{1}{4}  \label{Eq.InequalSquaresSpinExpnVals}
\end{eqnarray}%
and the final inequality (\ref{Eq.InequalBloch}) then follows by taking the
sum over particles $i$ and then using $\sum_{R}P_{R}=1$. If only the Schwarz
inequality is used instead of the more detailed consequences of Hermtiancy,
positiveness etc it can be shown that $\left\langle \widehat{S}%
_{x}^{i}\right\rangle _{R}^{2}+\left\langle \widehat{S}_{y}^{i}\right\rangle
_{R}^{2}+\left\langle \widehat{S}_{z}^{i}\right\rangle _{R}^{2}$ $\leq \frac{%
3}{4}$, which though correct is not useful.

Combining the inequalities in Eqs. (\ref{Eq.InequalSX2}), (\ref{Eq.IneqalSZ2}%
) and (\ref{Eq.InequalBloch}) into Eq. (\ref{Eq.VarianceSZ}) shows that 
\begin{eqnarray}
\left\langle \Delta \widehat{S}\,_{z}^{2}\right\rangle &=&\frac{N}{4}%
-\tsum\limits_{R}P_{R}\tsum\limits_{i}\left\langle \widehat{S}%
_{z}^{i}\right\rangle _{R}^{2}+\tsum\limits_{R}P_{R}\left\langle \widehat{S}%
_{z}\right\rangle _{R}^{2}-\left\langle \widehat{S}_{z}\right\rangle ^{2} 
\nonumber \\
&\geq &\frac{N}{4}-\tsum\limits_{R}P_{R}\tsum\limits_{i}\left\langle 
\widehat{S}_{z}^{i}\right\rangle _{R}^{2}  \nonumber \\
&\geq &\frac{N}{4}-\frac{1}{4}N+\sum_{R}P_{R}\,\tsum\limits_{i}\left(
\left\langle \widehat{S}_{x}^{i}\right\rangle _{R}^{2}+\left\langle \widehat{%
S}_{y}^{i}\right\rangle _{R}^{2}\right)  \nonumber \\
&\geq &\frac{1}{N}\left( \left\langle \widehat{S}\,_{x}\right\rangle
^{2}+\left\langle \widehat{S}\,_{y}\right\rangle ^{2}\right)
\label{Eq.FinalIequal}
\end{eqnarray}%
for the case of a non-entangled state. This result is that in Sorensen et al.%
\cite{Sorensen01a}. \pagebreak

\section{Appendix 9 - Revised Sorensen et al}

\label{Appendix - Revised Sorensen}

\subsection{Variance $\left\langle \Delta \widehat{S}\,_{x}^{2}\right\rangle 
$ for Single Mode Sub-Systems}

Here we will see if the modified approach to Sorensen et al can lead to a
useful inequality for $\left\langle \Delta \widehat{S}\,_{x}^{2}\right%
\rangle $ or.$\left\langle \Delta \widehat{S}\,_{y}^{2}\right\rangle $ that
applies when non-entangled states are those when \emph{all} the separate
modes $\widehat{a}_{i}$ and $\widehat{b}_{i}$ are the sub-systems . We will
attempt to follow the approach used for the simple two mode case in Section %
\ref{Section - Relationship Spin Squeezing & Entanglement}.

Firstly, the \emph{variance} for a Hermitian operator $\widehat{\Omega }$ in
a mixed state 
\begin{equation}
\widehat{\rho }=\sum_{R}P_{R}\,\widehat{\rho }_{R}
\end{equation}%
is always greater than or equal to the the average of the variances for the
separate components 
\begin{equation}
\left\langle \Delta \widehat{\Omega }\,^{2}\right\rangle \geq
\sum_{R}P_{R}\,\left\langle \Delta \widehat{\Omega }{}^{2}\right\rangle _{R}
\label{Eq.VarianceResultB}
\end{equation}%
where $\left\langle \Delta \widehat{\Omega }\,^{2}\right\rangle =Tr(\widehat{%
\rho }\,\Delta \widehat{\Omega }\,^{2})$ with $\Delta \widehat{\Omega }=%
\widehat{\Omega }-\left\langle \widehat{\Omega }\right\rangle $ and $%
\left\langle \Delta \widehat{\Omega }\,^{2}\right\rangle _{R}=Tr(\widehat{%
\rho }_{R}\,\Delta \widehat{\Omega }_{R}\,^{2})$ with $\Delta \widehat{%
\Omega }_{R}=\widehat{\Omega }-\left\langle \widehat{\Omega }\right\rangle
_{R}$ . The proof is straight-forward and given in Ref. \cite{Hoffmann03a}.

Next we calculate $\left\langle \Delta \widehat{S}\,_{x}^{2}\right\rangle
_{R}$, $\left\langle \Delta \widehat{S}\,_{y}^{2}\right\rangle _{R}$ and $%
\left\langle \widehat{S}_{x}\right\rangle _{R}$, $\left\langle \widehat{S}%
_{y}\right\rangle _{R}$, $\left\langle \widehat{S}_{z}\right\rangle _{R}$
for the case where 
\begin{equation}
\widehat{\rho }=\sum_{R}P_{R}\,\left( \widehat{\rho }_{R}^{a\,1}\otimes 
\widehat{\rho }_{R}^{b\,1}\right) \otimes \left( \widehat{\rho }%
_{R}^{a\,2}\otimes \widehat{\rho }_{R}^{b\,2}\right) \otimes \left( \widehat{%
\rho }_{R}^{a\,3}\otimes \widehat{\rho }_{R}^{b\,3}\right) \otimes .
\label{Eq.RevisedSorensenDensityOprNonEntB}
\end{equation}%
as is required for a \emph{general non-entangled} state \emph{all} $2N$
modes. Furthermore, the density operators for the individual modes must
represent possible physical states for such modes, so the super-selection
rule for atom number applies and we have 
\begin{eqnarray}
\left\langle (\widehat{a}_{i})^{n}\right\rangle _{a\,i} &=&Tr(\widehat{\rho }%
_{R}^{a\,i}(\widehat{a}_{i})^{n})=0\qquad \left\langle (\widehat{a}%
_{i}^{\dag })^{n}\right\rangle _{a\,i}=Tr(\widehat{\rho }_{R}^{a\,i}(%
\widehat{a}_{i}^{\dag })^{n})=0  \nonumber \\
\left\langle (\widehat{b}_{i})^{m}\right\rangle _{b\,i} &=&Tr(\widehat{\rho }%
_{R}^{b\,i}(\widehat{b}_{i})^{m})=0\qquad \left\langle (\widehat{b}%
_{i}^{\dag })^{m}\right\rangle _{b\,i}=Tr(\widehat{\rho }_{R}^{b\,i}(%
\widehat{b}_{i}^{\dag })^{m})=0  \nonumber \\
&&  \label{Eq.RevisedSorensenAveragesB}
\end{eqnarray}%
The Schwinger spin operators are%
\begin{eqnarray}
\widehat{S}_{x} &=&\sum_{i}(\widehat{b}_{i}^{\dag }\widehat{a}_{i}+\widehat{a%
}_{i}^{\dag }\widehat{b}_{i})/2=\sum_{i}\widehat{S}_{x}^{i}  \nonumber \\
\widehat{S}_{y} &=&\sum_{i}(\widehat{b}_{i}^{\dag }\widehat{a}_{i}-\widehat{a%
}_{i}^{\dag }\widehat{b}_{i})/2i=\sum_{i}\widehat{S}_{y}^{i}  \nonumber \\
\widehat{S}_{z} &=&\sum_{i}(\widehat{b}_{i}^{\dag }\widehat{b}_{i}-\widehat{a%
}_{i}^{\dag }\widehat{a}_{i})/2=\sum_{i}\widehat{S}_{z}^{i}
\label{Eq.NewSpinOprsB}
\end{eqnarray}%
where $\widehat{a}_{i}$, $\widehat{b}_{i}$ and $\widehat{a}_{i}^{\dag }$, $%
\widehat{b}_{i}^{\dag }$ respectively are mode annihilation, creation
operators. From Eqs. (\ref{Eq.NewSpinOprsB}) we find that

\begin{equation}
\widehat{S}\,_{x}^{2}=\tsum\limits_{i}(\widehat{S}_{x}^{i})^{2}+\tsum%
\limits_{i\neq j}\widehat{S}_{x}^{i}\widehat{S}_{x}^{j}
\end{equation}%
so that on taking the trace with $\widehat{\rho }_{R}$ and using Eqs. (\ref%
{Eq.RevisedSorensenDensityOprNonEntB}) we get after applying the commutation
rules $[\widehat{e},\widehat{e}^{\dag }]=\widehat{1}$ ($\widehat{e}=\widehat{%
a}$ or $\widehat{b}$)%
\begin{equation}
\left\langle \widehat{S}\,_{x}^{2}\right\rangle
_{R}=\tsum\limits_{i}\left\langle (\widehat{S}_{x}^{i})^{2}\right\rangle
_{R}+\tsum\limits_{i\neq j}\left\langle \widehat{S}_{x}^{i}\right\rangle
_{R}\left\langle \widehat{S}_{x}^{j}\right\rangle _{R}
\end{equation}

As we also have%
\begin{equation}
\left\langle \widehat{S}\,_{x}\right\rangle
_{R}=\tsum\limits_{i}\left\langle \widehat{S}_{x}^{i}\right\rangle
_{R}\qquad \left\langle \widehat{S}\,_{x}\right\rangle
_{R}^{2}=\tsum\limits_{i}\left\langle \widehat{S}_{x}^{i}\right\rangle
_{R}^{2}+\tsum\limits_{i\neq j}\left\langle \widehat{S}_{x}^{i}\right\rangle
_{R}\left\langle \widehat{S}_{x}^{j}\right\rangle _{R}
\end{equation}%
using Eqs. (\ref{Eq.RevisedSorensenDensityOprNonEntB}) and we see finally
that the variance $\left\langle \Delta \widehat{S}\,_{x}^{2}\right\rangle
_{R}$ is%
\begin{equation}
\left\langle \Delta \widehat{S}\,_{x}^{2}\right\rangle
_{R}=\tsum\limits_{i}\left\langle (\widehat{S}_{x}^{i})^{2}\right\rangle
_{R}-\tsum\limits_{i}\left\langle \widehat{S}_{x}^{i}\right\rangle _{R}^{2}
\end{equation}%
all the terms with $i\neq j$ cancelling out. and therefore from Eq. (\ref%
{Eq.VarianceResultB})%
\begin{equation}
\left\langle \Delta \widehat{S}\,_{x}^{2}\right\rangle \geq
\sum_{R}P_{R}\,\tsum\limits_{i}\left( \left\langle (\widehat{S}%
_{x}^{i})^{2}\right\rangle _{R}-\left\langle \widehat{S}_{x}^{i}\right%
\rangle _{R}^{2}\right)
\end{equation}

But using (\ref{Eq.RevisedSorensenAveragesB}) 
\begin{eqnarray}
(\widehat{S}_{x}^{i})^{2} &=&\frac{1}{4}(\widehat{b}_{i}^{\dag }\widehat{a}%
_{i}\widehat{b}_{i}^{\dag }\widehat{a}_{i}+\widehat{b}_{i}^{\dag }\widehat{a}%
_{i}\widehat{a}_{i}^{\dag }\widehat{b}_{i}+\widehat{a}_{i}^{\dag }\widehat{b}%
_{i}\widehat{b}_{i}^{\dag }\widehat{a}_{i}+\widehat{a}_{i}^{\dag }\widehat{b}%
_{i}\widehat{a}_{i}^{\dag }\widehat{b}_{i})  \nonumber \\
\left\langle (\widehat{S}_{x}^{i})^{2}\right\rangle _{R} &=&\frac{1}{4}%
(\left\langle (\widehat{b}^{\dag }\widehat{b})_{i}\right\rangle
_{R}+\left\langle (\widehat{a}^{\dag }\widehat{a})_{i}\right\rangle _{R})+%
\frac{1}{2}(\left\langle (\widehat{a}^{\dag }\widehat{a})_{i}\right\rangle
_{R}\left\langle (\widehat{b}^{\dag }\widehat{b})_{i}\right\rangle _{R})
\end{eqnarray}%
and 
\begin{equation}
\left\langle \widehat{S}_{x}^{i}\right\rangle _{R}=0
\end{equation}%
so that 
\begin{equation}
\left\langle \Delta \widehat{S}\,_{x}^{2}\right\rangle \geq
\sum_{R}P_{R}\,\tsum\limits_{i}\left( \frac{1}{4}(\left\langle (\widehat{b}%
^{\dag }\widehat{b})_{i}\right\rangle _{R}+\left\langle (\widehat{a}^{\dag }%
\widehat{a})_{i}\right\rangle _{R})+\frac{1}{2}(\left\langle (\widehat{a}%
^{\dag }\widehat{a})_{i}\right\rangle _{R}\left\langle (\widehat{b}^{\dag }%
\widehat{b})_{i}\right\rangle _{R})\right)
\end{equation}

Now using (\ref{Eq.RevisedSorensenAveragesB})%
\begin{equation}
\left\langle \widehat{S}_{z}^{i}\right\rangle _{R}=\frac{1}{2}(\left\langle (%
\widehat{b}^{\dag }\widehat{b})_{i}\right\rangle _{R}-\left\langle (\widehat{%
a}^{\dag }\widehat{a})_{i}\right\rangle _{R}))
\end{equation}%
\begin{eqnarray}
\left\langle \widehat{S}\,_{z}\right\rangle
&=&\sum_{R}P_{R}\,\tsum\limits_{i}\left\langle \widehat{S}%
_{z}^{i}\right\rangle _{R}  \nonumber \\
\frac{1}{2}|\left\langle \widehat{S}\,_{z}\right\rangle | &=&\frac{1}{2}%
\sum_{R}P_{R}\,|\tsum\limits_{i}\frac{1}{2}(\left\langle (\widehat{b}^{\dag }%
\widehat{b})_{i}\right\rangle _{R}-\left\langle (\widehat{a}^{\dag }\widehat{%
a})_{i}\right\rangle _{R}))|  \nonumber \\
&\leq &\sum_{R}P_{R}\,\frac{1}{4}\tsum\limits_{i}|(\left\langle (\widehat{b}%
^{\dag }\widehat{b})_{i}\right\rangle _{R}-\left\langle (\widehat{a}^{\dag }%
\widehat{a})_{i}\right\rangle _{R}))|  \nonumber \\
&\leq &\sum_{R}P_{R}\,\frac{1}{4}\tsum\limits_{i}(\left\langle (\widehat{b}%
^{\dag }\widehat{b})_{i}\right\rangle _{R}+\left\langle (\widehat{a}^{\dag }%
\widehat{a})_{i}\right\rangle _{R}))
\end{eqnarray}%
and thus%
\begin{eqnarray}
&&\left\langle \Delta \widehat{S}\,_{x}^{2}\right\rangle -\frac{1}{2}%
|\left\langle \widehat{S}\,_{z}\right\rangle |\;  \nonumber \\
&\geq &\sum_{R}P_{R}\,\tsum\limits_{i}\left( \frac{1}{4}(\left\langle (%
\widehat{b}^{\dag }\widehat{b})_{i}\right\rangle _{R}+\left\langle (\widehat{%
a}^{\dag }\widehat{a})_{i}\right\rangle _{R})+\frac{1}{2}(\left\langle (%
\widehat{a}^{\dag }\widehat{a})_{i}\right\rangle _{R}\left\langle (\widehat{b%
}^{\dag }\widehat{b})_{i}\right\rangle _{R})\right)  \nonumber \\
&&-\sum_{R}P_{R}\,\frac{1}{4}\tsum\limits_{i}(\left\langle (\widehat{b}%
^{\dag }\widehat{b})_{i}\right\rangle _{R}+\left\langle (\widehat{a}^{\dag }%
\widehat{a})_{i}\right\rangle _{R}))  \nonumber \\
&=&\sum_{R}P_{R}\,\frac{1}{2}\tsum\limits_{i}(\left\langle (\widehat{a}%
^{\dag }\widehat{a})_{i}\right\rangle _{R}\left\langle (\widehat{b}^{\dag }%
\widehat{b})_{i}\right\rangle _{R})  \nonumber \\
&\geq &0
\end{eqnarray}%
A similar proof shows that $\left\langle \Delta \widehat{S}%
\,_{y}^{2}\right\rangle -\frac{1}{2}|\left\langle \widehat{S}%
\,_{z}\right\rangle |\geq 0$ for the non-entangled state of all $2N$ modes.

This shows that for the general non-entangled state with all modes $\widehat{%
a}_{i}$ and $\widehat{b}_{i}$ as the sub-systems, the variances for two of
the spin fluctuations $\left\langle \Delta \widehat{S}\,_{x}^{2}\right%
\rangle $ and $\left\langle \Delta \widehat{S}\,_{y}^{2}\right\rangle $ are
both greater than $\frac{1}{2}|\left\langle \widehat{S}\,_{z}\right\rangle |$%
, and hence there is no spin squeezing for $\widehat{S}_{x}$ or $\widehat{S}%
_{y}$. Note that as $|\left\langle \widehat{S}\,_{y}\right\rangle |=0$, the
quantity $\sqrt{\left( |\left\langle \widehat{S}_{\perp \,1}\right\rangle
|^{2}+|\left\langle \widehat{S}_{\perp \,2}\right\rangle |^{2}\right) }$ is
the same as $|\left\langle \widehat{S}\,_{z}\right\rangle |$, so the
alternative criterion in Eq. (\ref{Eq.NewCriterionSpinSqueezing}) is the
same as that in Eq. (\ref{Eq.SpinSqueezingJXJY}) which is used here.

For the other spin fluctuation $\left\langle \Delta \widehat{S}%
\,_{z}^{2}\right\rangle $ since we have 
\begin{equation}
\left\langle \widehat{S}\,_{x}\right\rangle =\sum_{R}P_{R}\left\langle 
\widehat{S}\,_{x}\right\rangle _{R}=0\qquad \left\langle \widehat{S}%
\,_{y}\right\rangle =\sum_{R}P_{R}\left\langle \widehat{S}%
\,_{y}\right\rangle _{R}=0  \label{Eq.MeanNewSpinXYB}
\end{equation}%
then the other two uncertainty relationships just give $\left\langle \Delta 
\widehat{S}_{y}{}^{2}\right\rangle \left\langle \Delta \widehat{S}%
_{z}{}^{2}\right\rangle \geq 0$; $\left\langle \Delta \widehat{S}%
_{z}{}^{2}\right\rangle \left\langle \Delta \widehat{S}_{x}{}^{2}\right%
\rangle \geq 0$, so spin squeezing in $\widehat{S}\,_{z}$ is meaningless.

Hence we have shown that for a \emph{non-entangled} physical state for all
the $2N$ modes $\widehat{a}_{i}$ and $\widehat{b}_{i}$%
\begin{equation}
\left\langle \Delta \widehat{S}\,_{x}^{2}\right\rangle \geq \frac{1}{2}%
|\left\langle \widehat{S}\,_{z}\right\rangle |\quad and\quad \left\langle
\Delta \widehat{S}\,_{y}^{2}\right\rangle \geq \frac{1}{2}|\left\langle 
\widehat{S}\,_{z}\right\rangle |  \label{Eq.NonEntStateSpinSqCondnB}
\end{equation}%
so that spin squeezing in either $\widehat{S}_{x}$ or $\widehat{S}_{y}$
requires entanglement.

\subsection{Variance $\left\langle \Delta \widehat{S}\,_{z}^{2}\right\rangle 
$ for Two Mode Sub-Systems}

Here we will see if the modified approach to Sorensen et al can lead to a
useful inequality for $\left\langle \Delta \widehat{S}\,_{z}^{2}\right%
\rangle $ that applies when non-entangled states are those when the \emph{%
pairs} of modes $\widehat{a}_{i}$ and $\widehat{b}_{i}$ are the separate
sub-systems . We will attempt to follow the approach used by Sorensen et al
when identical particles $i$ were regarded as the sub-systems.

Now the general non-entangled state will be 
\begin{equation}
\widehat{\rho }=\sum_{R}P_{R}\,\widehat{\rho }_{R}^{1}\otimes \widehat{\rho }%
_{R}^{2}\otimes \widehat{\rho }_{R}^{3}\otimes ...
\end{equation}%
where the $\widehat{\rho }_{R}^{i}$ are now of the form given in Eq. (\ref%
{Eq.GeneralDensityOprModePair}) and the conditions in Eq. (\ref%
{Eq.RevisedSorensenAveragesB}) no longer apply. The Fock states are of the
form $\left\vert N_{ia}\right\rangle \otimes \left\vert N_{ib}\right\rangle $
for the pair of modes $\widehat{a}_{i}$ and $\widehat{b}_{i}$, and fior this
Fock state the total occupancy of the pair of modes is $N_{i}=N_{ia}+N_{ib}$%
. From the super-selection rule the density operator $\widehat{\rho }%
_{R}^{i} $ for the $i$th pair of modes $\widehat{a}_{i}$ and $\widehat{b}%
_{i} $ is diagonal in the total occupancy. For $N_{i}$ $=0$ there is one non
zero matrix element $(\left\langle 0\right\vert _{ia}\otimes \left\langle
0\right\vert _{ib})\,\widehat{\rho }_{R}^{i}\,(\left\vert 0\right\rangle
_{ia}\otimes \left\vert 0\right\rangle _{ib})$. For $N_{i}$ $=1$ there are
four non zero matrix elements, which may be written%
\begin{eqnarray}
(\left\langle 1\right\vert _{ia}\otimes \left\langle 0\right\vert _{ib})\,%
\widehat{\rho }_{R}^{i}\,(\left\vert 1\right\rangle _{ia}\otimes \left\vert
0\right\rangle _{ib}) &=&\rho _{aa}^{i}  \nonumber \\
(\left\langle 1\right\vert _{ia}\otimes \left\langle 0\right\vert _{ib})\,%
\widehat{\rho }_{R}^{i}\,(\left\vert 0\right\rangle _{ia}\otimes \left\vert
1\right\rangle _{ib}) &=&\rho _{ab}^{i}  \nonumber \\
(\left\langle 0\right\vert _{ia}\otimes \left\langle 1\right\vert _{ib})\,%
\widehat{\rho }_{R}^{i}\,(\left\vert 1\right\rangle _{ia}\otimes \left\vert
0\right\rangle _{ib}) &=&\rho _{ba}^{i}  \nonumber \\
(\left\langle 0\right\vert _{ia}\otimes \left\langle 1\right\vert _{ib})\,%
\widehat{\rho }_{R}^{i}\,(\left\vert 0\right\rangle _{ia}\otimes \left\vert
1\right\rangle _{ib}) &=&\rho _{bb}^{i}  \label{Eq.MatrixElemPairModes}
\end{eqnarray}%
For $N_{i}$ $=2$ there are nine non zero matrix element $(\left\langle
2\right\vert _{ia}\otimes \left\langle 0\right\vert _{ib})\,\widehat{\rho }%
_{R}^{i}\,(\left\vert 2\right\rangle _{ia}\otimes \left\vert 0\right\rangle
_{ib})$, $...$, $(\left\langle 0\right\vert _{ia}\otimes \left\langle
2\right\vert _{ib})\,\widehat{\rho }_{R}^{i}\,(\left\vert 0\right\rangle
_{ia}\otimes \left\vert 2\right\rangle _{ib})$ and the number increases with 
$N_{i}$.

If we restrict ourselves to general entangled states where $N_{i}$ $=1$ for
all pairs of modes, then the density operator $\widehat{\rho }_{R}^{i}$ is
of then form 
\begin{eqnarray}
\widehat{\rho }_{R}^{i} &=&\rho _{aa}^{i}(\left\vert 1\right\rangle
_{ia}\left\langle 1\right\vert _{ia}\otimes \left\vert 0\right\rangle
_{ib}\left\langle 0\right\vert _{ib})+\rho _{ab}^{i}(\left\vert
1\right\rangle _{ia}\left\langle 0\right\vert _{ia}\otimes \left\vert
0\right\rangle _{ib}\left\langle 1\right\vert _{ib})  \nonumber \\
&&+\rho _{ba}^{i}(\left\vert 0\right\rangle _{ia}\left\langle 1\right\vert
_{ia}\otimes \left\vert 1\right\rangle _{ib}\left\langle 0\right\vert
_{ib})+\rho _{bb}^{i}(\left\vert 0\right\rangle _{ia}\left\langle
0\right\vert _{ia}\otimes \left\vert 1\right\rangle _{ib}\left\langle
1\right\vert _{ib})  \label{Eq.PairModesdensityOpr}
\end{eqnarray}%
In addition Hermitiancy, positivity, unit trace $Tr(\widehat{\rho }%
_{R}^{i})=1$ and $Tr(\widehat{\rho }_{R}^{i})^{2}\leq 1$ can be used as in
Eq (\ref{Eq.ParamMatrixElements}) to parameterise the matrix elements in (%
\ref{Eq.MatrixElemPairModes}).%
\begin{eqnarray}
\rho _{aa}^{i} &=&\sin ^{2}\alpha _{i}\qquad \rho _{bb}^{i}=\cos ^{2}\alpha
_{i}  \nonumber \\
\rho _{ab}^{i} &=&\sqrt{\sin ^{2}\alpha _{i}\,\cos ^{2}\alpha _{i}}\,\sin
^{2}\beta _{i}\,\exp (+i\phi _{i})\qquad \rho _{ba}^{i}=\sqrt{\sin
^{2}\alpha _{i}\,\cos ^{2}\alpha _{i}}\,\sin ^{2}\beta _{i}\,\exp (-i\phi
_{i})  \nonumber \\
&&
\end{eqnarray}

The expectation values for the spin operators $\widehat{S}_{x}^{i}$, $%
\widehat{S}_{y}^{i}$ and $\widehat{S}_{z}^{i}$ associated with the $i$th
pair of modes are then 
\begin{eqnarray}
\left\langle \widehat{S}_{x}^{i}\right\rangle _{R} &=&Tr(\widehat{\rho }%
_{R}^{i}\,\frac{1}{2}(\widehat{b}_{i}^{\dag }\widehat{a}_{i}+\widehat{a}%
_{i}^{\dag }\widehat{b}_{i})  \nonumber \\
&=&\frac{1}{2}\left( \rho _{ab}^{i}+\rho _{ba}^{i}\right)  \nonumber \\
\left\langle \widehat{S}_{y}^{i}\right\rangle _{R} &=&\frac{1}{2i}\left(
\rho _{ab}^{i}-\rho _{ba}^{i}\right)  \nonumber \\
\left\langle \widehat{S}_{z}^{i}\right\rangle _{R} &=&\frac{1}{2}\left( \rho
_{bb}^{i}-\rho _{aa}^{i}\right)
\end{eqnarray}%
which are of exactly the same form as in Eq. (\ref%
{Eq.SpinExpnValuesMatrixElements}) as in the Appendix \ref{Appendix -
Sorensen Results} derivation of the original Sorensen et al \cite%
{Sorensen01a} results based on treating identical particles as the
sub-systems. The proof however is now different and rests on restricting the
states $\widehat{\rho }_{R}^{i}$ to each containing exactly one boson.

The remainder of the proof is exactly the same as in Appendix \ref{Appendix
- Sorensen Results} and we find that 
\begin{equation}
\left\langle \Delta \widehat{S}\,_{z}^{2}\right\rangle \geq \frac{1}{N}%
\left( \left\langle \widehat{S}\,_{x}\right\rangle ^{2}+\left\langle 
\widehat{S}\,_{y}\right\rangle ^{2}\right)
\end{equation}%
for non-entangled \emph{pairs} of modes $\widehat{a}_{i}$ and $\widehat{b}%
_{i}$. Thus when the interpretation is changed so that are the separate
sub-systems are these pairs of modes, it follows that spin squeezing
requires entanglement of all the mode pairs.

\pagebreak

\section{Appendix 10 - Heisenberg Uncertainty Principle Results}

\label{Appendix - Heisenberg Uncertainty Principle Results}

Here we derive the results in SubSection \ref{SubSection - Sorensen and
Molmer 2001} leading to inequalities for the variance $\left\langle \Delta 
\widehat{J}\,_{x}^{2}\right\rangle $ considered as a function of $%
|\left\langle \widehat{J}\,_{z}\right\rangle |$ for states where the spin
operators are chosen such that $\left\langle \widehat{J}\,_{x}\right\rangle
=\left\langle \widehat{J}\,_{y}\right\rangle =0$.

From the Schwarz inequality $\left\langle \widehat{J}\,_{z}\right\rangle
^{2}\leq \left\langle \widehat{J}\,_{z}^{2}\right\rangle $ so that 
\begin{equation}
\left\langle \widehat{J}\,_{x}^{2}\right\rangle +\left\langle \widehat{J}%
\,_{y}^{2}\right\rangle +\left\langle \widehat{J}\,_{z}\right\rangle
^{2}\leq \left\langle \widehat{J}\,_{x}^{2}\right\rangle +\left\langle 
\widehat{J}\,_{y}^{2}\right\rangle +\left\langle \widehat{J}%
\,_{z}^{2}\right\rangle =J(J+1)
\end{equation}%
giving Eq. (\ref{Eq.SchwarzResult}). Subtracting $\left\langle \widehat{J}%
\,_{x}\right\rangle ^{2}=\left\langle \widehat{J}\,_{y}\right\rangle ^{2}=0$
from each side gives

\begin{equation}
\left\langle \Delta \widehat{J}\,_{x}^{2}\right\rangle +\left\langle \Delta 
\widehat{J}\,_{y}^{2}\right\rangle +\left\langle \widehat{J}%
\,_{z}\right\rangle ^{2}\leq J(J+1)
\end{equation}

Substituting for $\left\langle \Delta \widehat{J}\,_{y}^{2}\right\rangle $
from the Heisenberg uncertainty principle result in Eq. (\ref{Eq.HUP}) gives 
\begin{equation}
\left\langle \Delta \widehat{J}\,_{x}^{2}\right\rangle ^{2}-\left(
J(J+1)-\left\langle \widehat{J}\,_{z}\right\rangle ^{2}\right) \left\langle
\Delta \widehat{J}\,_{x}^{2}\right\rangle +\frac{1}{4}\xi \left\langle 
\widehat{J}\,_{z}\right\rangle ^{2}\leq 0
\end{equation}%
The left side is a parabolic function of $\left\langle \Delta \widehat{J}%
\,_{x}^{2}\right\rangle $ and for this to be negative requires $\left\langle
\Delta \widehat{J}\,_{x}^{2}\right\rangle $ to lie between the two roots of
this function, giving 
\begin{eqnarray}
\left\langle \Delta \widehat{J}\,_{x}^{2}\right\rangle &\geq &\frac{1}{2}%
\left\{ \left( J(J+1)-\left\langle \widehat{J}\,_{z}\right\rangle
^{2}\right) -\sqrt{\left( J(J+1)-\left\langle \widehat{J}\,_{z}\right\rangle
^{2}\right) ^{2}-\xi \left\langle \widehat{J}\,_{z}\right\rangle ^{2}}%
\right\}  \nonumber \\
&& \\
\left\langle \Delta \widehat{J}\,_{x}^{2}\right\rangle &\leq &\frac{1}{2}%
\left\{ \left( J(J+1)-\left\langle \widehat{J}\,_{z}\right\rangle
^{2}\right) +\sqrt{\left( J(J+1)-\left\langle \widehat{J}\,_{z}\right\rangle
^{2}\right) ^{2}-\xi \left\langle \widehat{J}\,_{z}\right\rangle ^{2}}%
\right\}  \nonumber \\
&&
\end{eqnarray}%
which are the required inequalities in Eq. (\ref{Eq.HUPRestriction1}) and (%
\ref{Eq.HUPRestriction2}).

\pagebreak

\section{Figure Captions}

\bigskip

Figure 1. Bloch vector and spin fluctuations shown for original spin
operators.

\bigskip

Figure 2. Regions in the $<\Delta \widehat{J}_{x}^{2}>$ versus $|<\widehat{J}%
_{z}>|$ plane (shown shaded) for states that satisfy (a) the spin squeezing
inequality Eq. (\ref{Eq.SpinSqRestriction}) (b) the smaller Heisenberg
uncertainty principle inequality Eq. (\ref{Eq.HUPRestriction1}) and (c) the
larger HUP inequality Eq. (\ref{Eq.HUPRestriction2}). The case shown is for $%
J=1000$ and HUP factor $\xi =1$. Both $<\Delta \widehat{J}_{x}^{2}>$ and $|<%
\widehat{J}_{z}>|$ are in units of $J$. The spin operators are chosen so
that $<\widehat{J}_{x}>=<\widehat{J}_{y}>=0$.

\bigskip

Figure 3. As in Figure 2, but with $J=1000$ and HUP factor $\xi =10.0$.

\bigskip

Figure 3. As in Figure 2, but with $J=1$ and HUP factor $\xi =10.0$.
\pagebreak

\FRAME{ftbpF}{6.2587in}{4.2964in}{0pt}{}{}{spinsqgfig1.jpg}{\special%
{language "Scientific Word";type "GRAPHIC";display "USEDEF";valid_file
"F";width 6.2587in;height 4.2964in;depth 0pt;original-width
9.9998in;original-height 7.4996in;cropleft "0";croptop "1";cropright
"1";cropbottom "0";filename '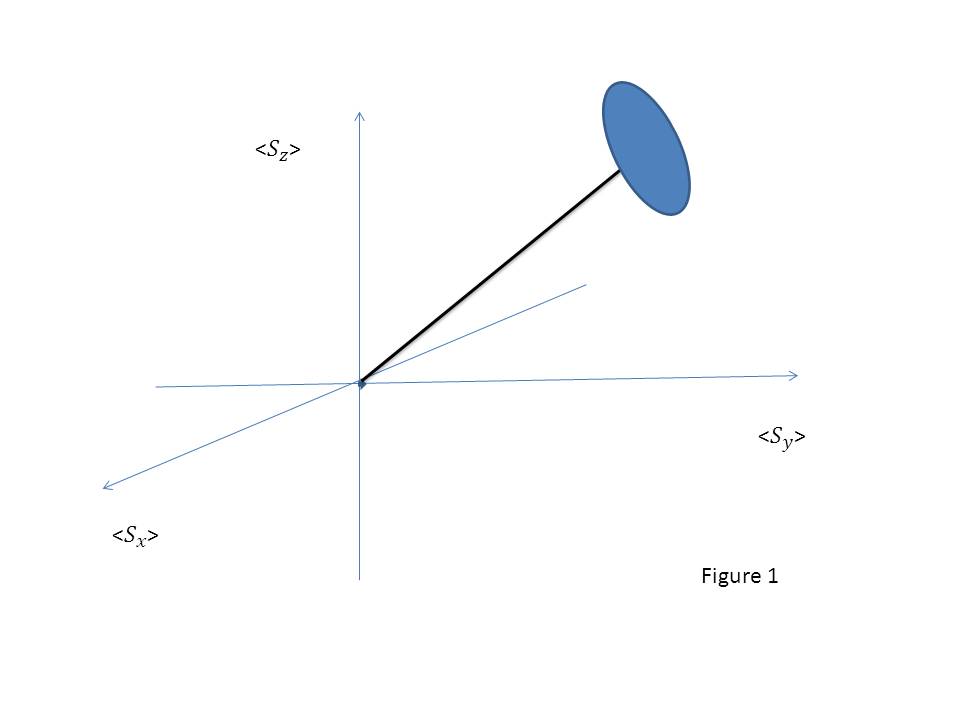';file-properties "NPEU";}}%
\pagebreak

\FRAME{ftbpF}{6.1021in}{5.5685in}{0pt}{}{}{spinsqgfig2.jpg}{\special%
{language "Scientific Word";type "GRAPHIC";display "USEDEF";valid_file
"F";width 6.1021in;height 5.5685in;depth 0pt;original-width
9.9998in;original-height 7.4996in;cropleft "0";croptop "1";cropright
"1";cropbottom "0";filename '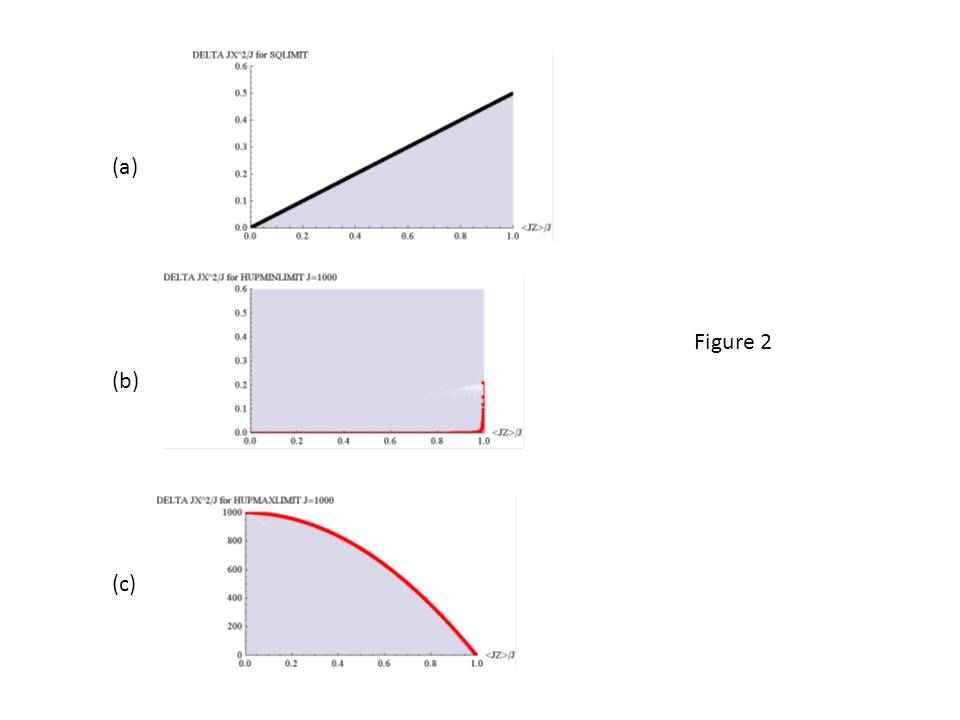';file-properties "NPEU";}}%
\pagebreak

\FRAME{ftbpF}{6.1756in}{4.7037in}{0pt}{}{}{spinsqgfig3.jpg}{\special%
{language "Scientific Word";type "GRAPHIC";display "USEDEF";valid_file
"F";width 6.1756in;height 4.7037in;depth 0pt;original-width
9.9998in;original-height 7.4996in;cropleft "0";croptop "1";cropright
"1";cropbottom "0";filename '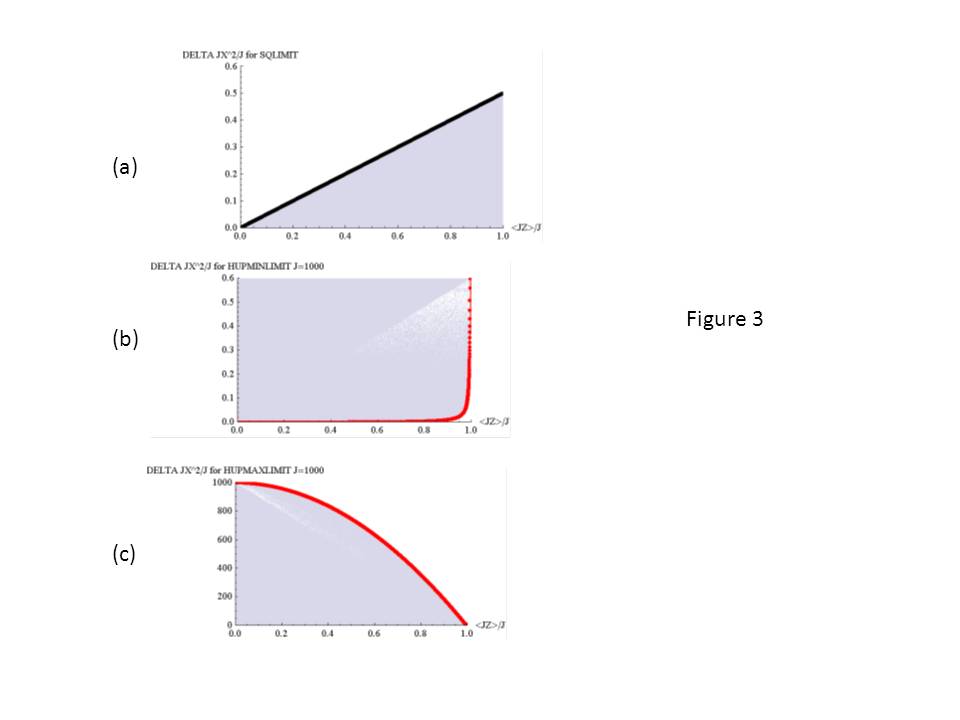';file-properties "NPEU";}}%
\pagebreak

\FRAME{ftbpF}{6.1964in}{5.0998in}{0pt}{}{}{spinsqgfig4.jpg}{\special%
{language "Scientific Word";type "GRAPHIC";display "USEDEF";valid_file
"F";width 6.1964in;height 5.0998in;depth 0pt;original-width
9.9998in;original-height 7.4996in;cropleft "0";croptop "1";cropright
"1";cropbottom "0";filename '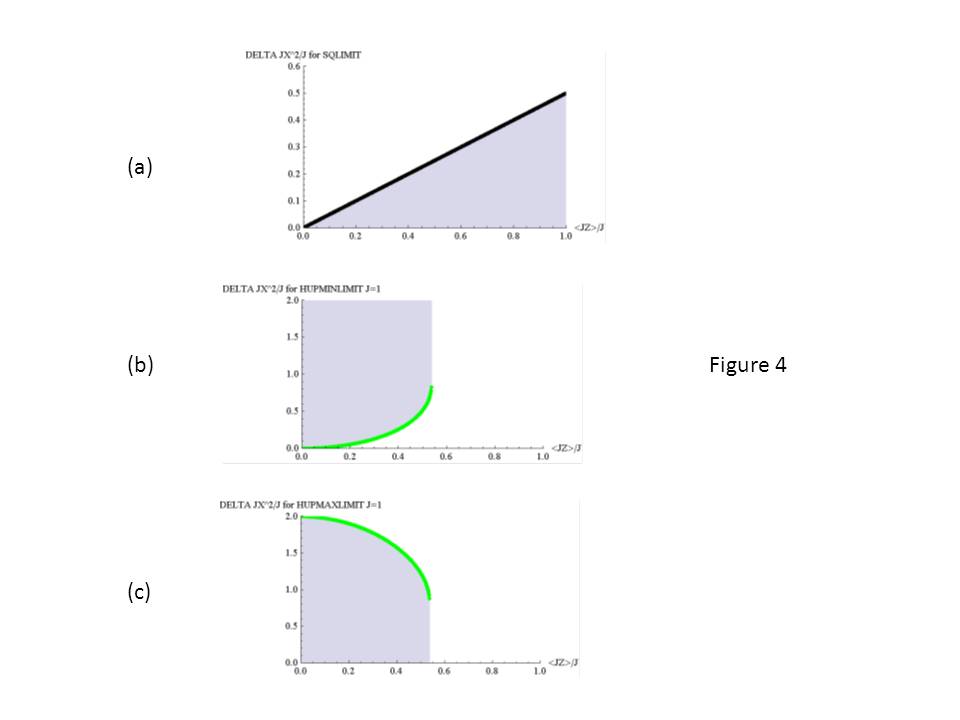';file-properties "NPEU";}}%
\pagebreak


\begin{thebibliography}{99}
\bibitem{Einstein35a} Einstein, A., Podolsky, B. and Rosen, N., \textit{%
Phys. Rev. }\textbf{1935}, \textit{47, }777.

\bibitem{Schrodinger35a} Schrodinger, E., \textit{Naturwiss. }\textbf{1935}, 
\textit{23}, 807. Translated in \textit{Quantum Theory and Measurement; }%
eds. J. A. Wheeler and W. Zurek, Princeton: 1983.

\bibitem{Bell65a} Bell, J. S., \textit{Physics }\textbf{1964}, \textit{1},
195.

\bibitem{Clauser69a} Clauser, J. F., Horne, M. A., Shimony, A. and Holt, R.
A., \textit{Phys. Rev. Letts. }\textbf{1969, }\textit{23, }880.

\bibitem{Werner89a} Werner, R. F., \textit{Phys. Rev. A }\textbf{1989}, 
\textit{40, }4277.

\bibitem{Peres93a} Peres, A. , \textit{Quantum Theory: Concepts and Methods; 
}Kluwer: Dortrecht, \textbf{1993}.

\bibitem{Nielsen00a} Nielsen, M. A. and Chuang, I. L., \textit{Quantum
Computation and Quantum Information; }Cambridge: Cambridge, \textbf{2000}.

\bibitem{Vedral07a} Vedral, V. , \textit{Introduction to Quantum Information
Science; }Oxford: Oxford, \textbf{2007}.

\bibitem{Barnett09a} Barnett, S. M., \textit{Quantum Information; }Oxford:
Oxford, \textbf{2009}.

\bibitem{Reid09a} Reid, M. D., Drummond, P. D., Bowen, W. P., Cavalcanti. E.
G. , Lam., P. K., Bachor, H. A., Anderson, U. L. and Leuchs, G., \textit{%
Rev. Mod. Phys. }\textbf{2009}, \textit{81}, 1727.

\bibitem{Reid12a} Reid, M. D., He, Q.-Y. and Drummond, P. D.,\textit{\
Front. Phys. }\textbf{2012}, \textit{7}, 72.

\bibitem{Simon02a} Simon, C. \textit{Phys. Rev. A }\textbf{2002, }\textit{%
66, }052323.

\bibitem{Hines03a} Hines, A.P., McKenzie, R.H. and Milburn, G.J., \textit{%
Phys. Rev. A }\textbf{2003}, \textit{67, }013609.

\bibitem{TerraCunha07a} Terra Cunha, M.O., Dunningham, J.A. and Vedral, V., 
\textit{Proc. Roy. Soc. A }\textbf{2007, }\textit{463, }2277.

\bibitem{Horodecki09a} Horodecki, R., Horodecki, P., Horodecki, M. and
Horodecki, K., \textit{Rev. Mod. Phys. }\textbf{2009}, \textit{81}, 865.

\bibitem{Guhne09a} Guhne, O. and Toth, G., \textit{Phys. Rep. }\textbf{2009}%
, \textit{474, }1.

\bibitem{Tichy11a} Tichy, M. C., Mintert, F. and Buchleitner, A. \textit{J.
Phys. B: At. Mol. Opt. Phys. }\textbf{\ 2011}, \textit{44, }192001.

\bibitem{Modi12a} Modi, K, Brodutch, A., Cable, H., Paterek, T. and Vedral,
V., \textit{Rev. Mod. Phys. }\textbf{2012}, \textit{84}, 1655.

\bibitem{Amico08a} Amico, L., Fazio, R., Osterloh, A. and Vedral, V., 
\textit{Rev. Mod. Phys. }\textbf{2008, }\textit{80}\textbf{, }517.

\bibitem{Wiseman03a} Wiseman, H. M. and Vaccaro, J. A., \textit{Phys. Rev.
Letts. }\textbf{2003, }\textit{91, }097902.

\bibitem{Dowling06b} Dowling, M. R., Doherty, A. C. and Wiseman, H. M., 
\textit{Phys. Rev. A }\textbf{2006}, \textit{73, }052323.

\bibitem{Benatti10a} Benatti, F., Floreanni, R. and Marzolino, U., \textbf{%
2010}, \textit{Ann. Phys.,325}, 924.

\bibitem{Benatti11a} Benatti, F., Floreanni, R. and Marzolino, U., \textbf{%
2011}, \textit{J. Phys. B: At. Mol. Opt. Phys.,44}, 1001.

\bibitem{Lunkes05a} Lunkes, C., Brukner, C. and Vedral, V., \textit{Phys.
Rev. Letts. }\textbf{2005, }\textit{95, }030503.

\bibitem{Bartlett06b} Bartlett, S. D., Doherty, A. C., Spekkens, R. W. and
Wiseman, H. W., \textit{Phys. Rev. A }\textbf{2006}, \textit{73}, 022311.

\bibitem{Jones07a} Jones, S. J., Wiseman, H. M. and Doherty, A. C. , \textit{%
Phys. Rev. A }\textbf{2007, }\textit{76, }052116.

\bibitem{Masanes08a} Masanes, L., Liang, Y.-C. and Doherty, A. C., \textit{%
Phys. Rev. Letts. }\textbf{2008, }\textit{100, }090403.

\bibitem{Verstraete03a} Verstraete, F. and Cirac, J. I. ,\textit{Phys. Rev.
Letts. }\textbf{2003, }\textit{91, }010404.

\bibitem{Schuch04a} Schuch, N., Verstraete, F. and Cirac, J. I., \textit{%
Phys. Rev. A }\textbf{2004}, \textit{70, }042310.

\bibitem{Wick52a} Wick, G. C., Wightman, A. S. and Wigner, E. P., \textit{%
Phys. Rev. }\textbf{1952}, \textit{88, }101.

\bibitem{Bartlett07a} Bartlett, S. D., Rudolph, T. and Spekkens, R. W. ,%
\textit{Rev. Mod. Phys. }\textbf{2007, }\textit{79}\textbf{, }555.

\bibitem{Hillery06a} Hillery, M. and Zubairy, M.S., \textit{Phys. Rev.
Letts. }\textbf{2006, }\textit{96, }050503.

\bibitem{Hillery09a} Hillery, M., Dung, H. T. and Niset, J. \textit{Phys.
Rev. A }\textbf{2009, }\textit{80, }052335.

\bibitem{Vaccaro08a} Vaccaro, J. A. , Anselmi, F., Wiseman, H. M. and
Jacobs, K., \textit{Phys. Rev. A }\textbf{2008}, \textit{77}, 032114.

\bibitem{White09a} White, G. A., Vaccaro, J. A. and Wiseman, H. M., \textit{%
Phys. Rev. A }\textbf{2009}, \textit{79}, 032109.

\bibitem{Paterek11a} Paterek, T., Kurzynski, P., Oi, D. K. L. and
Kaszlikowski, D., \textit{New J. Phys. }\textbf{2011}, \textit{13}, 043027.

\bibitem{Aharonov67a} Aharonov, Y. and Susskind, L., \textit{Phys. Rev. }%
\textbf{1967, }\textit{155, }1428.

\bibitem{Bartlett03a} Bartlett, S. D. and Wiseman, H. W., \textit{Phys. Rev.
Letts. }\textbf{2003, }\textit{91, }097903.

\bibitem{Sanders03a} Sanders, B. C., Bartlett, S. D., Rudolph, T. and
Knight, P. L., \textit{Phys. Rev. A }\textbf{2003, }\textit{68, }042329.

\bibitem{Kitaev04a} Kitaev, A., Mayers, D. and Preskill, J. , \textit{Phys.
Rev. A }\textbf{2004}, \textit{69, }052326.

\bibitem{van Enk05a} van Enk, S. J., \textit{Phys. Rev. A }\textbf{2005}, 
\textit{72, }064306.

\bibitem{Bartlett06a} Bartlett, S. D., Rudolph, T. and Spekkens, R. W. ,%
\textit{Int. J. Quant. Infn. }\textbf{2006}, \textit{4}, 17.

\bibitem{Wodkiewicz85a} Wodkiewicz, K. and Eberly, J. H., \textit{J. Opt.
Soc. Amer. B }\textbf{1985}, \textit{2}, 458.

\bibitem{Kitagawa93a} Kitagawa, M. and Ueda, M., \textit{Phys. Rev. A }%
\textbf{1993}, \textit{47, }5138.

\bibitem{Sorensen01a} Sorensen, A., Duan, L.-M., Cirac, J.I. and Zoller, P. 
\textit{Nature }\textbf{2001, }\textit{409, }63.

\bibitem{Micheli03a} Micheli, A., Jaksch, D., Cirac, J. I. and Zoller, P., 
\textit{Phys. Rev. A }\textbf{2003}, \textit{67, }013607.

\bibitem{Toth07a} Toth, G., Knapp, C., Guhne, O. and Briegel, H. J., \textit{%
Phys. Rev. Letts. }\textbf{2007, }\textit{99, }250405.

\bibitem{Hyllus12a} Hyllus, P., Pezze, L., Smerzi, A. and Toth, G. \textit{%
Phys. Rev. A }\textbf{2012}, \textit{86, }012237.

\bibitem{Pegg02a} Pegg, D. T., Barnett, S. M. and Jeffers, J., \textit{J.
Mod. Opt. }\textbf{2002}, \textit{49}, 913.

\bibitem{Pegg05a} Pegg, D. T. and Jeffers, J., \textit{J. Mod. Opt. }\textbf{%
2005}, \textit{52}, 1835.

\bibitem{Peres96a} Peres, A.,\textit{Phys. Rev. Letts. }\textbf{1996, }%
\textit{77, }1413.

\bibitem{Wiseman07a} Wiseman, H. M., Jones, S. J. and Doherty, A. C. , 
\textit{Phys. Rev. Letts. }\textbf{2007, }\textit{98, }140402.

\bibitem{Gisin91a} Gisin, N. , \textit{Phys. Letts. A }\textbf{1991, }%
\textit{154, }201.

\bibitem{Spreeuw98a} Spreeuw, R. J. C., \textit{Found. Phys. }\textbf{1998, }%
\textit{28, }361.

\bibitem{Borges10a} Borges, C. V. S., Hor-Meyl, M., Huguenin, J. A. O. and
Khoury, A. Z. , \textit{Phys. Rev. A }\textbf{2010}, \textit{82, }033833.

\bibitem{Isham95a} Isham, C. J., \textit{Quantum Theory; }Imperial College
Press: London, \textbf{1995}.

\bibitem{Reid03a} Reid, M. D., \textbf{2003}, In \textit{Quantum Squeezing},
P. D. Drummond and Z. Ficek, Eds. (Springer).

\bibitem{Greenberger03a} Greenberger, D. M., Horne, M. and Zeilinger, A., 
\textbf{2003}, In \textit{Bell's Theorem, Quantum Theory and Conceptions of
the Universe}, M. Kafatos, Ed. (Kluwer).

\bibitem{Rinner08a} Rinner, S. and Werner, E., \textit{Cent. Eur. J. Phys. }%
\textbf{2008}, \textit{6}, 178.

\bibitem{Haroche} Brune, M., Hagley, E., Dreyer, J., Maitre, X., Maali, C.,
Wunderlich, C., Raimond, J. M. and Haroche, S. \textit{Phys. Rev. Letts. }%
\textbf{1996, }\textit{77, }4887.

\bibitem{Pezze09a} Pezze, L. and Smerzi, A., \textit{Phys. Rev. Letts. }%
\textbf{2009, }\textit{102, }100401.

\bibitem{He12a} He, Q. Y., Drummond, P. D., Olsen, M. K. and Reid, M. D., 
\textit{Phys. Rev. A }\textbf{2012, }\textit{86, }023626.

\bibitem{Heaney10a} Heaney, L., Lee, S.-W. and Jaksch, D., \textit{Phys.
Rev. A }\textbf{2010, }\textit{82, }042116.

\bibitem{Cable05a} Cable, H., Knight, P. L. and Rudolph, T., \textit{Phys.
Rev. A }\textbf{2005}, \textit{71}, 042107.

\bibitem{Goold09a} Goold, J., Heaney, L., Busch, Th. and Vedral V., \textit{%
Phys. Rev. A }\textbf{2009}, \textit{80}, 022338.

\bibitem{Leggett01a} Leggett, A.J., \textit{Rev. Mod. Phys. }\textbf{2001, }%
\textit{73, }307.

\bibitem{Bach04a} Bach, R. and Rzazewskii, K, \textit{Phys. Rev. A }\textbf{%
2004}, \textit{70}, 063622.

\bibitem{Dalton12a} Dalton, B.J. and Ghanbari, S. \textit{J. Mod. Opt. }%
\textbf{2012}, \textit{59}, 287, \textit{ibid }\textbf{2013}, \textit{60},
(6).

\bibitem{Javainainen96a} Javanainen, J. and Yoo, S. M., \textit{Phys. Rev.
Letts. }\textbf{1996, }\textit{76, }161.

\bibitem{Stenholm02a} Stenholm, S., \textit{Phys. Scripta }\textbf{2002}, 
\textit{T102}, 89.

\bibitem{Molmer97a} Molmer, K., \textit{Phys. Rev. A }\textbf{1997}, \textit{%
55}, 4247.

\bibitem{Wiseman02a} Wiseman, H. W. and Vaccaro, J., \textit{Phys. Rev. A }%
\textbf{2002}, \textit{65}, 043605.

\bibitem{Wiseman02b} Wiseman, H. W. and Vaccaro, J., \textit{Phys. Rev. A }%
\textbf{2002}, \textit{65}, 043606.

\bibitem{Dowling06a} Dowling, M. R., Bartlett, S. D., Rudolph, T. and
Spekkens, R. W., \textit{Phys. Rev. A }\textbf{2006}, \textit{74}, 052113.

\bibitem{Dunningham11a} Dunningham, J. A., Rico Gutierrez, L. M. and Palge,
V., \textit{Optics and Spectroscopy, }\textbf{2011}, \textit{111}, 528.

\bibitem{Jaaskelainen06a} Jaaskelainen, M. and Meystre, P., \textit{Phys.
Rev. A }\textbf{2006}, \textit{73, }013602.

\bibitem{Rose57a} Rose, M.E. \textit{Elementary Theory of Angular Momentum; }%
Wiley: New York, \textbf{1957}.

\bibitem{Wineland94a} Wineland, D. J., Bollinger, J. J., Itano, W. M. and
Heinzen, D. J., \textit{Phys. Rev. A }\textbf{1994}, \textit{50, }67.

\bibitem{Toth09a} Toth, G., Knapp, C., Guhne, O. and Briegel, H. J., \textit{%
Phys. Rev. A }\textbf{2009, }\textit{79, }042334.

\bibitem{He11b} He, Q. Y., Peng, S.-G., Drummond, P. D.and Reid, M. D., 
\textit{Phys. Rev. A }\textbf{2011, }\textit{84, }022107.

\bibitem{Hoffmann03a} Hoffmann, H.F. and Takeuchi, S., \textit{Phys. Rev. A }%
\textbf{2003, }\textit{68, }032103.

\bibitem{Barnett89a} Barnett, S. M. and Pegg, D. T., \textit{Phys. Rev. A }%
\textbf{1989, }\textit{39, }1665.

\bibitem{He12b} He, Q. Y., Vaughan, T. G., Drummond, P. D.and Reid, M. D., 
\textit{New J. Phys. }\textbf{2012}, \textit{14}, 093012.

\bibitem{He11a} He, Q. Y., Reid, M. D., Vaughan, T. G., Gross, C.
Oberthaler, M., and Drummond, P. D., \textit{Phys. Rev. Letts. }\textbf{%
2011, }\textit{106, }120405.

\bibitem{Sorensen01b} Sorensen, A. and Molmer, K., \textit{Phys. Rev. Letts. 
}\textbf{2001, }\textit{86, }4431.

\bibitem{Duan00a} Duan, L-M., Giedke, G. Cirac, J. I. and Zoller, P., 
\textit{Phys. Rev. Letts. }\textbf{2000, }\textit{84, }2722.

\bibitem{Toth03a} Toth, G., Simon, C. and Cirac, J. I., \textit{Phys. Rev. A 
}\textbf{2003, }\textit{68, }062310.

\bibitem{Esteve08a} Esteve, J., Gross, C., Weller, A., Giovanazzi, S. and
Oberthaler, M. K., \textit{Nature }\textbf{2008}, \textit{455, }1216.

\bibitem{Gross10a} Gross, C., Zibold, T., Nicklas, E., Esteve, J. and
Oberthaler, M. K., \textit{Nature }\textbf{2010}, \textit{464, }1165.

\bibitem{Riedel10a} Riedel, M..F., Bohl, P., Li, Y., Hansch, T. W., Sinatra,
A. and Treutlein, P., \textit{Nature }\textbf{2010}, \textit{464, }1170.

\bibitem{Pegg91a} Pegg, D. T., \textit{J. Phys. A: Math. Gen., }\textbf{1991}%
, \textit{24}, 3031.

\bibitem{Tinkham64a} Tinkham, M. \textit{Group Theory and Quantum Mechanics; 
}McGraw-Hill: New York, \textbf{1964}.
\end{thebibliography}
\end{document}